\documentclass[aps, preprint, nofootinbib,preprintnumbers,eqsecnum,superscriptaddress,sort]{revtex4}
\pdfoutput=1
\usepackage{amssymb,amsmath,amsfonts,amsthm,latexsym,mathrsfs}
\usepackage{graphics, graphicx,color,epsfig,subfigure}
\usepackage[
      colorlinks=true,
      linkcolor=blue,
      urlcolor=blue,
      filecolor=black,
      citecolor=red,
      pdfstartview=FitV,
      pdftitle={},
        pdfauthor={Oscar Dias, Jorge Santos, Benson Way},
        pdfsubject={},
        pdfkeywords={},
        pdfpagemode=None,
        bookmarksopen=true
      ]{hyperref}
\usepackage{enumerate}
\def \be {\begin{equation}}
\def \ee {\end{equation}}
\def \bea {\begin{eqnarray}}
\def \eea {\end{eqnarray}}
\def \dd {\mathrm{d}}
\newcommand{\mbb}[1]{\mathbb{#1}}
\newcommand{\mrm}[1]{\mathrm{#1}}
\def\R{\mbb{R}}
\def\half{\frac{1}{2}}
\numberwithin{equation}{section}

\begin{document}

\title {Numerical Methods for \\ Finding Stationary Gravitational Solutions}

\author{\'Oscar J. C. Dias}
\email{ojcd1r13@soton.ac.uk}
\affiliation{STAG research centre and Mathematical Sciences, University of Southampton, UK}

\author{Jorge E. Santos}
\email{jss55@cam.ac.uk}
\affiliation{Department of Applied Mathematics and Theoretical Physics, University of Cambridge, Wilberforce Road, Cambridge CB3 0WA, UK \vspace{1 cm}}

\author{Benson Way}
\email{bw356@cam.ac.uk}
\affiliation{Department of Applied Mathematics and Theoretical Physics, University of Cambridge, Wilberforce Road, Cambridge CB3 0WA, UK \vspace{1 cm}}

\begin{abstract}\noindent{
The wide applications of higher dimensional gravity and gauge/gravity duality have fuelled the search for new stationary solutions of the Einstein equation (possibly coupled to matter). In this topical review, we explain the mathematical foundations and give a practical guide for the numerical solution of gravitational boundary value problems.  We present these methods by way of example: resolving asymptotically flat black rings, singly-spinning lumpy black holes in anti-de Sitter (AdS), and the Gregory-Laflamme zero modes of small rotating black holes in AdS$_5\times S^5$. We also include several tools and tricks that have been useful throughout the literature.}
\end{abstract}
\maketitle

\tableofcontents 
\newpage

\section{Introduction \label{Sec:Intro}}

The Einstein equation is intrinsically a dynamical system; it describes the evolution of spacetime and its interaction with matter.  As such, much effort has gone into the numerical time-evolution of gravitational systems (see \cite{Sperhake:2014wpa} for a review).   These include the characteristic formalism \cite{Winicour:2005ge,Chesler:2013lia}, the ADM, BSSN, Z4 formalisms \cite{Cook:2000vr,Baumgarte:2002jm,Gourgoulhon:2007ue,AlcubierreBook,BaumgarteShapiroBook,Centrella:2010mx,Heller:2012je}, and the generalized harmonic gauge formulation \cite{Pretorius:2004jg,Gundlach:2005eh,Lindblom:2005qh,Palenzuela:2006wp,Bantilan:2012vu,Hilditch:2015aba}. 

Yet, in this Topical Review, we focus instead on the non-dynamical aspects of the Einstein equation: stationary solutions\footnote{Here, we are using a loose definition of `stationary' that may include time-periodic solutions as well as out of equilibrium steady-state solutions. Precise statements will be given in Section \ref{Sec:statio}.}.  Though this is a more limited setting than the fully time-dependent scenarios, stationary solutions, especially black holes, are  arguably the most fundamental of all gravitational objects.   These solutions serve as possible endpoints to dynamical evolution, and  provide the basis for much of our understanding of general relativity, including topology, rigidity, and no hair theorems, as well as linear stability (see the many  motivations in the several contributions of \cite{HorowitzBook2012}).  From a more practical standpoint, because of the extra symmetry present in stationary solutions, they can be studied more systematically, and with significantly fewer computational resources than a full time-dependent simulation.  

More recently, gauge/gravity duality (also known as holography or AdS/CFT) has widened the applications of general relativity by formulating a dictionary between classical gravitational solutions and field theory states at strong coupling \cite{Maldacena:1997re,Gubser:1998bc,Witten:1998qj}  (see \cite{Aharony:1999ti,Maldacena:2011ut,Hubeny:2014bla,CasalderreySolana:2011us,Erdmengerbook} for reviews and textbooks). Applications of general relativity in anti-de Sitter (AdS) holography has placed greater emphasis on the phase diagram of stationary solutions, including in thermodynamic ensembles not suitable for dynamical evolution (see {\it e.g.} the reviews \cite{Horowitz:2010gk,Hubeny:2010ry,Hartnoll:2011fn,Marolf:2013ioa,Hubeny:2014bla}).  

But even with a stationary ansatz, the Einstein equation is a complicated set of coupled nonlinear PDEs that are difficult to solve. Nevertheless, there are methods to construct exact analytical solutions. These include inverse scattering methods, algebraically special solutions, Kerr-Schild techniques, applications of supersymmetry and generalised Harrison transformations \cite{Kramerbook,HorowitzBook2012}. However, these methods are restricted to situations with high symmetry.

Despite this, there are also a number of analytical approximation methods that can be used to construct nonlinear solutions.  These include the matched asymptotic expansion\footnote{In the context of time dependent perturbation theory, see \cite{Starobinsky:1973scalar,Starobinsky:1973,Unruh:1976fm,Detweiler:1980uk,Maldacena:1997ih,Dias:2007nj}} \cite{DEathbook}; the non-interacting thermodynamical model and the higher order matched asymptotic expansion for nonlinear solutions \cite{Basu:2010uz,Bhattacharyya:2010yg,Dias:2011tj,Dias:2011at,Stotyn:2011ns,Donos:2012ra,Cardoso:2013pza,Iizuka:2015vsa}; the fluid/gravity correspondence (a long-wavelength gradient expansion) \cite{Bhattacharyya:2008jc,Bhattacharyya:2008xc} reviewed in \cite{Hubeny:2010ry,Hubeny:2011hd}; the blackfold approach (for horizons with a separation of scales) \cite{Caldarelli:2008pz,Emparan:2009cs,Emparan:2007wm,Emparan:2009cs,Emparan:2009at,Caldarelli:2010xz,Armas:2010hz,Armas:2011uf,Camps:2012hw,Armas:2012bk,Armas:2013hsa,Armas:2014bia} reviewed in \cite{Emparan:2011br}; and the large $D$ limit of gravity \cite{Emparan:2013moa,Emparan:2013xia,Emparan:2013oza,Emparan:2014cia,Emparan:2014jca,Emparan:2014aba,Emparan:2015hwa,Emparan:2015rva,Bhattacharyya:2015dva,Suzuki:2015iha,Tanabe:2015hda}. 

While these methods are remarkable and extremely successful, they are only valid in their perturbative regimes.  To access regimes where these methods are not valid, or to test the extent to which these methods apply, one must resort to numerics.  In turn, these methods also supply a consistency check for numerical methods, and are often most accurate in regimes where numerical methods become practically difficult.  

We point out that numerical methods for stationary problems might also be of use to those more interested in time evolution.  The Einstein equation, formulated as a Cauchy problem, includes constraints that limit the set of initial data that can be considered; and the methods for finding suitable initial data are similar to those for finding stationary solutions.  As it stands, without matter, the issue of finding initial data is well understood  \cite{Cook:2000vr,AlcubierreBook,Okawa:2013afa}.  However, as discussed in \cite{Okawa:2014nda}, the situation is less satisfactory when matter fields are present.  

A numerical approach to finding stationary solutions requires a formulation of the Einstein equation that is suitable to be solved as a well-posed boundary value problem.  In this review, we shall focus on the Einstein-DeTurck formulation first introduced by Headrick, Kitchen and Wiseman \cite{Headrick:2009pv}.  This formulation follows from influential work on Ricci-DeTurck flow by DeTurck \cite{DeTurck1983,DeTurck2003} which in turn was motivated by Hamilton's Ricci flow\footnote{Ricci(-DeTurck) flow was introduced \cite{Hamilton1982,DeTurck1983,DeTurck2003} as a proposal to prove Thurston's geometrisation conjecture \cite{Thurston1982}, which includes  the Poincar\'e conjecture as a corollary. Indeed, Ricci-DeTurck flow with surgery turned out to be the key technical tool in Perelman's proof \cite{Perelman2002,Perelman2003}. Nice reviews and introductions to Ricci flow can be found in \cite{Morgan2005,Topping2006,ChowKnopfBook}.}  \cite{Hamilton1982}.  There is a review on this formulation by Wiseman \cite{Wiseman:2011by}, which includes a historical account and early applications and achievements. 

Our Topical Review will necessarily have some overlap with this review, but we will also cover subsequent technical and physical developments.  Our choice to focus on this method, at the expense of covering others in more detail, is mainly due to its flexibility, applicability to any cohomogeneity, and recent success. Another popular method for gravitational boundary value problems is to work in conformal gauge, which are restricted to cohomogeneity-2 problems. Initial applications and further details of this numerical method can be found in \cite{Wiseman:2002zc,Kol:2003ja,Kudoh:2004hs,Kudoh:2003ki,Aharony:2005bm,Kleihaus:1997mn,Kleihaus:1997ws,Kleihaus:2000kg,Hartmann:2001ic,Kleihaus:2002ee,Kleihaus:2005me,Kleihaus:2006ee,Kalisch:2015via}. For cohomogeneity-1 problems described by ODEs, the shooting method is a well established approach \cite{Press:1992zz}. As we mentioned before, in the Cauchy problem, one has to solve the elliptic constraint equations. Numerical methods traditionally used to solve these elliptic equations are reviewed in \cite{Cook:2000vr,Grandclement:2007sb} and references therein.
 
Once we have a formulation, the equations must be solved using some iterative algorithm (see \cite{Press:1992zz,HagemanYoungBook,VargaBook} for some standard treatments).  We will mostly describe Newton-Raphson, because of its basic importance, robustness, and efficiency; and Ricci flow because of its mathematical connection to the DeTurck method.  The implementation of these algorithms on a computer require some form of numerical discretisation, which we leave for the reader to choose. Those that are unfamiliar with these methods can see standard texts (\emph{e.g.} \cite{Press:1992zz,HagemanYoungBook,VargaBook,CanutoBook,Trefethen,Boyd,Grandclement:2007sb}) or appendix \ref{appendix:collocation}, which contains a short guide to collocation methods \cite{CanutoBook,Trefethen,Boyd,Grandclement:2007sb}.  

While in principle, this would supply the reader with all that is necessary for finding stationary solutions, the process of doing so is far from straightforward and requires much physical insight and guesswork.  For this, we include a number of additional tools and tricks that have been successfully used in the literature.  The most important of these involves finding zero modes of linear instabilities.  While this, in itself, is an important part of understanding the linear stability of systems, we present it here mostly as a tool for finding new black hole solutions.  Nevertheless, our review of this topic will be broad enough to cover some recent developments on linear stability.

The importance of zero modes might best be illustrated by the Gregory-Laflamme system\footnote{For a review, see \cite{Harmark:2007md,Horowitz:2011cq}; for connections to fluid instabilities, fluid/gravity, blackfolds, and the AdS/Ricci-flat correspondence see \cite{Cardoso:2006ks,Caldarelli:2008mv,Caldarelli:2008ze,Emparan:2009at,Camps:2010br,Caldarelli:2012hy,Caldarelli:2013aaa}.} \cite{Gross:1982cv,Horowitz:1991cd,Gregory:1993vy,Gregory:1994bj,Horowitz:2001cz,Gubser:2001ac,Reall:2001ag,Kol:2002xz,Wiseman:2002zc,Harmark:2002tr,Kudoh:2003ki,Kol:2003ja,Harmark:2003yz,Gorbonos:2004uc,Kudoh:2004hs,Sorkin:2004qq,Kleihaus:2006ee,Asnin:2006ip,Dias:2007hg,Headrick:2009pv,Monteiro:2009tc,Dias:2009iu,Dias:2010maa,Dias:2010eu,Dias:2011jg,Lehner:2010pn,Lehner:2011wc,Figueras:2012xj,Kalisch:2015via,Prestidge:1999uq,Dias:2010gk}, which shares many features with other higher dimensional black holes.  Recall that black strings and branes are unstable to gravitational perturbations along their extended directions \cite{Gross:1982cv,Gregory:1993vy,Gregory:1994bj}.  This is the \emph{Gregory-Laflamme instability}.  From a non-dynamical perspective, this system contains rich physics \cite{Gubser:2001ac,Reall:2001ag,Kol:2002xz,Wiseman:2002zc,Harmark:2002tr,Kol:2003ja,Harmark:2003yz,Gorbonos:2004uc,Kudoh:2004hs,Sorkin:2004qq,Sorkin:2006wp,Kleihaus:2006ee,Asnin:2006ip,Headrick:2009pv,Lehner:2010pn,Lehner:2011wc,Figueras:2012xj,Kalisch:2015via}.  The onset of the Gregory-Laflamme instability is a zero mode, which indicates the existence of a new branch of solutions.  These new solutions are non-uniform strings, which are static solutions that break the translation symmetry of the string \cite{Gubser:2001ac}. The family of non-uniform strings further connects in moduli space to spherical black holes through a topology-changing merger point  \cite{Kol:2002xz}.  This zero mode is therefore an important indicator for the existence of new solutions, and provides a guide for searching for them.  This property of zero modes is seen in many other systems, and is therefore an invaluable tool for finding stationary solutions.   

Let us now describe how knowledge of the stationary phase diagram might provide insight to the dynamical time-evolution of this system.  In five dimensions, the nonuniform strings have lower entropy (horizon area) than the unstable uniform strings.  The second law of black hole thermodynamics (otherwise known as the area theorem) thus forbids these nonuniform solutions from being the endpoint to the instability.  On the other hand, spherical black holes are entropically favoured, but such a dynamical transition from the uniform string would necessitate a topology change in the horizon and a violation of cosmic censorship \cite{Penrose:1969pc,HawkingEllisBook,waldbook,Christodoulou:1999,Senovilla:2014gza}. Indeed, a full time-dependent simulation of the five dimensional system \cite{Lehner:2010pn,Lehner:2011wc} has revealed that the horizon of the black string pinches off, leading to a violation of cosmic censorship.  However, in higher dimensions ($d\gtrsim14$), the nonuniform solutions have \emph{higher} entropy, and can serve as such an endpoint \cite{Sorkin:2004qq}.  A simulation in the large $D$ limit indeed finds that the system evolves towards a non-uniform black string \cite{Emparan:2015gva}.  

This Topical Review is structured as follows.  In the next section, we will review the status of various theorems associated with stationary solutions.  In the subsequent section, we describe linear perturbation theory and explain how zero modes are found.  At the end of this section, we give an example zero mode problem (the Gregory-Laflamme modes of rotating AdS$_5\times S^5$ black holes that was not studied previously).  Section \ref{Sec:DeTurcksection} contains an explanation of the Einstein-DeTurck formulation. The discussion of this formulation in the presence of matter fields is new to the literature. In Section \ref{Sec:BCs} boundary conditions are discussed in detail. We will then review Newton-Raphson and Ricci flow in section \ref{Sec:relaxation}, and then give a number of addition tools and tricks (including patching) in section \ref{Sec:tricks}.  The final two sections contains applications of these methods (black rings in 5, 6, and 7 dimensions in section \ref{Sec:rings} and AdS ultraspinning lumpy black holes in section \ref{Sec:RRR}; the latter study is novel). Appendix \ref{appendix:collocation} contains information on collocation methods and a rudimentary example of a boundary value problem. Appendix \ref{appendix:computecharges} reviews formalisms to compute asymptotic conserved charges and thermodynamic quantities.

\section{Review of Stationary Solutions \label{Sec:Revew}}

Prior to the beginning of the 21st century, stationary black holes were understood to be remarkably simple objects.  A number of black hole theorems, including topology, rigidity, no-hair, and stability theorems established that black holes are spherical in topology, uniquely specified by asymptotic charges, and stable.  Since an exact and general solution was already known, there was little motivation to search for new stationary solutions.  

But more recently, motivations from higher dimensions, string theory, holography, or simply a desire to understand general relativity more broadly, have lead to considerations that violate many of the assumptions of these black hole theorems.  Consequently, the physics of black holes is now far fuller and richer than previously believed.  This has fuelled the motivation for the numerical search for new stationary solutions. In this section, we review the various black hole theorems, explain how their assumptions are violated, and discuss the close connection between non-uniqueness and stability\footnote{Absent from our discussion are the singularity theorems that ultimately led to the formulation of the cosmic censorship conjecture \cite{Penrose:1969pc,HawkingEllisBook,waldbook,Christodoulou:1999} that we mention. The reader can find a recent review on these theorems in \cite{Senovilla:2014gza}.}.

\subsection{Black Hole Theorems \label{subsec:bhtheorems}} 

`The black hole uniqueness theorems' is a broad term that encompasses many theorems about spacetime topology, their symmetries (\emph{e.g.}, rigidity), and their asymptotic quantities.  Ultimately, these theorems strive to prove the uniqueness of a solution in a given theory.  Naturally, these theorems were first formulated in asymptotically flat 4-dimensional Einstein-Maxwell theory.  Reviews on the uniqueness theorems can be found in \cite{Heusler:1996,Chrusciel:2012jk,Robinson:2004zz} (for asymptotically flat $d=4$ spacetimes) and in \cite{Hollands:2012xy,Robinson:2004zz} (for higher dimensions and gravity with matter fields). These theorems are also thoroughly discussed in \cite{Emparan:2006mm,Emparan:2008eg,Emparan:2009at,Ionescu:2015dna}.

Note that these theorems apply to `stationary' solutions, meaning they have a Killing vector field that is timelike everywhere in the asymptotic region.  We will use this definition of stationary for now, until we begin to break many of the underlying assumptions of the theorems and subsequently require a broader notion of `stationary'.

Hawking's \emph{topological theorem} \cite{Hawking:1971vc,HawkingEllis:1973} constrains the topological properties of black hole horizons. It states that 4-dimensional  ($d=4$)  asymptotically flat stationary black holes obeying the dominant energy condition, must have horizons with spherical topology. 

Hawking's {\it rigidity theorem} \cite{Hawking:1971vc,HawkingEllis:1973,SudarskyW:1992,Chrusciel:1993cv,Friedrich:1998wq,Ionescu:2015dna} states that 4-dimensional non-extremal black holes with a compact bifurcate Killing horizon and a stationary Killing field ${\cal T}$ that is not normal to this horizon must also have commuting rotational Killing fields ${\cal R}=\partial_{\phi}$ that also commute with ${\cal T}$ and generates closed orbits with period $2\pi$.  Moreover, there is a linear combination $K = {\cal T} + \Omega_H {\cal R}$ that is normal to the horizon, where the constant $\Omega_H$ is the horizon angular velocity.  That is, $K$ is a Killing field that generates the horizon, and so the horizon is a Killing horizon.  The rigidity theorem guarantees that the black hole is time independent, axisymmetric, and must rotate along an isometry, and hence emits no gravitational radiation.  From a thermodynamic perspective, rigidity theorems guarantee that horizons have a constant well-defined temperature. Stationary rotating extremal black holes must also be axisymmetric \cite{Hollands:2008wn,Chrusciel:2008js}.

The topological and rigidity theorems are fundamental ingredients for the {\it uniqueness theorem} \cite{Israel:1967wq,Israel:1967za,Carter:1971zc,Wald:1971iw,Hawking:1971vc,Carter:1973,Carter:1986,Robinson:1975bv,Mazur:1982,Bunting:1983,Amsel:2009et,Figueras:2009ci}: all regular stationary, asymptotically flat (non-)degenerated black holes of the Einstein-Maxwell equations in $d=4$ dimensions are uniquely specified by their mass, angular momentum, and electric charge, and have horizon topology $S^2$.  The most general solution is the Kerr-Newman family \cite{Newman:1965my,Adamo:2014baa} which includes the Kerr \cite{Kerr:1963ud,Kerr:2007dk,Teukolsky:2014vca}, Reissner-Nordstr\"om \cite{Reissner:1916,Nordstrom:1918} and Schwarzschild \cite{Schwarzschild:1916} as special cases. This unique specification is an indication that black holes are featureless, and hence `have no hair'. If we allow for multi-black hole solutions there is the Majumdar-Papapetrou solution \cite{Majumdar:1947eu,Papaetrou:1947ib}, that describes a regular array of extremal black holes with charge equal to its mass. The uniqueness theorems of  \cite{Chrusciel:1994qa,Heusler:1996ex} show that this is the only electro-vacuum static solution with non-connected (degenerate) horizons. 

These uniqueness theorems led to the proposal of the Carter-Israel conjecture by Wheeler \cite{WheelerRuffiniNohair:1971}.  That is, that black holes formed through gravitational collapse should be fully described by its conserved charges, regardless of the field content of the initial data.  In particular, this posits that all measurable asymptotic quantities of black holes are dictated by a Gauss law. This conjecture is crucial for understanding the microscopic origins of black hole entropy, and has inspired the formulation of black hole thermodynamics \cite{Bekenstein:PhysTod1980,Bekenstein:1996pn,Wald:1999vt}. As highlighted in \cite{Reall:2002bh}, any microscopic or statistical description of black hole entropy is highly dependent on whether the macroscopic black hole is uniquely characterised by its asymptotic charges or not.   

Wheeler's Israel-Carter conjecture led to an early formulation of various {\it no-hair theorems} starting with \cite{Chase:1970,Penney:1968zz,Bekenstein:1972ny,Bekenstein:1971hc,Bekenstein:1972ky,Teitelboim:1972qx,Hartle1972,Heusler:1992ss,Bekenstein:1995un,Bekenstein:1996pn,Sudarsky:1995zg}. Most of these studies considered the simplest example of `hair' (asymptotic fields that are not captured by a Gauss law), namely hair that is sourced by a scalar field.  These situations include real or complex scalar fields, with possibly an additional Maxwell field.  These also include various self-interacting potentials  \cite{Heusler:1992ss,Sudarsky:1995zg,Hertog:2006rr}. Some of these permit a cosmological constant $\Lambda\neq 0$ (since this can be viewed as a constant massive scalar field), though most assume $\Lambda=0$. Reviews on no-hair theorems and on a variety of matter systems where they do not hold are reviewed in \cite{Bizon:1994dh,Bekenstein:1996pn,Heusler:1996,Volkov:1998cc,Ashtekar:2000nx,Chrusciel:2012jk}. 

The black hole theorems we have just discussed place limits on the existence of many stationary solutions.  Naturally, the full richness of gravitational physics only manifests when the assumptions of these theorems are violated or evaded.

\subsection{Evading No-Hair Theorems: Solitons and Hairy Black Holes \label{subsec:hair}} 

The weakest of these theorems are the no-hair theorems involving a scalar field, since they are limited to special theories.  Many `hairy' black holes can be found if one considers more general theories than the Einstein-Maxwell-Scalar family, such as those in supergravity \cite{Gibbons:1982ih,Gibbons:1987ps,Garfinkle:1990qj,Lee:1991qs,Horowitz:1991cd,Achucarro:1995nu}, Einstein-Yang-Mills-(dilaton)-(Higgs) \cite{Volkov:1989fi,Volkov:1990sva,Bizon:1990sr,Kuenzle:1990is,Breitenlohner:1991aa,Lavrelashvili:1992ia,Greene:1992fw,Hartmann:2001ic}, Einstein-Skyrme \cite{Bizon:1992gb,Droz:1991cx}.  More information can be found in the reviews \cite{Bizon:1994dh,Bekenstein:1996pn,Heusler:1996,Volkov:1998cc,Ashtekar:2000nx,Chrusciel:2012jk,Winstanley:2008ac}.  Even within the Einstein-Maxwell-Scalar(-AdS) family, hairy black holes can be constructed with a judicious choice of the self-interacting potential \cite{Torii:1998ir,Torii:2001pg,Zloshchastiev:2004ny,Martinez:2004nb,Martinez:2006an}. 

But in recent years, it was found that there are simpler ways of finding hairy black holes that evade the assumptions of existing no-hair theorems. Indeed, Einstein-Maxwell-Scalar theory with a negative cosmological constant (\emph{i.e.} AdS asymptotics) is now known to contain many hairy black hole solutions. In Poincar\'e AdS, this started with the hairy solutions of \cite{Gubser:2008px,Hartnoll:2008vx,Hartnoll:2008kx} (in the context of holographic superconductors), and in global AdS with the hairy black holes of \cite{Basu:2010uz,Dias:2011tj} (in the context of superradiance). Without the Maxwell field but still in AdS, Einstein-Scalar theory also has static hairy black holes including \cite{Hartnoll:2008kx,Faulkner:2010gj,Dias:2010ma} and rotating hairy black holes \cite{Dias:2011at,Stotyn:2011ns}. If we replace the AdS boundary of \cite{Dias:2011at} by a Dirichlet boundary sourced by the mass of a scalar in $\Lambda=0$, the solution of \cite{Dias:2011at} survives and describes a hairy Kerr black hole \cite{Herdeiro:2014goa}. Even just Einstein-AdS theory (with no matter fields) has black holes with gravitational hair \cite{Dias:2015rxy}.

An important and general way of creating hairy black holes is through \emph{near-horizon scalar condensation}, which was first noticed in  \cite{Gubser:2008px,Hartnoll:2008kx}. In AdS, there is a constraint on the mass of a scalar field due to normalisability (finiteness of energy) $\mu^2\geq\mu^2_{\mathrm{BF}}$, where $\mu^2_{\mathrm{BF}}<0$ is the Breitenl\"ohner-Freedman bound \cite{Breitenlohner:1982jf}.  Importantly, this bound is dependent on the dimension of the AdS space.  If the spacetime geometry interpolates between one `interior' AdS space to a different AdS space that defines the asymptotics, it is possible for the mass of the scalar field to be below the Breitenl\"ohner-Freedman bound in the interior AdS geometry, but above this bound in the asymptotic AdS geometry. If this happens, there is scalar condensation, and hairy black holes can exist \cite{Hartnoll:2008kx}. The easiest means of accomplishing this is to have an asymptotically AdS$_d$ solution with a near-extremal horizon, which has a near-horizon geometry that resembles AdS$_2\times (\text{transverse directions})$.  The most well-known examples are the holographic superconductors \cite{Gubser:2008px,Hartnoll:2008vx,Hartnoll:2008kx} (see also the review \cite{Hartnoll:2009sz,Herzog:2009xv,2010uqpt.book..701H,McGreevy:2009xe,Horowitz:2010gk}), which creates a near-horizon AdS$_2$ geometry via a charged near-extremal planar Reissner-Nordstr\"om black hole.  In that context, the asymptotic scalar field is interpreted as a superconducting condensate in the dual field theory.

After the holographic superconductors of \cite{Gubser:2008px,Hartnoll:2008vx,Hartnoll:2008kx}, many other hairy black holes with a  condensed matter dual interpretation have been constructed. Typically, depending on the asymptotic boundary condition, one gets gravitational solutions that holographically model different condensed matter systems. This area of research is quite broad and cannot be reviewed here. The reader can find reviews in \cite{Hartnoll:2009sz,Herzog:2009xv,2010uqpt.book..701H,McGreevy:2009xe,Horowitz:2010gk,Hartnoll:2011fn} and, for some of the developments that followed, on the Topical Review on `holographic lattices' that Classical and Quantum Gravity will soon publish \cite{DonosGauntlett:CQGreview2015}. Yet, for the purpose of the current Topical Review we should list numerical works on holographic superconductors that use methods of this review and where the reader can find applications \cite{Horowitz:2011dz,Horowitz:2012ky,GarciaGarcia:2012zd,Horowitz:2012gs,Donos:2012yu,Horowitz:2013jaa,Donos:2013wia,Withers:2013loa,Withers:2013kva,Ling:2013aya,Chesler:2013qla,Ling:2013nxa,Horowitz:2013mia,Dias:2013bwa,Hartnoll:2014cua,Hartnoll:2014gaa,Ling:2014saa,Mefford:2014gia,Withers:2014sja,Donos:2014yya,Hartnoll:2015faa,Rangamani:2015hka,Langley:2015exa,Hartnoll:2015rza}\footnote{We now take the opportunity to list other studies where numerical methods of this review were used. This includes the construction of plasma-balls  \cite{Aharony:2005bm,Figueras:2014lka} in the context of the gravity/Scherk-Schwarz correspondence \cite{Witten:1998zw,Aharony:2005bm,Lahiri:2007ae}, the construction of the black hole geometry dual to the deconfined phase of the BMN matrix model at strong 't Hooft coupling \cite{Costa:2014wya}, the construction of holographic duals of  localised defects in conformal field theories at strong coupling \cite{Dias:2013bwa,Horowitz:2014gva,Janik:2015oja}, the construction of bulk duals for generic holographic CFT states not described by a smooth near-horizon geometry \cite{Hickling:2015ooa} and Yang-Mills solutions in AdS$_4$ \cite{Kichakova:2014fta}.}.

Finally note that the near-horizon scalar condensation mechanism \cite{Gubser:2008px,Hartnoll:2008kx} can be generated from any other near-extremal horizons in AdS, such as global Reissner-Nordstr\"om, Kerr-AdS (or Myers-Perry) black holes, and even hyperbolic Schwarzschild-AdS black holes  \cite{Dias:2010ma}, although in these non-planar cases the hairy solutions are less for the gravity/condensed matter correspondence.

A more subtle way of evading the no-hair theorems is by breaking their underlying assumption that the scalar field has the same symmetries as the gravitational field. In particular, it is assumed that these fields are time independent and axisymmetric. Indeed, the gravitational field only needs to have the same symmetries as the stress tensor $T_{\mu\nu}$ coupled through the Einstein equation, not necessarily the matter fields themselves.  A simple example of this is a complex scalar field $\Phi\sim e^{-i\omega t+im\phi}$, which is neither axisymmetric nor time-independent, but its combination in the stress tensor $\Phi\bar\Phi$ and $\partial\Phi\partial\bar\Phi$ is. Such observations were first made in \cite{Yoshida:1997nd,Yoshida:1997qf}.

This observation led to the construction of \emph{boson stars}, which in some cases are gravitational back-reactions  of \emph{Q-balls} (see \cite{Lee:1991ax,Schunck:2003kk,Liebling:2012fv} for reviews of earlier works).  These are horizonless matter configurations which exist both in flat space and in AdS. As we have implied, fully time-independent boson stars with a real scalar field do not exist \cite{Friedberg:1986tp} (although time-dependent {\it oscillons} \cite{Seidel:1991zh,Maliborski:2013jca} do exist), but those with a complex scalar field do exist if confined in bound states by a (potential) well with Dirichlet boundary conditions. There are neutral, spherically symmetric ($m=0$) or planar-symmetric examples of boson stars \cite{Kaup:1968zz,Ruffini:1969qy,Yoshida:1997nd,Schunck:2003kk,Liebling:2012fv,Astefanesei:2003qy,Horowitz:2010jq,Buchel:2013uba}. There are also charged examples \cite{Jetzer:1989av,Basu:2010uz,Gentle:2011kv,Dias:2011tj}, which are sometimes simply called solitons since the harmonic time dependence of the scalar field can be removed by a gauge transformation of the Maxwell potential.  Non-axisymmetric cases ($m\neq 0$) can also be found \cite{Yoshida:1997qf,Schunck:1996book,Schunck:1996he,Astefanesei:2003rw,Hartmann:2010pm,Dias:2011at}, where all of these have metrics that are axisymmetric, though the scalar field is not. Notably, the examples in \cite{Hartmann:2010pm,Dias:2011at} in $d=5$ are fully dependent on time, a radial coordinate, and three angular coordinates, but the metric only depends on the radial coordinate, and so are well-suited as toy models for low-symmetry scenarios.   

As first observed in \cite{Schunck:1996book,Schunck:1996he,Yoshida:1997qf}, all the rotating boson stars \cite{Schunck:1996book,Schunck:1996he,Yoshida:1997qf,Astefanesei:2003rw,Hartmann:2010pm,Dias:2011at} preserve a Killing field $K=\partial_t + (\omega/m) \partial_\phi$, though $\partial_t$ and $\partial_\phi$ are individually non-Killing. That is, the solutions are not time-independent nor axisymmetric, but time-periodic, and are not, strictly speaking, stationary solutions (even though the metric is).  

With a boson star, a small black hole can often be added to it to obtain a hairy black hole.  This idea was first noticed not in the context of boson stars of Einstein-Scalar(-Maxwell) theory, but in the Einstein-Yang-Mills system. Indeed, a black hole can be added to the soliton \cite{Bartnik:1988am} of the theory leading to a black hole with non-abelian Yang-Mills hair \cite{Bizon:1990sr,Kuenzle:1990is,Volkov:1990sva} we briefly mentioned earlier.  These results were unexpected since vacuum Einstein and pure Yang-Mills in flat space do not contain solitonic solutions \cite{Lichnerowicz:1955,Coleman:1975,Deser:1976wq}.  Subsequent work \cite{Breitenlohner:1991aa,Lee:1991vy,Ortiz:1991eu,Aichelburg:1992st,Bizon:1992gb,Kastor:1992qy,Heusler:1992av} (see \cite{Bizon:1994dh,Volkov:1998cc,Ashtekar:2000nx} for a review) solidified the heuristic idea that small black holes can be placed in the core of solitonic solutions.

Exceptions to this idea were later pointed out in the form of no-go theorems in \cite{Kastor:1992qy,Pena:1997cy,Astefanesei:2003qy}. However, \cite{Dias:2011at} observed that these results occur essentially because boson stars have oscillatory time dependence $e^{-i\omega t}$, with $t\to\infty$ at the horizon of a static, uncharged, black hole. The scalar field thus oscillates infinitely near the horizon and cannot be smoothly continued inside.  These no-go theorems are evaded either by considering different theories (\emph{e.g.} the gravitational Abelian-Higgs model \cite{Basu:2010uz,Bhattacharyya:2010yg,Dias:2011tj}, possibly extended with massive Proca fields \cite{Brito:2015pxa}), or by adding rotation \cite{Dias:2011at,Herdeiro:2014goa}.

The analytic tools to demonstrate the existence of small black holes from the existence of a soliton chiefly come from an `interacting thermodynamic bound state model' \cite{Ashtekar:2000nx} (later refined into a computable `non-interacting thermodynamic model' in \cite{Basu:2010uz,Bhattacharyya:2010yg,Dias:2011at,Dias:2011tj}), and matched asymptotic expansions \cite{Basu:2010uz,Bhattacharyya:2010yg,Dias:2011at,Dias:2011tj,Stotyn:2011ns}. These show excellent agreement with numerical results in their regime of validity \cite{Dias:2011at,Dias:2011tj}.

There are also solitonic solutions that break time symmetry, even in the metric. Such configurations with real scalar fields are typically called oscillons \cite{Seidel:1991zh,Maliborski:2013jca}. Other configurations that additionally break rotational symmetries can occur in pure gravity within Einstein-AdS with `gravitational' hair.  These were coined \emph{geons} by Wheeler\footnote{The asymptotically flat geons of Wheeler \cite{Wheeler:1955zz} do not to exist because of dispersion at asymptotic infinity. For historical references on geons see \cite{Wheeler:1955zz,Brill:1957fx,Ernst:1957zza,Ernst:1957zz,Misner:1957mt,Melvin:1963qx,Brill:1964zz,Melvin:1965zza}.} \cite{Wheeler:1955zz} and were constructed perturbatively in \cite{Dias:2011ss} and fully nonlinearly in \cite{Horowitz:2014hja}.  Black holes can also be placed at the centre of a geon, leading to a family of hairy black holes with `gravitational' hair, which were called \emph{black resonators}. These were proposed and constructed perturbatively in \cite{Dias:2011ss} (see also details in \cite{Cardoso:2013pza}), and fully nonlineary in \cite{Dias:2015rxy} with numerics. Black resonators are pure gravitational solutions with a single helical Killing field whose existence was predicted in \cite{Reall:2002bh,Kunduri:2006qa}\footnote{Without spatial isometries, we can also have black holes with only one symmetry. One such example is the static and asymptotically flat black hole with charged vector meson hair \cite{Ridgway:1995ke,Ridgway:1995ac}, which can be interpreted as a magnetic monopole with a winding number and a black hole inside its core.  Other examples are \cite{Hartnoll:2014cua,Withers:2014sja,Hartnoll:2015faa}.}.

\subsection{Evading Topology Theorems \label{subsec:topology}} 

The most easily avoidable assumptions in Hawking's topological theorem are perhaps the assumptions of four dimensions, and asymptotic flatness.  Relaxing either of these assumptions will allow the existence of non-spherical black holes.  

The timelike boundary of AdS allows for a more general choice of asymptotics.  In the language of holography, this choice corresponds to choosing a `boundary metric', which serves as the spacetime background for the dual field theory\footnote{The asymptotic scaling symmetries of AdS correspond to a conformal symmetry of the field theory.  This implies that boundary metrics that are related by a conformal transformation share the same AdS asymptotic structure.  Because of this, extracting metric-dependent field theory quantities requires specifying one of these conformally equivalent metrics (\emph{i.e.} a `conformal frame' needs to be chosen).}. In AdS it is possible for black holes to have spherical, planar or hyperbolic symmetry \cite{Lemos:1994xp,Mann:1996gj,Cai:1996eg,Vanzo:1997gw,Mann:1997iz,Birmingham:1998nr}.  Indeed, each of these cases have boundary metrics that are conformal to $\mathbb R^t\times S^{d-2}$, $\mathbb R^t\times \mathbb R^{d-2}$, or $\mathbb R^t\times \mathbb H^{d-2}$.

Taking this idea further, one can choose a boundary metric that is conformal to a black hole spacetime (see \cite{Marolf:2013ioa} for a review). The black hole horizon on the boundary extends into the bulk, allowing for new horizon configurations with `droplet' or `funnel' shapes (see Fig.~3 of \cite{Marolf:2013ioa} for an illustration and a more thorough discussion of these solutions).   These solutions serve as tools to understand Hawking radiation at strong coupling and heat transport.  The existence of these solutions and their associated phase transitions were proposed in \cite{Hubeny:2009ru} which built on previous ideas of braneworld black holes \cite{Fitzpatrick:2006cd}. Concrete analytical examples of such solutions were found in \cite{Hubeny:2009ru,Hubeny:2009kz,Hubeny:2009rc,Caldarelli:2011wa,Haehl:2012tw,Haddad:2013tha,Emparan:2013fha,Emparan:2015hwa}, and numerically in \cite{Figueras:2011va,Fischetti:2013hja,Figueras:2013jja,Santos:2012he,Santos:2014yja}.  See also \cite{Emparan:1999wa,Figueras:2011gd,Abdolrahimi:2012qi,Abdolrahimi:2012pb} for related braneworld black hole constructions.

In higher dimensions, there are of course black strings and black-brane type solutions \cite{Horowitz:1991cd}, but these are not asymptotically flat.  If one also imposes asymptotic flatness, there are of course spherical black holes, including the higher dimensional versions of Kerr \cite{Kerr:1963ud,Kerr:2007dk,Teukolsky:2014vca} and Kerr-AdS  \cite{Carter:1968ks}, namely Myers-Perry black holes \cite{Myers:1986un,Myers:2011yc} and their AdS counterparts \cite{Hawking:1998kw,Gibbons:2004uw}.  But there are also five-dimensional black rings which have horizon topology $S^2\times S^1$ and rotation along the $S^1$ \cite{Emparan:2001wn} and doubly-spinning black rings \cite{Pomeransky:2006bd}. 
Multi-horizon solutions were also constructed in close form, namely the black Saturns \cite{Elvang:2007rd}, di-rings \cite{Iguchi:2007is} and bi-cycling rings \cite{Izumi:2007qx}. 
Some consequences of the existence of the black ring are reviewed in \cite{Emparan:2006mm,Emparan:2008eg}. 

All these solutions were later shown to be consistent with a generalisation of Hawking's result to higher dimensions \cite{Galloway:2005mf} that states that cross sections of  horizons in $d>4$ are of positive Yamabe type (\emph{i.e.} must admit metrics of positive curvature). In five dimensions, this theorem only allows for the horizon topologies $S^3$, $S^1\times S^2$, and Lens spaces $L_{p,q}$. By now, there are examples for each of these topologies \cite{Myers:1986un,Emparan:2001wn,Kunduri:2014kja}. In higher ($d\geq6$) dimensions, this theorem is less restrictive.

While the five-dimensional (Emparan-Reall) black ring is an exact analytic solution \cite{Emparan:2001wn,Emparan:2006mm}, higher dimensional rings with horizon topology $S^{d-2}\times S^1$ also exist.  For large angular momentum, these solutions exhibit a separation of scales and can be constructed by using the blackfold approach \cite{Emparan:2007wm,Armas:2014bia}.  Beyond this limit, numerical methods can be used \cite{Kleihaus:2012xh,Dias:2014cia}, which we will detail in section \ref{Sec:rings}. The blackfold approach agrees remarkably well with numerical results \cite{Dias:2014cia,Armas:2014bia}.  There are asymptotically AdS black rings as well, though no exact analytic solution is known.  In certain regimes, these can be constructed using the blackfold approach \cite{Caldarelli:2008pz,Armas:2010hz}, and in $d=5$ where constructed numerically \cite{Figueras:2014dta}.

Besides black rings, there are a large number of other horizon topologies that are consistent with the topology theorem \cite{Galloway:2005mf}.  The blackfold approach is particularly efficient at generating many of these \cite{Emparan:2007wm,Caldarelli:2008pz,Emparan:2009vd,Caldarelli:2010xz,Armas:2012bk,Armas:2015kra,Armas:2015nea}, which consist mostly of products of spheres.  But the blackfold approach is by no means exhaustive. For instance, ultraspinning spherical Myers-Perry black holes are connected to lumpy (also called bumpy, or rippled) black holes \cite{Dias:2014cia,Emparan:2014pra,Suzuki:2015iha}, which themselves are connected to black rings. None of these rippled solutions admit a blackfold approximation.

\subsection{Evading Rigidity Theorems \label{subsec:rigidity}} 

The rigidity theorems are perhaps the most difficult of these theorems to evade.  Previously, we already discussed that the extension of the topological theorem to AdS asymptotics and to higher dimensions is much less restrictive and allows for many new horizon topologies \cite{Galloway:2005mf}. On the other hand, the essentials of the rigidity theorem are kept unchanged in its extension to higher dimensions ($d>4$) and AdS asymptotics \cite{Hollands:2006rj,Moncrief:2008mr} (see also \cite{Morisawa:2004tc,Hollands:2007aj,Harmark:2009dh}). 

In higher dimensions (in asymptotically flat or global AdS), there could be up to $N=\left\lfloor\frac{d-1}{2}\right\rfloor$ independent angular momenta. The rigidity theorem states that a non-extremal black hole with a compact bifurcate Killing horizon generated by $K$ and a stationary Killing field $\mathcal T$ not normal to the horizon must also have at least one rotational Killing isometry that commutes with $\mathcal{T}$. As emphasised in \cite{Reall:2002bh}, the rigidity theorem only guarantees the existence of one rotational Killing isometry, even though all known exact black hole solutions contain $N$ rotational Killing isometries. However, solutions with the minimum number of rotational isometries do exist. One is the helical black ring constructed in the blackfold approximation \cite{Emparan:2009vd}. Moreover, \cite{Dias:2010eu,Durkee:2010ea} found that there are stationary ultraspinning perturbations of certain Myers-Perry black holes that generically break all but one of the rotational symmetries. These perturbations lead to new families of topologically spherical black holes that have a single rotational isometry.

The rigidity theorem does assume that there is a stationary Killing field that is not normal to the horizon in order to prove the existence of an additional rotational Killing field. It is therefore conceivable that the horizon generator $K$ is the only Killing field, and so we have a Killing horizon that is neither stationary nor axisymmetric, but time-periodic \cite{Reall:2002bh,Kunduri:2006qa,Dias:2011at}. Such black holes can even exist without matter, though they require AdS asymptotics \cite{Dias:2011ss,Dias:2015rxy}.  In fact, these are the same solutions as the \emph{black resonators} we have mentioned earlier. 

Another assumption of the rigidity theorem is compactness.  When this assumption is relaxed, there are black holes with non-Killing horizons.  This idea was first noticed within the context of holographic `shockwaves' \cite{Khlebnikov:2010yt,Khlebnikov:2011ka} and independently in the context of `droplet' and `funnel' solutions in \cite{Hubeny:2011yk}.  Horizons that extend to asymptotic regions can be required to have asymptotic horizon velocities via a prescribed boundary condition.  But different asymptotic regions may have different velocities, and so the horizon must `twist', leading to a non-rigid or non-Killing horizon. Similar boundary conditions can be imposed on the temperature instead.  Such non-Killing black holes were constructed in \cite{Fischetti:2012ps,Figueras:2012rb,Fischetti:2012vt,Emparan:2013fha}.  From a thermodynamic point of view, the rigidity theorem can be viewed as a property of heat transport on a horizon.  If a horizon is compact, it serves as its own heat source, and so the system equilibrates to a single temperature.  However, if the horizon is non-compact, heat can be sourced from some asymptotic boundary condition or from some singularity, and the horizon no longer needs to have a well-defined temperature. For a review of some of these ideas see \cite{Marolf:2013ioa}.

\subsection{Uniqueness and Instabilities \label{subsec:instabilities}} 

Most of the solutions we have mentioned also violate uniqueness.  That is, there are multiple solutions that share the same asymptotic charges.  Often, there is a close connection between such uniqueness-violating solutions, and the existence of instabilities. In this subsection, we will discuss the Gregory-Laflamme instability, the superradiant instability, and nonlinear instabilities, all of which are associated with the existence of new solutions\footnote{Absent from this topical review will be the Aretakis instability \cite{Aretakis:2011ha,Aretakis:2011hc,Aretakis:2011gz,Aretakis:2012ei,Aretakis:2012bm,Aretakis:2013dpa,Lucietti:2012sf} effecting extreme black holes. See \cite{Murata:2013daa} for a simulation of this instability. As far as we are aware, this instability is not related to the existence of any new stationary solutions.}.

We first point out that solutions where uniqueness theorems apply are believed to be linearly stable, and possibly also nonlinearly stable.  In particular, there is now overwhelming numerical evidence in favour of the linear mode stability of the Kerr-Newman black hole within Einstein-Maxwell theory. This process started with the organisation of the perturbation equations to a pair of master equations \cite{Regge:1957td,Zerilli:1970se,Newman:1961qr,Teukolsky:1972my,Geroch:1973am,Teukolsky:1973ha,Chandrasekhar:1978a,Chandrasekhar:1978b,Chandrasekhar:1985kt,Leaver:1985ax} that led to the proof of linear mode stability of the Kerr black hole \cite{Whiting:1988vc} (see the review \cite{Berti:2009kk}), and more recently to perturbative \cite{Pani:2013ija,Pani:2013wsa,Mark:2014aja} and numerical \cite{Zilhao:2014wqa,Dias:2015wqa} evidence in favour of the stability of Kerr-Newman black holes. Some of these results extend to higher dimensions and (A)dS including the mode stability of Tangherlini-Schwarschild  \cite{Tangherlini:1963bw,Kodama:2003jz,Ishibashi:2003ap,Kodama:2003kk}, and some classes of Myers-Perry black holes \cite{Kunduri:2006qa,Murata:2008yx,Kodama:2009bf}. However, as we review below, the stability properties of AdS black holes are typically different from their asymptotically flat counterparts \cite{Kodama:2003jz,Ishibashi:2003ap,Kodama:2003kk,Kovtun:2005ev,Friess:2006kw,Michalogiorgakis:2006jc,Berti:2009kk,Dias:2013sdc,Cardoso:2013pza}.

On the other hand, many of the solutions that violate uniqueness are unstable or connected to an instability. Again, a system that best illustrates this connection is the Gregory-Laflamme instability of the black string \cite{Harmark:2007md,Horowitz:2011cq}. The zero mode\footnote{We will discuss the definition of `zero mode' in Section \ref{subsec:zeromodes}.} of this instability \cite{Gross:1982cv,Gregory:1993vy,Gregory:1994bj,Gubser:2001ac,Reall:2001ag,Monteiro:2009tc,Dias:2009iu,Dias:2010maa,Dias:2010eu,Dias:2011jg,Prestidge:1999uq,Dias:2010gk} is associated with novel solutions that describe non-uniform strings and localized black holes \cite{Horowitz:2001cz,Kol:2002xz,Wiseman:2002zc,Harmark:2002tr,Kol:2003ja,Harmark:2003yz,Gorbonos:2004uc,Kudoh:2004hs,Sorkin:2004qq,Kleihaus:2006ee,Asnin:2006ip,Dias:2007hg,Headrick:2009pv,Figueras:2012xj,Kalisch:2015via}. As pointed out in \cite{Emparan:2003sy}, in $d\geq5$ dimensions the Gregory-Laflamme instability can cause even topologically spherical black holes with highly deformed horizons to break symmetries along their extended directions. This instability was confirmed in Myers-Perry black holes with sufficiently large angular momenta \cite{Dias:2009iu,Dias:2010maa,Dias:2010eu,Dias:2011jg,Shibata:2009ad,Shibata:2010wz,Dias:2014eua}, in black rings \cite{Arcioni:2004ww,Elvang:2006dd,Figueras:2011he,Santos:2015iua}, and in the global Schwarzschild-AdS$_5 \times S^5$ black hole \cite{Banks:1998dd,Peet:1998cr,Hubeny:2002xn,Dias:2015pda,Buchel:2015gxa} (see section \ref{subsec:glrot} for an extension of the latter to rotating black holes).

One particular class of these Myers-Perry instabilities is the \emph{ultraspinning instability}, seen in axisymmetric perturbations of singly-spinning Myers-Perry black holes in $d\geq 6$ dimensions.  For these perturbations, there is a zero mode that indicates new branches of solutions.  These are the lumpy (also called bumpy, or rippled) black holes we have mentioned earlier \cite{Harmark:2007md,Dias:2009iu,Emparan:2011ve,Dias:2014cia,Emparan:2014pra,Emparan:2014jca,Suzuki:2015iha}.  These lumpy black holes are connected in moduli space through topology-changing mergers \cite{Kol:2002xz,Emparan:2011ve} to black rings and some of their associated multi-horizon solutions \cite{Dias:2014cia,Emparan:2014pra}.  The ultraspinning instability is also present in $d\geq 6$ AdS spacetimes \cite{Dias:2010gk}, where zero modes again branch. The numerical construction of these AdS lumpy black holes will be presented for the first time in section \ref{Sec:RRR}.

This phenomenon of the onset of an instability leading to new solutions is very common, and we will review how to find these modes in the following section. However, we point out that there is no proof that this needs to be the case. For example, the \emph{bar-mode} instabilities of Myers-Perry black holes \cite{Emparan:2003sy,Shibata:2009ad,Shibata:2010wz,Dias:2014eua,PauBarM} do not lead to new asymptotically flat solutions since their onset does not have zero modes. That is, the onset is time dependent and has quadrupole momentum and thus emits gravitational radiation.

Many hairy black hole solutions branch from the zero mode of instabilities as well.  For example, the near-horizon scalar condensation mechanism \cite{Gubser:2008px,Hartnoll:2008kx,Dias:2010ma} we have mentioned earlier proceeds through a dynamical instability of a non-hairy solution\footnote{Though not every phase transition that generates a hairy solution proceeds via a dynamical instability. See for instance \cite{Hartnett:2012np}.}. In particular, the near-horizon scalar condensation instability on a planar Reissner-Nordstr\"om black hole \cite{Gubser:2008px,Hartnoll:2008kx} was shown to evolve in time \cite{Murata:2010dx} to the hairy black holes of \cite{Hartnoll:2008vx,Hartnoll:2008kx}. These are the very same black holes that, in a phase diagram of solutions, branch-off from the original Reissner-Nordstr\"om black hole at the zero mode of the instability.

The \emph{superradiant instability} is another instability that generates new solutions, and is distinct from the near-horizon scalar condensation instability above\footnote{Note that, unlike superradiant instabilities, near-horizon instabilities also exist in the context of neutral and static black holes \cite{Dias:2010ma}.}.  The superradiant instability has its origins in the Penrose process \cite{Penrose:1971uk} where energy can be extracted from the ergoregion of a black hole. (Stimulated) superradiance is the wave analogue of the Penrose process, and was proposed in \cite{Zeldovich:1971,Zeldovich:1972,Starobinsky:1973scalar,Starobinsky:1973,Teukolsky:1974yv}. Spontaneous superradiant emission can also occur and is usually mixed with Hawking thermal radiation \cite{Dias:2007nj}.  A detailed account of the historical evolution of superradiance can be found in \cite{Brito:2015oca}. Superradiance is present for scalar fields, electromagnetic waves, and gravitational waves.  For waves with a harmonic dependence $e^{-i\omega t + im\psi}$, superradiance occurs for horizon angular velocities with $\omega < m\Omega_H$ in Kerr \cite{Zeldovich:1971,Zeldovich:1972,Starobinsky:1973scalar,Starobinsky:1973,Teukolsky:1974yv}, or $\omega < m\omega_H + q\Phi_H $ in Kerr-Newman (and in spherical Reissner-Nordstr\"om) with charge chemical potential $\Phi_H$ and for a scalar field with charge $q$ \cite{Furuhashi:2004jk}.

These bounds on superradiance deserve further comment.  It turns out that the onset frequency of \emph{any} instability (defined by $\mathrm{Im}(\omega)=0$) in rotating black holes must satisfy the superradiant bound $0 \leq  \mathrm{Re}(\omega)  \leq m\Omega_H$ (and $0 \leq  \mathrm{Re}(\omega)  \leq q\Phi_H$ for a charged system).  The proof (see section V of \cite{Teukolsky:1974yv}) relies only on conservation of energy and angular momentum.  At the time, the only known instability was the superradiant instability, but this proof  applies much more broadly, such as to the Gregory-Laflamme instability of the (boosted) black string \cite{Gross:1982cv,Gregory:1993vy,Gregory:1994bj,Monteiro:2009tc,Dias:2009iu,Dias:2010maa,Dias:2010eu,Dias:2011jg,Prestidge:1999uq,Dias:2010gk,PhysRevD.73.084013} and black ring \cite{Santos:2015iua}, to the ultraspinning \cite{Emparan:2003sy,Dias:2009iu,Dias:2010maa,Dias:2010eu,Dias:2011jg,Dias:2010gk} and to the bar-mode instabilities \cite{Shibata:2009ad,Shibata:2010wz,Dias:2014eua}.  The proof, however, says nothing about what happens away from the onset of an instability where $\mathrm{Im}(\omega) > 0$ (see \cite{Teukolsky:1974yv} and chapter 12.4 of \cite{waldbook}).  The only known scenario (to our knowledge) where this bound is violated away from the onset is in the Gregory-Laflamme instability of the black ring \cite{Santos:2015iua}. 

A necessary condition for superradiance to be promoted to an instability is the existence of a boundary condition that prevents dispersion of waves at asymptotic infinity. Such systems are historically known as `black hole bombs' after \cite{Press:1972zz}.  This can be accomplished with a boundary condition that preserves the asymptotic charges of the system.  Dirichlet walls can be provided by a reflecting box \cite{Press:1972zz,King:1977,Cardoso:2004nk,Hod:2009cp,Rosa:2009ei,Hod:2009cw,Witek:2010qc,Lee:2011ez,Dolan:2012yt,Herdeiro:2013pia,Degollado:2013bha,Hod:2013fvl,Hod:2014pza,Li:2014gfg,Aliev:2014aba,Li:2014fna,DiMenza2015,Dolan:2015dha,Aliev:2015wla,Delice:2015zga}, by a massive scalar or Proca field potential well \cite{Damour:1976kh,Detweiler:1980uk,Zouros:1979iw,Furuhashi:2004jk,Dolan:2007mj,Strafuss:2004qc,Dolan:2012yt,Cardoso:2011xi,Yoshino:2012kn,Pani:2012bp,Witek:2012tr,Okawa:2014nda,Brito:2015pxa}, or by the AdS boundary with reflecting boundary conditions \cite{Basu:2010uz,Bhattacharyya:2010yg,Gentle:2011kv,Dias:2011tj,Hawking:1999dp,Cardoso:2004hs,Dias:2011at,Dias:2013sdc,Cardoso:2013pza,Dias:2015rxy}. 
Typically, the physical properties of these systems are independent on the particular setup used to impose the Dirichlet boundary condition. 

The studies of superradiance are complemented by mathematical proofs of (in)stability as reviewed in \cite{Dold:2015cqa}. A proof of the existence of exponentially growing massive scalar wave superradiant solutions in Kerr spacetimes was recently established in \cite{Dafermos:2008en,Dafermos:2010hb,Shlapentokh-Rothman:2013ysa,Dafermos:2014cua,Dafermos:2014jwa}. For Kerr-AdS, a similar statement was proven in \cite{Dold:2015cqa}, following the influential work \cite{Breitenlohner:1982jf,Ishibashi:2004wx,Vasy:2009,Holzegel:2009ye,Holzegel:2011uu,Holzegel:2011qj,Holzegel:2012wt,Warnick:2012fi,Holzegel:2012wt,2012arXiv1212.1907G,Holzegel:2013kna,Gannot:2014boa,Holzegel:2015swa}.

Just as the onset of the Gregory-Laflamme instability can lead to new solutions, the onset of superradiant instabilities can also lead to new solutions.  It so happens that many of these instabilities are connected to some of the hairy black holes that we have mentioned earlier.  Some of these hairy black holes arise by placing a small black hole in a solitonic solution, but as one increases the size of the black hole by varying parameters, eventually the hair disappears, and the hairy black hole joins with the hairless family of black holes in the theory.  This occurs at the onset of the superradiant instability. This effect can be seen in rotating black holes with neutral scalar hair \cite{Dias:2011at,Herdeiro:2014goa}, in charged black holes with charged scalar hair \cite{Basu:2010uz,Bhattacharyya:2010yg,Gentle:2011kv,Dias:2011tj}, and also in the black resonator/geon system \cite{Dias:2011ss,Horowitz:2014hja,Cardoso:2013pza,Dias:2015rxy} with just with gravitational hair. 

Although boson stars and geons (or other solitonic solutions) do not have horizons, it may still be possible for them to have ergoregions.  In \cite{friedman1978}, it was shown that asymptotically flat solitons with ergoregions can be unstable to the \emph{erogregion instability}. See \cite{cominsschutz1978,1996MNRAS.282..580Y,Cardoso:2005gj,Chowdhury:2007jx,Cardoso:2007az,Brito:2015oca} for applications.

Let us now address the endpoint of superradiant instabilities in AdS\footnote{Similar behaviour is expected for rotating or charged black holes inside a reflecting box. Note however, that for asymptotically flat massive scalars or Proca fields, we expect the system to radiate away its hair, with a yet unknown timescale.}.  For the static, charged (Reissner-Nordstr\"om) case, \cite{Basu:2010uz,Dias:2011tj} indicate that the resulting charged hairy black holes have higher entropy and are no longer superradiant unstable, suggesting that these solutions are the endpoint to the instability, though no explicit simulation has been carried out.  In the rotating case, the hairy black holes \cite{Dias:2011at,Dias:2015rxy} also have higher entropy, but are nevertheless still unstable to superradiant perturbations with higher wavenumbers (the charged system of \cite{Basu:2010uz,Dias:2011tj} does not have a similar fate because the charge $q$ of the scalar condensate is fixed by the theory). More precisely, the (yet unpublished) results of \cite{waldResonator} imply that any candidate endpoint must satisfy $\Omega_H L \leq 1$, and all back holes with scalar hair of  \cite{Dias:2011at} and all black resonators found in \cite{Dias:2015rxy} have $\Omega_H L > 1$. 
They are therefore not the endpoint of this instability, and the nature of the endpoint remains an open question.

There are (so far unsuccessful) attempts to address this question with numerical simulations \cite{Witek:2010qc,Bantilan:2012vu,East:2013mfa}.  However, some possibilities were offered in \cite{Dias:2011at,Dias:2015rxy,Niehoff:2015oga}.  One option would be for the endpoint to reach some yet undiscovered black hole with $\Omega_H L \leq 1$. However, \cite{Niehoff:2015oga} have ruled out the $\Omega_H L = 1$ possibility (in the pure gravitational case, but the result should extend to black holes with scalar hair).  We are left with two natural possibilities. Either a singular solution is reached in finite time leading to a violation of cosmic censorship \cite{Penrose:1969pc,HawkingEllisBook,waldbook,Christodoulou:1999,Senovilla:2014gza}, much as in the Gregory-Laflamme system \cite{Lehner:2010pn,Lehner:2011wc}. Alternatively, the system might evolve through a tower of metastable configurations, developing structure on smaller and smaller scales (\emph{i.e.} dominated by higher and higher wavenumbers). At a certain point  the Planck scale would be reached, quantum gravity effects would become relevant, leading to a violation of the spirit of cosmic censorship. 

Now let us address nonlinear (in)stabilities.  The nonlinear stability of de Sitter and Minkowski space was famously proved in \cite{Friedrich:1986}, and \cite{Christodoulou:1993uv} (see also new proofs and extensions \cite{Lindblad:2004ue,ChoquetBruhat:2006jc} and the reviews \cite{Dafermos:2008en,Ringstrom:2015jza}).  

Conspicuously missing from this proof is the nonlinear stability of AdS.  While AdS is linearly stable, it may be nonlinearly unstable as first conjectured in \cite{DafermosHolzegel2006}.  The reflecting boundary conditions of AdS prevent any energy dissipation.  It is therefore possible for an arbitrarily small energy excitation to continuously reflect off the AdS boundary and eventually form a black hole, leading to a nonlinear instability.  This conjecture was first tested numerically in \cite{Bizon:2011gg} by collapsing a spherically symmetric scalar field analogous to \cite{Choptuik:1992jv}. The results in \cite{Bizon:2011gg} indicate that a black hole forms for arbitrarily small amplitude of this scalar field.  In the dual field theory, black hole formation is dual to thermalisation.

A perturbative explanation for this instability was put forth in \cite{Bizon:2011gg}. At linear order in perturbation theory, the lack of dissipation in AdS yields a spectrum of evenly-spaced normal modes. At higher orders, resonances between these modes cause higher modes to be excited that grow linearly in time. In the generic case, this leads to a breakdown of perturbation theory, and is interpreted as the beginnings of a nonlinear instability.  Furthermore, the timescale for the breakdown of perturbation theory coincides with the time scale for black hole formation seen in the numerical simulations \cite{Bizon:2011gg}.  Since this argument implies a shift to shorter length scales, this phenomenon is called a \emph{weakly turbulent instability}\footnote{This is not related to recently shown behaviour of AdS horizons akin to (super)fluid turbulence.  For those, see \cite{Carrasco:2012nf,Adams:2012pj,Adams:2013vsa,Green:2013zba,Chesler:2014pka}.}. Irremovable resonances that drive the system unstable are also present in the purely gravitational sector \cite{Dias:2011ss}. 

The nature and mechanism that drive this instability, its detailed properties, the necessity of an exact resonant spectrum, and the question of whether the instability exists for all initial data or just a subset, are all ongoing matters that have yet to be fully resolved.  Some of these matters have been partially addressed, but is beyond the scope of this review, and we refer to the ongoing work in \cite{Bizon:2011gg,Dias:2011ss,Dias:2012tq,Buchel:2012uh,Buchel:2013uba,Maliborski:2013jca,Bizon:2013xha,Maliborski:2012gx,Maliborski:2013ula,Baier:2013gsa,Jalmuzna:2013rwa,Basu:2012gg,2012arXiv1212.1907G,Friedrich:2014raa,Maliborski:2014rma,Abajo-Arrastia:2014fma,Balasubramanian:2014cja,Bizon:2014bya,Balasubramanian:2015uua,daSilva:2014zva,Craps:2014vaa,Okawa:2014nea,Deppe:2014oua,Dimitrakopoulos:2014ada,Buchel:2014xwa,Craps:2014jwa,Yang:2015jha,Okawa:2015xma,Bizon:2015pfa,Green:2015dsa,Deppe:2015qsa,Craps:2015iia,Craps:2015xya,Evnin:2015gma,Menon:2015oda}.  Classical Quantum Gravity will soon publish a Topical Review on this subject \cite{Bizon:CQG2015}.

In this review, our attention is placed on the existence of normal modes. Though, as described above, any generic combination of these modes will lead to irremovable resonances and a consequent breakdown of perturbation theory, a single mode will not. If a single mode is excited, perturbation theory can be continued indefinitely, leading to the perturbative construction of a new solution.  For scalar fields this process leads to the oscillons \cite{Seidel:1991zh,Maliborski:2013jca}, charged boson stars \cite{Jetzer:1989av,Basu:2010uz,Gentle:2011kv,Dias:2011tj,Buchel:2012uh,Liebling:2012fv,Buchel:2013uba}, Proca stars \cite{Brito:2015pxa}, or rotating boson stars \cite{Yoshida:1997qf,Hartmann:2010pm,Dias:2011at}. For gravitational perturbations, this leads to the geons of \cite{Dias:2011ss,Horowitz:2014hja,Cardoso:2013pza,Dias:2015rxy}. These solitonic solutions, which we have mentioned in the context of superradiance, can therefore be also viewed as nonlinear normal modes of AdS. That is, these solitonic solutions and their nonlinear interactions connect the physics of superradiance to the weakly turbulent nonlinear instability of AdS or other confined Dirichlet systems with irremovable resonances \cite{Dias:2011at,Dias:2011ss,Dias:2015rxy}. 

\section{Linear Perturbation Theory and Zero Modes\label{Sec:linear}}

In this section, we review linear perturbation theory with an emphasis towards finding stationary solutions.  
Though linear perturbation theory is often used to study stability, it has also been an important tool for showing the existence of new stationary solutions, as well as providing a linear approximation to those solutions. 

\subsection{The General Problem}

Let the metric $\bar g$ be a solution of the Einstein equation (possibly coupled to matter).  Now consider infinitesimal fluctuations about this solution $g=\bar{g}+h$, and additional linear fluctuations of matter fields, if present.  Place $g$ in the Einstein equation and expand in powers of $h$.  The Ricci tensor linearises as \footnote{\label{foot:Rconventions}Our curvature convention is $R^\alpha{}_{\beta \mu \nu}=\partial_\mu \Gamma^\alpha{}_{\beta \nu}+\Gamma^\alpha{}_{\mu \sigma} \Gamma^\sigma{}_{\beta \nu}-(\mu\leftrightarrow \nu)$ and $R_{\alpha \beta}=R^\mu{}_{\alpha \beta \mu}$ such that the Ricci scalar $R=g^{\alpha \beta}R_{\alpha \beta}$ has the same sign as the cosmological constant $\Lambda$ for the (A)dS spacetime.}
\begin{equation}
R_{\mu\nu}[\bar g+h]=R_{\mu\nu}[\bar g]+\Delta_L h_{\mu\nu}+ \bar{\nabla}_{(\mu} v_{\nu)}+O(h^2)\;,
\end{equation} 
where  $\Delta_L$ is the usual quasi-linear second-order Lichnerowicz operator,
\begin{equation}
\Delta_L h_{\mu\nu} \equiv 
-\frac{1}{2}\bar{\nabla}^2h_{\mu\nu} - \bar{R}_\mu{}^\kappa{}_\nu{}^\lambda h_{\kappa\lambda} + \bar{R}_{(\mu}^{\phantom{(\mu}\kappa}h_{\nu)\kappa} \,,\quad \text{and}\quad  v_\nu \equiv \bar{\nabla}_\alpha {h^\alpha}_{{}\nu} - \frac{1}{2}\bar{\nabla}_\nu h\,,
\end{equation}
and all barred objects and indices are with respect to the background $\bar g$. At zeroth order, the Einstein equation is trivial in this expansion since $\bar g$ is a solution.  At first order, we have
\begin{equation}
\Delta_L h_{\mu\nu}+ \bar{\nabla}_{(\mu} v_{\nu)}-\frac{2\Lambda}{d-2}h_{\mu\nu}+(\text{linear matter terms}) =0\;.
\label{eq:linearjorge}
\end{equation}
We may additionally have linear equations in the matter fields.  Of course, this linear system is much easier to solve than the original fully nonlinear one.

Without further specifying a gauge, this problem is not well-posed.  For a general spacetime in the presence of matter, diffeomorphic gauge redundancy can be completely fixed using de-Donder gauge \cite{deDonder:1921}
\begin{equation}
v_\nu=0\,.
\label{eq:donder}
\end{equation}
On the other hand, if no matter fields are present, one can show that de-Donder gauge still leaves some gauge redundancy and one is able to additionally impose the traceless condition $h=0$. The simultaneous choice $\bar{\nabla}_\alpha {h^\alpha}_{{}\nu}=0$ and $h=0$ is called traceless-transverse gauge.

To simplify our discussion further, we will focus on backgrounds that admit a stationary Killing field $\partial_t$. We can then Fourier expand $h$ in eigenfunctions of $\partial_t$:
\begin{equation}
h_{\mu\nu}(t,x^i) = e^{-i\omega\,t}\,\widehat{h}_{\mu\nu}(x^i)\,.
\end{equation}
where lower case latin indices run over the $(d-1)$ coordinates that are not $t$. This reduces (\ref{eq:linearjorge}) and the gauge condition (\ref{eq:donder}) to a set of equations that take the form
\begin{equation}
\mathcal{H}_{\phantom{(0)}\mu\nu}^{(0)\phantom{\mu\nu}\rho\lambda}\,\widehat{h}_{\rho\lambda}-\omega\,\mathcal{H}_{\phantom{(1)}\mu\nu}^{(1)\phantom{\mu\nu}\rho\lambda}\,\widehat{h}_{\rho\lambda}-\omega^2\,\mathcal{H}_{\phantom{(2)}\mu\nu}^{(2)\phantom{\mu\nu}\rho\lambda}\,\widehat{h}_{\rho\lambda}=0\,,
\label{eq:generalH}
\end{equation}
where $\mathcal{H}_{\phantom{(i)}\mu\nu}^{(i)\phantom{\mu\nu}\rho\lambda}$, for $i\in\{0,1,2\}$, are second order differential operators that act on $\widehat{h}_{cd}$ and depend on derivatives of the coordinates $x^i$. The system \eqref{eq:generalH} is necessarily quadratic in $\omega$ since the Einstein equation is a second order differential equation.  

We caution that the linearised Einstein equation and gauge conditions contain much redundancy.  To reduce the system to the form (\ref{eq:generalH}), we must assume that a minimal set of equations was obtained.  That is, that this is a minimal set of independent equations that implies the entirety of the linearised Einstein equation and gauge conditions. This is in general not a straightforward task. While there are methods to deal with constrained eigenvalue problems (see for instance \cite{Gander1989815}), we will henceforth assume, as is often the case, that the system can always be brought to this form. 

To properly solve this system, we must supply boundary conditions.  These boundary conditions will depend upon the physical situation at hand.  For stability problems in flat space, one invariable chooses outgoing boundary conditions at future null infinity.  For AdS asymptotics, one typically imposes energy and angular momentum preserving boundary conditions\footnote{Incidentally, these are the boundary conditions for which the AdS/CFT correspondence is best understood.}, meaning that $h_{ab}$ cannot modify the boundary metric of $\bar{g}$.  Other options in AdS are possible, depending on the application. In the interior of the integration domain, regularity is almost always imposed, with black holes being regular in ingoing Eddington-Finkelstein coordinates.  See section \ref{Sec:BCs} for further discussion of boundary conditions.  

Once the boundary conditions are specified, (\ref{eq:generalH}) represents a quadratic St\"urm-Liouville-type problem (also known as a quadratic eigenvalue problem) in $\omega$, which can be readily solved using known methods, some of which we will describe in section \ref{subsec:solvequadeig}. These methods typically reduce (\ref{eq:generalH}) to a quadratic matrix eigenvalue problem which can be readily solved on a computer.  In general, the solution to (\ref{eq:generalH}) consists of a spectrum of eigenvalues $\omega$ and their corresponding eigenvectors $\widehat{h}_{\mu\nu}$ (and linear matter fields if present).  

Supposing a valid spectrum for $\omega$ is found, there are a number of conclusions that can be drawn from it.  Most important is perhaps $\mathrm{Im}(\omega)$, which determines stability.  If $\mathrm{Im}(\omega)>0$, then the fields are increasing exponentially in time, which indicates instability. If $\mathrm{Im}(\omega)<0$, then the system decays exponentially (showing dissipation) and is linearly mode stable to this particular perturbation.  Modes with $\mathrm{Im}(\omega)=0$ are marginal, and it is possible for these modes to lead to new stationary solutions.  We will discuss this possibility in the following subsection.

\subsection{Hunting for Zero Modes}\label{subsec:zeromodes}

As one varies the parameters (\emph{i.e.} scans the moduli space) of $\bar{g}$, it is sometimes possible for a perturbation to yield $\mathrm{Im}(\omega) = 0$.  This location in moduli space is an onset for an instability if it is a critical point between regions with $\mathrm{Im}(\omega) < 0$ and $\mathrm{Im}(\omega) > 0$.

If $\mathrm{Im}(\omega) = 0$, we can also have $\mathrm{Re}(\omega)=0$ or $\mathrm{Re}(\omega)\neq0$. The former is known as a \emph{zero frequency mode}, or sometimes a \emph{zero mode}. While the traceless-transverse gauge can be shown to fix all gauge freedom for $\omega\neq0$, the same is not necessarily true for $\omega = 0$.  If a zero mode is found, one therefore needs to verify if it is a pure gauge mode or not.  An elegant way to do this is to show that the associated Newman-Penrose scalars \cite{Newman:1961qr,Chandrasekhar:1985kt} (or its higher dimensional generalisation \cite{Godazgar:2012zq}) are non-vanishing.

If a non-gauge zero mode with $\omega =0$ is found, then there is a time-independent perturbation $h$ of our background metric $\bar{g}$ that is regular over all spacetime. In other words, the resulting full metric $\bar{g}+h$ is a linear approximation to a novel stationary solution of the Einstein equations that is perturbatively close, \emph{i.e.} connected, to the background metric $\bar{g}$.  Such a zero mode will indicate where to search for new solutions, and the perturbation could provide a natural seed for a Newton-Raphson method.

The $\mathrm{Re}(\omega)\neq0$ case can also lead to new solutions, but is more subtle since such a mode might radiate its energy away\footnote{For special cases, notably at the onset of superradiant instabilities, such modes are non-radiative.}.  So, in addition to checking that the mode cannot be pure gauge, one also needs to check that the mode is not radiative. There are two places where radiation can escape: a horizon and an asymptotic region. At a horizon, one must also verify that $\mathrm{Re}(\omega)$ will cancel the ingoing flux across the horizon. There are explicit examples with no fine-tuning where this is the case \cite{Dias:2011at,Cardoso:2013pza}. For an asymptotic region that is AdS, the reflecting boundary conditions ensure energy and angular momentum conservation, so no radiation leaks out. For asymptotically flat solutions, the boundary conditions at null infinity allow radiation to be lost, so one has to proceed more carefully. There are situations where this is not an issue, such as those where a potential well exists (say a massive scalar field), and for which $\mathrm{Re}(\omega)$ does not allow the wave to propagate out of this potential barrier.  

One way of finding onsets is to solve \eqref{eq:linearjorge} in de-Donder (or transverse-traceless) gauge and vary parameters until $\mathrm{Im}(\omega)=0$. If one is searching for a zero mode ($\omega=0$), there are simpler methods which involve setting $\omega=0$ from scratch and using a different parameter as an eigenvalue. An example of this approach will be presented in subsection \ref{subsec:glrot}.

One way this can be achieved is by searching for `negative modes' \cite{Gross:1982cv,Prestidge:1999uq,Monteiro:2009ke,Dias:2009iu,Dias:2010maa,Dias:2010eu,Dias:2011jg,Dias:2010gk}.  That is, consider the following problem in pure gravity:
\begin{equation}\label{eq:negmode}
\Delta_L h_{\mu\nu}=-k^2 h_{\mu\nu}\;,
\end{equation}
This problem arises in perturbations of higher-dimensional solutions of the form $(\mathcal M,\bar g )\times \Sigma$, for some manifold $\Sigma$, where $k^2$ is a wavenumber for harmonics on $\Sigma$.  One can set $\omega=0$ and view \eqref{eq:negmode} as a \emph{linear} eigenvalue problem in $k^2$.  Then solve for $k^2$ while varying the parameters until $k=0$, which puts us on a zero mode.  Often, (as for the black string), real, positive $k^2$ modes are connected to the zero mode, which can significantly simplify the problem. 

Once a non-gauge, non-dissipative onset is obtained, we can begin to search for new solutions.  We point out that it is possible for one or two solutions to branch from the onset of a particular mode.  This is due to the fact that if $h$ is a linear perturbation, so is $-h$.  Given sufficient symmetry, $h$ and $-h$ may describe the same physical situation, in which case only a single solution branches from the onset.  In other cases, these are not equivalent, and there will be two solutions that branch from the onset. 

As an example, consider a perturbation on a sphere that is proportional to $h_1=\sin\theta$, where $\theta\in[-\pi,\pi]$ is the polar angle.  Since the unperturbed sphere is symmetric under $\theta\to-\theta$, and the perturbations $h_1$ and $-h_1$ can be mapped to each other under this symmetry, they both describe the same physical situation. However, for $h_2=3\sin\theta^2-1$, we see that $h_2$ and $-h_2$ cannot be mapped to each other under the $\theta\to-\theta$ symmetry, so these are inequivalent perturbations.  This is the reason why the zero mode of black strings leads to one non-uniform branch \cite{Gubser:2001ac,Wiseman:2002zc}, while those of singly-spinning Myers Perry black holes lead to two \cite{Dias:2014cia,Emparan:2014pra}.  Perturbations of Schwarzschild-AdS$_5\times S^5$ show both behaviours depending on the wavenumber \cite{Dias:2015pda}. 

\subsection{Solving Quadradic Eigenvalue Problems}\label{subsec:solvequadeig}
Quadratic eigenvalue problems can be very difficult to solve. We will present below two methods to deal with nonlinear eigenvalue problems. The first will be specialised to quadratic (or more generally polynomial) eigenvalue problems, while the other can be applied to any nonlinear eigenvalue problem.

As an example we will solve the following eigenvalue problem in $\lambda$
\begin{subequations}
\begin{multline}
(1-x) \, x \, q''(x)+\left[1+3 \left(1-\frac{\lambda }{2}\right)\, x \right] q'(x)+ \\ + \left\{\left(\frac{4 x}{1-x}-\frac{1}{x}\right)+ 3 \left[1-\frac{1}{2 (1-x)}\right] \lambda-\frac{\lambda ^2}{2}\right\}q(x) = 0,
\label{eq:poleigeq}
\end{multline}
subject to the boundary conditions
\begin{equation}
q(0) = q(1) = 0.
\label{eq:polyeigbound}
\end{equation}
\label{eqs:polyeig}
\end{subequations}
The analytical solution of (\ref{eqs:polyeig}) is
\begin{equation}
q(x) \phantom{\,} = \phantom{\,}_2 F_1\left(\frac{\lambda}{2},\lambda,3,x \right)(1-x)\, x, \quad \text{with}\quad \lambda = -n, \, n\in\mathbb{N},
\label{eq:exactpoly}
\end{equation}
where $\phantom{\,}_2 F_1$ is the Gaussian hypergeometric function. Our aim is to recover this solution numerically.

Let us attempt to reduce this problem to a standard eigenvalue problem (more specifically the eigenvalue problem for a linear pencil).  For that, we introduce the following two-vector $\underline{Q}(x) = \{\lambda q(x), q(x)\}$. (\ref{eq:poleigeq}) can simply be written as
\begin{multline}
\left(
\left[
\begin{array}{c@{\hspace{1.5 cm}}c}
-\frac{3 x}{2}\partial_x +3\left[1-\frac{1}{2(1-x)}\right] &x(1-x)\partial_x^2+(1+3 x)\partial_x+\left(\frac{4 x}{1-x}-\frac{1}{x}\right)
\\
\\
1& 0
\end{array}
\right]
-
\right.
\\
\\
\left.
-\lambda
\left[
\begin{array}{ccc}
\frac{1}{2} & 0
\\
\\
0 & 1
\end{array}
\right]
\right)\underline{Q}(x) = 0.
\label{eq:polymatrix}
\end{multline}
This is now of the form $(A-\lambda B)\underline{Q}=0$ for some linear differential operators $A$ and $B$.  An equations of this form is called a generalised eigenvalue problem for the linear pencil $A-\lambda B$.  Using some appropriate numerical scheme to discretise $A$ and $B$ and to incorporate the boundary conditions, this would reduce to a matrix generalised eigenvalue problem which can be solved by a standard algorithm (such as QZ factorisation).  See appendix \ref{appendix:collocation} to see how to do this with collocation methods.  This procedure works for any quadratic eigenvalue problem and can be straightforwardly generalised to higher order polynomial eigenvalue problems (\emph{i.e.} higher order polynomials in $\lambda$).  

We note that the reduction to a linear pencil \eqref{eq:polymatrix} is not unique. It is usually good numerical practice to bring the operators $A$ and $B$ into some structured form (say, real, symmetric/anti-symmetric, Hermitian, etc) if possible. 

\begin{figure}
\centering
\subfigure[ $n=0$ eigenfunction.]{
\includegraphics[width = 5.5 cm]{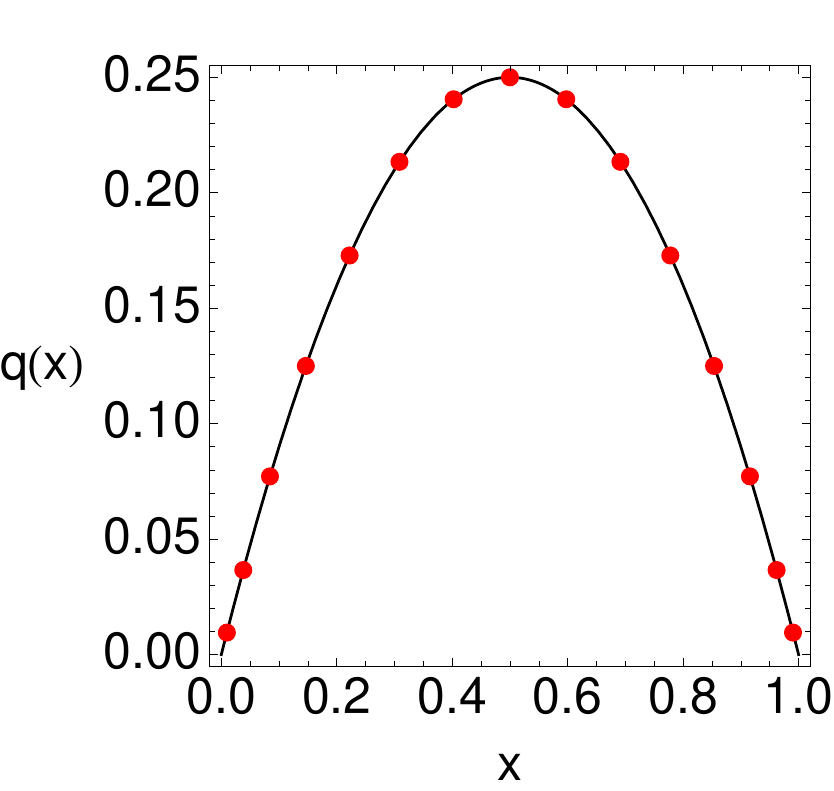}
\label{fig:eigena}
}
\subfigure[ $n=1$ eigenfunction.]{
\includegraphics[width = 5.5 cm]{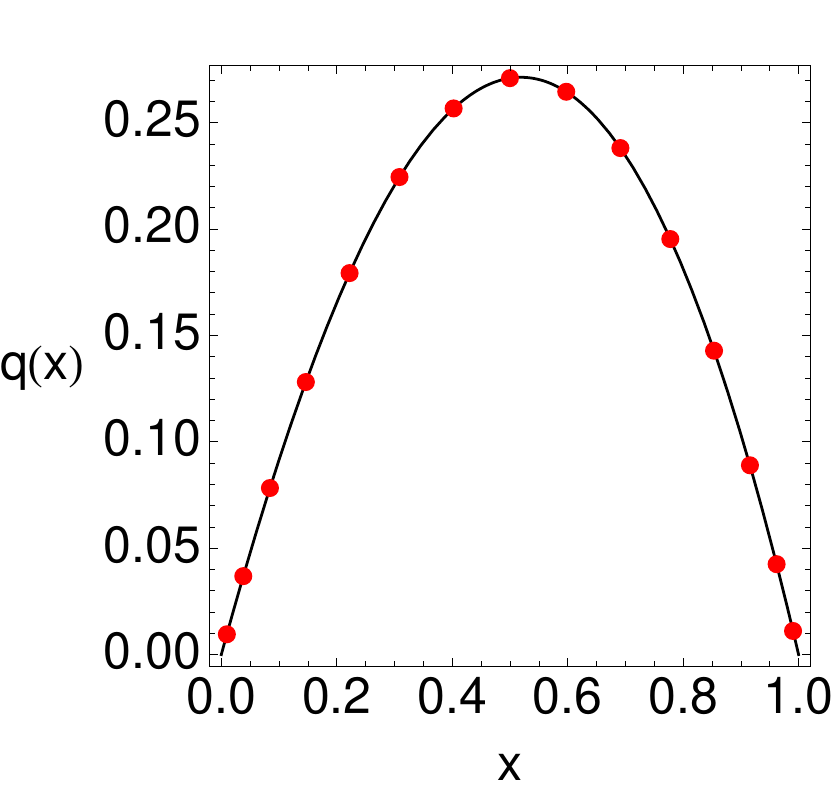}
\label{fig:eigenb}
}
\caption{\label{figs:eigen}Graphical representation of both the exact and numerical solutions of (\ref{eqs:polyeig}). The red dots represent the numerical result and the solid black line the exact result (\ref{eq:exactpoly}).}
\end{figure}

We have numerically solved (\ref{eq:polymatrix}) using pseudospectral collocation (see appendix \ref{appendix:collocation}) with $16$ points, and we find remarkable agreement between the analytical (\ref{eq:exactpoly}) and the numerical results (see Table \ref{tab:olare2}).
\begin{table}
\centering
\begin{tabular}{c|c}
& $|\lambda_{\mathrm{exact}}-\lambda_{\mathrm{num}}|$
\\
\hline
$n=0$ &$1.56615\times10^{-6}$
 \\
$n=1$ &$3.70670\times10^{-10}$
 \\
$n=2$ &$1.16275\times 10^{-4}$
 \\
$n=3$ &$2.96794\times 10^{-6}$
 \\
$n=4$ &$1.14458\times 10^{-4}$
 \end{tabular}
\caption{\label{tab:olare2} Numerical Vs exact result of (\ref{eqs:polyeig}).}
\end{table}
The numerical and exact eigenfunctions for the first two modes are shown in Figs.~\ref{figs:eigen}, and we confirm excellent agreement between them.

Eigenvalue problems are notoriously difficult numerical problems, and quadratic ones are worse.  The main reason for this is the appearance of spurious modes. These are modes that are numerical artefacts from the discretisation and not true eigenvalues to the continuous problem. In essence, the matrix version of the linear pencil $(A-\lambda B)\underline{Q}=0$ generically contains $N$ eigenvalues, where $N$ is the rank of the matrices $A$ and $B$, but the discretisation scheme is often only able to resolve $m\ll N$ modes.  There will therefore be $N-m$ unphysical spurious modes. This problem worsens considerably for high resolution grids/meshes, large number of functions, or a high dimension for the PDE. One can gain assurance that a mode is physical by solving the eigenvalue problem using many different resolutions, but this is not always possible if the number of spurious modes is too large or if the true mode is inaccurately resolved.   If only one or a few modes are desired, and one has a suitable guess for the eigenvalues and eigenfunctions, one can resort to a Newton-Raphson method, which we now describe. 

Consider the following more abstract nonlinear eigenvalue problem in $\{\mathfrak{f},\tilde{\lambda}\}$:
\begin{equation}
H({x,\tilde\lambda})\mathfrak{f}(x)=0\,\quad\text{with}\quad B_0(\tilde\lambda)\mathfrak{f}(0)=0\;,\qquad B_1(\tilde\lambda)\mathfrak{f}(1)=0\,,
\label{eq:nonlinearsturm}
\end{equation}
where $H(x,\tilde{\lambda})$, $B_0(\tilde\lambda)$, and $B_1(\tilde\lambda)$ are differential operators that can be nonlinear functions of $\tilde\lambda$. $B_i$ represent boundary conditions which can also be nonlinear in $\tilde\lambda$.  For the quadratic eigenvalue problem, $H$ takes the form $H \mathfrak{f}= H_0 \mathfrak{f}-\tilde{\lambda} H_1 \mathfrak{f}-\tilde{\lambda}^2 H_2\mathfrak{f}$, where each of the $H_i$ is a differential operator independent of $\tilde{\lambda}$, with the boundary conditions taking a similar form.  

Let us approach this problem via Newton-Raphson. We treat $\tilde\lambda$ as well as $\mathfrak f$ as unknown variables.  To have the proper number of equations and unknowns, we must supplement the problem with an additional normalisation condition $\mathcal N(\mathfrak f)=0$, for some functional $\mathcal N$.  The functional $\mathcal N$ can be something like $\mathfrak f(x_0)-1$, for some given point $x_0$ (provided it is known that the true solution satisfies $\mathfrak f(x_0)\neq 0$).  Another option is some integral condition such as $\int \mathfrak f^2 dx-1$ (provided a numerical scheme is capable of integration).  A third option is do choose a post-discretisation condition like $v_i\mathfrak f_i-1$ for some constant vector $v$.

In any case, following Newton-Raphson (see section \ref{subsec:newtonraphson} for more details on Newton-Raphson), we now linearise the system.  The result is the equation
\begin{equation}
\left[
\begin{array}{cc}
H(x,\tilde{\lambda}^{(n)}) & \frac{\partial H}{\partial\lambda}[x,\tilde\lambda^{(n)}] \mathfrak f^{(n)}\\
\frac{\delta \mathcal N}{\delta \mathfrak f}[\mathfrak f^{(n)}] & 0
\end{array}
\right]
\left[
\begin{array}{l}
\delta{\mathfrak{f}}^{(n)}\\
\delta{\tilde{\lambda}}^{(n)}
\end{array}
\right] = \,-\,\left[\begin{array}{l}
H(x,\tilde{\lambda}^{(n)})\mathfrak{f}^{(n)} \\
\mathcal N(\mathfrak f^{(n)})
\end{array}
\right]\,,
\end{equation}
which, given any $\mathfrak f^{(n)}$ and $\tilde\lambda^{(n)}$, is a linear equation in $\delta\mathfrak{f}^{(n)}$ and $\delta\tilde\lambda^{(n)}$.  Then we proceed with the usual Newton-Raphson algorithm.  That is, after a seed $\mathfrak f^{(0)}$ and $\tilde\lambda^{(0)}$ is given, we recursively iterate $\mathfrak f^{(n)}=\mathfrak f^{(n-1)}+\delta f^{(n-1)}$, $\tilde\lambda^{(n)}=\tilde\lambda^{(n-1)}+\delta\tilde\lambda^{(n-1)}$ by solving the linear equation above for $\delta f^{(n-1)}$, $\delta\tilde\lambda^{(n-1)}$ until some success or failure condition is met. The linear equations can be solved by any appropriate numerical scheme such as pseudospectral collocation (see appendix \ref{appendix:collocation}).

This method comes with all the usual drawbacks of Newton-Raphson.  It is only viable for finding specific modes, and requires a suitable seed solution.  But, if the seed is trustworthy, this method avoids the issue of spurious modes by assuring continuous connectedness to a physical mode\footnote{Unfortunately, it is still possible for Newton-Raphson to converge to a spurious mode.  Multiple resolutions should still be used for assurance.}.  It is therefore an ideal method for tracking the behaviour of a few modes as one varies parameters. This method has shown much success, for instance in \cite{Dias:2013sdc,Cardoso:2013pza}, where the quasinormal modes of the Kerr-AdS black hole were finally determined and in \cite{Dias:2015wqa} where the quasinormal spectrum of Kerr-Newman was investigated. See also \cite{Santos:2015iua} where this method was applied, together with patching techniques, to determine the instability growth rate of singly-spinning five-dimensional asymptotically flat black rings.

\subsection{\label{subsec:glrot}Application: Gregory-Laflamme Instability of Rotating Black Holes in AdS$_5\times$S$^5$}
Consider the following equations of motion in $d=10$ dimensions with a metric and a five-form flux field (derived from a four-form gauge potential):
\begin{equation}\label{eq:IIBsugra}
R_{MN}-\frac{1}{48}F_{MPQRS}F_N{}^{PQRS}=0\,,\qquad \nabla_MF^{MPQRS}=0\,,\qquad F_{(5)}=\star F_{(5)}\;.
\end{equation}
These are the equations of motion for type IIB supergravity with only the metric and Ramond-Ramond 5-form turned on.  Perhaps the most well-known solution to these equations is one where the metric is AdS$_5\times S^5$, and the form field is 
\begin{equation}
F_{\mu\nu\rho\sigma\tau}=\epsilon_{\mu\nu\rho\sigma\tau}\,\qquad F_{abde}=\epsilon_{abcde}\;,
\end{equation}
where $\epsilon_{\mu\nu\rho\sigma\tau}$ and $\epsilon_{abcde}$ are the volume forms of the base spaces AdS$_5$ and $S^5$, respectively. 

Without changing the five-form and the $S^5$ base space, any vacuum AdS$_5$ solution like Schwarzschild-AdS$_5$ is also a solution to \eqref{eq:IIBsugra}. Sufficiently small Schwarzschild-AdS$_5\times$S$^5$ black holes were shown to be unstable to a Gregory-Laflamme type instability in \cite{Hubeny:2002xn}. This solution is  a direct product of a Schwarzschild-AdS$_5$ black hole and a round five-sphere. Since the black hole can be made arbitrarily small, there will be a separation of scales between the round S$^3$ in Schwarzschild-AdS$_5$ and the five-dimensional sphere. This separation of scales causes the horizon geometry to resemble a black brane, which were shown to be unstable \cite{Gregory:1993vy}. 

While the growth rate of this instability has only been recently computed \cite{Buchel:2015gxa}, its onset has been determined some time ago \cite{Hubeny:2002xn}. We intend to generalise this computation, using the Newton-Raphson method described in the previous section, to five-dimensional rotating black holes with equal angular momenta along the two possible rotation planes -- the Hawking-Hunter-Taylor black hole \cite{Hawking:1998kw}. The full geometry, including the $S^5$, can be written as
\begin{equation}
\mathrm{d}s^2 = -\frac{f(r)}{h(r)}\mathrm{d}t^2+\frac{\mathrm{d}r^2}{f(r)}+r^2 \left[h(r)\left(\mathrm{d}\psi+\frac{\cos\theta}{2}\mathrm{d}\phi-\Omega(r)\mathrm{d}t\right)^2+\frac{1}{4}\left(\mathrm{d}\theta^2+\sin^2\theta\mathrm{d}\phi^2\right)\right]+L^2\,\mathrm{d}\Omega_5^2\,,
\end{equation}
where $L$ is the AdS length scale, $\mathrm{d}\Omega_5^2$ is a round five sphere of unit radius and
\begin{equation}
f(r) = \frac{r^2}{L^2}+1-\frac{r_0^2}{r^2}\left(1-\frac{a^2}{L^2}\right)+\frac{a^2r_0^2}{r^4}\,,\quad h(r) = 1+\frac{a^2r^2_0}{r^4}\,,\quad \text{and}\quad \Omega(r) = \frac{a r_0^2}{r^4\,h(r)}-\Omega_H\,.
\end{equation}

Asymptotically, the solution approaches AdS$_5\times$S$^5$ space, written in co-rotating coordinates\footnote{These are coordinates where the boundary metric rotates rigidly with angular velocity $\Omega_H$. In order to change to a static boundary metric, one performs a coordinate transformation $\tilde{\psi} = \psi-\Omega_H\,t$.}. The event horizon is located at $r=r_+$ (the largest real root of $f$). Requiring this solution to have a null tangent vector $\partial_t$, \emph{i.e.} choosing $(t,\psi)$ to be co-rotating coordinates, fixes the angular velocity $\Omega_H$ to be:
\begin{equation}
  \Omega_H = \frac{r_0^{2} a}{r_+^4+r_0^{2}a^2}
 \leq \Omega_H^{\rm ext}\,\qquad \hbox{where}\quad \Omega_H^{\rm ext}=\frac{1}{L}\sqrt{1+\frac{L^2}{2\,r_+^2}} \,. 
 \label{angvel}
\end{equation}
The solution saturating the bound in the angular velocity corresponds to an extreme black hole with a regular, but degenerate, horizon. Note that this upper bound in $\Omega_HL$ is always greater than one and tends to the unit value in the limit of large $r_+/L$.

It is convenient to parameterise the solution in terms of $(r_+,\Omega_H)$ instead of $(r_0,a)$ through the relations,
\begin{equation}
 r_0^{2}=\frac{r_+^{4} \left(r_+^2+L^2\right)}{r_+^2 L^2-a^2 \left(r_+^2+ L^2\right)}\,,
 \qquad a=\frac{r_+^2 L ^2 \Omega_H}{r_+^2+L^2}\,.
 \label{eq:MPrMa}
\end{equation}
Note that $r_+/L$ is completely gauge invariant, since it is related to the area of a spatial cross section of the horizon at constant $\psi$. 

Our task is to find the onset of the Gregory-Laflamme instability as a function of the dimensionless parameters $r_+/L$ and $\Omega_HL$. To do so, one can consider the full time-dependent linear perturbations and search for the parameters that yield $\mathrm{Im}(\omega)=0$. This is certainly possible, but is not what we are going to do here. Instead, we set $\omega = 0$ from the outset, and write down a generic metric perturbation that is stationary and respects the $t-\psi$ symmetry, \emph{i.e.} $\partial_t$ will be Killing and it will further have the discrete $(t,\psi)\to-(t,\psi)$ symmetry. We will then give $\Omega_H\,L$ and determine the value of $r_+/L$ at which the instability first appears.  That is, $r_+/L$ will take the place of the eigenvalue. 

Before we undertake this task, let us make the following technical remarks:
\begin{itemize}
\item The modes that generate the Gregory-Laflamme instability in this system deform the $S^5$, since these are the extended directions for small Schwarzschild-AdS$_5$ black holes.  Since there is full $SO(6)$ symmetry in the background solution, we expand our metric perturbation in scalar spherical harmonics on the five sphere. We denote these by $\mathbb{Y}_\ell$, and are such that
$$\Box_{S^5}\mathbb{Y}_\ell+\ell(\ell+4)\mathbb{Y}_\ell=0\,,$$
where $\ell \in \mathbb{Z}^+$. Note that for a given $\ell$, there could be many distinct harmonics. Each of these will have the same onset, but will lead to different nonlinear solutions.

\item These metric perturbations (related to massive Kaluza-Klein gravitons), only have components on the AdS$_5$ base space, which is possible only because our background metric is a direct product of a five-dimensional AdS solution and a round five sphere. 
\item These perturbations can be found in transverse-traceless gauge.  It is therefore impossible for the 5-form perturbations to couple to it. In essence, it is a spin-2 perturbation, which is impossible to generate through a gauge field perturbation. It can be verified that the metric perturbation completely closes the system without the need to introduce a 5-form perturbation.
\end{itemize}

Given the technical remarks made above, we take the following metric perturbation ansatz
\begin{multline}
h_{\mu\nu} \mathrm{d}x^\mu \mathrm{d}x^\nu\equiv -\frac{f(r)}{h(r)} q_1(r)\mathbb{Y}_\ell\mathrm{d}t^2+\frac{q_2(r)\mathbb{Y}_\ell\mathrm{d}r^2}{f(r)}+r^2 \Bigg[q_3(r)\mathbb{Y}_\ell\,h(r)\left(\mathrm{d}\psi\frac{\cos\theta}{2}\mathrm{d}\phi-\Omega(r)\mathrm{d}t\right)^2\\
+\frac{q_4(r)\mathbb{Y}_\ell}{4}\left(\mathrm{d}\theta^2+\sin^2\theta\mathrm{d}\phi^2\right)-2 h(r)q_5(r)\Omega(r)\mathbb{Y}_\ell\mathrm{d}t\,\left(\mathrm{d}\psi+\frac{\cos\theta}{2}\mathrm{d}\phi-\Omega(r)\mathrm{d}t\right)\Bigg]\,
\end{multline}
where the five variables $q_i$, with $i\in\{1,\ldots,5\}$ are to be determined numerically.

Now we put this ansatz into the linearised equation \eqref{eq:linearjorge}, and obtain a minimum set of equations. The transverse-traceless gauge allows $q_1$ and $q_3$ to be written in terms of the remaining quantities and their first derivatives:
\begin{subequations}
\begin{align}
&q_1 = -q_2-q_3-2 q_4
\\
&q_3 = \frac{2 q_4\left[r h f^\prime-f \left(r h^\prime+2 h\right)\right]+2r f h\,q_2^\prime+2r^3 h^3 q_5\,\Omega \Omega^\prime+q_2\,\left[2 r h f^\prime+f \left(6 h-r h^\prime\right)\right]}{2 f \left(r h^\prime+h\right)-r h f^\prime}\,,
\end{align}
\end{subequations}
where $^\prime$ denotes differentiation with respect to $r$. We are thus left with $q_2$, $q_4$ and $q_5$ to be determined by the equations of motion. 

We now change coordinates to a compact radial direction $y$
\begin{equation}
r = \frac{r_+}{1-y^2}\,,
\end{equation}
which places the horizon at $y=0$ and asymptotic infinity at $y=1$. The remaining equations of motion can be solely expressed in terms of $y$ and the two dimensionless parameters $\alpha \equiv \Omega_H \,L$ and $\beta \equiv r_+^2/L^2$.

For boundary conditions (see section \ref{Sec:BCs}), we impose regularity in both the past and future horizons, which amounts to $q_2^\prime(0)=q_4^\prime(0)=q_5^\prime(0)$. At infinity, we have to proceed more carefully. We perform the  change of variables:
\begin{equation}
q_2 = Q_1\,,\quad q_4 = Q_2\quad\text{and}\quad q_5 = \frac{Q_3}{(1-y^2)^2}\,,
\end{equation}
where our final equations can be written in the following schematic vectorial form
\begin{equation}
(1-y)^2 \mathbb{A}(y)\cdot\frac{\partial^2 \mathbf{q}}{\partial y^2}+(1-y) \mathbb{B}(y)\cdot\frac{\partial \mathbf{q}}{\partial y}+\mathbb{C}(y)\cdot \mathbf{q}=0\,,
\nonumber
\end{equation}
where $\mathbb{A}$, $\mathbb{B}$ and $\mathbb{C}$ are $3\times3$ matrices with coefficients that depend on $y$, $\alpha$, and $\beta$; and $\mathbf{q}$ is a vector with components $\mathbf{q} = \{Q_1,Q_2,Q_3\}$. Near asymptotic infinity ($y=1$), $\mathbb{A}$, $\mathbb{B}$ and $\mathbb{C}$ attain regular values. In particular
\begin{align}
\mathbb{A}(1) = \left[
\begin{array}{ccc}
1 &0 &0
\\
0&1&0
\\
0&0&1
\end{array}
\right]\,,\quad
\mathbb{B}(1) = \left[
\begin{array}{ccc}
7 & 0 &0
\\
2 & 3 & 0
\\
0 & 0 & 7
\end{array}
\right]\,\quad \text{and}
\\
\mathbb{C}(1) = \left[
\begin{array}{ccc}
-(\ell+6)(\ell-2) &0 &0
\\
12&-\ell(\ell+4)&0
\\
0&0&-(\ell+6)(\ell-2) 
\end{array}
\right]\,.
\end{align}
This then allows us to determine the most general solution at asymptotic infinity, by inspecting the auxiliary equation
\begin{equation}
(1-y)^2 \mathbb{A}(1)\cdot\frac{\partial^2 \mathbf{q}_0}{\partial y^2}+(1-y) \mathbb{B}(1)\cdot\frac{\partial \mathbf{q}_0}{\partial y}+\mathbb{C}(1)\cdot \mathbf{q}_0=0\,.
\label{eq:generaleq}
\end{equation}
The general solution to this equation can be  determined by noting that if we further change variables to $\chi = -\log(1-y)$, the system above reduces to a system of coupled ODEs with constant coefficients, which can be  analytically solved using standard eigenvalue methods. The most general solution $\mathbf{q}_0$ compatible with normalisability takes the following form
\begin{align}
&\mathbf{q}_0^1(y) = C_1 (1-y)^{\ell+6} \nonumber
\\
&\mathbf{q}_0^2(y) =C_2 (1-y)^{\ell+4}+\frac{C_1 \ell  (1-y)^{\ell +6}}{2 (\ell +3)}\,,
\\
&\mathbf{q}_{0}^3(y) =C_3 (1-y)^{\ell +6}\nonumber
\end{align}
where $C_1$, $C_2$ and $C_3$ are constants. Note that the most general solution to (\ref{eq:generaleq}) depends of course on six integration constants, but the remaining three are not compatible with normalisability\footnote{To be precise, if we are interested in the holographic dual of $\mathcal{N}=4$ SYM, then these are the relevant decays. In particular, for sufficiently low $\ell$, there are other integration constants that might also appear allowed by normalisability, but they correspond to double-trace deformations of $\mathcal{N}=4$ SYM that we will not consider here.}.

Motivated by the discussion above, we then perform one final change of variables:
\begin{align}
&Q_1 = (1-y^2)^{\ell+6} \hat{q}_1 \nonumber
\\
&Q_2 = (1-y^2)^{\ell+4} \hat{q}_2
\\
&Q_3 = (1-y^2)^{\ell+6} \hat{q}_3 \nonumber
\end{align}
The boundary conditions are now simply given by $\hat{q}_i^\prime(0)=\hat{q}_i^\prime(1)=0$.

\begin{figure}[ht]
\centering
\includegraphics[width=0.5\textwidth]{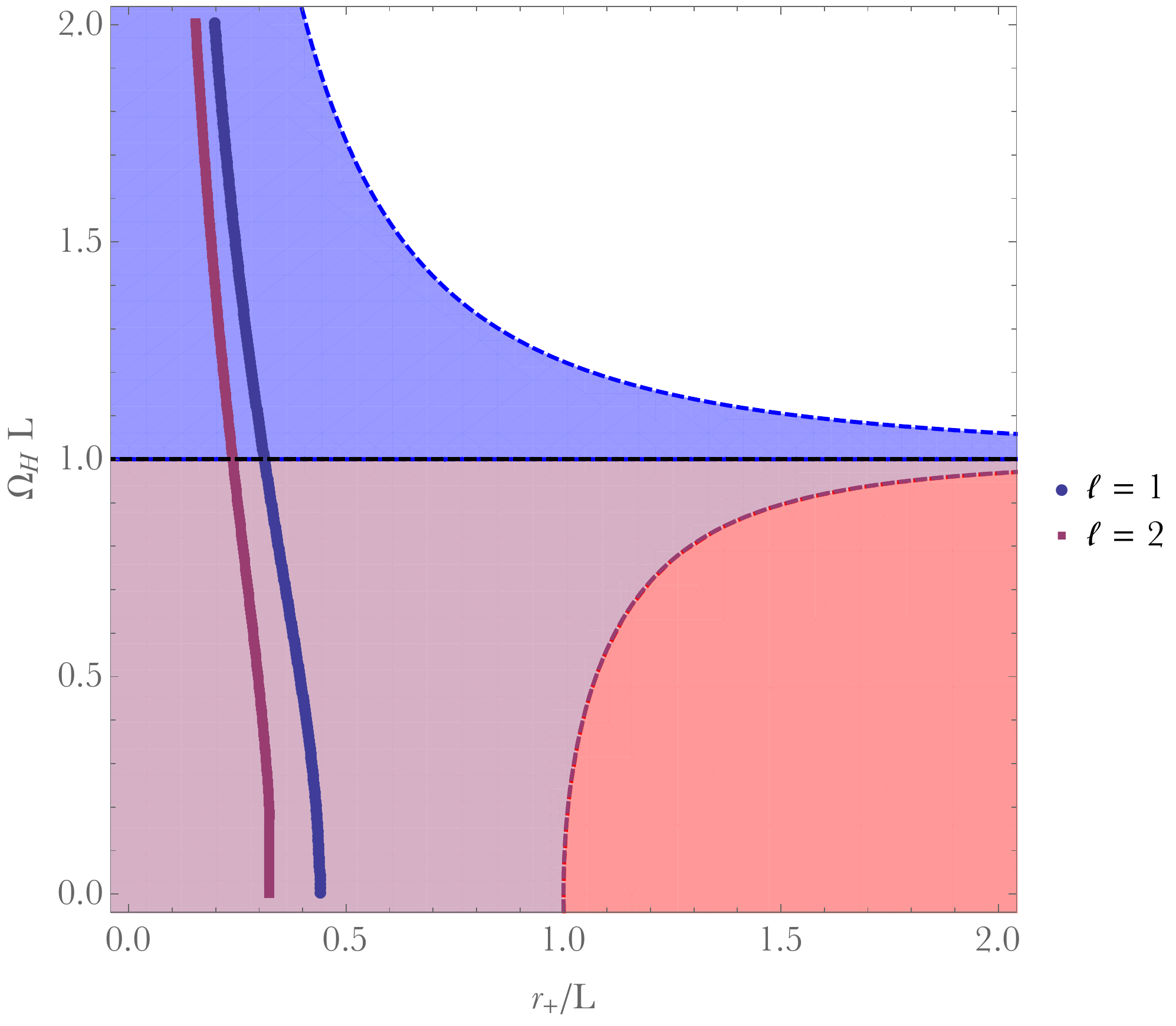}
\caption{\label{fig:rotGL}Onset of the rotating Gregory-Laflamme instability. The dotted line below $\Omega_H L = 1$ is the Hawking-Page transition and the dotted line above $\Omega_H L = 1$ is extremality. The left (right) vertical line is the onset of the $\ell = 1$ ($\ell = 2$) mode. We expect that the left side of each of these curves to be the unstable region. For $\Omega_H L = 0$, our results reproduce those in \cite{Hubeny:2002xn,Dias:2015pda}.}
\end{figure}

Given an $\alpha\equiv\Omega_H L$, the equations of motion are a 9\textsuperscript{th} order polynomial eigenvalue problem in $\beta\equiv{r_+}^2/L^2$. We solve this problem using the Newton-Raphson method described in the previous section. We expect the Gregory-Laflamme mode for rotating solutions to be connected to that of Schwarzschild AdS$_5\times S^5$.  Therefore, we can use the known value of $\beta$ in \cite{Hubeny:2002xn,Dias:2015pda} for $\alpha=0$ as a seed for small $\alpha$ perturbations. Our findings are shown in Fig.~\ref{fig:rotGL}, where we can see that increasing rotation shuts down the instability, since we need to move towards smaller values of $r_+/L$.

The results of Fig.~\ref{fig:rotGL} are physically interesting. We first note that the line with $\Omega_H L = 1$ separates two distinct regions of the moduli space of the Hawking-Hunter-Taylor black hole \cite{Hawking:1998kw}. Solutions with $\Omega_H L>1$ were shown to be unstable to the superradiant instability \cite{Cardoso:2013pza}, whereas solutions with $\Omega_H L<1$ are expected to be linearly stable to the same instability\footnote{Note that solutions with $\Omega_H L =1$ are likely to be nonlinearly unstable.}. One could envisage a scenario where the onset of the rotating Gregory-Laflamme, presented in Fig.~\ref{fig:rotGL}, would asymptote from below to $\Omega_H L =1$ as $r_+/L$ increases. This would shield these black holes from the superradiant instability by first being unstable to the Gregory-Laflamme instability. We see that this is not the case. Indeed, we can find regions of parameter space where small Hawking-Hunter-Taylor black holes are stable to the rotating Gregory-Laflamme instability, and yet unstable to the superradiant instability, and \emph{vice versa}.

We expect new families of black holes with deformed $S^3\times S^5$ horizons to branch from either of the onsets displayed in Fig.~\ref{fig:rotGL}. Let us focus on solutions which preserve the isometries $\mathbb{R}_t\times SO(4)\times SO(5)$. Due to discrete symmetries, we expect one such family from the $\ell = 1$ onset, and two families from the $\ell = 2$. The construction of these novel black hole solutions remains an open problem.

\section{The DeTurck Method\label{Sec:DeTurcksection}}

 In 1952, Choquet-Bruhat  \cite{FouresBruhat:1952zz,Bruhat:1967}, and later Fischer and Marsden in 1972 \cite{Fischer:1972} and Choquet-Bruhat and Geroch \cite{ChoquetBruhat:1969cb}, famously proved that the Einstein equation is a hyperbolic system of differential equations, \emph{i.e.} proved the well-posedness of the Cauchy problem for the Einstein equation \cite{Dafermos:2008en,Ringstrom:2015jza}\footnote{This result was built on earlier influential work on the harmonic gauge by de Donder \cite{deDonder:1921}, and on the existence and uniqueness theorems for general quasilinear wave equations in the 1930's and Leray's notion of global hyperbolicity \cite{LerayBook}, as reviewed in the lecture notes of \cite{Dafermos:2008en}.}. On the other hand, once symmetries are assumed in a static or stationary ansatz, the Einstein equation typically becomes a mixed elliptic-hyperbolic system of differential equations.  This fact complicates the numerical search for static or stationary solutions, which requires a well-posed boundary value problem.  Rather than this mixed elliptic-hyperbolic system, we would like an elliptic set of equations to guarantee well-posedness.
 
 For cohomogeneity-2 problems, it is possible to work in conformal gauge.  That is, if the system depends on the coordinates $x$ and $y$, assume an ansatz of the form
 \begin{equation}
\mathrm ds^2= \Omega(x,y)(dx^2+dy^2)+g_{ij}(x,y)dz^idz^j\;,
 \end{equation}
 where the $i,j$ indices run over all coordinates except for $x$ and $y$, $\Omega>0$ and $g_{ij}$ has Lorentzian signature. The fact that any two-dimensional metric is conformally flat allows us to write an ansatz of this form.  There is residual conformal symmetry that is fixed by specifying the integration domain. The Einstein equation will yield a set of elliptic equations of motion for $\Omega$ and $g_{ij}$, as well as a set of constraint equations.  The constraint equations are used to supply consistent boundary conditions. 
 
Conformal gauge has been used successfully in a number of cases \cite{Wiseman:2002zc,Kol:2003ja,Kudoh:2004hs,Kudoh:2003ki,Aharony:2005bm,Kleihaus:1997mn,Kleihaus:1997ws,Kleihaus:2000kg,Hartmann:2001ic,Kleihaus:2002ee,Kleihaus:2005me,Kleihaus:2006ee,Kalisch:2015via}. But it is restricted to cohomogeneity-2 problems, and the application of boundary conditions is not always straightforward. 

In this section, we describe an alternative method  that attempts to reformulate the Einstein equation into a manifestly elliptic form. This is the Einstein-DeTurck formulation, first introduced in the seminal work of \cite{Headrick:2009pv} and further developed in \cite{Adam:2011dn,Figueras:2011va,Figueras:2012rb,Fischetti:2012vt}. This method has greater flexibility than conformal gauge; it works in any cohomogeneity and for a broad range of different types of solutions.  Of course, the method also has its drawbacks which will also be discussed in this section.  

\subsection{Stationarity \label{Sec:statio}}
We start with a brief preamble about what we mean by stationary solutions of the Einstein equation. One operational way of defining stationarity is the following: a stationary solution is a solution to a well-posed boundary value problem.

In general, finding whether an equation can be solved as a boundary value problem is not an easy task. There are cases, however, where one can prove this. In particular, if a solution is stationary in the more restrictive sense, \emph{i.e.} in the sense that it admits a Killing field that is asymptotically timelike, then one can prove that the problem is elliptic, and thus a well-posed boundary value problem. Known black hole solutions such as the Kerr-(AdS) black hole are stationary in this restricted sense. On the other hand, solutions with a single helical Killing vector field, such as the ones numerically constructed in \cite{Horowitz:2014hja,Dias:2015rxy}, or the flowing geometries of \cite{Figueras:2012rb,Fischetti:2012vt}, do not fall under this more restrictive definition, and yet can be found with the methods outlined in this review.
 
\subsection{Ellipticity and the Harmonic Einstein Equation \label{Sec:DeTurck}}

\subsubsection{Ellipticity and DeTurck Gauge-Fixing \label{Sec:DeTurck1}}
The Einstein equation with a cosmological constant $\Lambda$  on a $d$-dimensional spacetime $({\cal M},g)$ can be written as (we postpone the inclusion of matter fields for later on) 
\begin{equation}\label{dT:Einstein}
R_{\mu\nu}-\frac{2 \Lambda}{d-2}\, g_{\mu\nu}=0,
\end{equation} 
where $R_{\mu\nu}$ is the Ricci tensor.

This is a quasi-linear second-order differential equation. To determine the character of this equation we linearise the Einstein equation about fluctuations of the metric, $g_{\mu\nu}\to g_{\mu\nu}+h_{\mu\nu}$
\begin{equation}\label{dT:EinsteinLinear}
\Delta_Rh_{\mu\nu}\equiv\delta R_{\mu\nu}-\frac{2 \Lambda}{d-2}\, h_{\mu\nu}=0  \qquad \Leftrightarrow  \qquad  \Delta_L h_{\mu\nu}+ \nabla_{(\mu} v_{\nu)} -\frac{2 \Lambda}{d-2}\, h_{\mu\nu}=0,
\end{equation} 
where  $\Delta_L$ is the usual quasi-linear second-order Lichnerowicz operator (as we have defined in the previous section),
\begin{equation}\label{dT:Lichn}
\Delta_L h_{\mu\nu} \equiv 
-\frac12\nabla^2h_{\mu\nu} - R_\mu{}^\kappa{}_\nu{}^\lambda h_{\kappa\lambda} + {R_{(\mu}}^\kappa h_{\nu)\kappa} \,,\quad \hbox{and}\quad  v_\nu \equiv \nabla_\alpha {h^\alpha}_{{}\nu} - \frac12\partial_\nu h\,.
\end{equation}
To determine the character of the differential system of equations \eqref{dT:EinsteinLinear} we examine the {\it principal symbol} of $\Delta_R$, \emph{i.e.} the highest derivative terms in $\Delta_R$ (in particular, note that the cosmological constant term is not relevant to determine the character of the equation),
\begin{equation}\label{dT:PricSymb1}
P_g h_{\mu\nu} = \frac{1}{2} g^{\alpha \beta} \left(- \partial_\alpha \partial_\beta h_{\mu\nu}+ 2 \partial_\alpha \partial_{(\mu}  h_{\nu) \beta}   - \partial_\mu  \partial_\nu h_{\alpha\beta} \right).
\end{equation}
Short wavelength perturbations have large second derivatives, so they govern the principal symbol operator. Indeed, when moving from \eqref{dT:EinsteinLinear} into \eqref{dT:PricSymb1} we effectively discard the Riemann and Ricci curvature terms, which means that we keep only perturbations with wavelength much smaller than any curvature scale of the metric.
Replacing the derivatives $\partial_\alpha$ by a covector $\zeta_\alpha$, \eqref{dT:PricSymb1} reads
\begin{equation}\label{dT:PricSymb2}
P_g(\zeta) h_{\mu\nu} = \frac{1}{2} \left(-  \zeta^2 h_{\mu\nu} +2 \zeta^\alpha \zeta_{(\mu} h_{\nu) \alpha} -  \zeta_\mu  \zeta_\nu h \right)
\end{equation}
and the {\it linear} operator $\delta R_{\mu\nu}$ (and thus the {\it non-linear} operator $R_{\mu\nu}$) is said to be {\it elliptic} if and only if $P_g(\zeta) h_{\mu\nu}\neq 0$ for all non-vanishing vectors $\zeta \in \mathbb{R}^d$  and for every point $x$ in ${\cal M}$. 

An inspection of \eqref{dT:PricSymb2} concludes that $\delta R_{\mu\nu}$ is not elliptic. The main reason being that any perturbation of the form $h_{\mu\nu}=\zeta_{(\mu}\xi_{\nu)}$, which translates into $h_{\mu\nu}=\partial_{(\mu}\xi_{\nu)}$ in a coordinate basis,  yields $P_g(\zeta) h_{\mu\nu}= 0$ for any vector $\xi$. 
A perturbation of this type is a short wavelength (local) diffeomorphism generated by $\xi$. Indeed, recall that an infinitesimal diffeomorphism generated by a gauge vector field $\xi$ is given by $h_{\mu\nu} = \nabla_{(\mu} \xi_{\nu)}$. This reduces to $ \nabla_{(\mu} \xi_{\nu)} \sim \partial_{(\mu} \xi_{\nu)}$ for a gauge parameter $\xi$ that varies on very short wavelength scales.
Therefore, we arrive to the important conclusion that the lack of ellipticity of the Einstein equation is a direct consequence of gauge invariance. 

In other words, any short wavelength perturbation can be decomposed as $h_{\mu\nu} = \hat h_{\mu\nu} + \partial_{(\mu}\xi_{\nu)}$ where $\hat h_{\mu\nu}$ is the transverse (physical) part ($\partial_\nu\hat h^\nu{}_\mu -\frac12\partial_\mu\hat h = 0$)  and $ \partial_{(\mu}\xi_{\nu)}$ is a longitudinal (pure gauge) contribution. One has  $P_g  \hat{h}_{\mu\nu} =-\frac{1}{2} \partial^\alpha \partial_\alpha \hat{h}_{\mu\nu}$, while $P_g$ annihilates the longitudinal fluctuation. To move towards a elliptical formulation of the equations of motion, we would like to eliminate the longitudinal pure gauge mode perturbations by imposing the so-called {\it de-Donder} or {\it harmonic gauge}, 
\begin{equation}\label{dT:harmonicgauge}
\partial_\nu {h^\nu}_\mu - \frac12\partial_\mu h = 0,
\end{equation}
\emph{i.e.} we want to find a gauge covector $\xi_\mu$ such that, for short wavelengths, one has $\delta \xi_\mu =\partial_\nu {h^\nu}_\mu - \frac12\partial_\mu h$. The simplest choice is  
\begin{equation}\label{dT:harmonicgauge}
\xi_\mu = g^{\lambda\nu}\left(
\partial_\lambda g_{\nu\mu} - \frac12\partial_\mu g_{\lambda\nu} \right) \qquad \Leftrightarrow \qquad 
 \xi^\mu = g^{\alpha\nu}\Gamma^\mu_{\alpha\nu}= -\nabla^2x^\mu\,.
\end{equation}
where $x^\mu$ is some coordinate chart (viewed as scalar functions), $\Gamma^\mu_{\alpha\nu} $ is the Levi-Civita affine connection associated to $g$ and $\nabla^2$ is the scalar Laplacian.

The reader familiar with time evolution in numerical relativity will recognise that the local gauge fixing choice $\xi^\mu =  g^{\alpha\nu}\Gamma^\mu_{\alpha\nu} =  -\nabla^2x^\mu=0$ is nothing but the usual \emph{harmonic gauge} (and associated harmonic coordinates $x^\mu$) often made in dynamical hyperbolic problems where Einstein equation is solved as a Cauchy problem. Indeed, with the local gauge choice $\xi^\mu = 0$, the principal symbol of the Einstein operator \eqref{dT:Einstein} becomes simply
$P_g h_{\mu\nu} =-\frac{1}{2} \partial^\alpha \partial_\alpha h_{\mu\nu}$
which is manifestly hyperbolic for Lorentzian solutions, \emph{i.e.} it describes propagating waves along a light-cone.

The local harmonic gauge choice \eqref{dT:harmonicgauge} breaks gauge invariance as desired but also breaks covariance, which is not so desirable. To recast the system in a covariant form, we promote the coordinate partial derivatives in \eqref{dT:harmonicgauge} to covariant derivatives with respect to an arbitrary but fixed background $\bar{g}$ on ${\cal M}$,
\begin{equation}\label{dT:harmonicgaugeCov}
\xi_\mu = g^{\alpha\nu}\left(
\bar\nabla_\alpha g_{\nu\mu} - \frac12\bar\nabla_\mu g_{\nu\alpha}\right) \qquad
\Leftrightarrow \qquad \xi^\mu = g^{\alpha\nu}(\Gamma^\mu_{\alpha\nu} - \bar\Gamma^\mu_{\alpha\nu}),
\end{equation}
where $\bar\Gamma^\mu_{\alpha\nu}$ is the Levi-Civita connection for $\bar g_{\mu\nu}$ and $\bar\nabla_\mu$ is the associated covariant derivative. Usually, $\bar{g}$ and $\bar\Gamma$ are denoted by the {\it reference metric} and {\it reference connection}, respectively. Since the difference between two connections is a tensor, $\xi$ is a globally defined covariant vector field.

At this point, we have just made a local gauge-fixing that casts the Einstein equations as a manifestly hyperbolic system rather than elliptic. In particular, the hyperbolic character of $R_{\mu\nu}$, expressed in the principal symbol $P_g h_{\mu\nu} =-\frac{1}{2} \partial^\alpha \partial_\alpha h_{\mu\nu}$, is just an explicit restatement of the well-posedness of the Cauchy problem in the Einstein equation first proved by Choquet-Bruhat \cite{FouresBruhat:1952zz,Bruhat:1967,ChoquetBruhat:1969cb,Fischer:1972}.  To recover an elliptic system instead, we must impose certain symmetries.  We will do this in the following subsection, but let us now finish this subsection with a few asides.

The DeTurck vector \eqref{dT:harmonicgaugeCov} can be viewed as a global version of a local vector used to define generalised harmonic coordinates \cite{Friedrich,Garfinkle:2001ni}. Harmonic coordinates obey the homogeneous wave equation $\nabla^2x^\mu=0$, \emph{i.e.}  $g^{\alpha\nu}\Gamma^\mu_{\alpha\nu} =0$ while the generalised harmonic coordinates obey the inhomogeneous wave equation $\nabla^2x^\mu=H^\mu$  in local coordinates for some choice of $H^\mu$. Generalised harmonic coordinates are commonly employed in time-dependent problems, and initial contributions and reviews can be found in \cite{Garfinkle:2001ni,Szilagyi:2001fy,Szilagyi:2002kv,Pretorius:2004jg,Gundlach:2005eh,Pretorius:2005gq,Lindblom:2005qh,Szilagyi:2006qy,Palenzuela:2006wp,Pretorius:2007nq,Szilagyi:2009qz,Hilditch:2015aba}.

Working in the generalised harmonic gauge amounts to choosing the gauge vector $\xi^\mu = g^{\alpha\nu} \Gamma_{\alpha\nu}^{\mu} + H^{\mu}$. Locally, we can establish an equivalence between the generalised harmonic and the DeTurck gauge-fixing by setting $H^\mu = - g^{ \alpha\nu} \bar{\Gamma}_{\alpha\nu}^{\mu}$. But in the generalised harmonic formulation, $H^\mu$ is not a global vector field as it is in the DeTurck formulation. Additionally, in generalised harmonic coordinates $H$ is fixed locally while in the DeTurck formulation one fixes $\bar{\Gamma}$ instead. These are inequivalent since the metric $g$ appears in the relation between $H$ and $\bar{\Gamma}$.  With these caveats in mind, the vanishing of the DeTurck vector $\xi^\mu=0$ can be seen as a generalised harmonic gauge condition.

The choice of vector field $\xi$ \eqref{dT:harmonicgaugeCov} was first introduced in the mathematical context of Riemannian geometry to show that weakly parabolic Ricci flow \cite{Hamilton1982} is  diffeomorphic to the  strongly parabolic Ricci-DeTurck flow \cite{DeTurck1983,DeTurck2003}.  We will describe this further in Section \ref{Sec:RicciFlow}.  Physical applications of Ricci flow include string theory with the one-loop approximation to the renormalization group flow of sigma models \cite{Friedan1985}, and numerical relativity \cite{Garfinkle:2003an,Headrick:2006ti,Holzegel:2007zz,Holzegel:2007ud,Headrick:2007fk,Headrick:2009pv,Figueras:2011va}.

\subsubsection{The Einstein-DeTurck Equation \label{Sec:DeTurck2}}
	
We are now in a position to introduce the Einstein-DeTurck formulation \cite{Headrick:2009pv}, which uses the DeTurck gauge-fixing and symmetries to cast the Einstein equation in elliptic form. Following \cite{Headrick:2009pv}, let us add the covariant gauge-fixing DeTurck term $\nabla_{(\mu}\xi_{\nu)}$ to the Einstein equation \eqref{dT:Einstein} to get the {\it Einstein-DeTurck equation} or {\it harmonic equation} (hereafter, the superscript `$H$'  refers to the harmonic equation).
\begin{equation}
\label{dT:EinsteinDeTurck}
R^H_{\mu\nu}-\frac{2 \Lambda}{d-2}\, g_{\mu\nu}=0,\qquad \hbox{with} \quad R^H_{\mu\nu}\equiv R_{\mu\nu} - \nabla_{(\mu} \xi_{\nu)}\,, \qquad \xi^\alpha \equiv g^{\mu\nu} \left( \Gamma^{\alpha}_{\mu\nu} - \bar{\Gamma}^{\alpha}_{\mu\nu} \right) \,,
\end{equation}
where again $\bar\Gamma$ is the Christoffel connection for a reference metric $\bar g$. 

Now we linearise the Einstein-DeTurck equation to get the principal symbol.
\begin{equation}\label{dT:dTEinsteinLinear}
\Delta_H h_{\mu\nu} =0  \qquad \Leftrightarrow  \qquad  \Delta_R h_{\mu\nu} -\frac12\pounds_\xi h_{\mu\nu}=0\,,
\end{equation}
where $ \Delta_R h_{\mu\nu}= \Delta_L h_{\mu\nu}+ \nabla_{(\mu} v_{\nu)}$ (with $\Delta_L$ being the Lichnerowicz operator) which was introduced in \eqref{dT:EinsteinLinear}, and  ${\cal L}_\xi g_{\mu\nu} = 2\nabla_{(\mu} \xi_{\nu)}$ is the Lie derivative of the metric with respect to the DeTurck vector field $\xi$. As desired, the fundamental property of the Einstein-DeTurck equation is that the principal symbol of the associated linearised operator about a background $g$ is simply,
\begin{equation}
\label{dT:PricSymbH}
P^H_g h_{\mu\nu} = - \frac{1}{2}  \partial^\alpha \partial_\alpha h_{\mu\nu}, 
\end{equation}
which is manifestly elliptic (hyperbolic) for a Riemannian (Lorentzian) background $g$, as oppose to the principal symbol \eqref{dT:PricSymb1} of the Einstein equation.
Relevant for the programme of constructing generic gravitational solutions, we shall see that it is also manifestly elliptic for Lorentzian geometries with certain symmetries. 

The Einstein-DeTurck method involves solving the equation \eqref{dT:EinsteinDeTurck} rather than the Einstein equation \eqref{dT:Einstein}. But a solution of the Einstein-DeTurck equation \eqref{dT:EinsteinDeTurck} is only a solution of the Einstein equation if $\xi=0$.  In some situations, it is indeed possible to have solutions of \eqref{dT:EinsteinDeTurck} with $\xi\neq0$.  Such solutions are called Ricci solitons, and must be avoided when searching for numerical solutions of the Einstein equation.

Under certain symmetries and asymptotic boundary conditions, it is possible to prove mathematically that Ricci solitons do not exist.  We will discuss these cases in Section \ref{Sec:RicciSolitons}.  In practice, when the the existence of Ricci solitons cannot be ruled out, we can still solve the Einstein-DeTurck equation numerically and monitor $\xi$. If the Einstein-DeTurck equation is elliptic, then local uniqueness of solutions is guaranteed.  That is, a Ricci soliton is distinguishable from a true solution, and can be practically ruled out by verifying that $\xi=0$ to machine precision. 

We pause momentarily to comment on the way the gauge-fixing  condition $\xi^\mu = 0$ is implemented in \eqref{dT:EinsteinDeTurck}. This is not an algebraic gauge-fixing condition (this would be the case, \emph{e.g.} if we imposed specific conditions on the metric components like in conformal gauge). Instead, the DeTurck gauge-fixing is itself a differential equation for the (unknown) metric $g$ given a choice of reference background $\bar{g}$. The differential DeTurck gauge-fixing condition $\xi^\mu = 0$ is only realised after solving the full set of equations.  Indeed, the possibility of Ricci solitons means there may be solutions that violate this gauge condition.  Because the full Einstein-DeTurck equations are solved, there are $d(d+1)/2$ independent components of the metric (unless extra symmetries are assumed), rather than the usual $d(d-1)/2$ independent components in the Einstein equation where the Bianchi identity supplies $d$ extra conditions.

This means of gauge-fixing has advantages and disadvantages.  One of its advantages is that the gauge $\xi^\mu = 0$ is dependent on the reference metric.  One can therefore change the reference metric to better adapt to the solution as a means of obtaining better numerical accuracy without altering the physical solution.  On the other hand, it is sometimes difficult to control unwanted gauge artefacts that can appear.  An example of this is a pure-gauge logarithmic term that can appear in a power-series expansion off an AdS boundary.  These gauge terms may make it more difficult to resolve the solution accurately.  Additionally, since we do not impose the gauge conditions explicitly, there will be more metric functions to solve for.

The DeTurck method relies on a choice of the reference metric $\bar g$.  There is much freedom in choosing this metric, but there are still some restrictions. While $R_{\mu\nu}$ shares the same isometries as $g$, the full harmonic operator $R_{\mu\nu}^H$ has the same symmetries of $g_{\mu\nu}$ if and only if the reference metric $\bar{g}$ preserves the same symmetries and causal structure as the desired metric $g$.  Said another way, the boundary conditions on $g$ are only consistent with $\xi=0$ if $\bar g$ satisfies the same boundary conditions. The criteria for the choice of reference metric will be further elucidated in Sections \ref{Sec:DeTurck5} and \ref{Sec:seed1}.

\subsubsection{Geometries with and without Manifestly Elliptic Einstein-DeTurck Equations\label{Sec:DeTurck5}}

The Einstein-DeTurck formulation attempts to yield an elliptic formulation of the Einstein equation.  But actually, this is only possible in certain circumstances.  This should be unsurprising since the Einstein equation is intrinsically hyperbolic.  We will discuss three classes of symmetric Lorentzian ans\"atze that have so far been used to find Einstein solutions using the Einstein-DeTurck equations: stationary (and static) geometries, geometries with helical Killing vector fields, and flowing geometries.  We have been using the term `stationary' to loosely mean all of these cases, but for this section, we will refer to a more precise notion of `stationary'. We will show that the Einstein-DeTurck equations are manifestly elliptic for most stationary geometries, possibly elliptic for geometries with helical Killing vector fields, and mixed elliptic-hyperbolic for flowing solutions.

\vspace{0.2cm}
{\underline{ 1. Stationary Geometries}}
\vspace{0.2cm}

In its restricted sense, a spacetime geometry is stationary if it admits a Killing vector field that is timelike everywhere in the asymptotic region.  This class includes static spacetimes (which admit global timelike Killing vectors), rotating geometries, and boosted black branes. 

Many stationary black holes are governed by the rigidity theorems \cite{Hawking:1971vc, Hawking:1971vc,HawkingEllis:1973,SudarskyW:1992,Chrusciel:1993cv,Friedrich:1998wq,Hollands:2006rj,Moncrief:2008mr} (see subsections \ref{subsec:bhtheorems} and \ref{subsec:rigidity}). If there is a Killing field $\partial_t$ that is not normal to the horizon, then compact, non-extremal black holes must have at least one rotational Killing vector field $\Omega_H^{(\alpha)}\mathcal{R}_{(\alpha)}$, where $\mathcal{R}_{(\alpha)}$ are asymptotic rotational Killing fields and the constants $\Omega_H^{(\alpha)}$ are the horizon angular velocities. One can form the Killing field $\partial_t+\Omega_H^{(\alpha)}\mathcal{R}_{(\alpha)}$, which serves as a horizon generator.  This ultimately guarantees that any rotation or motion of the horizon is generated by an isometry and does not radiate.

Though black branes are not compact, they still often admit Killing horizons, and these $\mathcal{R}_{(\alpha)}$ Killing fields describe asymptotic translation symmetries.  It is also possible to have more than one disconnected Killing horizons, $\mathcal{H}_1 , \ldots , \mathcal{H}_k$. In this case each component is a Killing horizon generated by a linear combination of ${\cal T}$ and ${\cal R}_{(\alpha)}$, $K_{\mathcal{H}_k} = {\cal T} + \Omega_{H_k}^{(\alpha)} {\cal R}_{(\alpha)}$ for some constants $\Omega_{H_k}^{(\alpha)}$ (that can differ for each horizon).

These considerations motivated by rigidity suggest that it is appropriate to write the line element in a coordinate frame that is adapted to the isometries of the stationary system. Let us assume that $\partial_t$ and $\partial_{z^a}$ are commuting Killing vector fields. Define  $z^A = \{ t, z^a \}$ to be the coordinates associated to the Killing fields  $\partial_t$ and $\partial_{z^a}$. For rotational $\partial_{z^a}$, we normalise the associated compact orbit to have period to $2\pi$ and thus $z^a$ is periodic with $z^a \sim z^a + 2 \pi$. Consider the following class of Lorentzian stationary line elements
\begin{equation}
\label{dT:StationaryAnsatz}
\mathrm{d}{s}^2 = {g}_{\mu\nu} \mathrm{d}X^\mu \mathrm{d}X^\nu = {G}_{AB}(x) \left[ \mathrm{d}z^A + {A}^A_i(x) \mathrm{d}x^i \right] \left[\mathrm{d}z^B +{A}^B_j(x) \mathrm{d}x^j \right] +{h}_{ij}(x) \mathrm{d}x^i \mathrm{d}x^j\;,
\end{equation}
where $G_{AB}$ is Lorentzian and $h_{ij}$ is Euclidean.  While this is a rather general form, all known stationary solutions so far have an extra discrete symmetry that would imply $A^A_i=0$.  We will not need to assume this symmetry to prove ellipticity, but it considerably simplifies the numerical problem.

Next we have to choose our reference metric.  This metric must take the same form as above, so we write
\begin{equation}
\label{dT:StationaryAnsatzRef}
\mathrm{d}{s}^2 = \bar{g}_{\mu\nu} \mathrm{d}X^\mu \mathrm{d}X^\nu = \bar{G}_{AB}(x) \left[ \mathrm{d}z^A + \bar{A}^A_i(x) \mathrm{d}x^i \right] \left[ \mathrm{d}z^B +\bar{A}^B_j(x) \mathrm{d}x^j \right] +\bar{h}_{ij}(x) \mathrm{d}x^i \mathrm{d}x^j\;.
\end{equation}

With these considerations we are in position to finally prove that the Einstein-DeTurck equation for the symmetric ansatz \eqref{dT:StationaryAnsatz} is manifestly elliptic. 
First note that the principal symbol \eqref{dT:PricSymbH}  of the Lorentzian Einstein-DeTurck operator reads 
\begin{eqnarray}\label{dT:PricSymbStationary}
P^H_g g_{AB} & = & -\frac{1}{2} g^{\alpha\beta} \partial_ \alpha \partial_\beta {g}_{AB}  = -\frac{1}{2} h^{mn} \partial_m\partial_n {G}_{AB} \,, \nonumber \\
P^H_g g_{ij} & = & -\frac{1}{2} g^{\alpha\beta} \partial_ \alpha \partial_\beta {g}_{\mu\nu}  = -\frac{1}{2} h^{mn} \partial_m \partial_n {h}_{ij} \,.
\end{eqnarray}
Moreover, in going from \eqref{dT:PricSymbH} to \eqref{dT:PricSymbStationary} for the ansatz \eqref{dT:StationaryAnsatz}, we used the fact that $g_{\mu\nu}$ is independent of the isometric directions $z^A = \{ t, z^a \}$ (\emph{i.e.} $\partial_{z^A} g_{\mu\nu}=0$).
Consequently, the character of the Einstein-DeTurck equation for the stationary ansatz \eqref{dT:StationaryAnsatz} is controlled only by the metric $h^{ij}$. Though $g$ is Lorentzian, the metric  $g^{ij}=h^{ij}$ is Euclidean.   This means that  $h^{ij}$ is positive definite and thus it follows from \eqref{dT:PricSymbStationary} that the Einstein-DeTurck equation is manifestly elliptic. 

\vspace{0.2cm}
{\underline{ 2. Geometries with a Helical Killing Vector Field}}
\vspace{0.2cm}

As we have mentioned above, the rigidity theorems state that the existence of a Killing field $\partial_t$ that is not normal to a compact, non-extremal, bifurcate horizon implies the existence of a additional Killing field $\partial_{\varphi}$.  The combination $\partial_t+\Omega_H \partial_\varphi$ is then null at the horizon and is therefore the horizon-generating Killing field.  However, the theorem still allows for the existence of horizon-generating Killing field $K=\partial_t+\Omega_H \partial_\varphi$ such that $\partial_t$ and $\partial_\varphi$ are not Killing fields. It is then conceivable that there are black holes with Killing horizons, but only a single Killing field. This Killing field is not usually timelike at asymptotic infinity, so these are not, strictly speaking, stationary solutions. Rather, they are time-periodic.  Such black holes indeed arise as rotating black holes in global AdS$_4$.  These solutions were recently constructed by numerically solving boundary value problems, and were coined {\it black resonators}.

We also point out that there are also smooth, horizonless solutions with a helical horizon-generating Killing field.  The zero-size limit of black resonators are {\it geons}, which have this property.  Like black resonators, geons are time-periodic.

An ansatz for such a solution can be written in a similar way to \eqref{dT:StationaryAnsatz}. 
\begin{eqnarray}
\label{dT:HelicalAnsatz0}
\mathrm{d}{s}^2 = {G}_{AB}(t,\varphi,x) \left[ \mathrm{d}z^A + {A}^A_i(t,\varphi,x) \mathrm{d}x^i \right] \left[ \mathrm{d}z^B +{A}^B_j(t,\varphi,x) \mathrm{d}x^j \right] +{h}_{ij}(t,\varphi,x)\mathrm{d}x^i \mathrm{d}x^j  \nonumber \\\;.
\end{eqnarray}
where  $z^A=\{t,z^a\}=\{t,\varphi\}$.  We can write a reference metric in a similar form.  An important difference between \eqref{dT:StationaryAnsatz} and  \eqref{dT:HelicalAnsatz0} is that in the latter, $\partial_t$ and $\partial_\varphi$ are not Killing, so all the metric components depend also on $t$ and $\varphi$.  

Now we need an ansatz for a geometry with a helical Killing field. To exploit the fact that $K=\partial_t+\Omega_H \partial_\varphi$ is a Killing vector, perform the change of variables: 
\begin{eqnarray}
\mathrm{d}\tau = \mathrm{d}t\,, \qquad \mathrm{d}\phi=\mathrm{d}\varphi+\Omega_H \mathrm{d}t\,.
\end{eqnarray}
In these new coordinates, we have $K=\partial_\tau$. An ansatz that accommodates the helical isometry generated by $K=\partial_\tau$ of the solution is 
\begin{eqnarray}
\label{dT:HelicalAnsatz}
\mathrm{d}{s}^2 &=& {g}_{\mu\nu} \mathrm{d}X^\mu \mathrm{d}X^\nu \nonumber\\
&=& {\cal N}(\phi,x) \left[ \mathrm{d}\tau + {A}_i(\phi,x) \mathrm{d}x^i \right]^2 +{h}_{ij}(\phi,x) \mathrm{d}x^i \mathrm{d}x^j\,.
\end{eqnarray}
A reference metric can be put in a similar form. But note that since the Killing field $K=\partial_\tau$ is not a globally timelike, we cannot guarantee that ${\cal N}(\phi,x)$ is everywhere positive and finite. Accordingly, we cannot assume that ${\rm det}\, h_{ij}>0$, \emph{i.e.} we cannot assume that $({\cal M}, h)$ is a smooth Riemannian manifold with Euclidean metric signature. Consequently, we cannot straightforwardly prove that the Einstein-DeTurck equations for the helical ansatz \eqref{dT:HelicalAnsatz} yield a manifestly elliptic system of equations. 

But without theory, it is still possible to rely on practice.  One can still solve the Einstein-DeTurck equation, and {\it \`a posteriori} verify that the equations of motion are elliptic in a neighbourhood of the solution.  If this is the case, we have confirmed that we have solved a well-posed boundary value problem and we further rely on local uniqueness to eliminate the possibility of a Ricci soliton. This strategy was used in \cite{Horowitz:2014hja,Dias:2015rxy} to find the geons and black resonators.

\vspace{0.2cm}
{\underline{ 3. Flowing Geometries with Non-Killing Horizons}}
\vspace{0.2cm}

The assumptions of the rigidity theorems can be evaded in another way: by having black holes with non-compact horizons. In this case, it is possible to have black holes with non-Killing horizons.  These solutions are regular in the future horizon but not the past horizon, and are sometimes called {\it flowing} geometries.  By being non-compact, these horizons extend to some asymptotic regions, where boundary conditions can be imposed.  If one has two such asymptotic regions, these boundary conditions can necessitate a solution with a non-Killing horizon.  That is, the boundary conditions demand that the horizon must be non-rigid.  For example, one can impose that the black hole has two different temperatures in each of these regions, or impose different horizon velocities.  

Such geometries have been constructed using the Einstein-DeTurck method \cite{Figueras:2012rb,Fischetti:2012vt}. Unfortunately, the Einstein-DeTurck equations reduce to a mixed hyperbolic-elliptic PDE system.  In spite of this, the DeTurck vector $\xi$ was verified to vanish to machine precision.  This suggest that the results in \cite{Figueras:2012rb,Fischetti:2012vt} are valid solutions of the Einstein equation, but we do not (yet) have the added guarantee of local uniqueness of solutions.  Nevertheless, these examples demonstrate the potentially far-reaching utility of the Einstein-DeTurck method.

\subsubsection{Einstein-DeTurck Equation in the Presence of Matter Fields\label{Sec:DeTurck6} }

Our discussion thus far has focused on the vacuum Einstein-DeTurck equation, possibly with a cosmological constant.  Let us now generalise this discussion to matter fields.  Consider the action
\begin{equation}
S = \int_\mathcal{M} \mathrm{d}^{d}x\sqrt{-g}\left[R-2\Lambda+\mathcal L_m\right]\,,
\end{equation}
for some matter Lagrangian $\mathcal L_m$, which yields the standard Einstein equation
\be
R_{\mu\nu}-\frac{R}{2}g_{\mu\nu}+\Lambda g_{\mu\nu}= T_{\mu\nu}\,,
\ee
for a matter stress tensor $T_{\mu\nu}$, along with some additional equations of motion for the matter fields.  To get the Einstein-DeTurck equations with matter, we must write this in trace-reversed form.  That is, we take the trace of this equation, solve with respect to $R$ and substitute back in.  The result is
\be
R_{\mu\nu}=\frac{2\Lambda}{d-2}g_{\mu\nu}+T_{\mu\nu}-\frac{1}{d-2}T g_{\mu\nu}\;.
\ee
Now to fix the principal symbol for the metric, we add the DeTurck term, that gives
\be
R_{\mu\nu}-\nabla_{(\mu}\xi_{\nu)}=\frac{2\Lambda}{d-2}g_{\mu\nu}+T_{\mu\nu}-\frac{1}{d-2}T g_{\mu\nu}\;,
\ee
which is the Einstein-DeTurck equation with matter.  The principal symbol of this equation is the same as before.  

But we would like to have the same principal symbol in the equations of motion for the matter fields as well.  This is already guaranteed for scalar fields
\be
\mathcal L_{\Phi}=-2 \nabla_\mu \Phi \nabla^\mu\Phi-4 V(\Phi)\;,
\ee
which gives the equation of motion 
\begin{equation}
\Box \Phi-V^\prime(\Phi)=0\,.
\end{equation}
This should be unsurprising since scalar fields contain no gauge freedom. This procedure also holds for complex scalar fields.

Now let us consider Maxwell fields, where the extra gauge freedom will spoil the principal symbol. The Lagrangian is
\begin{equation}
\mathcal L_{EM} =-F^{\mu\nu}F_{\mu\nu}\;,
\end{equation}
where $F = \mathrm{d}A$ and $A$ is the one-form potential. The Maxwell equation reads
\begin{equation}
\nabla_\rho F^{\rho\phantom{\mu}}_{\phantom{\rho}\mu}=0\,.
\label{eq:maxwell1}
\end{equation}
Note that in the absence of torsion, $F_{\mu\nu} = 2 \partial_{[\mu}A_{\nu]} =  2 \nabla_{[\mu}A_{\nu]}$ and so (\ref{eq:maxwell1}) reduces to
\begin{equation}
\Box A_{\mu} - \nabla_\mu \nabla_\rho A^{\rho}-R_{\mu\rho}A^{\rho}=0\,,
\label{eq:maxwella}
\end{equation}
where we have used the Ricci identities for vector fields. The first term has the desired form for the principal symbol, but the second term spoils the principal symbol, since it also involves second derivatives. In order to solve this, we add to the Maxwell equation (\ref{eq:maxwell1}) the following covariant gauge fixing term
\begin{equation}
\nabla_\rho F^{\rho\phantom{\mu}}_{\phantom{\rho}\mu}-\nabla_\mu \chi=0\,,
\label{eq:maxwell2}
\end{equation}
where we choose $\chi = \nabla_\rho A^\rho -\nabla_\rho \bar{A}^\rho$, and $\bar{A}$ is a reference one-form gauge field of our choice, that must satisfy the same asymptotic boundary conditions and regularity conditions as the gauge field we wish to find. This procedure was first outlined in \cite{Rozali:2013fna}. Note that since we are dealing with torsion-free spacetimes, (\ref{eq:maxwell2}) automatically implies
\begin{equation}
\Box \chi = 0\,.\nonumber
\end{equation}
This is a consequence of the following mathematical identity $\nabla_\mu \nabla_\nu F^{\mu\nu}\equiv0$, which itself follows from the Ricci identities for rank two tensors. With our choice of $\chi$, (\ref{eq:maxwell2}) can be solely expressed in terms of $A_\mu$ and $\bar A_\mu$:
\begin{equation}
\Box A_{\mu} -\left(\frac{2\Lambda}{d-2}g_{\mu\rho}+2 F_{\mu\phantom{\lambda}}^{\phantom{\mu}\lambda}F_{\rho\lambda}-\frac{g_{\mu\rho}}{d-2}F^{\lambda\beta}F_{\lambda\beta}\right)A^{\rho}=-\nabla_\mu \nabla_\rho \bar{A}^{\rho}\,,
\end{equation}
where we see that the principal symbol of this equation is governed by $g^{\mu\nu}\partial_\mu \partial_\nu$, as desired. Note also that we have substituted $R_{\mu\nu}$ from the Einstein equation (without DeTurck term). The reason for this is simple: after this substitution, the only second order differential operator in this equation is only acting on $A_\mu$. After the addition of this novel term to the Maxwell equation, we can restore the DeTurck term in the Einstein-DeTurck equation in the usual way. The final system of equations is then given by
\begin{subequations}
\begin{equation}
R_{\mu\nu}-\nabla_{(\mu}\xi_{\nu)} = \frac{2\Lambda}{d-2}g_{\mu\nu}+2 F_{\mu\phantom{\rho}}^{\phantom{\mu}\rho}F_{\nu\rho}-\frac{g_{\mu\nu}}{d-2}F^{\lambda\rho}F_{\lambda\rho}\,,
\label{eq:ricci2}
\end{equation}
\begin{equation}
\Box A_{\mu} -\left(\frac{2\Lambda}{d-2}g_{\mu\rho}+2 F_{\mu\phantom{\lambda}}^{\phantom{\mu}\lambda}F_{\rho\lambda}-\frac{g_{\mu\rho}}{d-2}F^{\lambda\beta}F_{\lambda\beta}\right)A^{\rho}=-\nabla_\mu \nabla_\rho \bar{A}^{\rho}\,.
\label{eq:maxwell4}
\end{equation}
\end{subequations}
These equations simplify in some special cases. If the spacetime is stationary with respect to a Killing vector field $\kappa$, and we seek solutions for which the only component of the Maxwell field that is non-vanishing is $\kappa^\rho A_\rho$, then it is trivial to show that $\nabla_\rho A^\rho$ is always zero. In this case, we can freely set $\bar{A}=0$ and the situation is similar to the scalar field case.

One might wonder why this simple procedure for the gauge field works in general. The reason has to do with the fact that the Einstein equation (minimally coupled to any type of matter that is not fluid-like) is quasi-linear. As such, we can envisage adding gauge fixing terms that are linear in the corresponding gauge transformations, and these should always render the equations governed solely by the usual principal symbol $g^{\mu\nu}\partial_\mu \partial_\nu$.

In the language of differential forms, and for a generic $p-$form, say $F_{(p)} = \mathrm{d}C_{(p-1)}$, the gauge fixing term above generalises to
\begin{equation}
\star \mathrm{d} \star F_{(p)}-\mathrm{d}P_{(p-2)}=0\,,
\end{equation}
where $P_{(p-2)} = \star \mathrm{d}\star C_{(p-1)}-\star \mathrm{d}\star \bar{C}_{(p-1)}$ and $\bar{C}_{(p-1)}$ is a reference form field of our choice.

\subsection{Non-existence of Ricci Solitons \label{Sec:RicciSolitons}}

So far we have seen how to deform the Einstein equation, possibly coupled to matter, into the Einstein-DeTurck equation. The hope is that solutions to the Einstein-DeTurck equation necessarily coincide with those of the Einstein equation.  That is, we would like to show that Ricci solitons (solutions with $\xi\neq0$) do not exist.  While this is not possible to prove in general, if the Einstein-DeTurck equation is elliptic, then there are local uniqueness theorems that guarantee that solutions with $\xi\neq0$ are distinguishable from $\xi=0$.  That is, Ricci solitons cannot be arbitrarily close to Einstein solutions, except for a measure zero set in moduli space.  Therefore, Ricci solitons can be practically ruled out (on a case by case basis) for elliptic Einstein-DeTurck equations by verifying that $\xi^\mu \xi_\mu=0$ to machine precision.

There are, however, certain exceptional circumstances where it is possible to prove that Ricci solitons do not exist, so any Einstein-DeTurck solution is also an Einstein solution.  We will only consider the cases of static geometries, \emph{i.e.} geometries that admit a global everywhere timelike Killing field $\mathcal{T}$. Furthermore, we introduce coordinates $z^\mu = \{ t, z^a \}$ adapted to the Killing field $\mathcal{T}=\partial_t$. Note that since we want to focus on static solutions, we also demand that the line element must be invariant under the discrete symmetry $t\to-t$. This means we can Euclideanise our line element, by setting $t = -i\tau$, in which case our $d$-dimensional metric becomes manifestly Euclidean. It is in this context that the non-existence proof of Ricci-solitons is best understood.

We start with the Einstein equation coupled to a conserved stress energy tensor
\begin{equation}
R_{\mu\nu} -\frac{R}{2}g_{\mu\nu} = T_{\mu\nu}\,.
\end{equation}
As in the previous section, we move to the trace-reversed version of the Einstein equation
\begin{equation}
R_{\mu\nu}= T_{\mu\nu} - \frac{g_{\mu\nu}}{d-2}T\equiv \widetilde{T}_{\mu\nu}\,.
\end{equation}
This is the form of the Einstein equation that we augment by the covariant DeTurck term $\nabla_{(\mu}\xi_{\nu)}$. The Einstein-DeTurck equation becomes
\begin{equation}
R_{\mu\nu}-\nabla_{(\mu}\xi_{\nu)}= \widetilde{T}_{\mu\nu}\,.
\label{eq:general}
\end{equation}
$\xi$ does not contain additional degrees of freedom because of the contracted Bianchi identities.  Taking the divergence of (\ref{eq:general}), gives
\begin{equation}
\nabla^\mu R_{\mu\nu}-\frac{1}{2}\Box \xi_\nu-\frac{1}{2}R_{\nu\mu}\xi^\mu-\frac{1}{2}\nabla_\nu \nabla_\mu \xi^\mu = -\frac{1}{d-2} \nabla_\nu T\,,
\end{equation}
where we have used that $T_{\mu\nu}$ is covariantly conserved. We now note that by taking the trace of (\ref{eq:general}) we can express $\nabla_\mu \xi^\mu$ as a function of $R$ and $T$ only. Substituting back into the previous equation, yields the following equation
\begin{equation}
\Box \xi_\nu+R_{\nu\mu}\xi^\mu=0\,,
\label{eq:xi}
\end{equation}
where we have used the contracted Bianchi identity. 

In the case of a static geometry, $\xi$ only has nontrivial components in the $a$ index (\emph{i.e.} the non-$t$ components of $\mu$), and so that $\lambda = \xi^\mu \xi_\mu = \xi^a \xi_a$ is positive definite. Using (\ref{eq:xi}), one can obtain the following scalar partial differential equation
\begin{equation}
\Box \lambda + \xi^\mu \nabla_\mu \lambda = -2 \widetilde{T}_{\mu\nu} \xi^\mu \xi^\nu+2\nabla^\mu \xi^\nu\nabla_\mu \xi_\nu\,.
\label{eq:sacred}
\end{equation}
Any non-trival Ricci soliton must have $\lambda \neq0$ and, since after Euclideanisation we are working in Riemannian geometry with positive Euclidean signature, we also have $\lambda = 0 \Leftrightarrow \xi = 0$. 

Now consider the cases for which the right hand side of (\ref{eq:sacred}) is non-negative. This occurs, for instance, if $T_{\mu\nu} = 2 \Lambda/(d-2)g_{\mu\nu}$ and $\Lambda\leq0$, \emph{i.e.} a non-positive cosmological constant.  Under this assumption, if a solution $\phi$ of (\ref{eq:sacred}) exists, then there must be a solution $f>0$ to
\begin{equation}
\Box f + \xi^\mu \nabla_\mu f \geq0
\label{eq:aux}
\end{equation}
on a fixed background $(\mathcal{M},g,\xi)$. Note that the converse is not necessarily true. That is to say, the existence of solutions to (\ref{eq:aux}) does not imply the existence of solutions to (\ref{eq:sacred}). 

When the right hand side of (\ref{eq:aux}) is positive, such an equation admits a local maximum principle\footnote{See for instance \cite{Figueras:2011va} and references therein.}, which states the following: (1) $f$ can only attain a maximum at the boundary of $\mathcal{M}$; and (2) at the maximum, $\partial_n f>0$ for outward-pointing normal $\partial_n$. But since $f>0$, and $f$ can only reach a maximum at $\partial \mathcal{M}$, it suffices to show that $f$ must be zero at the boundaries, to show that $f$ is everywhere zero.  This completes the proof that, in the static case for which the right hand side of (\ref{eq:sacred}) is positive definite, no Ricci solitons exist if appropriate boundary conditions are given at the boundary of $\mathcal{M}$. In the next sections, we will show how such boundary conditions can sometimes be given such that $\xi$ is zero at all asymptotic boundaries, and thus is zero everywhere.

\section{Boundary Conditions \label{Sec:BCs}}

Having a well-posed boundary value problem also requires properly imposing boundary conditions.  Here, we discuss the allowable boundary conditions in the Einstein-DeTurck equation \eqref{dT:EinsteinDeTurck}, some of which can also be applied to finding zero modes of linear perturbations \eqref{eq:linearjorge}.  These boundary conditions will be different from those arising from time-dependent linear perturbations. For those, we refer the reader to  \cite{Press:1973zz,Teukolsky:1974yv,Dias:2010maa,Dias:2014eua,Santos:2015iua} for asymptotically flat backgrounds and to \cite{Kovtun:2005ev,Friess:2006kw,Michalogiorgakis:2006jc,Dias:2009ex,Dias:2013sdc} for asymptotically AdS backgrounds.

Boundary conditions are typically imposed on a coordinate hyperslice, say at $Z=0$.  In a neighbourhood of this slice, the metric takes the form 
\begin{equation}\label{dT:bdrymetric}
\mathrm{d}s^2 = n^2 \mathrm{d}Z^2 + \gamma_{ij} (\mathrm{d}x^i + \alpha^i \mathrm{d}Z) ( \mathrm{d}x^j + \alpha^j \mathrm{d}Z). 
\end{equation}
Without imposing extra symmetries, the Einstein-DeTurck equation \eqref{dT:EinsteinDeTurck} has $d(d+1)/2$ independent components and $d(d+1)/2$ unknown metric functions.  We must therefore have $d(d+1)/2$ boundary conditions. This line element \eqref{dT:bdrymetric} invites us to fix $d(d-1)/2$ of these boundary conditions by imposing conditions on either  the induced metric $\gamma_{ij}{\bigl |}_{Z=0}$, or the extrinsic curvature $K_{ij} = \frac{1}{2 n} \left( \partial_Z \gamma_{ij} - 2 \nabla_{(i} \alpha_{j)} \right){\bigl |}_{Z=0}$.  The remaining $d$ boundary conditions should come from agreement with the Einstein equation, which requires  $\xi^\mu=0$, so it is natural to draw these remaining boundary conditions from $\xi^\mu=0{\bigl |}_{Z=0}$.  For linear perturbations, a similar set of boundary conditions can be drawn from the de-Donder gauge conditions or the transverse-traceless conditions.  

Now let us discuss boundary conditions more specifically.  The integration domain typically extends to some singular point in the line element, which is usually a coordinate singularity.  Boundary conditions are imposed at these singular points.  First, we distinguish {\it asymptotic} boundaries, {\it asymptotic extremal} boundaries, and {\it fictitious}  boundaries.  As the name suggests, asymptotic boundaries or asymptotic extremal boundaries are those where the proper distance from any other point is infinite. Boundary conditions at asymptotic infinity are typically chosen so that the asymptotic structure is preserved. Extremal horizons are more subtle, and we will address these later in this section.

On the other hand, fictitious boundaries are a finite proper distance from other points.  Typical examples of fictitious boundaries are an axis and a non-degenerate horizon. They are useful to fully exploit the symmetries of the problem at hand, but require boundary conditions that enforce regularity.  That is, that there exists a coordinate chart where the fictitious boundary is manifestly regular.  These may be coordinates similar to the familiar Cartesian coordinates or Eddington-Finkelstein coordinates. 

Besides the distinction between asymptotic and fictitious boundaries, let us also distinguish between a {\it defining} boundary condition and a {\it derived} boundary condition. For any (well-posed) second-order differential equation, an expansion off a hyperslice (say $Z=0$) would typically yield two unknown coefficients, say $A(x)$ and $B(x)$.  All remaining terms in the expansion are determined if $A(x)$ and $B(x)$ are known.  Together, $A(x)$ and $B(x)$ give two functional degrees of freedom, one of which must be fixed by a boundary condition, which we call a {\it defining} boundary condition.  Once a defining boundary condition is imposed, any other conditions derived from the equations of motion and this defining boundary condition are {\it derived} boundary conditions.

Let us illustrate this concept with a simple example. Consider the simple ODE
\be
f''(z)+f(z)=0\,.
\ee
Expanding this equation in a power series about $z=0$, we find that $f(0)$ and $f'(0)$ are undetermined, with all remaining terms in the power series determined by these two.  Let us impose the boundary condition $f(0)=0$, which is a defining boundary condition because it fixes one of the two undetermined coefficients.  The equation of motion and this boundary condition imply $f''(0)=0$, which is a derived boundary condition.  

This concept is useful when performing function redefinitions.  For example, the defining boundary condition $f(0)=0$ above allows the function redefinition $f(z)=z g(z)$, which automatically satisfies this condition so long as $g$ remains finite, which will always be  true when performing numerics.  This means that it is possible to impose defining boundary condition at our convenience (to improve numerical accuracy) merely by defining functions in a certain way.  If we continue to expand the equations of motion, $g'(0)=0$ is a derived boundary condition, and $g(0)$ is left undetermined until the full system is solved.  Often in numerical computations, function redefinitions of this kind are performed, and derived boundary conditions are imposed.  

From a mathematical point of view, imposing defining boundary conditions is all that is necessary for a well-posed boundary value problem.  If one imposes defining boundary conditions by an appropriate definition of functions, it is not necessary to further impose derived boundary conditions, since they are usually a direct consequence of the equations of motion.  Nevertheless, these derived conditions are usually imposed anyway to improve numerical accuracy.  Furthermore, it is a check of well-posedness to perform a series expansion and find only one undetermined coefficient.  

For the Einstein-DeTurck method, we have both a metric $g$ and reference metric $\bar g$.  As we have mentioned, $\bar g$ must have the same symmetries and causal structure as $g$ in order to be compatible with $\xi=0$.  In particular, this means that $\bar g$ must have the desired asymptotic boundaries, as well as regular ficticious boundaries.  

Since $\bar g$ is usually held fixed, the singular points in $\bar g$ will affect the series expansion for the metric $g$.  For fictitious boundaries, $\bar g$ must be regular, which implies that some coordinate transformation is known where the line element is manifestly regular (or, at a minimum that the metric is invertible). This coordinate transformation can then be used to obtain regularity conditions on $g$.  The series expansion in $g$ typically contains two undetermined coefficients, one of which causes $g$ to be non-invertible under the same coordinate transformation that makes $\bar g$ manifestly invertible.  Regularity demands that this non-invertible coefficient vanishes.  In other words, the defining boundary condition for regularity is for the metric $g$ to be invertible under the same coordinate transformation that makes $\bar g$ invertible.  Smoothness, a more stringent condition than invertibility, is typically enforced by the equations of motion as derived boundary conditions.

For the remainder of this section, we will discuss more specific cases and how their boundary conditions are imposed. For the cases discussed in subsections \ref{Sec:BCsH}, \ref{Sec:BCsAxis} and \ref{Sec:BCext}, the reader might also benefit from the exposition given in \cite{Adam:2011dn}.

\subsection{Asymptotic Boundaries \label{Sec:BCs1}}

As explained in Section \ref{Sec:DeTurck2} we must choose the reference background $\bar{g}$ to be such that it preserves the same symmetries and causal structure as the desired geometry $g$. In particular, this means that such a reference geometry shares the same asymptotic boundary as $g$. For example, if we desire $g$ to be an asymptotically flat metric, then $\bar g$ must also be asymptotically flat.  One typically imposes a Dirichlet type boundary condition here, so that $g$ matches $\bar g$,  
 \begin{equation}\label{dT:BCasymp}
g{\bigl |}_{\partial M}=\bar{g}{\bigl |}_{\partial M}\,.
\end{equation}

\subsection{Non-Extremal Killing Horizons \label{Sec:BCsH}}

Suppose we seek the boundary conditions at a non-extremal bifurcate Killing horizon generated by the Killing field $K = \partial_t + \Omega_H^{(a)} \partial_{z_{(a)}}$, where $\partial_t$ is a Killing vector field that is asymptotically timelike and $\partial_{z_{(a)}}$ are rotational Killing vector fields. $K$ has isometry group $\mathbb{R}$ with a fixed point at the bifurcation surface and its orbits close on the future and past horizons. This bifurcation surface is a fictitious boundary and regularity conditions could be determined by moving to Eddington-Finkelstein coordinates.  Alternatively, one could also Euclideanise the metric under a Wick rotation and demand regularity in Cartesian coordinates, as we will soon demonstrate.

As we have mentioned, the procedure is to find a coordinate transformation where $\bar g$ is invertible, then apply the same coordinate transformation to $g$ and demand that it too is invertible. Here, we will go slightly farther and obtain one of the derived boundary conditions.

Begin by writing the metric ansatz $g$ and reference background $\bar{g}$ as in \eqref{dT:StationaryAnsatz} and \eqref{dT:StationaryAnsatzRef}, respectively. Next, we want to write these line elements in a frame that is adapted to the helical isometry of the horizon generator.  This is achieved with a Wick-rotation (Euclideanisation) of a rescaled time coordinate and with the introduction of new azimuthal coordinates $\psi^a$,
\begin{eqnarray}\label{dT:shiftHorizon}
\tau = i \,\kappa \, t\,, \qquad \psi^a=z^a-\Omega_H^{(a)} t\,,
\end{eqnarray}
such that $\tau \sim \tau+2\pi$ and $ \psi^a\sim  \psi^a+2\pi$. These coordinates are adapted to the Killing symmetry of the horizon because the horizon generator now reads simply $K=\partial_\tau$. If we require the norm of $K$ at the asymptotic boundary to be $1$ we further find that the constant $\kappa$ is the surface gravity of the horizon, related to its temperature by $\kappa=2\pi T_H$. 

We can always complete our coordinate chart by introducing a radial coordinate $\rho$ such that the horizon boundary is at $\rho=0$, \emph{i.e.} $|K|_{\rho=0}=0$. This also means that $x^i=\{\rho,x^{\hat{\imath}} \}$. A Taylor expansion (to lowest order) of the reference background \eqref{dT:StationaryAnsatzRef} around the horizon boundary $\rho=0$ can always be written as 
\begin{eqnarray}
\label{dT:HorizonExpansion}
\dd s^2 &=& \bar{B}\, \dd \rho^2 +\bar{A} \, \rho^2 \dd\tau^2 +  \bar{C}\, \rho^3  \dd \rho\,\dd \tau+  \bar{F}_{\hat{\imath}} \, \rho  \,\dd\rho \,\dd x^{\hat{\imath}} +  \bar{G}_{\hat{\imath}} \,\rho^2\,\dd \tau\,\dd x^{\hat{\imath}} +  \bar{f}_a\,  \rho\,  \dd \rho \,\dd \psi^a + \bar{g}_a \, \rho^2 \dd \tau \,\dd \psi^a  
\nonumber\\
&&  + \bar{h}_{ \hat{\imath} \hat{\jmath}} \dd {x}^{\hat{\imath}} \dd {x}^{ \hat{\jmath}}  + \bar{G}_{ab} \left( \dd\psi^a + \bar{A}^{a}_{\hat{\imath}} \dd x^{\hat{\imath}} \right) \left( \dd \psi^b +\bar{A}^b_{\hat{\jmath}} \dd x^{\hat{\jmath}} \right)\;,
\end{eqnarray}
where the reference metric functions $\bar{A},\bar{B},\bar{C},\bar{F}_{\hat{\imath}},\bar{G}_{\hat{\imath}},\bar{f}_a,\bar{g}_a,\bar{h}_{ \hat{\imath} \hat{\jmath}},\bar{A}^{a}_{\hat{\imath}}$ and $\bar{G}_{ab}$  are independent of $\tau$. The metric takes a similar form with bars removed from the functions.  We have factored-out certain powers of $\rho$ in some of the metric components for reasons that will be clear soon. 

Now, we need a coordinate transformation that ensures this line element is regular.  Of course, such a transformation would depend on the form of the various metric functions.  A particularly simple form would be for the reference metric functions to be smooth functions of $\rho^2$, with $\bar A=\bar B$ on the horizon.  It is always possible for a regular reference metric with a non-extremal horizon to be brought in such a form.  In this case, we can move to a Cartesian coordinate chart  with
\begin{equation}\label{dT:HorizonExpansion1}
X=\rho \cos \tau\, , \qquad Y=\rho \sin \tau\,.
\end{equation}
Now introduce manifestly regular and smooth 1-forms
\begin{equation}\label{dT:HorizonExpansion2}
E^{\rho}=\rho \,\dd\rho =X \,\dd X+Y\,\dd Y \, , \qquad E^{\tau}=\rho^2 \dd\tau =X \,\dd Y-Y\,\dd X.
\end{equation}
The expansion \eqref{dT:HorizonExpansion} about the horizon can now be written as
\begin{eqnarray}
\label{dT:HorizonExpansion3}
\dd s^2 &=& \bar{B} \left(\dd \rho^2 +\frac{ \bar{A}}{ \bar{B}} \, \rho^2 \dd \tau^2\right) +   \bar{C}\, E^{\rho} E^{\tau} +   \bar{F}_{\hat{\imath}}\,\dd x^{\hat{\imath}} E^{\rho} +  \bar{G}_{\hat{\imath}}\,\dd x^{\hat{\imath}}E^{\tau} +   \bar{f}_a\, \dd \psi^a E^{\rho} +  \bar{g}_a \,\dd \psi^a  E^{\tau}
\nonumber\\
&&  +  \bar{h}_{ \hat{\imath} \hat{\jmath}} \dd {x}^{\hat{\imath}} \dd {x}^{ \hat{\jmath}}  +  \bar{G}_{ab} \left( \dd\psi^a +  \bar{A}^{a}_{\hat{\imath}} \dd x^{\hat{\imath}} \right) \left( \dd\psi^b + \bar{A}^b_{\hat{\jmath}} \dd x^{\hat{\jmath}} \right)\;,
\end{eqnarray}
which is manifestly regular except for possibly the first two terms.  But since $\bar A=\bar B$ at $\rho=0$, the first two terms become $\bar B(\mathrm{d}X^2+\mathrm{d}Y^2)$, plus additional higher order terms that go as $\rho^2=X^2+Y^2$, which is regular. 

Next, we must impose boundary conditions that guarantee that the metric $g$ is regular at the horizon.  Suppose the metric is written in the form \eqref{dT:HorizonExpansion} (with bars removed).  Also suppose that the reference metric satisfies the conditions we mentioned earlier: $\bar A=\bar B$ and reference metric functions being smooth functions of $\rho^2$. Then one can show that
\begin{equation}\label{dT:horizonBCs}
\hspace{-0.5cm}A{\bigl |}_{\rho=0}=B{\bigl |}_{\rho=0} \, , \qquad \{\partial_\rho A, \partial_\rho B,\partial_\rho C,\partial_\rho F_{\hat{\imath}},\partial_\rho G_{\hat{\imath}},\partial_\rho f_a,\partial_\rho g_a,\partial_\rho h_{ \hat{\imath} \hat{\jmath}},\partial_\rho G_{ab} \}{\bigl |}_{\rho=0}=0
\end{equation}
are derived boundary conditions on the metric that assures regularity.  These can be confirmed by expanding the equations of motion about the horizon in a polynomial series expansion. In taking this series expansion, we have already discarded independent solutions that diverge (typically, but not necessarily, logarithmic ones), and where demanding that these diverging terms vanish is the defining boundary condition. The boundary conditions \eqref{dT:horizonBCs} then follow from the equations of motion as derived boundary conditions.

Note that we have a Neumann condition for every metric function along with $A=B$ at $\rho=0$.  Thus we have obtained one more (derived) boundary condition than there are metric functions. One is free to impose any independent combination of these since these are derived boundary conditions.  Our experience has shown that the choice here makes little difference in the end. 

One might arrive at different boundary conditions with a different coordinate choice.  But those will amount to reworking the discussion above in a different set of coordinates.  It is, however, still necessary to choose the reference background $\bar{g}$ to contain a horizon that is generated by the same Killing field as that of $g$, and thus have the same temperature and angular velocities.  

\subsection{Axes of Symmetry \label{Sec:BCsAxis}}

Assume now that the solution we seek is axisymmetric, \emph{i.e.} it is periodic in a coordinate $\phi$. The axis of symmetry, say $\rho=0$, occurs when the $U(1)$ symmetry generated by the Killing vector field $\partial_\phi$ has a fixed point, $|\partial_\phi |_{\rho=0}=0$, and this is a fictitious boundary.  The most familiar version of this is perhaps the origin of polar coordinates.  

We will present the boundary conditions for the fixed point of a $U(1)$ symmetry, but it generalises for that of an $SO(n)$ symmetry via a projection to a $U(1)$ subgroup.  

Boundary conditions at an axis of symmetry are similar to that of a Euclideanised horizon. For a reference metric with a regular axis, it is always possible to bring the reference metric into the form 
\begin{eqnarray}
\label{dT:PolarAxis}
\dd s^2 &=& \bar{B}\, \dd\rho^2 +\bar{A} \, \rho^2 \dd \phi^2 +  \bar{C}\, \rho^3 \dd\rho \,\dd \phi+  \bar{F}_{\hat{\imath}} \, \rho  \,\dd\rho\, \dd {x}^{\hat{\imath}} + \rho^2 \bar{G}_{\hat{\imath}} \dd \phi \,\dd {x}^{\hat{\imath}} +  \bar{f}_{\hat{A}}\,  \rho\,  \dd\rho \,\dd z^{\hat{A}} + \bar{g}_{\hat{A}} \, \rho^2 \dd\phi \,\dd z^{\hat{A}}  
\nonumber\\
&&  + \bar{h}_{ \hat{\imath} \hat{\jmath}} \dd {x}^{\hat{\imath}} \dd {x}^{ \hat{\jmath}}  + \bar{G}_{ \hat{A} \hat{B} }\left( \dd z^{\hat{A}} + \bar{A}^{\hat{A}}_{\hat{\imath}} \dd x^{\hat{\imath}} \right) \left( \dd z^{\hat{B}} +\bar{A}^{\hat{B}}_{\hat{\jmath}}\dd x^{\hat{\jmath}} \right),
\end{eqnarray}
where the metric functions $\bar{A},\bar{B},\bar{C},\bar{F}_{\hat{\imath}},\bar{G}_{\hat{\imath}},\bar{f}_a,\bar{g}_a,\bar{h}_{ \hat{\imath} \hat{\jmath}},\bar{A}^{a}_{\hat{\imath}}$ and $\bar{G}_{ab}$ are independent of $\phi$, which is a periodic coordinate with period $2\pi$. All of the reference metric functions are smooth functions of $\rho ^2$, and $\bar A=\bar B$ at $\rho=0$. 

To demonstrate that this line element is regular, we move to a Cartesian coordinate chart $\{X,Y, z^{\hat{A}},x^{\hat{\imath}}\}$ defined via 
\begin{equation}\label{dT:PolarAxis1}
X=\rho \cos \phi \, , \qquad Y=\rho \sin \phi\,, 
\end{equation}
and define the smooth and regular 1-forms 
\begin{equation}\label{dT:PolarAxis2}
E^{\rho}=\rho \,\dd\rho =X \,\dd X+Y\,\dd Y \, , \qquad E^{\phi}=\rho^2 \dd\phi =X \,\dd Y-Y\,\dd X,
\end{equation}
which allow us to rewrite \eqref{dT:PolarAxis} as
\begin{eqnarray}
\label{dT:PolarAxis3}
\dd s^2 &=& \bar{B} \left(\dd\rho^2 +\frac{\bar{A}}{\bar{B}} \, \rho^2 \dd \phi^2\right) +  \bar{C}\, E^{\rho} E^{\phi} +  \bar{F}_{\hat{\imath}}  \dd {x}^{\hat{\imath}} E^{\rho} + \bar{G}_{\hat{\imath}}  \dd {x}^{\hat{\imath}} E^{\phi} +  \bar{f}_{\hat{A}} \dd z^{\hat{A}} E^{\rho} + \bar{g}_{\hat{A}} \dd z^{\hat{A}}  E^{\phi}  \nonumber\\
&&  + \bar{h}_{ \hat{\imath} \hat{\jmath}} \dd {x}^{\hat{\imath}} \dd {x}^{ \hat{\jmath}} 
+ \bar{G}_{ \hat{A} \hat{B} } \left( \dd z^{\hat{A}} + \bar{A}^{\hat{A}}_{\hat{\imath}} \dd x^{\hat{\imath}} \right) \left( \dd z^{\hat{B}} +\bar{A}^{\hat{B}}_{\hat{\jmath}} \dd x^{\hat{\jmath}} \right)\;,
\end{eqnarray}
which is guaranteed to be regular by the requirements on the reference metric functions.  

Now, writing the metric $g$ in the same form as $\bar g$, but with bars removed, we find the derived boundary conditions
\begin{equation}\label{dT:axesBCs}
\hspace{-0.5cm}A{\bigl |}_{\rho=0}=B{\bigl |}_{\rho=0} \, , \qquad \{\partial_\rho A,\partial_\rho B,\partial_\rho C,\partial_\rho F_{\hat{\imath}},\partial_\rho G_{\hat{\imath}},\partial_\rho f_{\hat{A}},\partial_\rho g_{\hat{A}},\partial_\rho h_{ \hat{\imath} \hat{\jmath}},\partial_\rho G_{\hat{A} \hat{B}} \}{\bigl |}_{\rho=0}=0\,.
\end{equation}

Let us also note that these boundary conditions can be used even if there is a conical singularity.  Since one can freely rescale the angular coordinate $\phi\to\alpha\phi$ without affecting the equations of motion and derived boundary conditions, these singularities can always be removed locally around any $U(1)$ axis.  Local boundary conditions of this sort are therefore unaware of the existence of any conical singularity. 

\subsection{Extremal Killing Horizons \label{Sec:BCext}}

Extremal horizons are amongst the most relevant types of horizons in applications of AdS/CFT to condensed matter systems. These describe ground states of the theory, and possibly novel types of matter.  In some cases, they exhibit universal criticality, which on the gravity side manifests itself through the existence of uniqueness theorems of geometries near regular extremal horizons \cite{Kunduri:2007vf,Figueras:2008qh,Kunduri:2008rs,Kunduri:2008tk,Kunduri:2009ud,Kunduri:2013ana}.

However, unlike non-extremal horizons, there is no known universal method for handling boundary conditions of all extremal horizons.  For now, let us restrict ourselves to smooth, static, simply connected, extremal horizons.

For such horizons, ingoing Eddington-Finkelstein coordinates can be found.  Furthermore, there is a Killing vector field $\partial_t$ that is timelike outside the horizon (as is the case for static black holes). One can then show that a coordinate chart $(t,\rho,x^a)$ with $a  = 1,\ldots,d-2$ can be found such that \cite{Figueras:2011va}
\begin{equation}
\mathrm{d}s^2 = -T(\rho,x)\rho^2 \mathrm{d}t^2 +R(\rho,x)\left[\frac{\mathrm{d}\rho}{\rho}+\rho\,\omega_a(\rho,x) \mathrm{d}x^a\right]^2+\gamma_{ab}(\rho,x)\mathrm{d}x^a\mathrm{d}x^b\,.
\label{eq:extgeneral}
\end{equation}
for $\rho>0$, where $T,\,R>0$ are smooth functions of $\rho$ and $x$ near $\rho = 0$ such that
\begin{equation}
\lim_{\rho\to0} T(\rho,x) = T_0(x)\,,\quad \lim_{\rho\to0}R(\rho,x) = T_0(x)\,,\quad \lim_{\rho\to0}\gamma_{ab}(\rho,x) = \gamma_{ab}^0(x) \quad \text{and}\quad\lim_{\rho\to0}\omega_a(\rho,x) <+\infty\,.
\end{equation}
In addition, one can also deduce relations between $\left.\partial_\rho T\right|_{\rho=0}$ and $\left.\partial_\rho R\right|_{\rho=0}$, but they will not be useful in what follows. In general, both these quantities are non-vanishing. The general line element (\ref{eq:extgeneral}) admits a scaling limit, in which we set $\rho = \epsilon \hat{\rho}$ and $t = \hat{t}/\epsilon$ as then take $\epsilon\to0$. This is the so-called near-horizon limit. The resulting line element
\begin{equation}
\mathrm{d}s^2_0 = T_0(x)\left(-\hat{\rho}^2 \mathrm{d}\hat{t}^2 +\frac{\mathrm{d}\hat{\rho}^2}{\hat{\rho}^2}\right)+\gamma^0_{ab}(x)\mathrm{d}x^a\mathrm{d}x^b\,,
\label{eq:extgeneral0}
\end{equation}
is itself a solution of the Einstein equation and is often called the near-horizon limit of an extremal black hole. To actually solve for $T_0(x)$ and $\gamma^0(x)$, one has to input the above line element in the Einstein equation, and determine the corresponding solutions. In general, the smooth solution to these near-horizon equations depends on a number of real parameters, which cannot be fixed via any local calculation. Let us represent these by $C^i$. For instance, the near-horizon geometry of an extremal Reissner-Nordstr\"om black hole in global AdS depends on the total charge of the black hole measure in units of the AdS radius, but that number cannot be fixed just by considering the near-horizon geometry.

In a few special cases with sufficient foresight, these $C^i$ are known explicitly.  In this case, one could choose a reference metric with the desired near-horizon geometry as \eqref{eq:extgeneral0} and impose a Dirichlet condition. (This would fix $\xi^\mu\xi_\mu = 0$ at $\rho = 0$; see \cite{Figueras:2011va} for more details.)  This, for instance, has been used in \cite{Figueras:2011va,Horowitz:2014gva}.  

In general, we do not know what these constants $C^i$ are.  One way to proceed is to consider a family of Dirichlet boundary conditions parametrised by these $C^i$, then vary these constants until a solution is found.  A simpler method which as been successfully applied in \cite{Horowitz:2014gva} is to perform the coordinate transformation $\rho \to y^2$.  With these coordinates, the extremal horizon is located at $y=0$, and $\mathrm{d}t^2$ now vanishes as $y^4$. Since a regular solution admits a power series expansion in $\rho$ \cite{Figueras:2011va}, this means that the first nontrivial term in the expansion in $y$ will be proportional to $y^2$, and so a simpler boundary condition can be imposed at the horizon, namely $\left.\partial_y T \right|_{y=0}= 0$, and similarly for the remaining metric functions. We still have $T(0,x)\propto R(0,x)$ as an additional regularity condition (the constant of proportionality is dictated by the reference metric).

Note that in the $y$ coordinates, the boundary conditions are the same as those of a static, non-extremal horizon written in the form \eqref{dT:HorizonExpansion}. Indeed, in some cases, one can obtain extremal horizons by taking the finite-temperature solutions with these boundary conditions, and parametrically reducing the temperature to zero.  

But extremal horizons are fundamentally different from non-extremal horizons.  Recall that an expansion of a PDE about a hyperslice should yield two free coefficients that are functions of the transverse directions. A defining boundary condition fixes one of these, leaving a full function free. This is what happens in all of the non-extremal boundary conditions we have seen so far.  The fact that we are instead left with \emph{constants} $C^i$ might seem strange from a PDE standpoint.  Indeed, this is an indication that demanding regularity is an over-constraining boundary condition for extremal horizons. In fact, demanding regularity for extremal horizons might be too restrictive from a physical standpoint as well.  There are many examples where the zero-temperature limit of regular horizons is singular (see for example \cite{Horowitz:2009ij,Horowitz:2002ym,Horowitz:2014gva,Hickling:2015ooa}). Unfortunately, we do not know how to impose that an extremal horizon is the limit of a regular finite temperature horizon.  Worse, we do not even know a general form for the series expansion about an extremal horizon that yields the two free coefficients.  

\subsection{Non-Killing Horizons\label{Sec:BCnonKil}}
As discussed in Section \eqref{Sec:DeTurck5}, we might be interested in flowing geometries that have a {\it non}-Killing horizon, \emph{i.e.} a horizon that is regular in the future horizon $\mathcal{H}^+$  but not in the past horizon $\mathcal{H}^-$. All solutions that have been constructed in this class of geometries are stationary, in the sense that an asymptotically timelike Killing vector $\mathcal T$ field exists.

Introduce coordinates that are adapted to this Killing field, \emph{i.e} consider a coordinate chart $\{t,x^a\}$ where $\mathcal{T}=\partial_t$. In such coordinates, the most general line element compatible with the above reads
\begin{equation}
\mathrm{d}s^2 = -T(x) \mathrm{d}t^2 +\gamma_{ab}(x) [\mathrm{d}x^a+\omega^a(x) \mathrm{d}t][\mathrm{d}x^b+\omega^b(x) \mathrm{d}t]\,.
\label{eq:generalflow}
\end{equation}
The reference metric takes a similar form, but with $T$ and $\gamma_{ab}$ replaced by $\bar{T}$ and $\bar{\gamma}_{ab}$, respectively. Note that the principal symbol of the Einstein-DeTurck equation associated to this line element is controlled by $g^{ab}$ which is in general not a positive definite matrix, and so the Einstein-DeTurck equations for this class of geometries is not necessarily elliptic. In fact, it can be shown that in flowing geometries, $\omega^a(x)$ is non-zero on the horizon, yielding a $g^{ab}$ that is necessarily non-positive. The Einstein-DeTurck is thus of the mixed elliptic-hyperbolic type for this class of line elements (see subsection \ref{Sec:DeTurck5}). In general, little is know about such systems of equations, and it does not come as a surprise that the boundary conditions are poorly understood in this case as well.

There are two inequivalent methods that seem to work, in the sense that the components of the DeTuck vector (after the calculation is done) vanishes to machine precision. Note that without elliptic equations, we do not have the added guarantee of local uniqueness to distinguish Einstein solutions from Ricci solitons.  We will describe these methods briefly and refer the readers to \cite{Figueras:2012rb} and \cite{Fischetti:2012vt} for more details.

One method \cite{Figueras:2012rb} is to work in ingoing Eddington-Finkelstein coordinates and impose a fictitious boundary condition in the interior of the horizon. But since the horizon is now an output of the computation and is not known \emph{a priori}, the integration domain must be chosen to be sufficiently large to cover the horizon. 

Another method \cite{Fischetti:2012vt} is to write a reference metric for which a coordinate transformation to regular ingoing Eddington-Finkelstein coordinates is known. This coordinate transformation is then applied to the general metric ansatz \eqref{eq:generalflow} to derive a set of regularity conditions that are imposed as boundary conditions.  In this method, the integration domain stops at the future horizon, and the Eddington-Finkelstein coordinates are used only to determine boundary conditions.

\subsection{Non-Symmetric Axes\label{Sec:BCnonaxisym}}
In certain circumstances, an axis (say, the fixed point of some rotation $\partial_\varphi$) is not an axis of symmetry (\emph{i.e.}, $\partial_\varphi$ is not a Killing field). For the sake of presentation we restrict ourselves to codimension-$2$ (polar-like coordinates). The procedure here can be straightforwardly generalised to higher codimensions.  Let us place such an axis at the coordinate $\rho=0$ and introduce a coordinate chart $\{\rho,\varphi,x^a\}$, with $a=1,\ldots,d-2$. Under these assumptions, a coordinate system can be found such that
\begin{multline}
\mathrm{d}s^2 = G_{ab}(x,\rho,\varphi)\mathrm{d}x^a\mathrm{d}x^b+M(x,\rho,\varphi)\left[\mathrm{d}\rho+\omega_a(x,\rho,\varphi)\mathrm{d}x^a\right]^2+\\
S(x,\rho,\varphi)\,\rho^2\,\left[\mathrm{d}\varphi+\kappa_a(x,\rho,\varphi) \mathrm{d}x^a+\frac{\eta(x,\rho,\varphi)\mathrm{d}\rho}{\rho}\right]^2\,.
\end{multline}
If the axis is regular, then it must permit some general polynomial expansion in some Cartesian coordinate system around $\rho = 0$. Since we are focusing on a codimension-$2$ axis, we will take these cartesian coordinates to be labeled by $X_1$ and $X_2$. That is to say, the line element above close to $\rho = 0$ can always be brought to the following simple form
\begin{multline}
\mathrm{d}s^2 = \tilde{G}_{ab}(x,X_1,X_2)\mathrm{d}x^a\mathrm{d}x^b+\Phi_1(x,X_1,X_2)\left[\mathrm{d}X_1+\tilde{\omega}_a(x,X_1,X_2)\mathrm{d}x^a\right]^2+\\
\Phi_2(x,X_1,X_2)\,\left[\mathrm{d}X_2+\tilde{\kappa}_a(x,X_1,X_2) \mathrm{d}x^a+\tilde{\eta}(x,X_1,X_2) \mathrm{d}X_1\right]^2\,.
\label{eq:expansion}
\end{multline}
where all functions of $\{x,X_1,X_2\}$ are smooth functions around $X_1=X_2 =0$.

The line element (\ref{eq:expansion}) is not singular at any point, which means that we should be able to obtain regularity conditions at $\rho=0$, by equating both line elements and determining the coordinate transformation between $\{\rho,\varphi\}$ in a perturbative expansion around $X_1$, $X_2\,\sim0$. That is to say, one needs to find the coordinate transformation
\begin{equation}
X_i = \sum_{n=0}^{+\infty}F^n_i(x,\varphi)\rho^n\quad\text{for}\quad i\in\{1,2\}\,.
\label{eq:expansion}
\end{equation}

In general, this is a difficult and tedious task, but it can be simplified with an appropriate choice of reference metric. Choose a reference metric where the transformation between both line elements is simple, at least to second order in $\rho$. One then uses this transformation to construct the aforementioned coordinate map and determine regularity. Note that the coordinate transformation that one gets from the reference metric is not the one used to determined regularity of the actual metric we want to find. In terms of the expansion (\ref{eq:expansion}), only the $n=0$ terms are fixed by the reference metric.
Let us consider for example a situation where the reference metric is chosen to be
\begin{equation}
\mathrm{d}s^2 = \bar{G}_{ab}(x)\mathrm{d}x^a\mathrm{d}x^b+\bar{S}(x)(\mathrm{d}\rho^2+\rho^2\,\mathrm{d}\varphi^2)+\mathcal{O}(\rho^3)\,.
\end{equation}

In this case, the regularity conditions are simply
\begin{align}
&\left.\partial_\rho G_{ab}\right|_{\rho=0}=0\,,\quad \left.\partial_\rho M\right|_{\rho=0}=0\,,\quad \left.\partial_\rho S\right|_{\rho=0}=0\,,\quad \left.\partial_\rho \omega_a\right|_{\rho=0}=0 \nonumber
\\
&\left.\partial_\rho \eta\right|_{\rho=0}=0\,,\quad \left.\partial_\rho \kappa\right|_{\rho=0}=0\,.
\end{align}

In the special case where the line element is static, it is easy to show that these boundary conditions, together with the above reference metric, imply $\left.\partial_\rho (\xi^\mu \xi_\mu)\right|_{\rho=0}=0$.

\subsection{Boundary Conditions and the DeTurck Vector \label{Sec:BCs4}}

We still need to address whether the boundary conditions that we imposed above for the metric components are consistent with the requirement that Einstein solutions must have DeTurck vector field $\xi=0$.

For that, note that the Bianchi identity for the Einstein-DeTurck equation imply that $\xi$ obeys an elliptic linear second order differential equation,
\begin{equation}\label{dT:eqxi}
\nabla^\mu R^H_{\mu\nu} - \frac12\partial_\nu R^H 
=-\frac12\left(\nabla^2\xi_\nu + {R_\nu}^\mu\xi_\mu\right) = 0\,.
\end{equation}
We must confirm that the boundary conditions applied to $\xi$ are such that \eqref{dT:eqxi} is a well-posed elliptic PDE for $\xi^\mu$ and consistent with $\xi=0$ as required to get an Einstein solution.

Let us first confirm that this is the case for the asymptotic boundary. 
It follows from the definition \eqref{dT:harmonicgaugeCov} of the DeTurck vector field that the BCs for the metric \eqref{dT:BCasymp} at the asymptotic boundary impose 
\begin{equation}\label{dT:asympBCxi}
\xi^\mu|_{\partial M}=0
\end{equation}
which is indeed consistent with $\xi=0$. If we further assume the spacetime is static and has a negative cosmological constant, this implies that (provided this is the only asymptotic end) that the DeTurck vector is zero in the interior.

Consider now a fictitious boundary located at $\rho=0$, \emph{e.g.} a horizon or an axes of symmetry.  In these cases, we can explicitly check that the horizon boundary conditions \eqref{dT:horizonBCs} and the axes boundary conditions \eqref{dT:axesBCs} individually imply:
\begin{eqnarray}\label{dT:fictitiousBCxi}
&&\xi^\rho  {\bigl |}_{\rho = 0} = 0 \, , \quad \partial_\rho \xi^{\tilde{i}}  {\bigl |}_{\rho = 0}  = 0 \, , \quad \partial_\rho \xi^A  {\bigl |}_{\rho = 0} = 0\;.
\end{eqnarray}
In these cases, the Neumann boundary conditions for the tangential components of $\xi^\mu$ do not imply that $\xi$ must vanish but are certainly compatible with $\xi=0$. Moreover, in the static case and for a negative cosmological constant, we can see from $\left.\partial_\rho (\xi^\mu \xi_\mu)\right|_{\rho=0}=0$. That fictitious boundaries cannot be where the maximum of $\xi^\mu \xi_\mu$ is located, since the maximum principle dictates $\left.\partial_\rho(\xi^\mu \xi_\mu)\right|_{\rho=0}>0$.

\subsection{Boundary Conditions for Matter Fields\label{Sec:BCmatter}}

Let us finish this section by discussing boundary conditions for matter fields.  We will restrict ourselves to scalar fields and Maxwell fields.  These boundary conditions apply generally and are not specific to the Einstein-DeTurck method.

\subsubsection{Asymptotic Boundary \label{Sec:BCmatterI}}

Consider first a nonlinear solution with a real or complex Klein-Gordon scalar field $\Phi$ with mass $\mu$ and charge $q$.  Let $t$ be an asymptotic time coordinate, $r$ a radial coordinate, and $x$ the remaining asymptotic coordinates.  For complex solutions, there is a Fourier decomposition $\Phi\sim\Psi e^{-i\omega t}$, with possibly other spatial wavenumbers.  

The asymptotic behaviour of this scalar field must asymptotically solve the Klein Gordon equation in AdS or Minkowski backgrounds. As usual, we have two independent solutions and we must choose boundary conditions that ensure the solution is normalisable (has finite energy). We must treat the asymptotically AdS and flat cases separately. 

For asymptotically flat backgrounds, defined as
\begin{equation}
\mathrm{d}s^2 = -\mathrm{d}t^2+\mathrm{d}r^2+r^2\,\mathrm{d}\Omega_{d-3}^2+\mathcal{O}(r^{-\delta})
\end{equation}
with $\delta>0$, the asymptotic behaviour of the scalar field must solve the Klein-Gordon equation. For zero modes and the nonlinear problem with a stationary scalar cloud, the frequency $\omega$ is a real number satisfying $\omega\leq\mu$ so that bound states are trapped by the massive potential barrier\footnote{At linear order this frequency is a normal mode of massive bound states in Kerr, which is then corrected at higher order in perturbation theory until we get the full nonlinear result.}. In this case, asymptotically we have the general solution
\begin{equation} \label{BCm:flatdecay}
 \Psi(r,x)\simeq   A_{-}(x)\, r^{c_{-}} \,e^{-\sqrt{\mu^2-\omega^2}\,r} +A_{+}(x) \,r^{c_{+}} \,e^{\sqrt{\mu^2-\omega^2}\,r}+\ldots\,,\qquad\text{with}\qquad c_{\pm}=c_{\pm}(d,\mu,\omega)\;,
 \end{equation}
and we eliminate the divergent term by choosing the boundary condition,
 \begin{equation}\label{BCm:flatdecay1}
A_{+}=0.
 \end{equation}

For asymptotically AdS backgrounds, and in Fefferman-Graham coordinates $\{z,x\}$ (see appendix \ref{appendix:computecharges}), one instead has the behaviour \cite{Breitenlohner:1982jf,Mezincescu:1984ev}
\begin{equation}
\label{BCm:AdSdecay}
 \Phi(r,x)\simeq
A_{+}(x) \, z^{\Delta_+}+A_{-}(x) \, z^{\Delta_-}+\cdots\,, \quad
 \hbox{with} \quad \Delta_\pm=\frac{d-1}{2}\pm
 \sqrt{\frac{(d-1)^2}{4}+\mu^2 L^2}\;,
 \end{equation}
 where $L$ is the AdS length scale. Stability of the AdS background (\emph{i.e.} the demand that energy is finite) requires $\Delta_\pm$ to be real. The mass of the scalar field must then obey the Breitenl\"ohner-Freedman (BF) bound \cite{Breitenlohner:1982jf,Mezincescu:1984ev}
 \begin{equation}
\label{BCm::AdSdecay1}
\mu^2 \ge \mu^2_{\mathrm{BF}}\equiv -\frac{(d-1)^2}{4 L^2}.
 \end{equation}
Another special value is the unitarity bound 
 \begin{equation}
\label{BCm::AdSdecay2}
 \mu^2_{\mathrm{unit}}\equiv \mu^2_{\mathrm{BF}} +1/L^2.
\end{equation}
We have to distinguish three windows for the scalar mass: $\mu^2 \geq \mu^2_{\mathrm{unit}}$, $\mu^2_{\mathrm{BF}} < \mu^2 < \mu^2_{\mathrm{unit}}$, and $\mu^2 = \mu^2_{\mathrm{BF}}$.

For scalars with $\mu^2 \geq \mu^2_{\mathrm{unit}}$, the mode $A_{+}$ with faster fall-off is normalisable and $A_{-}$ is the non-normalisable mode. It is therefore customary to choose $A_{-}=0$.  According to the AdS/CFT dictionary, the coefficient $A_{+}(x)$ is then proportional to the expectation value $\langle {\cal O}(x)\rangle$ of the boundary operator $\mathcal{O}$ that has dimension $\Delta_+$.

For scalars with $\mu^2_{\mathrm{BF}} < \mu^2 < \mu^2_{\mathrm{unit}}$ both modes in \eqref{BCm:AdSdecay} are normalisable and one has more freedom to choose normalisable boundary conditions. For example, one can impose either the standard boundary condition $A_{+}=0 $, or the alternative boundary condition $A_{-}=0$. In the AdS/CFT correspondence, this choice dictates whether the operator ${\cal O}$ dual to $\Phi$ has dimension $\Delta_+$ or $\Delta_-$ and in both cases they are not sourced, \emph{i.e.} the boundary theory is not deformed \cite{Klebanov:1999tb}. These two choices are the only ones that respect the AdS symmetries at large radius \cite{Hertog:2004rz}.

One can also impose the mixed condition $A_{+}=\kappa A_{-}$ which is also known as double-trace boundary condition since it corresponds to a deformation of the dual theory by adding the term $-\kappa \int d^{d-1} x\;{\cal O}^{\dagger}{\cal O}$ to its action $S_{\mathrm{bdry}}$ \cite{Witten:2001ua,Sever:2002fk}. If $\kappa < 0$, a bulk `positive energy theorem' under this double trace boundary condition was proved in \cite{Faulkner:2010fh}. 

Yet another possibility is to impose inhomogeneous boundary conditions, such as $A_{-}=A_0 \cos(k x)$, for some constant $A_0$ and $k$. Of course, one can also impose more general functions $A_{-}=A(x)$.  This yields the expectation value $\langle {\cal O}(x)\rangle \propto A_{+}(x)$. 

To summarise, we can choose the boundary conditions
\begin{equation}
\label{BCm::AdSdecay4}
\mu^2_{\mathrm{BF}} < \mu^2 < \mu^2_{\mathrm{unit}}: \quad
\left\{
\begin{array}{l}
A_{+}=0, \quad \hbox{standard} \\
A_{-}=0, \quad \hbox{alternative} \\
A_{+}=\kappa A_{-}, \quad \hbox{double-trace} \\
A_{-}=A(x), \quad \hbox{inhomogeneous} 
\end{array}
\right.
\end{equation}

Precisely at the Breitenl\"ohner-Freedman bound, $\mu^2 = \mu^2_{\mathrm{BF}}$, the asymptotic behaviour \eqref{BCm:AdSdecay} does not hold because one of the appearance of logarithmic terms,
 \begin{equation}
\label{BCm:AdSdecay5}
 \Phi(z,x)\simeq
z^{\Delta_{\mathrm{BF}}}\left[ A_{+}(x) +A_{-}(x) \, \log\,z  \right]\,.  \end{equation}
 This logarithmic term is a non-normalizable mode that causes AdS to be unstable \cite{Hubeny:2004cn} so we must impose the boundary condition 
\begin{equation}
\label{BCm::AdSdecay6}
A_{-}=0\;,\qquad (\text{for}\;\; \mu^2 = \mu^2_{\mathrm{BF}} )\;.  
\end{equation}
This concludes our discussion of asymptotic boundary conditions for scalar fields.

A Maxwell gauge field in AdS asymptotically takes the form:
  \begin{equation}
\label{BCm:AdSdecayMax}
 A_\mu(z,x)\simeq  A_{(0)\mu}(x)+ J_\mu(x)z^{d-3}+\cdots\;,
 \end{equation}
 where $A_{(0)\mu}(x)$ is the boundary gauge potential, $J^\mu(x)$ is a boundary current and the expansion above is only valid in the radial gauge, \emph{i.e.} $A_z = 0$. Here, the index $\mu$ only runs through coordinates on the boundary metric.  For example, the time component $A_{(0)t}=\mu_A$ is the chemical potential, and $J_t$ is related to the charge density $\rho_A$. The possible choice of boundary conditions for Maxwell fields is very similar to those of scalar fields with $\mu^2_{\mathrm{BF}}<\mu^2<\mu^2_{\mathrm{unit}}$.

\subsubsection{Fictitious Boundaries \label{Sec:BCmatterH}}
At fictitious boundaries, matter fields must be manifestly regular in the same coordinates where the metric is manifestly regular.  For gauge fields, this refers to the field strength tensor $F=\mathrm{d}A$, rather than the gauge potential.

\section{Numerical Algorithms to Solve the Gravitational Equations \label{Sec:relaxation}}

\subsection{The Newton-Raphson Algorithm}\label{subsec:newtonraphson}

In subsections \ref{Sec:DeTurck} and \ref{Sec:BCs}, we have completed the formulation of our boundary value problem to find gravitational solutions. Assuming we have a proper reference metric and boundary conditions, the Einstein-DeTurck equations will yield a set of nonlinear boundary value PDEs to be solved by some numerical method.  The most commonly used method for solving nonlinear boundary value problems is Newton-Raphson.

Let us therefore begin with a rudimentary introduction to Newton-Raphson.  In its most basic form, Newton-Raphson is a root-finding algorithm.  Let us begin by solving a one-dimensional root problem.  Given a function $f$ and its derivative $f'$, find an $x_s$ such that $f(x_s)=0$.  The procedure begins with a guess $x_0$.  Near $x_0$, the function behaves like 
\be
f(x)=f(x_0)+f'(x_0)(x-x_0)+O((x-x_0)^2)\;.
\ee
If $x_0$ is sufficiently close to a root, we can attempt to get closer by finding the root of the Taylor series above, truncated to linear order.  Then 
\be\label{newton1d}
x_1=x_0-\frac{f(x_0)}{f'(x_0)}
\ee
should be closer to a root than $x_0$.  We can continue to iterate this process until we reach the desired accuracy.  

Let us analyze the speed of convergence for this method.  Let $x_s$ be the true root $f$, and let the error after the $n$-th step be $\epsilon_n=x_s-x_n$.  Then a Taylor series about $x_n$ gives
\be
0=f(x_s)=f(x_n+\epsilon_n)=f(x_n)+\epsilon_n f'(x_n)+\frac{\epsilon_n^2}{2}f''(x_n)+O(\epsilon_n^3)\;.
\ee
Solving for $f(x_n)$ and then dividing by $f'(x_n)$ gives
\be
\frac{f(x_n)}{f'(x_n)}=-\epsilon_n-\frac{\epsilon_n^2}{2}\frac{f''(x_n)}{f'(x_n)}+O(\epsilon_n^3)\;.
\ee
Then from the Newton-Raphson method,
\be
\epsilon_{n+1}=x_s-x_{n+1}=x_s-\left[x_n-\frac{f(x_n)}{f'(x_n)}\right]=-\frac{\epsilon_n^2}{2}\frac{f''(x_n)}{f'(x_n)}+O(\epsilon_n^3)\;.
\ee
Therefore, Newton-Raphson converges quadratically.  This means, roughly, that near a root the number of significant digits will double at each step.  This rapid convergence makes Newton-Raphson a powerful method for finding roots.  

Now we can generalise Newton-Raphson to higher dimensions.  We wish to find a root of the function $F:\R^k\rightarrow\R^k$.  In general, this is a difficult problem and cannot be solved without sufficient insight.  Unlike the root of a single one-dimensional function, there are also very few numerical methods available to accomplish this task.  Following the same procedure as before, we expand a Taylor series about the vector $x_n$:
\be
F(x)=F(x_n)+J_F(x_n)\cdot(x-x_n)+O((x-x_n)^2)\;,
\ee
where $J_F$ is the Jacobian of $F$.  Then the higher dimensional analogue of \eqref{newton1d} should be rewritten as
\be
J_F(x_n)\cdot(x_{n+1}-x_n)=-F(x_n)\;,
\ee
where now $x_n\in\R^k$.  Rather than invert the matrix $J_F(x_n)$, it is usually more efficient and accurate to solve the linear system of equations for $x_{n+1}-x_n$.  There are many standard and efficient algorithms for solving a linear system (such as LU decomposition). 

From here, we move to a functional version of Newton-Raphson that is useful for PDEs. We begin with some set of differential equations $E_i[x,f_1,\ldots,f_N]=0$, where $i=1,\ldots N$ and $f_i$ are functions of $x$, which can stand for any number of coordinates. Here, $E_i$ can be a function of the $f$'s as well as their derivatives.   Expanding about a particular set of functions $f_j^{(n)}$ to linear order gives us
\be
E_i[x,f_1,\ldots,f_N]=E_i[x, f_1^{(n)},\ldots, f_N^{(n)}]+\frac{\delta E_i}{\delta f_j}[x,f_1^{(n)},\ldots, f_N^{(n)}]\delta f_j+O(\delta f^2)\;,
\ee
where $\delta f_j=f_j-f_j^{(n)}$.  Note that here, $\frac{\delta E_i}{\delta f_j}$ is a (second-order) differential operator on $\delta f_j$.  The whole expression $\frac{\delta E_i}{\delta f_j}\delta f_j$ can be computed by setting $f_j\rightarrow f_j^{(n)}+\epsilon \delta f_j^{(n)}$ in $E_i$, differentiating with respect to $\epsilon$, then setting $\epsilon\rightarrow0$.  Boundary conditions (of the form $B_i=0$) can be treated in a similar way.  The only difference is that $B_i$ are not full functions of $x$, but evaluated at the boundaries of the integration domain.  

If the set of $f_j^{(n)}$ is sufficiently close to a solution $E[x,f_1,\ldots,f_N]=0$, we can write, approximately,
\be\label{functionalnewton}
\frac{\delta E_i}{\delta f_j}[x,f_1^{(n)},\ldots, f_N^{(n)}]\delta f_j=E_i[x, f_1^{(n)},\ldots, f_N^{(n)}]\;.
\ee
Newton-Raphson then amounts to taking a seed $f_1^{(0)},\ldots,f_N^{(0)}$ and solving the above linear equation (subject to the similarly linearised boundary conditions) for the $\delta f_j$'s and then setting $f_j^{(n)}=f_j^{(n-1)}+\delta f_j$, repeating the process as necessary.  These linear equations may be solved using standard PDE methods.  See appendix \ref{appendix:collocation} for an introduction to collocation methods which can be used to solve these systems.

There are a number general difficulties with Newton-Raphson.  The first difficulty may be in the evaluation of the root derivative (\emph{i.e.} $f'$, the Jacobian $J_F$, or the functional Jacobian $\frac{\delta E}{\delta f}$).  In certain cases, computing this derivative may be overly costly or inaccurate.  For boundary value problems, $E$ is usually known explicitly, and functional derivatives can be taken analytically. 

The other issue is that the convergence of the solution depends critically on the initial guess (or seed).  If the seed is not carefully chosen, the iterations can diverge, or possibly enter an infinite cycle.  The starting points that allow convergence (the basin of attraction) can be extremely intricate, even in the simplest of equations.  In fact, the basin of attraction for an equation can define a fractal set (these are called Newton fractals).  

In some cases, the failure of convergence can be mitigated with line searches. Parametrise the Newton step as
\be
f_{\mrm{new}}=f_{\mrm{old}}+\lambda\delta f\qquad\lambda\in\R\;,
\ee
where we will suppress the subscripts (in $f_j$, etc.) for this discussion.  We choose $\lambda$ in a manner that would improve convergence.  The idea is to prevent the iterations from wandering too far and possibly out of the basin of attraction.  The standard line search is to choose $\lambda$ such that $E$ is minimised.  This can be done, for instance, with a bisection search within the interval $\lambda\in(0,1]$ (\emph{i.e.} repeatedly halving this interval).  In certain cases, there may be certain known restrictions on $f$ (say, it must be positive-definite), in which case $\lambda$ can be controlled to ensure that $f$ remains positive.  

Finally, Newton-Raphson only finds a single solution.  One will need a separate, well-motivated seed to find other solutions and there is no systematic way to find all of them (see appendix \ref{appendix:collocation} for an example).  When one root dominates the basin of the attraction, the other ones may be difficult to find.

\subsection{Ricci Flow \label{Sec:RicciFlow}}
Besides Newton-Raphson, there is another method to solve the Einstein equation which is based on Ricci-flow.  It has a number of drawbacks which makes it less appealing than Newton-Raphson.  Nevertheless, it can also possibly serve as a means for obtaining seeds for Newton-Raphson.  It is also more geometrical than Newton-Raphson, and is interesting from a mathematical standpoint.
  
Given the vacuum Einstein equation, $R_{\mu\nu}=0$, we can always form the so-called {\it Ricci flow equation}  \cite{Hamilton1982}
\begin{eqnarray}\label{dT:Ricciflow}
\frac{\dd}{\dd\tau} \,g_{\mu\nu} = -2 R_{\mu\nu},
\end{eqnarray}
where $\tau$ is to be viewed as a  (fictitious) flow time. {\it Ricci flow} is the one-parameter family of metrics $g(\tau)$ that satisfy this parabolic equation. The Ricci-Flow method involves solving \eqref{dT:Ricciflow} as a geometric evolution equation for the metric $g$ after some initial guess for the metric $g_{\mu\nu}{\bigl |}_{\tau=0}$ is given.  The hope is that the system will evolve towards a fixed point $g$ that solves the Einstein equation $R_{\mu\nu}(g)=0$. 

Standard parabolic theory contains short-time existence (and uniqueness) theorems for strictly parabolic system of PDEs (see \emph{e.g.} \cite{Hamilton1982}).\footnote{Recall that a second order PDE on ${\cal M} \subset \mathbb{R}^n$ for a function $g:{\cal M} \to \mathbb{R}$ of the form $\frac{\dd g}{\dd\tau}=A_{ij}\,\partial_i \partial_j g+B_{i}\,\partial_i g +C\,g$ with smooth coefficients $A_{ij},B_{i},C$ is said to be strictly (or strongly) parabolic if $A_{ij}$ is uniformly positive definite, \emph{i.e.} if the operator $A$ is elliptic. (A familiar example is the heat equation $\frac{\dd g}{\dd\tau} = \nabla^2  g$). If $\frac{\dd g}{\dd\tau}= G(g)$ is stricty parabolic at $g^{(s)}$, then there is a short-time existence and uniqueness theorem: it states that there exist $\epsilon>0$ and a smooth one-parameter family $g(\tau)$ for $\tau\in [0,\epsilon]$ such that $\frac{\dd g}{\dd\tau}= G(g)$ for $\tau\in [0,\epsilon]$ and $g(0)=g^{(s)}$.}  However, \eqref{dT:Ricciflow} is only a weakly parabolic system of PDEs and thus this theorem does not immediately apply.  An equation of the form \eqref{dT:Ricciflow} is strictly parabolic only if the operator $R_{\mu\nu}$ is elliptic. The strategy is therefore to replace this with the Einstein-DeTurck operator to get the  {\it Ricci-DeTurck flow equation}  \cite{DeTurck1983,DeTurck2003}
\begin{eqnarray}\label{dT:dTRicciflow}
\frac{\dd}{\dd\tau}\,g_{\mu\nu} = -2 R^H_{\mu\nu}\quad \Leftrightarrow \quad \frac{\dd}{\dd \tau} \,g_{\mu\nu} = -2R_{\mu\nu}+2\nabla_{(\mu} \xi_{\nu)},
\end{eqnarray}
which can be a strictly parabolic system of PDEs, under certain conditions, as we have discussed earlier.  As parabolic equations tend to describe diffusion or heat dissipation, the Ricci(-DeTurck) flow equation can be seen as a non-linear heat equation for the metric. 

The Ricci-DeTurck flow \eqref{dT:dTRicciflow} is particularly useful because it is diffeomorphic to Ricci flow \eqref{dT:Ricciflow}. Indeed the DeTurck term $\nabla_{(\mu} \xi_{\nu)}$ just introduces an infinitesimal diffeomorphism at each point along the flow. Moreover, if the reference metric $\bar{g}$ preserves the same isometries of $g$, the Ricci-DeTurck flow $g(\tau)$ will also preserve them. Consequently, although the time evolution in the space of metrics depends explicitly on the choice of reference metric $\bar{g}$, the flow on the space of geometries (\emph{i.e.}, metrics up to diffeomorphisms) is independent of $\bar{g}$. In equivalent words, fixed points of the flow are geometric invariants. 
 
The above description suggests that the diffusion properties of the Ricci-DeTurck flow can be used as an efficient algorithm to find a numerical solution to gravitational equations. This method solves a cohomogeneity-$n$ problem by solving a $n+1$ parabolic differential equation, of which there are many standard approaches.  Often, one chooses a spatial discretisation and a temporal discretisation (time-stepping).  Options for spatial discretisation include those that are used for numerical boundary value problems such as finite differences, finite elements, spectral, etc.  Options for time-stepping include the Runge-Kutta family, or implicit methods like Crank-Nicolson or Backwards differencing.  Since the aim is to resolve late-time behaviour, implicit methods may provide extra numerical stability (and hence efficiency if large time steps can be used).  Some standard time-stepping methods can be found in \cite{Press:1992zz,Boyd}.
 
However, Ricci-DeTurck flow is particularly sensitive to whether or not the desired solution has a Lichnerowicz linear operator $\Delta_L$, defined in \eqref{dT:Lichn}, with a negative mode. If this is the case, the time evolution of the Ricci-DeTurck flow will drive the system \emph{away} from the fixed point rather than towards it. This is a major drawback of the Ricci-DeTurck method. Many black hole solutions have a Lichnerowicz operator that yields negative modes. The primary example is the Schwarzschild black hole that has the famous Gross-Yaffe-Perry negative mode \cite{Gross:1982cv} (that also signals the onset of the Gregory-Laflamme instability in the associated black string \cite{Reall:2001ag}). This negative mode is not eliminated when rotation is turned on to get the Kerr black hole \cite{Monteiro:2009tc} nor for moderate values of the electric charge \cite{Monteiro:2008wr}. Moreover it persists in higher dimensions \cite{Monteiro:2009tc,Dias:2009iu,Dias:2010maa,Dias:2010eu,Dias:2011jg} and in certain regions of moduli space of AdS black holes \cite{Prestidge:1999uq,Gubser:2000ec,Monteiro:2009ke,Gubser:2000mm,Dias:2010gk}. 

Despite this drawback, Ricci-DeTurck flow may still prove useful.  First, there are instances where negative modes do not arise, \emph{e.g.} see \cite{Figueras:2011va,Santos:2014yja}. Second, even though a desired solution has negative modes, these negative modes often do not dominate the spectrum. Therefore, Ricci-DeTurck flow may move towards a fixed point before being driven away by the negative mode.  This provides a means of obtaining a seed for a Newton-Raphson algorithm.  

\section{Other Tools and Tricks \label{Sec:tricks}}
\subsection{Finding a Seed \label{Sec:seed1}}
One of the most difficult tasks in solving boundary value problems with Newton-Raphson is finding a good seed.  Here, we attempt to give a few strategies for finding seeds.

The easiest case to find new solutions is when nearby solutions are known.  When solutions are parametrised by some parameter $\lambda$, the well-posedness of the boundary value problem ensures that functions will change continuously with continuous changes in $\lambda$.  Then there is the clear strategy of `marching'.  Take small steps $\delta\lambda$, and find new solutions at $\lambda+\delta\lambda$ by using the solution at $\lambda$ as a seed.  This is typically the strategy that is employed to explore parameter space.

Another case where nearby solutions are known is when a zero mode has been found.  As explained in section \ref{subsec:zeromodes}, a zero mode often indicates that new branch of solutions exist.  From a zero mode, there are several options one can use to find this new branch.  If perturbative functions are known (that is, the eigenfunctions), they can be used to approximate a seed.  Otherwise, one may have to guess, as we do for the example in section \ref{Sec:RRR}.  It can sometimes be the case that the basin of attraction for the background solution is too large to find these new solutions in this way.  For this, one can attempt to use a change of parameters.  For example, suppose a background contains a $U(1)$ symmetry $\partial_{\phi}$, but the new branch of solutions does not contain this symmetry and breaks it as the usual Fourier expansion $e^{m\phi}$.  Let $\lambda$ be a parameter for these solutions.  We can promote this parameter to an unknown and introduce the new parameter
\be
\epsilon\equiv\int f(\phi)\sin(m\phi)\dd\phi\;,
\ee
where $f$ is some metric function evaluated at some curve parametrised by $\phi$.  This definition of the new parameter can be added as a new equation, with $\lambda$ as a new unknown. Any solution with $\epsilon\neq0$ has $\phi$ dependence, and hence cannot go back to the background solution.  This was the strategy used in \cite{Horowitz:2014hja,Dias:2015rxy}.

The cases where nearby solutions are not known is much more difficult.  With sufficient insight, one may be able to create a reference metric that is sufficiently close to the true solution and use the reference metric as a seed.  To aid in this process, one can create, two (or several) metrics that approximate `near' and `far' regions and join them together with an interpolating function $I(r)$:
\be
\dd s^2=[1-I(r)]\dd s^2_\mathrm{far}+I(r)\dd s^2_\mathrm{near}\;,
\ee
for some `radial' coordinate $r$.  This strategy was successfully employed in \cite{Figueras:2014lka,Figueras:2014dta}.

We finish this subsection with one final trick that can be used.  Suppose we wish to solve $G_{\mu\nu}[g]=0$ (where $G_{\mu\nu}$=0 could be the Einstein-DeTurck equation) for the metric $g$. Then consider the equation
\be
G_{\mu\nu}[g]-\delta G_{\mu\nu}[\bar g]=0\;,
\ee
where $\bar g$ is the reference metric and $\delta\in[0,1]$.  By construction, $\delta=0$ is our original equation, and $\delta=1$ has the solution $g=\bar g$.  The strategy is to first set $\delta=1$ and $g=\bar g$, and then slowly march $\delta$ down to $\delta=0$.  This method is akin to introducing a stress tensor in the Einstein-DeTurck equations that is a solution on the reference metric, then slowly turning off the stress tensor.  Equations with matter can be solved in a similar fashion, either by introducing a similar equation for the matter fields, or by first solving the matter equations on a fixed background $\bar g$ (with $\delta=1$) before lowering $\delta$.  This was done successfully in \cite{Horowitz:2014gva}.

\subsection{Turning Points}\label{subset:turning}
While varying parameters, it is possible to reach a turning point.  For example, suppose $\lambda$ parametrises a family of solutions.  It might be the case that as one increases $\lambda$, a limit is reached near $\lambda=\lambda_\mathrm{max}$.  If there are no indicators that some singular behaviour is occurring, this may be a turning point.  That is, one may be able to find new solutions by \emph{decreasing} $\lambda$.  There are multiple solutions for a given value of $\lambda$ that happen to meet at $\lambda_\mathrm{max}$.  

To find these new solutions, one has to decrease $\lambda$ without simply back-tracking on the already known solutions.  The strategy is to adjust the Newton-Raphson seed in such a way that will land us on these new solutions.  This might be accomplished as follows.  Let $\lambda_0<\lambda_1<\lambda_\mathrm{max}$, with $\delta\lambda\equiv\lambda_1-\lambda_0$ and $\lambda_1+\delta\lambda>\lambda_\mathrm{max}$.   That is, we are increasing the parameter $\lambda$ in steps $\delta\lambda$ in Newton-Raphson and have reached a limit where there fails to be a solution at $\lambda_1+\delta\lambda$.  If there is a turning point, we may expect there to be multiple solutions at $\lambda_1$.  Now let us assume that in going from $\lambda_0$ to $\lambda_1$, the change in the functions is similar to going from $\lambda_1$ to the alternate solution at $\lambda_1$.  This suggests seed
\be
\Sigma_1+\Delta(\Sigma_1-\Sigma_0)\;,
\ee
where $\Sigma_0$ and $\Sigma_1$ are the known solutions at $\lambda_0$ and $\lambda_1$, respectively, and $\Delta$ is some number of our choosing.  Choosing $\Delta=0$ would just recover the same solution we already have, namely $\Sigma_1$, but a large enough $\Delta$ might kick the solution enough to give us a new solution.  For this trick to work, $\delta\lambda$ may have to be sufficiently small. This procedure was successfully applied in \cite{Santos:2014yja,Dias:2014cia}.  We will give an example where this trick is applied in section \ref{Sec:rings}.

\subsection{Increasing the Dimension of Spheres \label{Sec:seed2}}
Suppose we have a $d$-dimensional solution with $S^n$ spherical symmetry.  It is possible to obtain $d+1$ dimensional solutions with $S^{n+1}$ spherical symmetry.  One means of doing this is to rewrite the equations of motion for any $S^k$ symmetry, and then treat $k$ as a free real parameter.  Though non-integer $k$ does not have any physical meaning, they still yield well-posed boundary value problems.  One can then slowly deform $k$ from $n$ to $n+1$ by repeated application of Newton-Raphson.  Unfortunately, there is an issue with this method.  Consider the Schwarzschild-Tangherlini metric
\be
\dd s^2=-\left(1-\frac{r_0^{k-1}}{r^{k-1}}\right)\dd t^2+\frac{dr^2}{1-\frac{r_0^{k-1}}{r^{k-1}}}+r^2\dd\Omega_{k}\;,
\ee
where $k=d-2$.  For any noninteger $k$, there are fractional powers in the fall-off of the metric components.  That is, the metric components are non-smooth, which may pose a difficulty to a numerical method, particularly (pseudo-)spectral methods.  

Instead, one could take the following alternative approach.  Consider the Einstein-DeTurck equation in $d$ dimensions $(G^{H}_{d})^\mu{}_\nu=0$.  If there is spherical symmetry, the components related to the spherical coordinates will satisfy $(G^{H}_d)^i{}_j\propto\delta^{i}{}_j$.  Since these components contain much redundant information, let the indices $\bar\mu$, $\bar\nu$ represent just one of the spherical components and the remaining non-spherical components.  Note that by definition, if we increase the dimension of the sphere, the number of components in the indices $\bar\mu$ and $\bar\nu$ remain the same.  Therefore, we can write an equation of the form
\be
(G^{H}_{d})^{\bar\mu}{}_{\bar\nu}(1-\delta)+\delta (G^{H}_{d+1})^{\bar\mu}{}_{\bar\nu}=0\;.
\ee
By construction, $\delta=0$ is the equation of motion in $d$ dimensions with, say, $S^k$ spherical symmetry, while $\delta=1$ is the equation of motion in $d+1$ dimensions with $S^{k+1}$ spherical symmetry.  Furthermore, terms that are the same among $G^{H}_{d}$ and $G^{H}_{d+1}$ do not get modified with the above construction.  This includes the principal symbol, so ellipticity is not lost.  Given a solution in $d$ dimensions, we can then solve the above equation, slowly moving $\delta$ from $0$ to $1$ with repeated application of Newton-Raphson.  

Unlike the first case where the dimension of the sphere is the parameter, this process is less prone to producing fractional powers.  Instead, it tends to yield sums of two different powers and attempts to change the coefficients between these terms. That is, something roughly of the form $(1-\delta)/r^{k-1}+\delta/r^{k}$.  This is how black rings in higher dimensions were constructed in \cite{Dias:2014cia}.

\subsection{Patching\label{Sec:Patching}}
Many boundary value problems of interest require an integration domain with more than four natural boundaries.  These include Kaluza-Klein black holes \cite{Kudoh:2003ki,Kudoh:2004hs,Headrick:2009pv}, black rings \cite{Kleihaus:2012xh,Dias:2014cia,Figueras:2014dta}, hovering black holes \cite{Horowitz:2014gva}, AdS domain wall and plasma ball solutions \cite{Aharony:2005bm,Figueras:2014lka}, and black droplets \cite{Santos:2014yja}.  Since most numerical methods work on domains which are rectangular\footnote{We will assume our problems are cohomogeneity-$2$, but all of the methods herein can be extended to higher dimensions.}, finding these solutions with such methods requires some way of dealing with the extra boundary.  

One way (such as those in \cite{Kudoh:2003ki,Kudoh:2004hs,Kleihaus:2012xh}) is to find a new set of coordinates where two boundaries are mapped to one.  This new coordinate system typically contains coordinate singularities, which can be an issue for numerics.  For gravitational problems, if one works in conformal gauge, this choice of gauge can ensure that these singularities do not show up in the numerical metric functions.  There are fewer such guarantees in DeTurck gauge.

Another method is to use patching.  Patching works much like the construction of manifolds, where the integration domain is covered by various `patches', each in their own coordinate system.  The patches are then joined together in a suitable way.  

Numerical codes that implement patching differ depending on whether the patches overlap or meet only on patch boundaries. Where patches overlap, interpolation is typically used to ensure that the functions are the same on the two patches.  For patches that do not overlap, extra patching conditions that match functions and their derivatives must be imposed.  Although the former method of overlap patching has been successfully employed with the DeTurck method in \cite{Headrick:2009pv,Figueras:2014dta,Figueras:2014lka}, it is with the later method of patching that we will be primarily concerned with here\footnote{In this section, `patching' is a concept for collocation methods, where the equations of motion are solved directly.  There are also finite element based approaches to (non-overlap) patching, which solve an integral form of the equation of motion and impose continuity through a condition of a surface term after an integration by parts.  Though these methods are used extensively in scientific and engineering computations, they have (so far) seen little use in numerical relativity.  We therefore do not comment on them further.}.  We will henceforth assume all patches to be non-overlapping.

Besides accommodating a more complicated integration domain, there are other reasons why one might attempt to use patching.  Often, to increase the accuracy in extracting physical quantities, one would increase the resolution of the grid.  For large enough grids, the computational resources required can become prohibitively expensive.  However, if one only needs a higher resolution in a particular region (say, near one of the boundaries of the integration domain), one can include patches with a finer grid to increase the accuracy \cite{Figueras:2012xj,Emparan:2014pra,Kalisch:2015via}.

Another application of patching allows one to use higher-order methods with functions that are less smooth.  Higher order methods (like spectral methods, or high-order finite differencing), often have rapid convergence with increasing grid size, but requires functions that are sufficiently smooth.  If the location of non-smoothness is understood, one can patch a lower-order grid that covers the regions that are less smooth \cite{Horowitz:2012ky,Horowitz:2012gs,Horowitz:2013jaa,Hartnoll:2014cua,Hartnoll:2014gaa,Hartnoll:2015faa,Hartnoll:2015rza}.  With this method and a carefully chosen grid, it is possible to keep the rapid convergence of the higher-order method.  If there is a discontinuity in some (second or higher) derivative of a function, one can also patch two high-order grids together, keeping the non-smooth location on a patch boundary.  

\subsubsection{Transfinite Interpolation}

While the implementation of patching is relatively straightforward (demand the agreement of fundamental fields and their derivatives on patch boundaries), there is still an issue that remains to be resolved: many integration domains cannot be broken up into rectangles.  But, one can still divide them into `warped' rectangular regions (loosely speaking, these are regions that have four `corners' and four possibly curved `edges'), and then patch these regions together.  The task of placing grids on these warped rectangles can be accomplished by using transfinite interpolation.

\begin{figure}[t]
\centerline{\includegraphics[width=.35\textwidth]{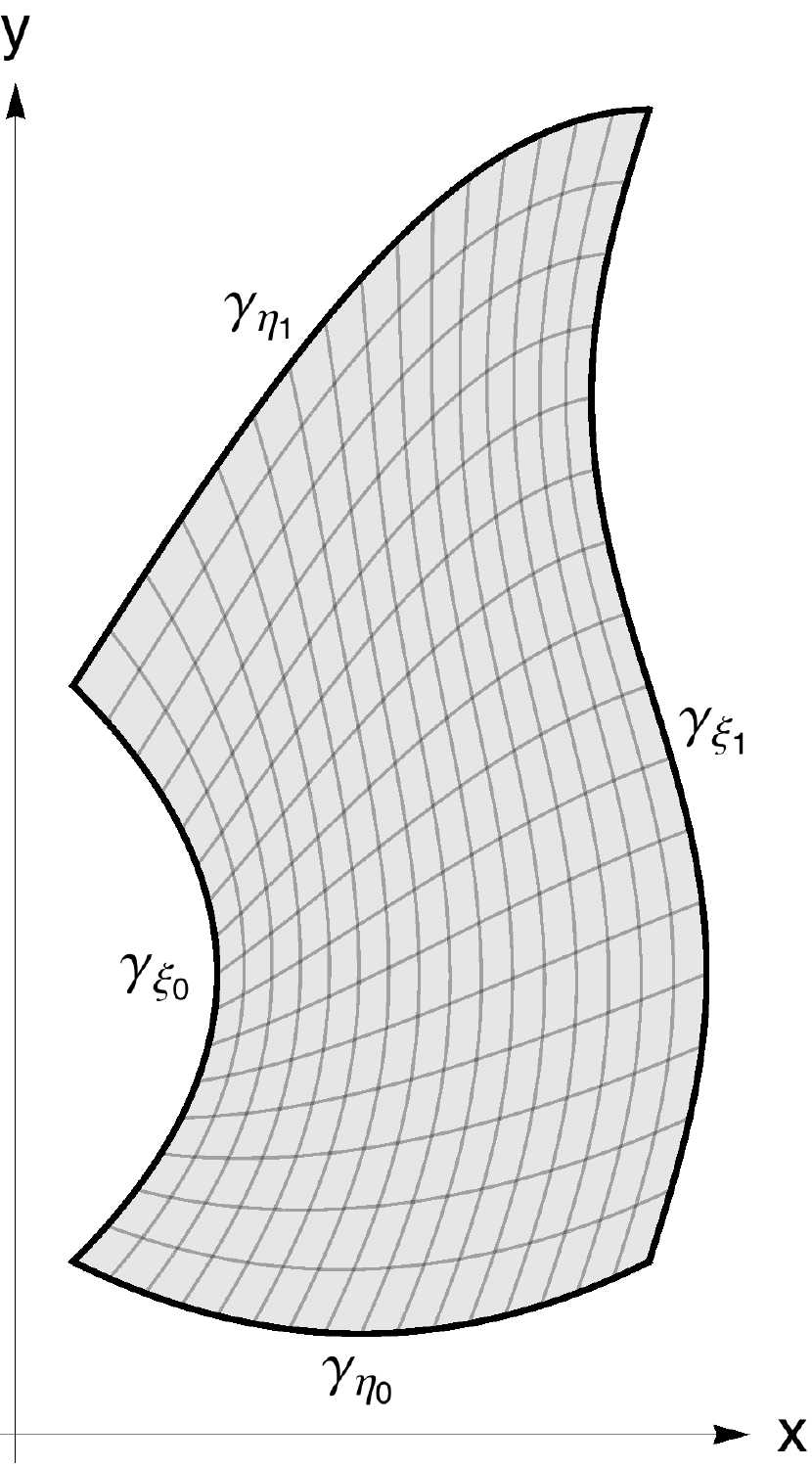}
\hspace{1cm}\includegraphics[width=.45\textwidth]{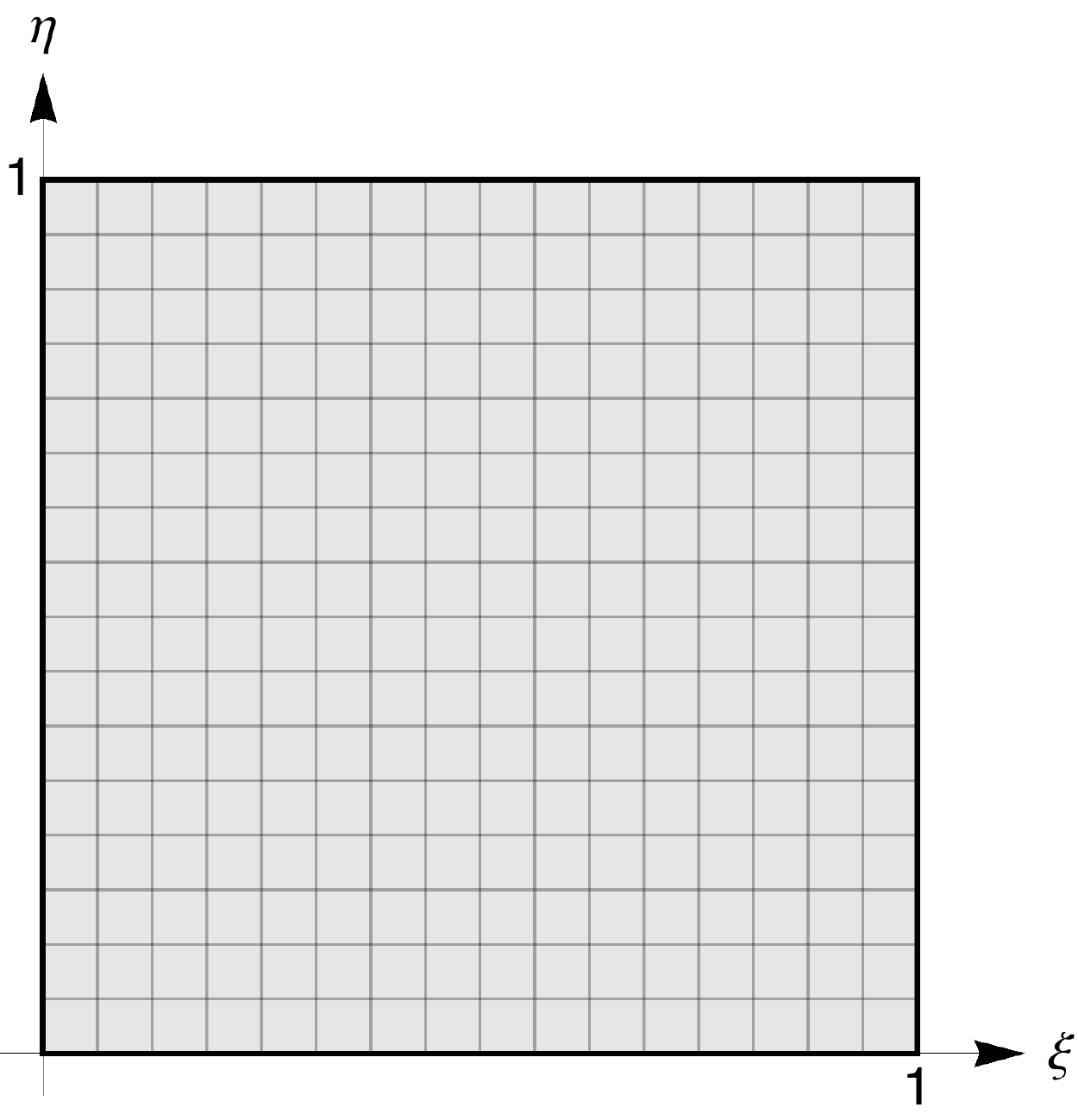}}
\caption{Warped rectangle in physical space and logic space.}
\label{fig:tfimaps}
\end{figure}

Consider a warped rectangular region such as the one in Fig.~\ref{fig:tfimaps}.   This region is defined in some set of coordinates $x$ and $y$ appropriate for specifying the boundary value problem, which we refer to as `physical space'.  However, we wish to do numerical computations in a rectangular domain where we can easily place a grid.  So we will map this region into a new set of coordinates $\xi$ and $\eta$, which we call `logic space' where numerics will be carried out.  

Let us attempt to use logic space to parametrise the four curves of the edges in physical space as follows
\begin{align}\label{transparam}
\vec\gamma_{\xi_0}(\eta)&=(x_{\xi_0}(\eta), y_{\xi_0}(\eta))\nonumber\\
\vec\gamma_{\xi_1}(\eta)&=(x_{\xi_1}(\eta), y_{\xi_1}(\eta))\nonumber\\
\vec\gamma_{\eta_0}(\xi)&=(x_{\eta_0}(\xi), y_{\eta_0}(\xi))\nonumber\\
\vec\gamma_{\eta_1}(\xi)&=(x_{\eta_1}(\xi), y_{\eta_1}(\xi))\;,
\end{align}
where the $\vec\gamma$'s are curves in physical space, and the coordinates $\xi$ and $\eta$ lie in the unit interval $[0,1]$.  The various functions $x_k(\lambda)$ and $y_k(\lambda)$ are any functions of our choosing such that the parametrised curves satisfy the following consistency conditions on the `corners':
\begin{align}
\vec\gamma_{\xi_0}(0)&=\vec\gamma_{\eta_0}(0)\nonumber\\
\vec\gamma_{\xi_1}(0)&=\vec\gamma_{\eta_0}(1)\nonumber\\
\vec\gamma_{\xi_0}(1)&=\vec\gamma_{\eta_1}(0)\nonumber\\
\vec\gamma_{\xi_1}(1)&=\vec\gamma_{\eta_1}(1)\;.
\end{align}
These conditions are all that is necessary to generate a coordinate map. If one requires a regular coordinate transformation, we also require that the derivatives at the corners never line up.  This is equivalent to the condition
\be
|\vec\gamma_i'\cdot\vec\gamma_j'|\neq|\vec\gamma_i'||\vec\gamma_j'|,\qquad\text{(at corners with $i \neq j$)}\;.
\ee

To generate a coordinate transformation $\vec x(\xi,\eta)$ from logic space to physical space, we use the transfinite interpolation formula, also known as the Coon map.
\begin{align}
\vec x(\xi,\eta)&=(1-\xi)\gamma_{\eta_0}(\xi)+\eta \gamma_{\eta_1}(\xi)+(1-\xi)\gamma_{\xi_0}(\eta)+\xi\gamma_{\xi_1}(\eta)\nonumber\\
&\qquad-\Big[\xi\eta\gamma_{\eta_1}(1)+\xi(1-\eta)\gamma_{\eta_0}(1)+\eta(1-\xi)\gamma_{\eta_1}(0)+(1-\xi)(1-\eta)\gamma_{\eta_0}(0)\Big]\;.
\end{align}
Any grid points on logic space can then be mapped to points in physical space using this map.  Derivatives in physical space can be computed from derivatives in logic space via the chain rule and inverse function theorem.  Note that for these purposes, this map does not need to be easily invertible. This was used with great success in \cite{Santos:2014yja,Dias:2014cia,Horowitz:2014gva,Santos:2015iua}.  

\section{Application: Black Rings\label{Sec:rings}}

Now let us apply the various tools in this review to the construction of black rings in \cite{Dias:2014cia}.  These are asymptotically flat singly-spinning solutions to vacuum Einstein gravity with $S^1\times S^{d-3}$ horizon topology.  In this section, we numerically construct rings in $d=5$, $6$, and $7$ dimensions.  The integration domain for black rings naturally have five boundaries: three axes, a horizon, and asymptotic infinity.  This is therefore a natural example to demonstrate patching and transfinite interpolation as was done in section \ref{Sec:Patching}.  We will aim to be thorough since the details in \cite{Dias:2014cia} are sorely lacking.  We note that the numerical construction of black rings in $d=6$ has been done previously with different methods in \cite{Kleihaus:2012xh}\footnote{Black rings with $d=7$ were also constructed by the authors of \cite{Kleihaus:2012xh}, but the results of that calculation were not presented.}.
	
Black rings in $d=5$ are known analytically where the solution can be written as \cite{Emparan:2001wn}
\begin{align}\label{analyticring}
\dd s^2&=-\frac{F(\tilde y)}{F(\tilde x)}\left(\dd \tilde t-C \tilde R\frac{1+\tilde y}{F(\tilde y)}\dd\tilde\psi\right)^2\nonumber\\
&\qquad +\frac{\tilde R^2}{(\tilde x-\tilde y)^2}F(\tilde x)\left[-\frac{G(\tilde y)}{F(\tilde y)}\dd\tilde\psi^2-\frac{\dd \tilde y^2}{G(\tilde y)}+\frac{\dd \tilde x^2}{G(\tilde x)}+\frac{G(\tilde x)}{F(\tilde x)}\dd\tilde\phi^2\right]\;,
\end{align}
where
\be
F(\xi)=1+\lambda\xi\;,\qquad G(\xi)=(1-\xi^2)(1+\nu\xi)\;, \qquad C=\sqrt{\lambda(\lambda-\nu)\frac{1+\lambda}{1-\lambda}}\;.
\ee
The coordinates range in $x\in[-1,1]$ and $y\in (-\infty,-1]$, and the parameters $\lambda$ and $\nu$ satisfy $0<\nu\leq\lambda<1$.  
	
In a certain limit, this metric resembles that of a black string.  This can easily be seen from the redefinitions
\be
r=\frac{-\tilde R}{\tilde y},\qquad \cos\theta=\tilde x\;,\qquad \nu=\frac{r_0}{\tilde R}\;,\qquad \lambda=\frac{r_0\cosh^2\sigma}{\tilde R}\;,
\ee
after which \eqref{analyticring} becomes
\begin{align}\label{analyticring2}
\dd s^2&=-\frac{\hat f}{\hat g}\left(\dd t-r_0\sinh\sigma\cosh\sigma\sqrt{\frac{\tilde R+r_0\cosh^2\sigma}{\tilde R-r_0\cosh^2\sigma}}\frac{\frac{r}{\tilde R}-1}{r\hat f}\tilde R\,\dd\psi\right)^2\nonumber\\
&\qquad+\frac{\hat g}{(1+\frac{r\cos\theta}{\tilde R})^2}\left[\frac{f}{\hat f}\left(1-\frac{r^2}{\tilde R^2}\right)\tilde R^2\,\dd\psi^2+\frac{\dd r^2}{(1-\frac{r^2}{\tilde R^2})f}+\frac{r^2}{g}\dd\theta^2+\frac{g}{\hat g}r^2\sin^2\theta\,\dd\phi^2\right]\;,
\end{align}
where
\be
f=1-\frac{r_0}{r}\;,\qquad\hat f=1-\frac{r_0\cosh^2\sigma}{r}\;,\qquad g=1+\frac{r_0}{\tilde R}\cos\theta\;,\qquad\hat g=1+\frac{r_0\cosh^2\sigma}{\tilde R}\cos\theta\;.
\ee
In the limit
\be\label{nearhorizon}
r,\,r_0,\,r_0\cosh^2\sigma\ll \tilde R\;,
\ee
and redefinition $\psi=z/\tilde R$, the line element \eqref{analyticring2} becomes that of the boosted black string with boost parameter $\sigma$.  From this picture of the $S^1\times S^2$ ring, $R$ can be viewed as the size of the $S^1$ and $r_0$ can be viewed as the size of the $S^2$.  This also suggests that the parameter $\nu=r_0/R$ can be used as a measure of how thin or fat the ring is.
	
In order to avoid a conical singularity, we would set 
\be
\lambda=\frac{2\nu}{1+\nu^2}\;.
\ee
But let us instead consider the static ring with $\lambda=\nu$.  Let us make a number of further redefinitions for numerical convenience:
\begin{align}
\tilde x &=1-2(1-x^2)^2\;,\qquad \tilde y=-\frac{1-(1-\nu)(1-y^2)^2}{\nu}\;,\nonumber \\
\tilde t &=\frac{2\nu\,t}{\sqrt{1-\nu^2}}\;,\qquad \tilde\psi=\frac{\psi}{\sqrt{1-\nu}}\;,\qquad \tilde\phi=\frac{\phi}{\sqrt{1-\nu}}\;,\nonumber\\
 \nu& =\frac{\beta^2}{2+\beta^2}\;,\qquad \tilde R=\frac{\sqrt{1+\beta^2}R}{\beta^2}\;,
\end{align}
the static ring becomes
\begin{align}\label{staticring}
\dd s^2=R^2\bigg[-\frac{(1-y^2)^2\,\dd t^2}{g_x}+\frac{(1+\beta^2)g_x}{h^2}&\bigg(\frac{4\,\dd y^2}{(2-y^2)g_y}+\frac{y^2(2-y^2)g_y\,\dd\psi^2}{\beta^4}\nonumber\\
&\qquad+\frac{4\,\dd x^2}{(2-x^2)g_x}+x^2(2-x^2)(1-x^2)^2\,\dd\phi^2\bigg)\bigg]\;,
\end{align}
where
\be
g_x=1+\beta^2x^2(2-x^2)\;,\qquad g_y=\beta^2+y^2(2-y^2)\;,\qquad h=\beta^2x^2(2-x^2)+y^2(2-y^2)\;.
\ee
As we shall see, this choice of coordinates will make imposing boundary conditions particularly simple.
	
Let us make a few comments about the static line element \eqref{staticring} since it will serve as a starting point for our numerical construction\footnote{We could have instead begun with the regular rotating black ring, but doing so gave us poorer numerical results.}.  This is a one-parameter family with $\beta>0$.  Here, the coordinate range is $x\in[0,1]$ and $y\in[0,1]$, with the topologically $S^1\times S^2$ horizon at $y=1$, the axis of the $S^1$ at $y=0$, the outer axis of the $S^2$ at $x=0$, and the inner axis at $x=1$.  The temperature is fixed to be $T=1/(2\pi)$, and the periods of the angles $\psi$ and $\phi$ are fixed to be $2\pi$.  There is a conical singularity at $x=1$ with a conical excess with factor $\sqrt{1+\beta^2}$.  All other circles close off smoothly.  

Asymptotic infinity is at the coordinate point $x=y=0$.  To see this more explicitly, we can define the new coordinates
 \be\label{rhoximap}
 \rho=\sqrt h=\sqrt{\beta^2 x^2(2-x^2)+y^2(2-y^2)}\;,\qquad \xi=\sqrt{1-\frac{\beta x\sqrt{2-x^2}}{\sqrt{\beta^2 x^2(2-x^2)+y^2(2-y^2)}}}\;,
 \ee
whose inverse is
\be\label{xymap}
x=\sqrt{1-\sqrt{1-\frac{\rho^2(1-\xi^2)^2}{\beta^2}}}\;,\qquad y=\sqrt{1-\sqrt{1-\rho^2 \xi^2(2-\xi^2)}}\;.
\ee
Taking the limit $\rho\rightarrow0$ in these new coordinates, the metric becomes
\be
\dd s^2=-\dd t^2+\frac{1+\beta^2}{\beta^2}\left[\frac{\dd\rho^2}{\rho^4}+\frac{1}{\rho^2}\left(\frac{4\,\dd\xi^2}{2-\xi^2}+\xi^2(2-\xi^2)\dd\psi^2+(1-\xi^2)^2\dd\phi^2\right)\right]+\mathcal O(\rho^{-1})\dd\rho\,\dd\xi\;.
\ee
Which is asymptotically the line element of Minkowski space.  To get this in a more familiar form, set
\be
r=\frac{\sqrt{1+\beta^2}}{\beta}\frac{1}{\rho}\;,\qquad \cos\theta=1-\xi^2\;,
\ee
which yields
\be
\dd s^2=-\dd t^2+\dd r^2+r^2 d\Omega_{3}^2+\mathcal O(r^{-1})\dd r\dd\xi\;.
\ee
	
From the line element \eqref{staticring}, we would like to write down a suitable reference metric for the construction of regular black rings in five and higher dimensions.   We have chosen
\begin{align}\label{ringref}
\dd s^2=R^2\bigg\{-\frac{(1-y^2)^2\,\dd t^2}{g_x}+\frac{(1+\beta^2)g_x}{h^2}&\bigg[\frac{4\,\dd y^2}{(2-y^2)g_y}+\frac{y^2(2-y^2)g_y}{\beta^4}\left(\dd\psi-\omega \left(\frac{h}{g_x}\right)^{\lceil \frac{d-1}{2}\rceil}\dd t\right)^2\nonumber\\
&\qquad+\frac{4\,\dd x^2}{(2-x^2)g_x}+\frac{x^2(2-x^2)(1-x^2)^2}{f_x}\,\dd\Omega_{d-4}^2\bigg]\bigg\}\;,
\end{align}
where
\be
f_x=1+\alpha^2\, x^2(2-x^2)\;.
\ee
Note that we have introduce the parameters $\alpha$ and $\omega$, and that this reference metric reduces to the static ring \eqref{staticring} when $d=5$, $\alpha=0$ and $\omega=0$.  Note also that the additional functions do not spoil asymptotic infinity which sits at $x=y=0$.  
	
The parameter $\alpha$ now controls the conical excess, with the conical singularity disappearing when $\alpha=\beta$.   If $\alpha=\beta$, $\beta$ no longer determines any physical parameters and is reduced to pure gauge.  Since it originally came from the parameter $\nu$, it roughly controls the thinness/fatness of the ring in the reference metric, and it can be used as a numerical means of adapting our gauge to the specific ring at hand (larger $\beta$ for fatter rings and smaller $\beta$ for thinner rings).  We find that varying $\beta$ as we vary physical parameters can significantly improve the numerics.
	
The parameter $\omega$ is the angular frequency of the horizon.  The power $\lceil \frac{d-1}{2}\rceil$ was chosen so that this factor decays sufficiently fast asymptotically.  More will be said about this in a moment.

With a reference metric in hand, we can now write down a metric ansatz.  The one we have chosen is given by
\begin{align}\label{ringansatz}
\dd s^2=R^2&\bigg\{-\frac{(1-y^2)^2T\,\dd t^2}{g_x}\nonumber\\
&+\frac{(1+\beta^2)g_x}{h^2}\bigg[\frac{4A\,\dd y^2}{(2-y^2)g_y}+\frac{y^2(2-y^2)g_y}{\beta^4}S_1\left(\dd\psi-W \left(\frac{h}{g_x}\right)^{\lceil \frac{d-3}{2}\rceil}\dd t\right)^2\nonumber\\
&\qquad\qquad\qquad+\frac{4B}{(2-x^2)g_x}\left(\dd x-\frac{x(2-x^2)(1-x^2)y(1-y^2)}{h}F\,\dd y\right)^2\nonumber\\
&\qquad\qquad\qquad\qquad\qquad\qquad\qquad\qquad+\frac{x^2(2-x^2)(1-x^2)^2}{f_x}S_2\,\dd\Omega_{d-4}^2\bigg]\bigg\}\;,
\end{align}
where $T$, $A$, $B$, $S_1$, $S_2$, $W$, and $F$ are functions of $x$ and $y$.  Note the replacement of $\omega$ by the function $W$ and the change in the power $\lceil \frac{d-1}{2}\rceil$ to $\lceil \frac{d-3}{2}\rceil$. 
	
The ansatz \eqref{ringansatz} and reference metric \eqref{ringref} are written in a coordinate system where asymptotic infinity is at the coordinate point $x=y=0$.  This is not well controlled for asymptotics.  Our approach is to patch a grid around $x=y=0$ in a different coordinate system where the asymptotics are under better numerical control.  The $\rho$, $\xi$ coordinates defined by the transformations \eqref{rhoximap} and \eqref{xymap} serve this purpose well.  Actually, we can sometimes do better.  For odd dimensions, one can show that any asymptotic data occurs in even powers of the standard radial coordinate, so we can use $\tilde\rho=\rho^2$ instead of $\rho$.  This effectively halves the number of derivatives needed to extract physical quantities from infinity.  
	
At this point, we can also explain our choice of powers $\lceil \frac{d-1}{2}\rceil$ and $\lceil \frac{d-3}{2}\rceil$.  The angular momentum is read off infinity from this term, which asymptotically goes as
\be
\frac{1}{\rho^2}(\dd\psi-(\sim J) \rho^{d-1}\dd t)^2\;.
\ee
The particular power of $\lceil \frac{d-3}{2}\rceil$ (recall $\rho=\sqrt h$) was chosen so that the angular momentum could be read from just one derivative (in $\rho$ for even $d$ and $\tilde\rho$ for odd $d$) of $W$.  We will need to impose $W=0$ as a boundary condition, so to be compatible with this, we require the power $\lceil \frac{d-1}{2}\rceil$ in the reference metric.
	
In the new $(\rho,\xi)$ or $(\tilde\rho,\xi)$ coordinate system, we chose to keep the same metric functions used in \eqref{ringansatz} rather than define new ones.  Indeed, the particular factors in the $\dd x\,\dd y$ cross term in \eqref{ringansatz} were chosen so that the ansatz in the $(\rho,\xi)$ or $(\tilde\rho,\xi)$ coordinates would be as simple as possible (in this case, this means no square roots appear). In our case, this would be the simplest choice since the asymptotic boundary condition at $\rho=0$ or $\tilde\rho=0$ ($x=y=0$ in the other coordinates) is specified by Dirichlet data.  If the boundary conditions are more complex, it may be advantageous to define new metric functions more closely adapted to the geometry in the new coordinates, in which case patching is handled by demanding that the metrics, rather than the metric functions, are equivalent.  This was done, for example, in \cite{Santos:2014yja,Horowitz:2014gva} in order to handle multi-horizon geometries.

\begin{figure}[t]
\centerline{\includegraphics[width=.45\textwidth]{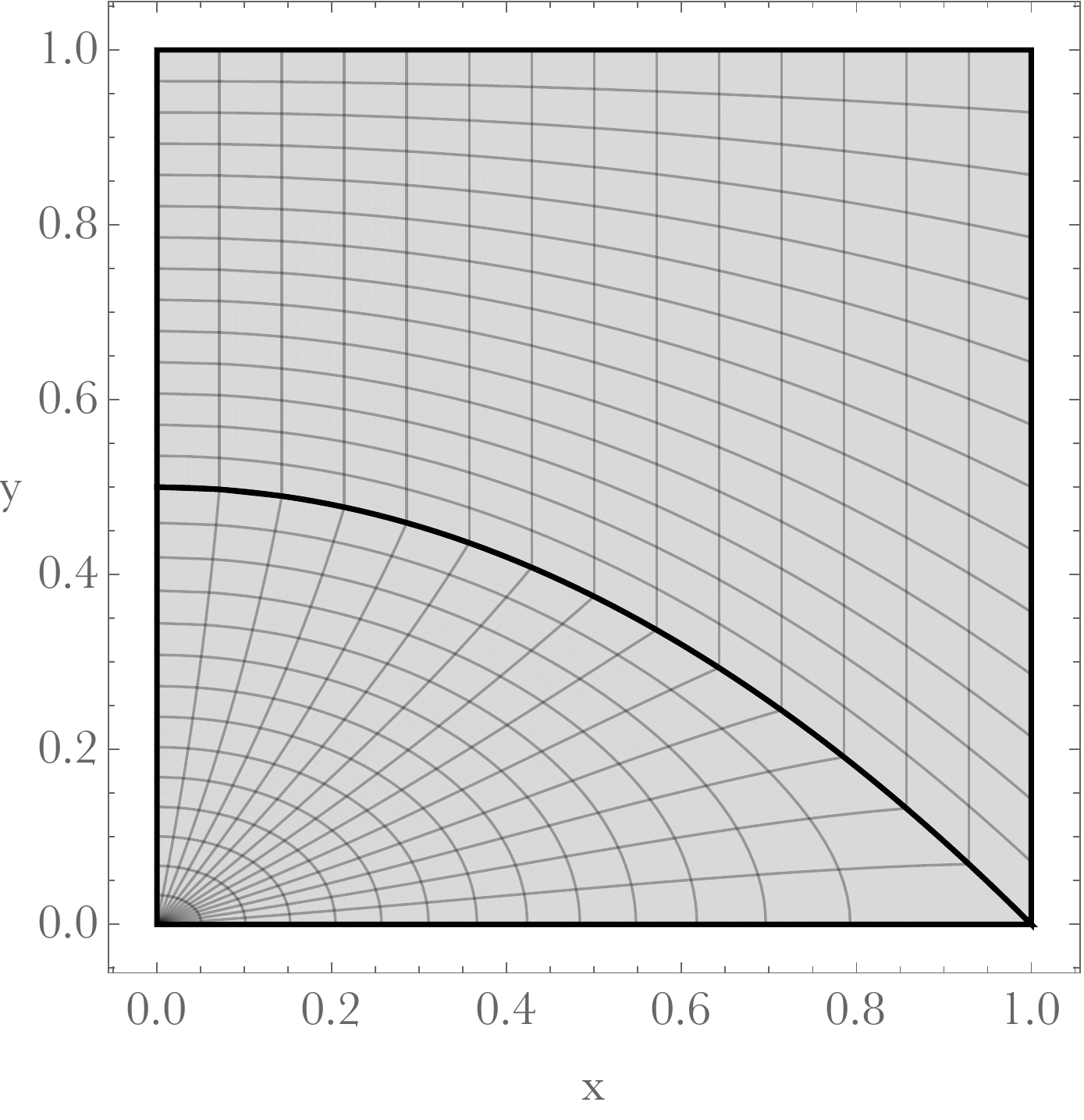}
\hspace{1cm}\includegraphics[width=.45\textwidth]{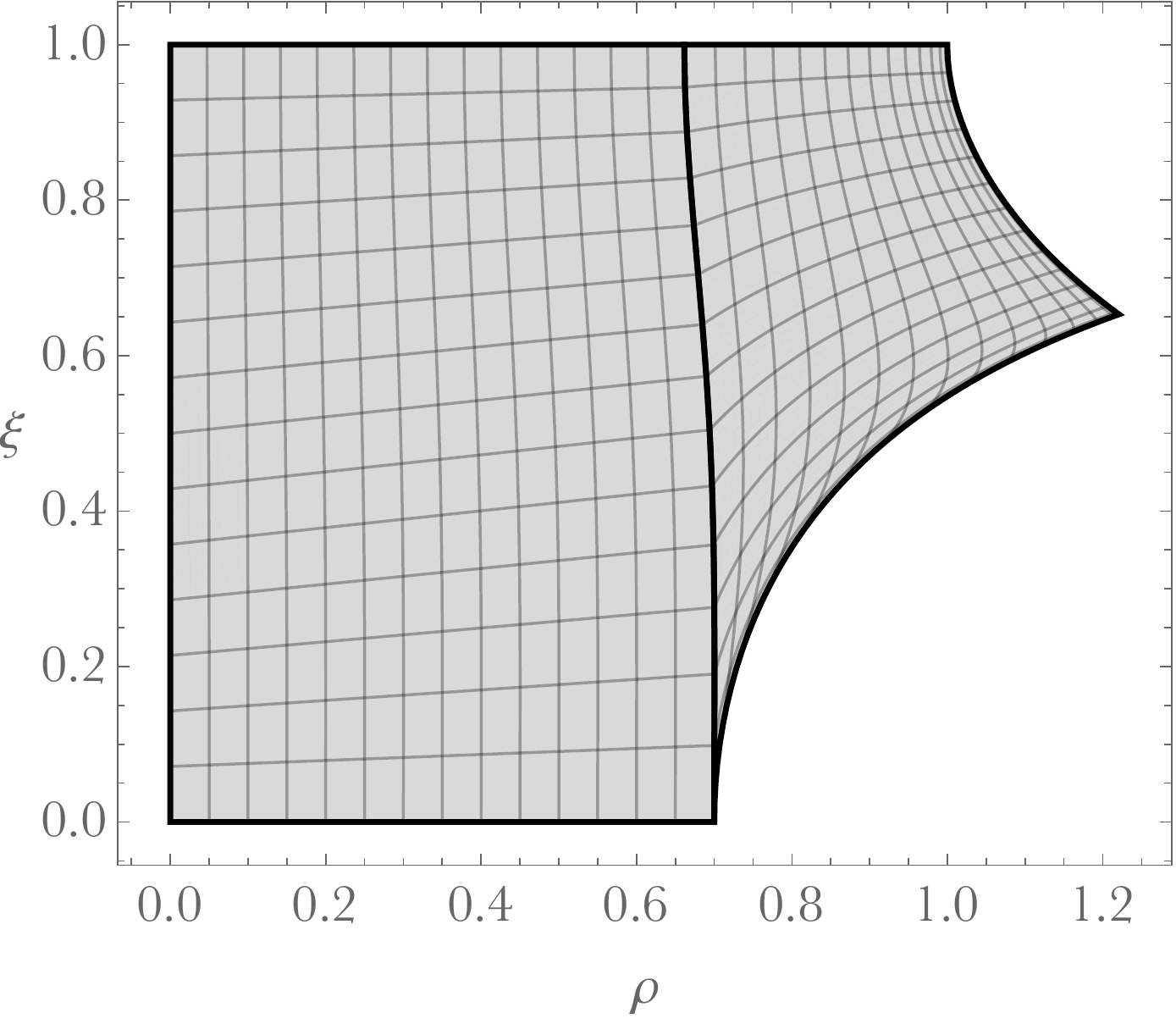}}
\caption{Patches and coordinates for black rings.}
\label{fig:ringcoord}
\end{figure}

Now we can choose patches and place grids on these patches using transfinite interpolation.  An example of such a construction is shown in Fig.~\ref{fig:ringcoord}.  Note that we have chosen the entirety of asymptotic infinity to be covered by a patch in the $(\rho, \xi)$ coordinate system.  There is still a significant amount of freedom here in the choice of patch boundaries and parametrisation in \eqref{transparam}.  These are usually chosen just so the grids look sensible.  As a numerical check, we can demonstrate that our results do not change when the patch boundary or parameterisations \eqref{transparam} are varied.  
	
Now let us proceed with a discussion of boundary conditions.  At asymptotic infinity, we must recover Minkowski space.  This gives us the Dirichlet conditions
\be
(T,A,B,S_1,S_2,W,F)|_{\rho=0,\text{ or } \tilde\rho=0}=(1,1,1,1,1,0,0)\;.
\ee
We only impose this boundary condition in the $(\rho,\xi)$  or $(\tilde\rho,\xi)$ coordinate system where we have numerical control.
	
The remaining boundary conditions require regularity.  Our choice of coordinates for the ansatz and reference metric were such that the metric functions are even in $x$, $1-x^2$, $y$, and $1-y^2$.  Our choice for the coordinates $\rho$ and $\xi$ are such that we get even functions of $1-\xi^2$ and $\xi$.  This choice yields simple Neumann boundary conditions for the metric functions.  

From our choice of patch seen in Fig.~\ref{fig:ringcoord}, the boundary condition at the axis $S^1$ ($y=0$ or $\xi=0$) is only imposed in the $(\rho,\xi)$  or $(\tilde\rho,\xi)$ coordinate system.  Nevertheless, the boundary conditions are simple enough to present here in both coordinate systems:
\begin{align}
&A|_{y=0\text{ or }\xi=0}=S_1|_{y=0\text{ or }\xi=0}\;,\nonumber\\
&(\partial_y T,\partial_y A,\partial_y B,\partial_y S_1,\partial_y S_2,\partial_y W,\partial_y F)|_{y=0}=(\partial_\xi T,\partial_\xi A,\partial_\xi B,\partial_\xi S_1,\partial_\xi S_2,\partial_\xi W,\partial_\xi F)|_{\xi=0}=0\;.
\end{align}
This appears to be one more condition than is required.  In particular, there appears to be a choice of two among three conditions that involve $A$, and $S_1$.  Here, it makes little difference which two of these are imposed, since these are derived from the equations of motion directly. 
	
Similarly, at the outer axis ($x=0$ or $\xi=1$), we have
\begin{align}
&B|_{y=0\text{ or }\xi=0}=S_2|_{y=0\text{ or }\xi=0}\;,\nonumber\\
&(\partial_x T,\partial_x A,\partial_x B,\partial_x S_1,\partial_x S_2,\partial_x W,\partial_x F)|_{x=0}=(\partial_\xi T,\partial_\xi A,\partial_\xi B,\partial_\xi S_1,\partial_\xi S_2,\partial_\xi W,\partial_\xi F)|_{\xi=1}=0\;.
\end{align}
From our choice of patches, this is a boundary shared by two patches.  
	
The remaining two boundaries are only imposed in the $x$, $y$ coordinates.  At the inner axis $x=1$, we have
\begin{align}
&B|_{x=1}=S_2|_{x=1}\;,\nonumber\\
&(\partial_x T,\partial_x A,\partial_x B,\partial_x S_1,\partial_x S_2,\partial_x W,F)|_{x=1}=0\;.
\end{align}
Here, there is a Dirichlet-type condition $F=0$. Recall that the inner axis can contain a conical singularity.  These conditions ensure that the conical deficit can be locally removed by a trivial rescaling of angular coordinates, are equivalent to regularity when there is no conical singularity.  

At the horizon $y=1$, we have
\begin{align}
&T|_{y=1}=A|_{y=1}\;,\nonumber\\
&(\partial_y T,\partial_y A,\partial_y B,\partial_y S_1,\partial_y S_2,W,\partial_y F)|_{y=1}=(0,0,0,0,0,\omega,0)\;.
\end{align}
There are Dirichlet-type conditions $W=\omega$, which determines the angular velocity of the horizon, and also $F=0$.  
	
Finally, in addition to the physical boundary conditions, there are also the artificial patching conditions.  These are simply that the metric functions are equal on patch boundaries, as well as their normal derivatives off the patch boundary.
		
Since we are using Newton-Raphson, we require a good seed solution.  Fortunately, there is already the analytic solution \eqref{staticring}.  Our strategy then is to slowly deform this solution until we arrive at the desired solution.  Let us first tally the number of parameters available to us, excluding those associated to the choice of patches.  These are the dimension $d$, a parameter $\alpha$ that dials the conical defect, a parameter $\beta$ that controls the thinness/thickness of the ring (this is pure gauge when $\alpha=\beta$),  and the angular frequency $\omega$. The radius $R$ was only used to set a scale drops out of the equations of motion.  
	
Now let us describe our general strategy for finding black rings numerically.  First, set $d$ in the ansatz \eqref{ringansatz} and reference metric \eqref{ringref} to the desired dimension, except for the sphere $S^{d-4}$, which we set to an $S^1$.  The point of this is mainly to fix the power $\lceil\frac{d-3}{2}\rceil$ while we increase the dimension of the sphere.  Now set $\alpha=0$ and $\omega=0$, which just reproduces the static ring solution given in \eqref{staticring}.  Now we attempt to remove the conical deficit while spinning up the ring.  With some fixed $\beta$ ($\beta=1$ worked well for us), increase $\omega$ and $\alpha$ until $\alpha=\beta$.  This gives us a regular black ring in five dimensions.  Now we increase the dimension of the sphere from an $S^1$ to $S^{d-4}$ using the method described in section \ref{Sec:seed2}.  Some adjustments might have to be made in $\omega$ for this to work.  This gives us a black ring in the desired dimension.  	
Afterwards, we fix $\alpha=\beta$ while exploring the physical parameter space of the black rings.  At this point, the temperature is fixed, so $\omega$ parameterises the solutions, and $\beta$ is a pure gauge parameter that is used to adapt the coordinate system to the black ring.  As $\omega$ is varied, changes are made to $\beta$ and possibly the patching parameters to keep good numerical accuracy.  Asymptotic charges $E$ and $J$ can be computed from a Komar integral at infinity
\begin{align}\label{KomarCharges}
E&=\frac{1}{8\pi \mathcal G}\frac{d-2}{d-3}\int_{\partial\Sigma} \dd x^{d-2}\sqrt{\gamma}n_{\mu}\sigma_{\nu}\nabla^\mu K^\nu\;,\nonumber\\
J&=-\frac{1}{8\pi \mathcal G}\int_{\partial\Sigma} \dd x^{d-2}\sqrt{\gamma}n_{\mu}\sigma_{\nu}\nabla^\mu R^\nu\;,
\end{align}
where $\partial\Sigma$ is a sphere at spatial infinity with induced metric $\gamma_{ij}$, $\sigma^\mu$ is an outward-pointing normal vector, $n^\mu$ is the unit normal vector to the space like surface $\Sigma$, $K^\mu$ is the timeline Killing vector and $R^\mu$ is the rotational Killing vector.   Other thermodynamic quantities like the temperature, entropy, and angular velocity can be read near the horizon in the standard way. We use the Smarr law $\frac{d-3}{d-2}M=T_{H}S_H+\Omega_HJ$, the first law of thermodynamics $\dd M=T_H\dd S_H+\Omega_H\dd J$, and the vanishing of the DeTurck vector $\xi^2=0$ as monitors of numerical accuracy.  As an additional check, we verify that small changes to $\beta$ and the patching parameters do not change our numerical results.  More quantitative values for these checks can be found in \cite{Dias:2014cia}.

\begin{figure}[t]
\centerline{\includegraphics[width=.45\textwidth]{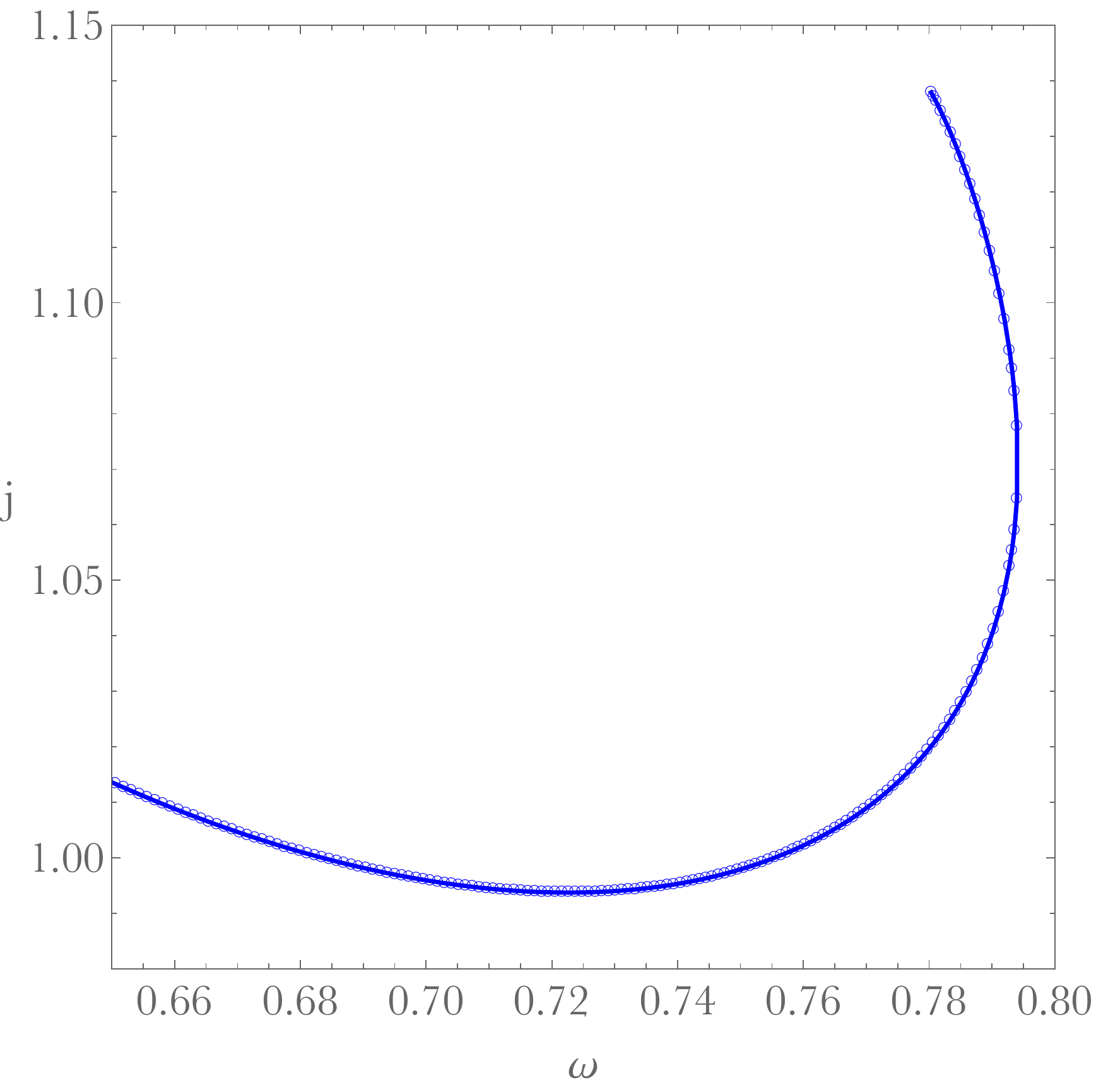}
\hspace{1cm}\includegraphics[width=.45\textwidth]{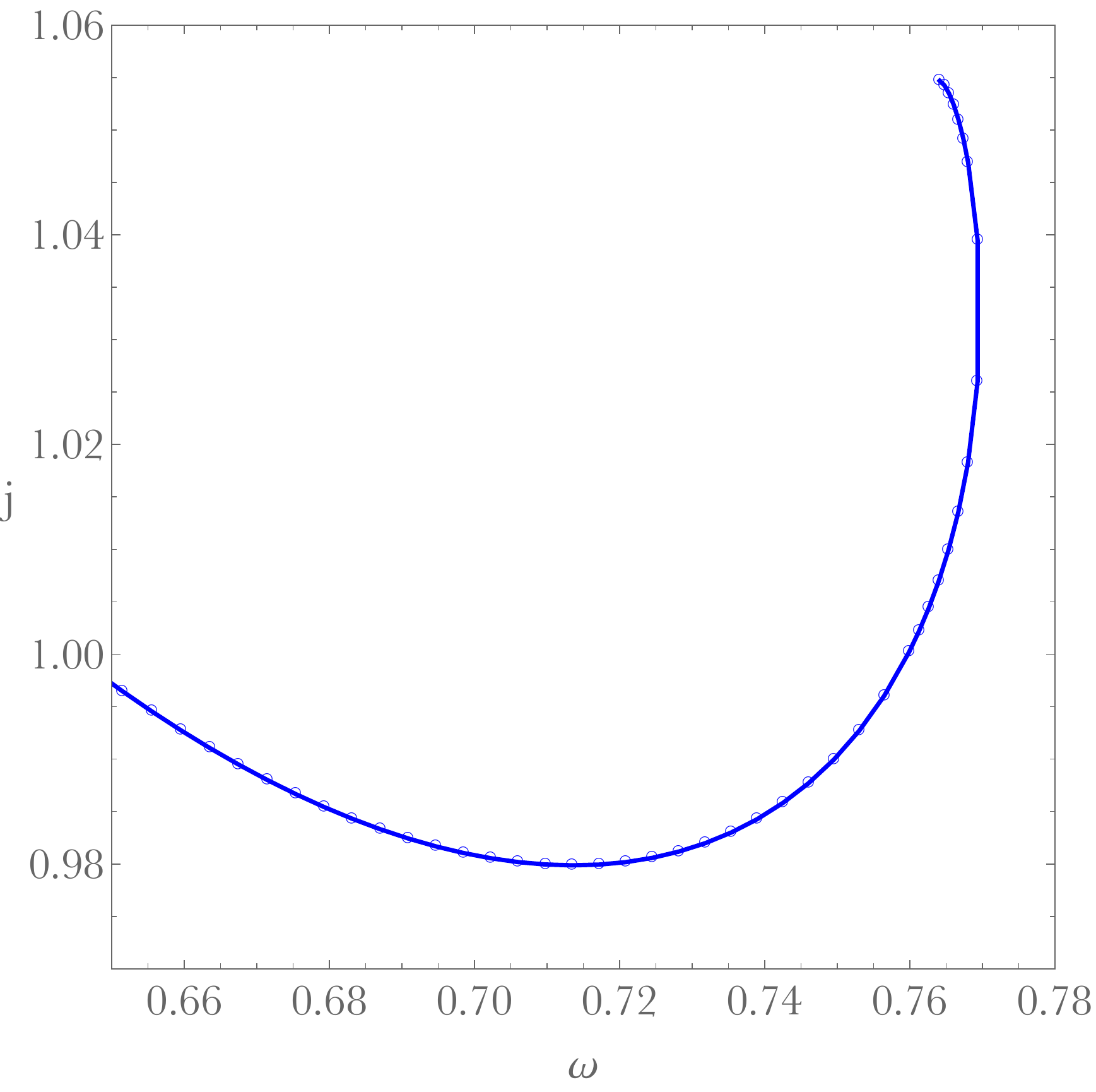}}
\caption{Dimensionless angular momentum $j$ (in mass units) as a function of the angular frequency $\omega$ for $d=6$ (left) and $d=7$ (right).  In both cases, $\omega$ reaches some maximum value, and must be decreased to continue the family of solutions.}
\label{fig:ringdata}
\end{figure}

In the $d=6$ and $d=7$ rings, it turns out that a single $\omega$ can give two physically distinct solutions, and there is a turning point somewhere.  This phenomenon can be seen in Fig.~\ref{fig:ringdata}.  As we follow a line of solutions by increasing $\omega$, we will find a maximum $\omega_{\max}$.  We must decrease $\omega$ in order to continue finding solutions, but we must also avoid backtracking on the solutions we already have.  The strategy for going around this turning point was outlined in section \ref{subset:turning}.

\section{Application: Ultraspinning Lumpy Black Holes in AdS\label{Sec:RRR}}

Now let us apply these methods to a find new solution that has never been constructed in the literature.  We are interested in solutions of Einstein(-AdS) gravity in dimensions $d\geq 6$. Singly-spinning Myers-Perry black holes in $d\geq 6$, including those in AdS have no bound on its angular momentum. For large rotation, these black holes becomes highly deformed and resemble black branes.  Black branes are unstable, so these highly spinning black holes should also be unstable, as first pointed out by Emparan and Myers \cite{Emparan:2003sy}. This instability was confirmed (both in Minkowski  \cite{Emparan:2003sy,Dias:2009iu,Dias:2010maa,Dias:2010eu,Durkee:2010ea,Dias:2011jg,Emparan:2014jca} and AdS \cite{Dias:2010gk}) by solving for linearised perturbations of the Einstein equation, and is called the {\it ultraspinning instability} (see section \ref{subsec:instabilities}).

In a phase diagram of stationary solutions that are asymptotically AdS (or flat), the onset of the instability is associated to a bifurcation to a new branch of axisymmetric black holes that preserve the same $SO(d-3)\times SO(2)$ symmetry as the linear unstable mode \cite{Harmark:2007md,Dias:2009iu,Emparan:2011ve}. Such black holes were coined {\it lumpy} or {\it bumpy} black holes since their horizon is distorted by ripples along the polar direction.  Asymptotically flat lumpy black holes in $d=6$ and $d=7$ were explicitly constructed in \cite{Dias:2014cia,Emparan:2014pra} using the Einstein-DeTurck method with a pseudospectral collocation grid and Newton-Raphson's relaxation, as described in previous sections.

Here, we want to construct some examples of asymptotically AdS$_7$ lumpy black holes that merge with the Myers-Perry-AdS$_7$ black hole at the onset of the AdS ultraspinning instability. We choose to illustrate the $d=7$ case because of its potential relevance for the AdS$_7\times S^4/CFT_6$ correspondence.  For this review, we will focus on technical/numerical details rather than on the physical properties of the system. The latter will be discussed elsewhere. 
 
The Myers-Perry-AdS$_7$ black hole is a solution of Einstein-AdS gravity \eqref{dT:Einstein} that is usually written in Boyer-Lindquist coordinates as
\begin{eqnarray}\label{RRR:ds2MPAdS}
\hspace{-1.5cm}\mathrm{d}s^2&=& 
-\frac{\Delta_r}{\Sigma}\left(\mathrm{d}t-\frac{a\,\sin^2\theta}{\Xi} \mathrm{d}\varphi \right)^2+
\frac{\sin^2\theta\, \Delta_\theta}{\Sigma }\left(\frac{r^2+a^2}{\Xi }\,\mathrm{d}\varphi-a \,\mathrm{d}t\right)^2 
+\frac{\Sigma }{\Delta _r}\,\mathrm{d}r^2 +\frac{\Sigma }{\Delta_\theta}\,\mathrm{d}\theta^2 \nonumber\\
&&+ r^2\,\cos^2\theta \,\mathrm{d}\Omega_3^2,
\end{eqnarray}
where $\Delta_r = \left(r^2+a^2\right) \left(1+\frac{r^2}{L^2}\right)-\frac{r_+^2}{r^2} \left(r_+^2+a^2\right) \left(1+\frac{r_+^2}{L^2}\right)$, $\Delta_\theta=1-\frac{a^2}{L^2}\,\cos^2(\theta)$, $\Sigma=1-\frac{a^2}{L^2}$ and $\Xi=1-\frac{a^2}{L^2}$. It depends on two parameters: the horizon radius $r_+$ and the rotation parameter $a$. The AdS radius $L$ sets the scale of the system.

As it stands, this solution rotates at the boundary of AdS$_7$ with angular velocity $\Omega_\infty=-\frac{a}{L^2}$. This rotating frame is not adequate to discuss the thermodynamics of the solutions \cite{Caldarelli:1999xj,Gibbons:2004ai,Papadimitriou:2005ii}. Therefore, we  perform the coordinate transformation
\begin{equation}\label{RRR:coordtransfPhi}
\varphi = \phi - \frac{a}{L^2}\,t\,,
\end{equation}
so that in the new frame $\{t,r,\theta,\phi,x^i_{S^3}\}$ there is no rotation at the boundary. Indeed, the solution has a boundary metric conformal to the Einstein static universe $R_t\times S^5$,
\begin{equation}\label{RRR:bdy}
\mathrm{d}s^2_{\mathrm{bdy}} = -\mathrm{d}t^2+\mathrm{d}\theta^2+\sin^2\theta \mathrm{d}\phi^2+\cos^2\theta \,\mathrm{d}\Omega_3^2 \,.
\end{equation}

The lumpy AdS$_7$ black hole we wish to construct is also conformal to the Einstein static universe and preserves the same $SO(4)\times SO(2)$ rotational symmetries as the Myers-Perry-AdS$_7$ black hole. Let us define the new coordinates $\{x,y\}$
\begin{equation}\label{RRR:newcoord}
\cos\theta=x \sqrt{2-x^2}\,, \qquad  r=\frac{r_+}{\sqrt{y}}\,,
\end{equation}
which have range $0\leq x\leq 1$ (with an axis of the $S^5$ at $x=0$ and another axis at $x=1$) and $0\leq y\leq 1$ (with the AdS boundary being at $y=0$ and the horizon located at $y=1$). We shall see why this choice of coordinates is convenient when we discuss boundary conditions. Now introduce the dimensionless horizon radius and rotation parameters,
\begin{equation}\label{RRR:adimensional}
y_+=\frac{r_+}{L}\,, \qquad  \alpha =\frac{a}{r_+}\,.
\end{equation}
Note that since the AdS radius $L$ sets the scale of the problem, a more natural pair of dimensionless quantities would be $r_+/L$ and $a/L$. Instead, we choose \eqref{RRR:adimensional}  because these were the quantities used in the linearised search of the ultraspinning instability in \cite{Dias:2010gk}, and we will need the results of \cite{Dias:2010gk} for a Newton-Raphson seed. Moreover,  $\alpha =\frac{a}{r_+}$ was also the quantity used in the construction of the asymptotically flat lumpy black holes (where the scale $L$ is absent). Therefore, with the choice \eqref{RRR:adimensional} the procedure below with $L\to \infty$ outlines the construction of the asymptotically flat lumpy black holes of \cite{Dias:2014cia}  (where many technical details were omitted). 

We want to use the Einstein-DeTurck method, so we require an ansatz and a reference metric.  A general ansatz that is compatible with a boundary metric \eqref{RRR:bdy} that has a horizon at $y=1$ and preserves $SO(4)\times SO(2)$ rotational symmetry is
\begin{eqnarray}\label{RRR:ds2}
&& \hspace{-1.5cm}\mathrm{d}s^2= r_+^2 {\biggl[}-\frac{\Delta_x}{y_+^2 \left(1-y_+^2\alpha^2\right)^2} \, \frac{\Sigma \, \Delta_y}{\rho \, y}\,A (1-y) \mathrm{d}t^2+\left(1-x^2\right)^2\frac{\rho}{\Sigma}\,\frac{S_2}{y}\left(\mathrm{d}\phi-\frac{y^3 \Delta_x}{\rho \,y_+} \,\Omega \, \mathrm{d}t \right)^2\nonumber\\
&& \hspace{0.3cm}+\frac{ \Sigma }{4 \Delta _y}\,\frac{B}{ y^2 (1-y)} \,\mathrm{d}y^2+\frac{4 \Sigma}{\left(2-x^2\right) \Delta _x}\,\frac{S_1}{y} (\mathrm{d}x+F\mathrm{d}y)^2+x^2 \left(2-x^2\right)\,\frac{S_3}{y}\,\mathrm{d}\Omega_3^2  {\biggr]}
\end{eqnarray}
with  
\begin{eqnarray}\label{RRR:ds2aux}
&& \Delta _x=1-y_+^2\alpha ^2 x^2 \left(2-x^2\right)   \,, \qquad \Delta _y=y_+^2 \left[1+\left(1+\alpha ^2\right) y (y+1)\right]+y \left[1+\left(1+\alpha^2\right) y\right],
\nonumber\\
&& \Sigma =1+\alpha ^2 x^2 \left(2-x^2\right) y\,, \qquad 
\rho =\frac{\Delta _x \left(1+\alpha ^2 y\right)^2-\alpha ^2 \left(1-x^2\right)^2 (1-y) \Delta _y}{\left(1-\alpha ^2 y_+^2\right)^2},
\end{eqnarray}
and functions ${\cal F}=\{A,B,\Omega,F,S_1,S_2,S_3\}$ that depend on the radial and polar coordinates, ${\cal F}={\cal F}(x,y)$. The horizon of this solution, at $y=1$, is a Killing horizon generated by the linear combination of the stationary Killing field ${\cal T}=\partial_t$ and the rotational Killing field ${\cal R}=\partial_\phi$, $K=\partial_t +\Omega_H \partial_\phi$, where the horizon angular velocity is
\begin{equation}\label{RRR:OmH}
\Omega_H=-\frac{g_{t\phi}}{g_{\phi\phi}}{\biggl |}_{y=1}=\frac{\alpha \left(1+y_+^2\right)}{1+\alpha^2}\,.
\end{equation}
Furthermore, (anticipating that regularity at the horizon requires the boundary condition $A(x,1)=B(x,1)$) the temperature of this horizon is ($\kappa$ is the surface gravity) 
\begin{equation}\label{RRR:TH}
T_H =\frac{\kappa}{2 \pi}{\biggl |}_{y=1}=\frac{2+\alpha^2+y_+^2 \left(3+2 \alpha^2 \right)}{2 \pi \left(1+\alpha^2\right)}\,.
\end{equation}
This ansatz has the nice property that it reduces to the Myers-Perry-AdS$_7$ black hole when we set  
$A=1, \: B=1, \:  S_1=1,  \:   S_2=1, \:  S_3=1, \:  \Omega\equiv \Omega_0=\frac{\alpha  \left(1+\alpha^2\right) \left(1+y_+^2\right)}{\left(1-\alpha ^2 y_+^2\right)^2}$  and  $F=0$.  

The reference geometry $\bar{g}$ must preserve the same symmetries and must have the same asymptotics and horizon as the lumpy black hole geometry $g$ we wish to construct. A natural choice in the problem at hand is to take the Myers-Perry-AdS$_7$ black hole as the reference geometry. So the reference geometry is given by \eqref{RRR:ds2} with the replacement ${\cal F} \to \bar{\cal F}=\{\bar{A},\bar{B},\bar{\Omega},\bar{F},\bar{S}_1,\bar{S}_2,\bar{S}_3\}$ with 
\begin{equation}\label{RRR:ref}
\bar{A}=1, \:\: \bar{B}=1,\:\:  \bar{S}_1=1, \:\:  \bar{S}_2=1,\:\:  \bar{S}_3=1, \:\:  \bar{\Omega}\equiv \Omega_0=\frac{\alpha \left(1+\alpha^2\right) \left(1+y_+^2\right)}{\left(1-\alpha ^2 y_+^2\right)^2}\,, \:\:  \bar{F}=0 \,.
\end{equation}
As required, this reference geometry has the same  $SO(4)\times SO(2)$ isometry group and the same asymptotics as the lumpy black hole \eqref{RRR:ds2} we will search for. Moreover, it also has a horizon located at $y=1$ with the same angular velocity \eqref{RRR:OmH} and temperature  \eqref{RRR:TH} as the lumpy black hole to be found. 

Our task now is to solve the Einstein-DeTurck equation \eqref{dT:EinsteinDeTurck}. Since  \eqref{RRR:ds2} and the reference metric are of the form of the stationary ansatz \eqref{dT:StationaryAnsatz} adapted to the isometries generated by ${\cal T}=\partial_t$ and ${\cal R}=\partial_\phi$, we know section \ref{Sec:DeTurck5} that the Einstein-DeTurck equation is a manifestly elliptic system of seven PDEs for the unknowns ${\cal F}=\{A,B,\Omega,F,S_1,S_2,S_3\}$. 

It is convenient to introduce new functions related to the original ${\cal F}=\{A,B,\Omega,F,S_1,S_2,S_3\}$ as
\begin{equation}\label{RRR:redefF}
q_1=A, \:\: q_2=B, \:\:  q_3=x(1-x)F, \:\:  q_4=S_1, \:\:  q_5=S_1, \:\:  q_6=S_3, \:\:  q_7=y \,\Omega\,.
\end{equation}
Most of these are trivial relabelings. The factors of $x$ and $1-x$ in the redefinition of $F$ are introduced so that the boundary condition for $q_3$ $x=0$ and $x=1$ will be Neumann, rather than Dirichlet.  The Neumann boundary condition is convenient because it makes closer contact with the analysis made in Section  \ref{Sec:BCs}, but is equally as good as a Dirichlet condition on $F$ instead.  The redefinition $q_7$ needs a justification that we now address while discussing boundary conditions.

At the asymptotic boundary, $y=0$, our metric must reduce to the (Myers-Perry-AdS$_7$) reference metric \eqref{RRR:ref}. Thus we impose the Dirichlet boundary conditions,
\begin{equation}\label{RRR:bdyBC}
\{q_1,\,q_2,\,q_3,\,q_4,\,q_5,\,q_6,\,q_7\}{\bigl |}_{y=0}=\{1,1,0,1,1,1,0\}\,.
\end{equation}
Note that with the choice $q_7=y\, \Omega$ was needed to impose the asymptotic boundary conditions $q_7{\bigl |}_{y=0}=0$. Had we chosen to define $q_7=\Omega$, we would instead impose $q_7{\bigl |}_{y=0}=\Omega_0$ with $\Omega_0$ defined in \eqref{RRR:ref}. We find that the former option yields, in practice, better numerics.

The Boyer-Lindquist-like `isometric'  chart of coordinates $\{t,y,x,\phi\}$ does not cover the fixed point $y=1$ (the Killing horizon) of the isometry generated by the Killing vector field $K=\partial_t +\Omega_H \partial_\phi$. As explained in detail in  section \ref{Sec:BCsH},  we can treat the horizon at $y=1$ as a fictitious boundary where we impose boundary conditions that guarantee that the solution is smooth. This requires the boundary conditions,
\begin{equation}\label{RRR:HorBC}
q_1{\bigl |}_{y=1}=q_2{\bigl |}_{y=1}\,, \qquad q_7{\bigl |}_{y=1}=\Omega_0 \,,
\end{equation}
which, as described in section \ref{Sec:BCsH}, guarantee that the solution has angular velocity and temperature given by \eqref{RRR:OmH} and \eqref{RRR:TH}, respectively. In addition, solving the Einstein-DeTurck equation in a Taylor expansion around $y=1$ we find that that the other functions need to obey Robin (\emph{i.e.}, mixed Dirichlet-Neumann) boundary conditions. These expressions too lengthy to warrant their inclusion here. These boundary conditions are not exactly like those in section \ref{Sec:BCsH} since we are using different coordinates.

The compact Killing field $\partial_\phi$  generates $U(1)$ orbits with period $2\pi$ and has a fixed point at the rotation axis $x=1$. Here, the boundary conditions must ensure regularity at the axis. As explained in section \ref{Sec:BCsAxis}, this implies
\begin{equation}\label{RRR:axisBC}
q_4{\bigl |}_{x=1}=q_5 {\bigl |}_{x=1}\,, \qquad   \{\partial_x q_1,\,\partial_x q_2,\,\partial_x q_3,\,\partial_x q_5,\,\partial_x q_6,\,\partial_x q_7\}{\bigl |}_{x=1}=\{0,0,0,0,0,0\}\,.
\end{equation}
In particular, note the reason we have chosen to work with the angular coordinate $x$ defined in \eqref{RRR:newcoord}: in the vicinity of the axis the $x,\phi$ piece of the line element  reads $ds^2{\bigl |}_{y=1}\sim q_4 \left(d\Theta^2+\frac{q_5}{q_4}\,\Theta^2d\phi^2 \right)$, \emph{i.e.} $g_{\phi\phi}{\bigl |}_{y=1}\propto \Theta^2$ which makes direct contact with the regularity analysis done when discussing \eqref{dT:PolarAxis3}  in section  \ref{Sec:BCsAxis}. (Had we chosen instead $\cos\theta=\tilde{x}$ for instance, we would have $g_{\phi\phi}{\bigl |}_{y=1}\propto \tilde \Theta \equiv (1-\tilde{x})$ which tends to yield more complicated boundary conditions).

There is another axis at $x=0$, where we must also impose regularity.  Just like the other axis, we get the boundary conditions
\begin{equation}\label{RRR:equatorBC}
q_4{\bigl |}_{x=0}=q_6 {\bigl |}_{x=0}\,, \qquad   \{\partial_x q_1,\,\partial_x q_2,\,\partial_x q_3,\,\partial_x q_5,\,\partial_x q_6,\,\partial_x q_7\}{\bigl |}_{x=0}=\{0,0,0,0,0,0\}\,.
\end{equation}

With a set of equations, and boundary conditions, we are now in a position to search for the lumpy black holes numerically, parametrised by $y_+$ and $\alpha$ (which is equivalent to a parametrisation by $\Omega_H$ and $T_H$.  We do so using Newton-Raphson (see section \ref{subsec:newtonraphson}) with pseudospectral collocation on a Chebyshev grid (see appendix \ref{appendix:collocation}).  

But in order for this method to be successful, we require a satisfactory seed solution.  Typically, these trial functions $q_j^{(s)}$ are chosen using an educated guess. In the present case, a good strategy for this guess is as follows. Near the ultraspinning merger the lumpy black hole is perturbatively close to the Myers-Perry-AdS$_7$ black hole, which is also the reference background \eqref{RRR:ref}. Therefore, we can try a seed that is the  Myers-Perry-AdS$_7$ black hole plus a deformation, proportional to an amplitude ${\cal A}$, that depends on $x$ and $y$ and obeys the boundary conditions \eqref{RRR:HorBC}, \eqref{RRR:axisBC} and \eqref{RRR:equatorBC}. We should try different choices for the $x,y$-dependence of this deformation and  for the value of the amplitude. Hopefully, with the right deformation as a seed, the Newton-Raphson solution will converge to a new solution.  This kind of trial and error is a well-known limitation of Newton-Raphson.  

To follow the strategy just outlined, it is necessary to pinpoint the merger curve where the lumpy black holes branch-off from the Myers-Perry-AdS$_7$ black hole. That is, we would like to have the critical curve $\alpha(y_+)$ that describes the onset of the ultraspinning instability. Fortunately, this linear study was done in \cite{Dias:2010gk} where the linearised Einstein equations were solved to look for the zero-modes of the ultraspinning instability. For example, for the Myers-Perry-AdS$_7$ black hole with $y_+=0.3$, the ultraspinning instability is present for any dimensionless values of the rotation $\alpha\geq \alpha_{\rm merger}$ where $\alpha_{\rm merger} \simeq 2.627$. Fixing $y_+=0.3$, we should try a seed that has $\alpha$ around this critical value.   

As we have alluded to in section \ref{subsec:zeromodes}, the sign of the deformation amplitude ${\cal A}$ can matter, with different signs giving different solutions. This matters when, given a linear perturbation $\delta h$ near an onset, there is no discrete symmetry of the background spacetime that can map linear perturbations $\delta h$ to $-\delta h$.  This is the case for the Myers-Perry-AdS$_7$ black holes.  In this case, fixing $y_+$, solutions with $\alpha>\alpha_{\rm merger} $  correspond to positive deformation amplitudes ${\cal A}$.  The resulting lumpy black holes span a space of solutions that joins\footnote{Here we are borrowing ideas from the asymptotically flat system where the black rings were explicitly constructed \cite{Kleihaus:2012xh,Dias:2014cia} and assuming that similar physics applies in AdS. In particular, AdS black rings were construct numerically in \cite{Figueras:2014dta}  for $d=5$.} Myers-Perry-AdS$_7$ black holes, which have spherical $S^5$ topology, to AdS$_7$ black rings which have horizon topology $S^1 \times S^4$. On the other hand, solutions with $\alpha<\alpha_{\rm merger} $ ($\alpha_{\rm merger} \sim 2.627$) correspond to deformation amplitudes ${\cal A}<0$ and describe a second branch of lumpy black holes that are not connected to black rings. In this respect, lumpy AdS black holes are  similar to the asymptotically flat lumpy black holes of \cite{Dias:2014cia,Emparan:2014pra}.

We are now ready to give a specific set of trial functions for the seed. Recalling that for $y_+=0.3$ the ultraspinning onset occurs at $\alpha \simeq 2.627$, we find that the trial functions\footnote{In the simulations and results presented here, we will always describe lumpy black holes with $y_+=0.3$ for definiteness. But these methods can be used for other values of $y_+$.}   
\begin{eqnarray}\label{RRR:seed}
&& q_1^{(s)}=1+{\cal A} \left(1-y^2\right) x^2 \left(2-x^2\right), \nonumber \\
&& q_2^{(s)}=1+{\cal A} \left(1-y^2\right) x^2 \left(2-x^2\right), \nonumber \\
&& q_3^{(s)}={\cal A} \, y \left(1-y^2\right) (1-y) x^2 \left(1-x^2\right)^2, \nonumber \\
&& q_4^{(s)}=1+{\cal A}  \left(1-y^2\right) x^2 \left(2-x^2\right), \nonumber \\
&& q_5^{(s)}=1+{\cal A}  \left(1-y^2\right) x^2 \left(2-x^2\right), \nonumber \\
&& q_6^{(s)}=1+{\cal A}  \left(1-y^2\right) x^2 \left(2-x^2\right), \nonumber \\
&& q_7^{(s)}=\Omega_0 \, y \left[1+ {\cal A}  \left(1-y^2\right) x^2 \left(2-x^2\right)\right],
\end{eqnarray}
with $\alpha=2.680$ and ${\cal A}=1/5$ provide a good seed that causes the Newton-Raphson algorithm to converge.  We emphasise that the choice of these functions is by no means unique, and is just an educated guess.  To give the reader a small idea of the practical implementation of this step, typically, for amplitudes much smaller than ${\cal A}\sim 1/5$, Newton-Raphson converges to a Myers-Perry-AdS$_7$ black hole rather than a new lumpy solution.  In this case the seed is too close to the Myers-Perry-AdS$_7$ black hole's basin of attraction and insufficiently close to that of the lumpy black hole's. Values much larger than ${\cal A}\sim 1/5$ give a seed that is too far from any solution, and Newton-Raphson fails to converge.

Now, once we have a lumpy black hole solution with $\{ y_+=0.3,\alpha=2.680\}$, we can now search for solutions closer to the merger (if $2.627<\alpha<2.680$) or farther away (if $\alpha>2.680$).  We simply use this solution as a seed for a problem with a small change in $\alpha$ ($\delta \alpha \sim10^{-3}$).  In this way, we can explore the parameter-space of solutions.  Sufficiently small steps must be taken here to ensure that the seeds are close enough for Newton-Raphson.

The solutions just described are those that eventually connect to black rings.  We can also construct the other lumpy black hole branch with $\alpha<2.627$ (and $y_+=0.3$) that is not connected to black rings. We can repeat the  procedure described above with $\alpha<2.627$, this time taking a negative amplitude, ${\cal A}<0$, and $\alpha<2.627$ in \eqref{RRR:seed}, or in a similar set of trial functions.  

However, since we already have access to some of the lumpy black hole solutions, we can take an alternative approach to reduce the amount of guesswork.  Take the metric functions $q_j$ of one of the solutions that we already have with $\alpha>2.627$. Call these auxiliary functions $q_j^{(aux)}$. Now consider the construction
\begin{eqnarray}\label{RRR:seedRing}
&& q_j^{(s)}=1+{\cal A}\left( q_j^{(aux)}-1 \right)\,, \quad  \hbox{if} \:\: j\neq 3, \nonumber \\
&& q_3^{(s)}=\tilde{\cal A} \, q_3^{(aux)}\,,
\end{eqnarray}
with a negative amplitude $\cal A$. This should give a good seed for a $\alpha<\alpha_{\rm merger}$ lumpy black hole for some pair of $\alpha,{\cal A}$ after some trial and error. As an example (out of many), choosing $q_j^{(aux)}$ to be the solution with $\alpha=2.658>2.627$, ${\cal A}=-1/9$, and $\tilde{\cal A}=-1$ gives a good seed. Having our first lumpy black hole with $\alpha<2.627$ we can now march in $\alpha$ towards   (if $2.620<\alpha<2.627$) and away ($\alpha<2.620$) from the ultraspinning merger, again with a step of $\delta\alpha=10^{-3}$. This generates the full family of lumpy black holes with $y_+=0.3$ that are not connected to the black ring.    

Near the merger, we find that $q_j(x,y)$ becomes $1$ or $0$, as expected.  Further from the merger, this system is expected to reach various singularities, and curvature invariants grow large.  This makes it increasingly difficult to find numerical solutions.

Once we have solutions, we need to extract physical quantities from them, such as the energy $E$, angular momentum $J$, entropy $S$, and curvature invariants.  To accurately obtain these quantities, we need sufficient numerical resolution.  The resolution is increased steadily until these quantities do not change significantly (for our results we choose $\sim10^{-6}$ for the energy, $\sim10^{-10} $ for the angular momentum, and $\sim10^{-12}$ for the entropy).  The difference in error between these quantities is related to the difficulty of their computation, which we will address below.

Having described in detail the methods we used to construct the lumpy black holes, we are finally ready to present the results. Given the nature of this review, we will not focus much on the physical aspects of the results (to be presented elsewhere) but more on their numerical aspects. 

\begin{figure}
\centering
\includegraphics[width=1.0\textwidth]{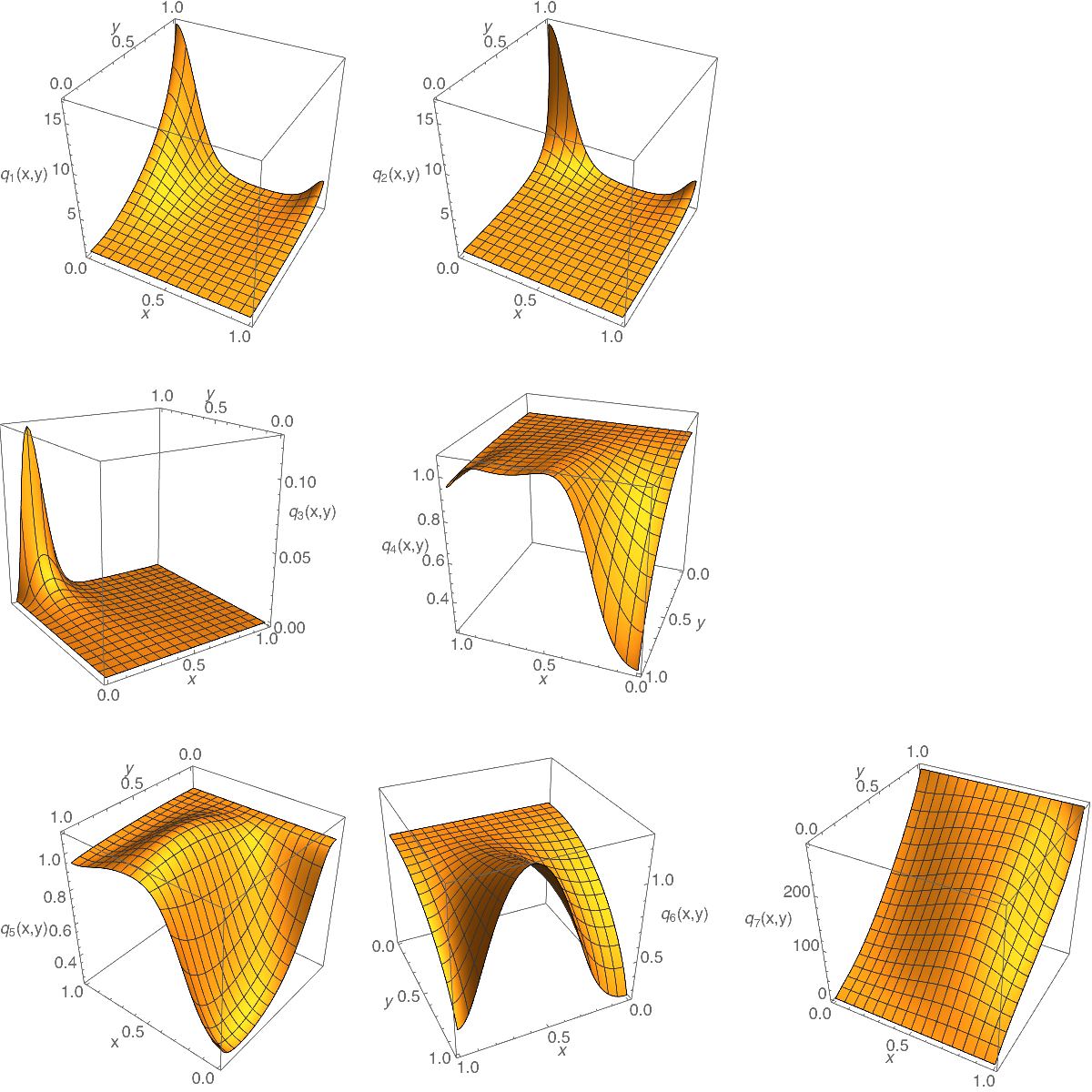}
\caption{Metric functions for the most deformed lumpy black hole whose branch is expected to be connected to the black ring. It has $y_+=0.3$ and $\alpha=2.775$ (\emph{i.e.} with $E=1.457943$ and angular momentum $J=0.805340$). Recall that the Myers-Perry-AdS$_7$ black hole has $q_{1,2,4,5,6}(x,y)=1$, $q_3(x,y)=0$ and $q_7(x,y)=y \Omega_0$. This data uses the resolution $N\times N =57\times 57$.}\label{Fig:RRRfunctionsRing}
\end{figure}

\begin{figure}
\centering
\includegraphics[width=1.0\textwidth]{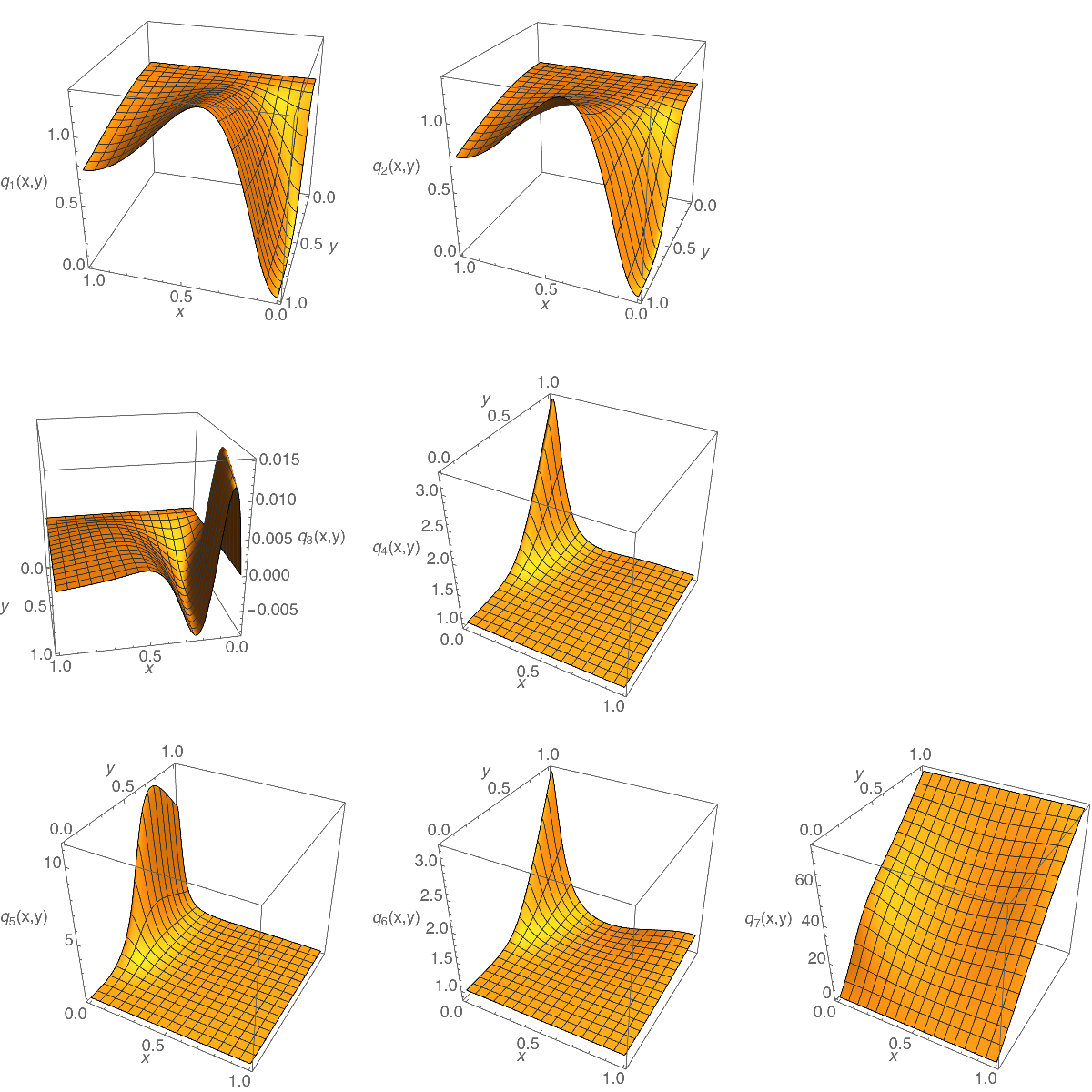}
\caption{Metric functions for the most deformed lumpy black hole whose branch is not connected to the black ring. It has $y_+=0.3$ and $\alpha=2.400$ (\emph{i.e.} with $E=0.547864$ and angular momentum $J=0.225911$). This data uses a resolution $N\times N =61\times 61$.}\label{Fig:RRRfunctionsOther}
\end{figure}

\begin{figure}[ht]
\centering
\includegraphics[width=.6\textwidth]{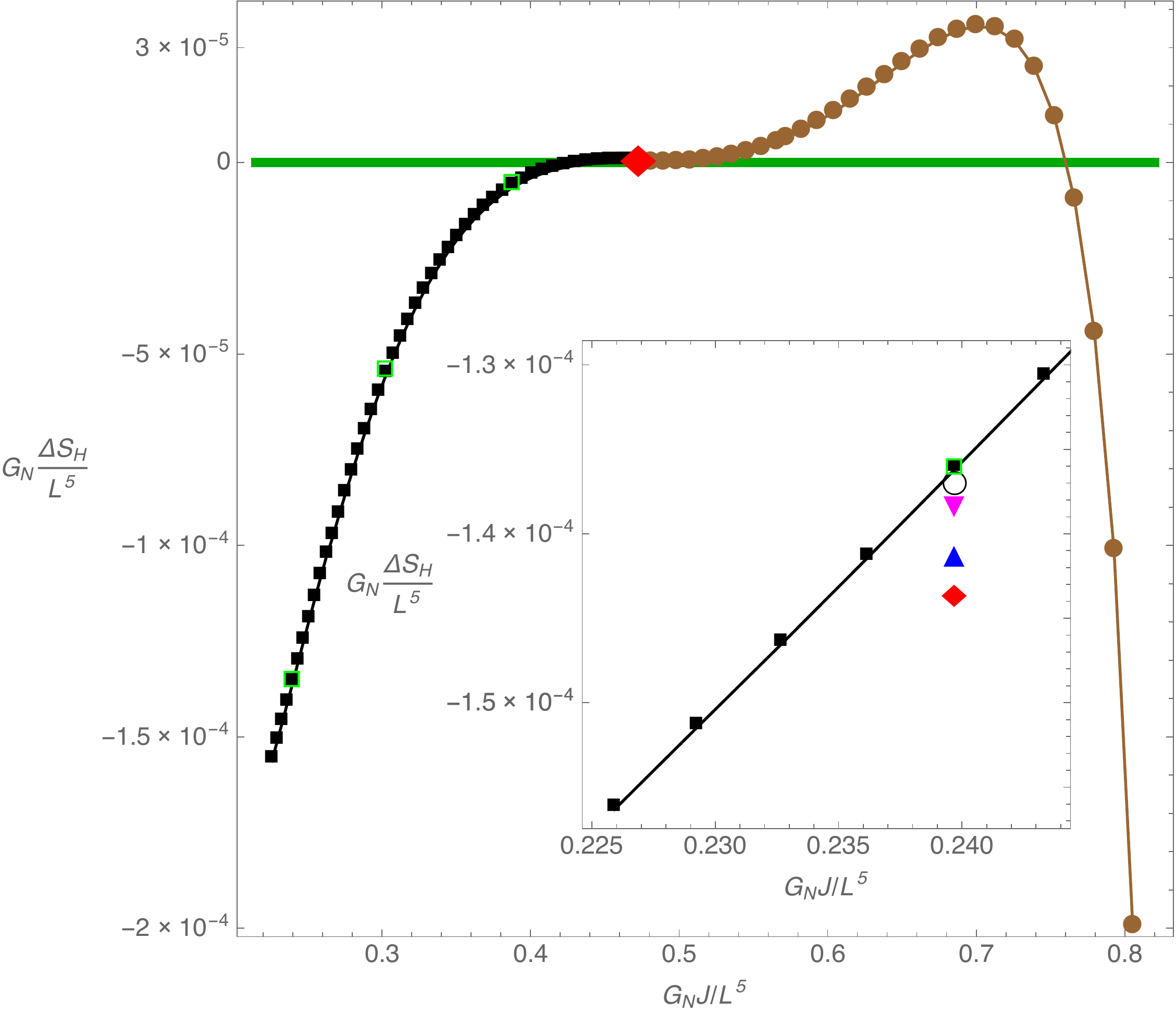}
\caption{Phase diagram of AdS$_7$ black hole solutions in the microcanonical ensemble. We plot the difference in entropy $\Delta S_H$ between a given AdS$_7$ black hole solution and the Myers-Perry-AdS$_7$ black hole with the same energy $E$ and same angular momentum $J$, as a function of $J$ in units of $L$. The horizontal green curve is the singly-spinning Myers-Perry-AdS$_7$ black holes, the red diamond is the onset of the ultraspinning instability,  and the brown dots and black squares describe the two branches of lumpy black holes.  The brown dots are expect to join a black ring family. These used a resolution $N\times N=51 \times 51$ except for the last 9 black squares (on the left) which used $N\times N=61 \times 61$. The inset plot is zoomed in around one of the solutions. The points, in increasing $\Delta S_H$ use resolutions $N=27,31,41,51,59,61$. The $N=59$ and $N=61$ points overlap.}\label{Fig:RRRdSJ}
\end{figure}

The output of the numerical code are the metric functions $q_j$'s that, via the redefinitions \eqref{RRR:redefF}, appear in the ansatz \eqref{RRR:ds2}. These metric functions are not usually presented since what is physically more relevant are the thermodynamic and gauge invariant quantities that can be computed from them. However, the reader may be curious as to how they look. For that we pick the most deformed lumpy black hole of the branch that is expected to be connected to the ring (this is the lumpy black hole with the highest angular momentum that will be later on presented in the phase diagram of Fig.~\ref{Fig:RRRdSJ}). Its seven metric functions are displayed in Fig.~\ref{Fig:RRRfunctionsRing}. For reference, recall that the Myers-Perry-AdS$_7$ black hole has $q_{1,2,4,5,6}(x,y)=1$, $q_3(x,y)=0$ and $q_7(x,y)=y \Omega_0$. On the other hand, and for comparison, the metric functions of the  most deformed lumpy black hole of the other branch (this is the lumpy black hole with the lowest angular momentum in the phase diagram of Fig.~\ref{Fig:RRRdSJ}) are shown in  Fig.~\ref{Fig:RRRfunctionsOther}.

To compute the physical conserved charges for these solutions, we use the holographic renormalisation method \cite{deHaro:2000xn,Henningson:1998gx}, which is specific to asymptotically AdS solutions. First we need to take the line element \eqref{RRR:ds2} and expand it in Fefferman-Graham coordinates $\{z,X\}$ with $X=\{t,\chi,\phi\}$  around the boundary at $z=0$,
\begin{equation} \label{RRR:FGds2}
\dd s^2= \frac{L^2}{z^2} \left[ \dd z^2 +  g_{ij}(z,X)  \dd X^i \dd X^j\right]
\end{equation}
where the $6$-dimensional metric reads
\begin{equation} \label{RRR:FGds2}
g(z,X)=g_{(0)} + z^2 g_{(2)} + z^4 g_{(4)} +z^6 g_{(6)} + h_{(6)} z^{6} \log z^2 +
\mathcal{O}(z^{7})
\end{equation}
with $g_{(n)}$ and $h_{(6)}$ being functions of $X$ (\emph{i.e.} of $\chi$ since $t$ and $\phi$ are isometric directions). To expand \eqref{RRR:ds2} asymptotically as in \eqref{RRR:FGds2} on needs a Fefferman-Graham coordinate transformation $\{y,x\} \to \{z,\chi\}$ of the type 
\begin{eqnarray} \label{RRR:FGds2}
&& y=a_0(\chi)\left( z^2+\sum_{k=3}^n a_k(\chi) z^k \right), \nonumber \\
&& x=b_0(\chi) \left( 1+ \sum_{k=1}^n b_k(\chi) z^k \right),
\end{eqnarray}
where the functions $a_k$ and $b_k$ are determined requiring that at each order one has $g_{z X}=0$ and $g_{zz}=1$. One also chooses the normalization $g_{tt}=-1 + \mathcal{O}(z^2)$.

In these conditions, it follows from the holographic renormalisation procedure \cite{deHaro:2000xn,Henningson:1998gx} that the holographic stress tensor is given by
\begin{equation} \label{RRR:holoTab}
\left\langle T_{ij} \right\rangle={3  \over 8 \pi G_N}\, 
\left(g_{(6) ij}-A_{(6)ij}+{1\over 24}S_{ij} \right),
\end{equation}
with \footnote{Recall from footnote \ref{foot:Rconventions} that our curvature convention is such that the AdS Ricci scalar is negative. It is thus opposite to the convention used in \cite{deHaro:2000xn,Henningson:1998gx}. Accordingly, in our expressions, the terms proportional to the curvature have opposite sign to those derived in \cite{deHaro:2000xn,Henningson:1998gx}. Moreover, all contractions and all indices are raised and lowered with the metric $g_{(0)}$.} 
\begin{eqnarray} \label{RRR:holoTabaux}
\hspace{-1cm} A_{(6) ij} &=& {1 \over 3} {\biggl \{}
2\left(g_{(2)} g_{(4)}\right)_{ij}+\left(g_{(4)} g_{(2)}\right)_{ij}-\left(g_{(2)}^3\right)_{ij} 
+ {1 \over 8}\,\left[ \hbox{Tr}\, g_{(2)}^2 - \left( \hbox{Tr}\, g_{(2)}\right)^2\right]\, g_{(2) ij} \nonumber \\
\hspace{-1cm} &&\hspace{0.5cm}-\hbox{Tr}\, g_{(2)}\,\left[ g_{(4)ij} - \frac{1}{2} \left( g_{(2)}^2 \right)_{ij} \right] \nonumber \\
\hspace{-1cm} && \hspace{0.5cm}
-\left[{1 \over 8} \hbox{Tr}\, g_{(2)}^2 \hbox{Tr}\, g_{(2)} - {1 \over 24} \left(\hbox{Tr}\, g_{(2)}\right)^3
-{1 \over 6} \hbox{Tr}\, g_{(2)}^3
+{1\over 2} \hbox{Tr} \, (g_{(2)}g_{(4)})\right]\,g_{(0) ij}  {\biggr \}}, \nonumber \\
\hspace{-1cm} 
S_{ij}&=& \nabla^2C_{ij}+2R^{k \ l}_{\ i \ j} C_{kl}
+4\left(g_{(2)}g_{(4)}-g_{(4)}g_{(2)}\right)_{ij}
+{1\over 10}\left(\nabla_i\nabla_j B
-g_{(0)ij}\nabla^2 B\right) \nonumber \\
\hspace{-1cm} &&
+{2\over 5}g_{(2)ij}B+g_{(0)ij}\left(-{2\over 3}\hbox{Tr} \, g_{(2)}^3
-{4\over 15}(\hbox{Tr} \,g_{(2)})^3+
{3\over 5}\hbox{Tr} \, g_{(2)}\hbox{Tr} \, g^2_{(2)}\right)\,, \nonumber\\
\hspace{-1cm} 
 C_{ij}&=&\left( g_{(4)}-{1\over 2}g^2_{(2)}+{1\over 4}g_{(2)}\hbox{Tr} \,g_{(2)}\right)_{ij}+ {1\over 8}g_{(0)ij}B \,, \qquad \quad
B=\hbox{Tr} \, g^2_2-(\hbox{Tr} \, g_2)^2\,,
\end{eqnarray}
where $\hbox{Tr} \, g_{(2)} =g_{(0)}^{ij} g_{ij}^{(2)}$, $\hbox{Tr} [ g_{(2)}^2 ]=g_{(2)}^{ij} g_{ij}^{(2)}$ and $\left( g_{(2)} g_{(4)}\right)_{ij}  =g_i^{(2) k} g_{kj}^{(4)}$.  The holographic stress is covariantly conserved, $\nabla_i \left\langle T^{ij} \right\rangle =0$ (with the covariant derivative taken with respect to $g_{(0)}$), and its trace is proportional to the conformal anomaly $a_{(6)}$ (see (28) of \cite{Henningson:1998gx}) which vanishes in our case, $\left\langle T_{i}^{\:\:i} \right \rangle=-\frac{a_{(6)}}{8 \pi G_N}=0$.

The energy $E$ and angular momentum $J$ can be computed by integrating the holographic stress tensor. This is done by pulling-back $\langle T_{ij} \rangle$ to a 5-dimensional spatial hypersurface  $\Sigma_t$ (\emph{i.e.} with $z=0$ and $t=constant$), with unit normal $n$ and induced metric $\sigma^{ij}=g_{(0)}^{ij}+n^{i}n^{j}$. To get the energy (angular momentum) we contract this quantity with the Killing vector ${\cal T}=\partial_t$ (${\cal R}=\partial_\phi$) that generates time (rotational) translations. The integral of this contraction gives the energy (angular momentum) 
\begin{eqnarray} \label{RRR:lumpyThermo0}
E&=&-\int_{\Sigma_t}\sqrt{\sigma} \langle T_i^j \rangle \, {\cal T}^i \,n_j\,, \nonumber\\
J&=&\int_{\Sigma_t}\sqrt{\sigma} \langle T_i^j \rangle {\cal R}^i \, n_j\,.
\end{eqnarray}

On the other hand, the entropy is just the horizon area divided by $4\pi$. Altogether, the entropy $S_H$, energy $E$ and angular momentum $J$ of the lumpy AdS$_7$ black hole are given by,
\begin{eqnarray} \label{RRR:lumpyThermo}
S_H &=& \frac{L^5}{G_N}\,\frac{2 \pi^3 y_+^5 \left(1+\alpha ^2\right)}{1-y_+^2\alpha^2}
\int _0^1\dd x\,x^3\left(2-3 x^2+x^4 \right)\sqrt{q_4(x,1)\,q_5(x,1)}\,q_6(x,1){}^{\frac{3}{2}}\,,\nonumber\\
E&=& \frac{L^4}{G_N}\,\frac{\pi ^2}{8}\, \left(1-y_+^2 \alpha^2\right){}^2 
\int _0^1 \dd x \,\frac{x^3 \left(2-3 x^2+x^4 \right)}{\left[1-y_+^2 \alpha ^2\, x^2 \left(2-x^2\right) \right]{}^3}\,  
{\biggl \{} -\frac{5}{2}\nonumber\\
&&\hspace{0.5cm} +\frac{4 y_+^4 \left[1- y_+^2\alpha^2 x^2 \left(2-x^2\right)\right]^3}{\left(1-y_+^2\alpha^2\right)^4}{\biggl [} \left(1+y_+^2\right)\left(1+\alpha^2\right) {\biggl (} 5+y_+^2 \alpha ^2\left[1-6 x^2 \left(2-x^2\right)\right] {\biggr )} \nonumber\\
&&\hspace{0.5cm}-y_+^2 \left(1-y_+^2\alpha^2\right) \partial_y^3 q_1(x,0)
{\biggr ]}
{\biggr \}}
\nonumber\\
J&=& \frac{L^5}{G_N}\, 3 \pi ^2 y_+^5 \int _0^1 \dd x\, x^3 \left(2-x^2\right) \left(1-x^2\right)^3 \partial_y q_7(x,0)\,.
\end{eqnarray}
If we set $q_1=1,q_4=1,q_5=1,q_6=1$ and $q_7=y \Omega_0$, with $\Omega_0$ defined in \eqref{RRR:ref}, we recover the entropy, energy and angular momentum of the Myers-Perry-AdS$_7$ black hole,
\begin{eqnarray}\label{RRR:MPthermo}
&& S_{\rm{MP}}=\frac{L^5}{G_N}\, \frac{\pi^3 \, y_+^5 \left(1+\alpha^2 \right)}{4 \left(1-y_+^2 \alpha^2 \right)} \nonumber\\
&& E_{\rm{MP}}=\frac{L^4}{G_N} \, \frac{\pi^2 \,y_+^4 \left(1+y_+^2\right) \left(1+\alpha^2\right) \left(5-3 y_+^2 \alpha^2 \right)}{16 \left(1-y_+^2\alpha^2 \right)^2} \nonumber\\
&& J_{\rm{MP}}=\frac{L^5}{G_N}\,\frac{\pi^2 \, \alpha\, y_+^5 \left(1+\alpha^2\right) \left(1+y_+^2\right)}{8 \left(1-y_+^2\alpha^2\right)^2}
\end{eqnarray}
and its angular velocity and temperature are given by  \eqref{RRR:OmH} and \eqref{RRR:TH}. 

Fig.~\ref{Fig:RRRdSJ} describes the phase diagram of the asymptotically AdS$_7$ stationary black hole solutions. This is the phase diagram in the microcanonical ensemble, whereby we fix the energy $E$ and angular momentum $J$ and compute the entropy $S_H$. The thermodynamically favoured solution is the one with higher entropy. The comparison is made in terms of dimensionless quantities, namely $G_N E/L^4$, $G_N J/L^5$, $G_N S_H/L^5$. The difference in entropy between the lumpy and Myers-Perry-AdS$_7$ black hole is extremely small. Therefore, to better illustrate the results,  we take the difference between the entropies of the lumpy AdS$_7$ black hole and the  Myers-Perry-AdS$_7$ black hole at the same energy and angular momentum. To find the entropy of the Myers-Perry-AdS$_7$ black hole at the same $E$ and $J$ as the lumpy black hole, we just need to solve the two last relations of \eqref{RRR:MPthermo}  with respect to $y_+$ and $\alpha$ and then insert these in the relation \eqref{RRR:MPthermo} for the entropy. 

To illustrate how crucial it is to increase the grid resolution $N$ (recall we use Chebyshev grids) and to identify the resolution where we can stop, in the inset plot of Fig.~\ref{Fig:RRRdSJ} we pick one of the solutions and we show how its thermodynamic quantities change as the resolution $N$ is varied. Note that the results for $N=59$ and $N=61$ already match which signals that we can conclude our numerical runs.  

\begin{figure}[th]
\centering
\includegraphics[width=.49\textwidth]{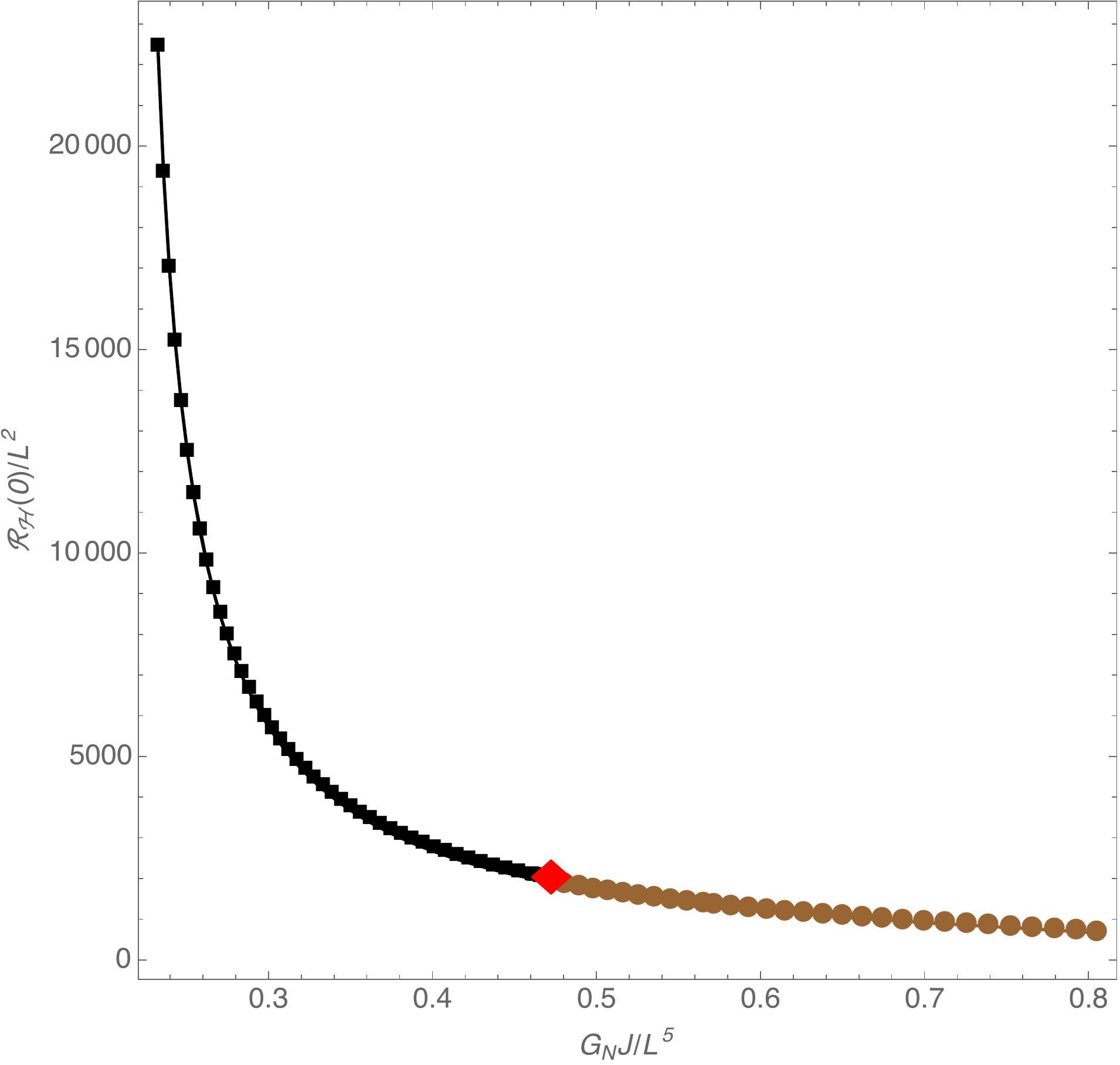}\hspace{0.5cm}
\includegraphics[width=.47\textwidth]{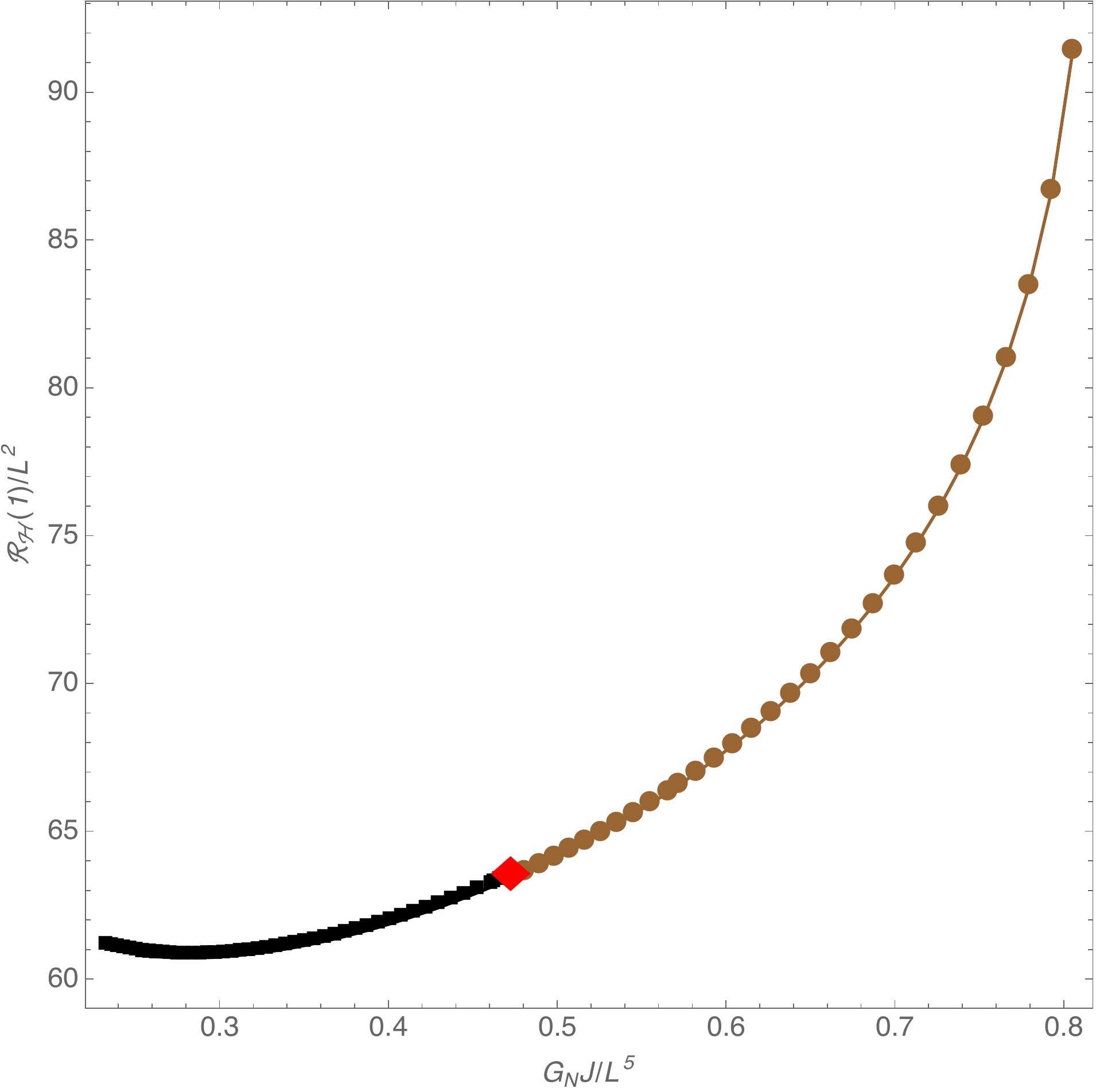}
\caption{Pull-back of the Ricci scalar evaluated at the horizon and at $x=0$ (left panel) and $x=1$  (right panel).}\label{Fig:RRRricci}
\end{figure}

A non-trivial check of our non-linear code is the fact that the merger of the lumpy black holes with the Myers-Perry-AdS$_7$ black hole occurs (red diamond in  Fig.~\ref{Fig:RRRdSJ}) precisely at the critical rotation $\alpha=2.627$ (for $y_+=0.3$) that was found using linear perturbation theory in  \cite{Dias:2010gk}. Moving away from this merger, in either directions, the lumpy black holes become increasingly deformed along the polar direction. Eventually, continuing to find solutions becomes difficult because curvature invariants are getting large. For the branch that connects to the black ring (brown disks), the pull-back of the Ricci scalar evaluated at the horizon ($y=1$) and at $x=1$ grows large (see right panel of Fig.~\ref{Fig:RRRricci}). On the other hand, for the other lumpy branch (black squares) it is the  Ricci scalar evaluated at the horizon and at the axes $x=0$ that grows large (see left panel of Fig.~\ref{Fig:RRRricci}). The fact that this curvature invariant diverges at different polar locations distinguishes the two branches of lumpy black holes. 

There are a number of other numerical checks that can be done.  One is to verify that the first law of thermodynamics $dE=T_H dS_H +\Omega_H J$ is satisfied. Since we are working with solutions at constant $y_+=0.3$, the first law is obeyed if,
\begin{equation}\label{RRR:firstlaw}
1-T_H\, \frac{\partial_\alpha S_H}{\partial_\alpha E}-\Omega _H \,\frac{\partial_\alpha J}{\partial_\alpha E}=0\,.
\end{equation}
In the worst cases, we find that \eqref{RRR:firstlaw} is satisfied to an error of $10^{-6}$. \footnote{We mention that for a given spacetime dimension, the asymptotic charges in AdS are more difficult to extract than in flat space \cite{Dias:2014cia}. Indeed, for the same grid resolution of $N \times N =51\times 51$, the asymptotically flat case satisfies the first law two orders of magnitude better than the AdS case. Note that for asymptotically flat solutions, we can also test the numerical results against the Smarr relation, $\frac{d-3}{d-2}\,E=T_H\,S_H+\Omega_H J$.}

\begin{figure}[ht]
\centering
\includegraphics[width=.5\textwidth]{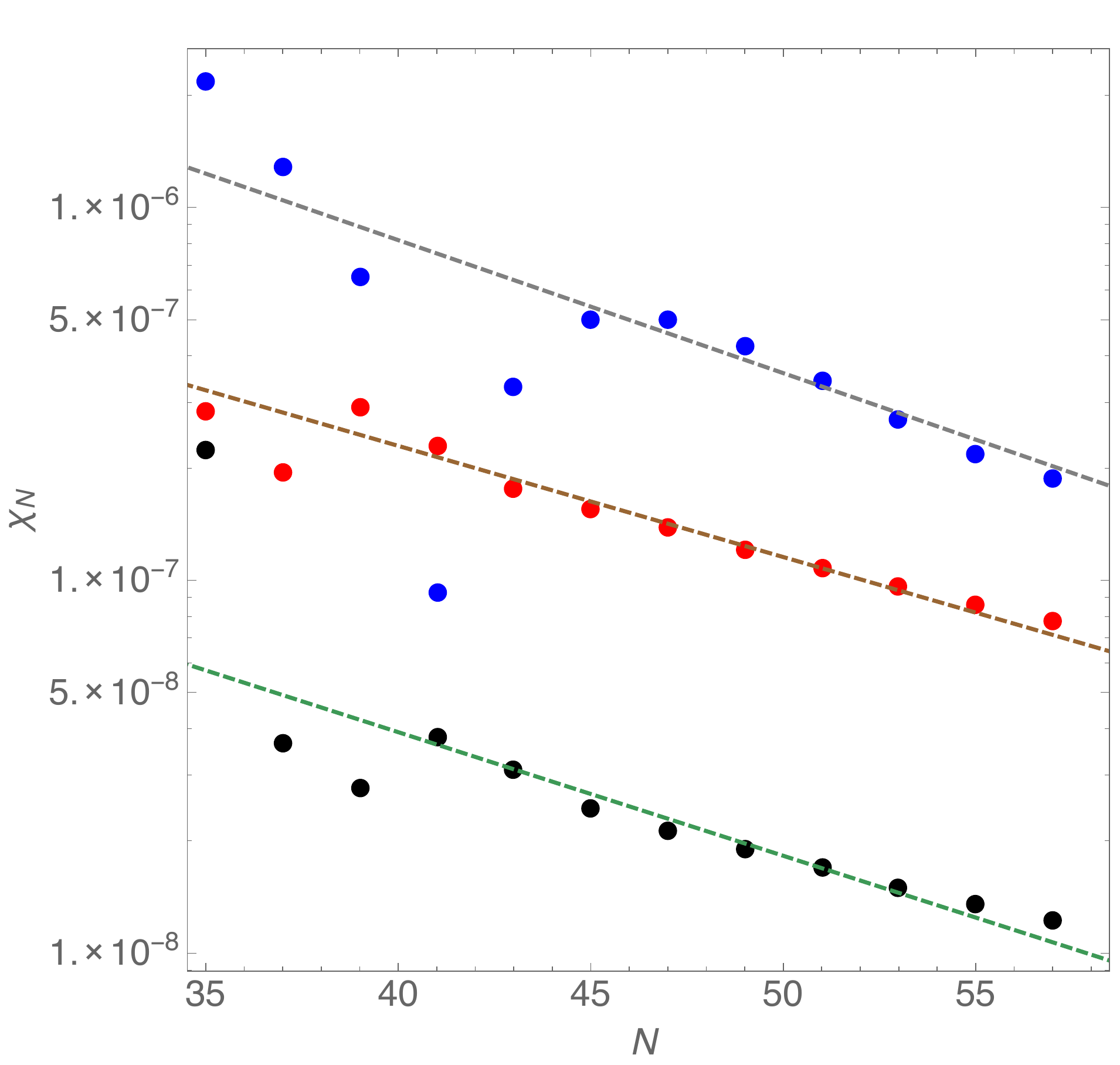}\hspace{0.5cm}
\includegraphics[width=.45\textwidth]{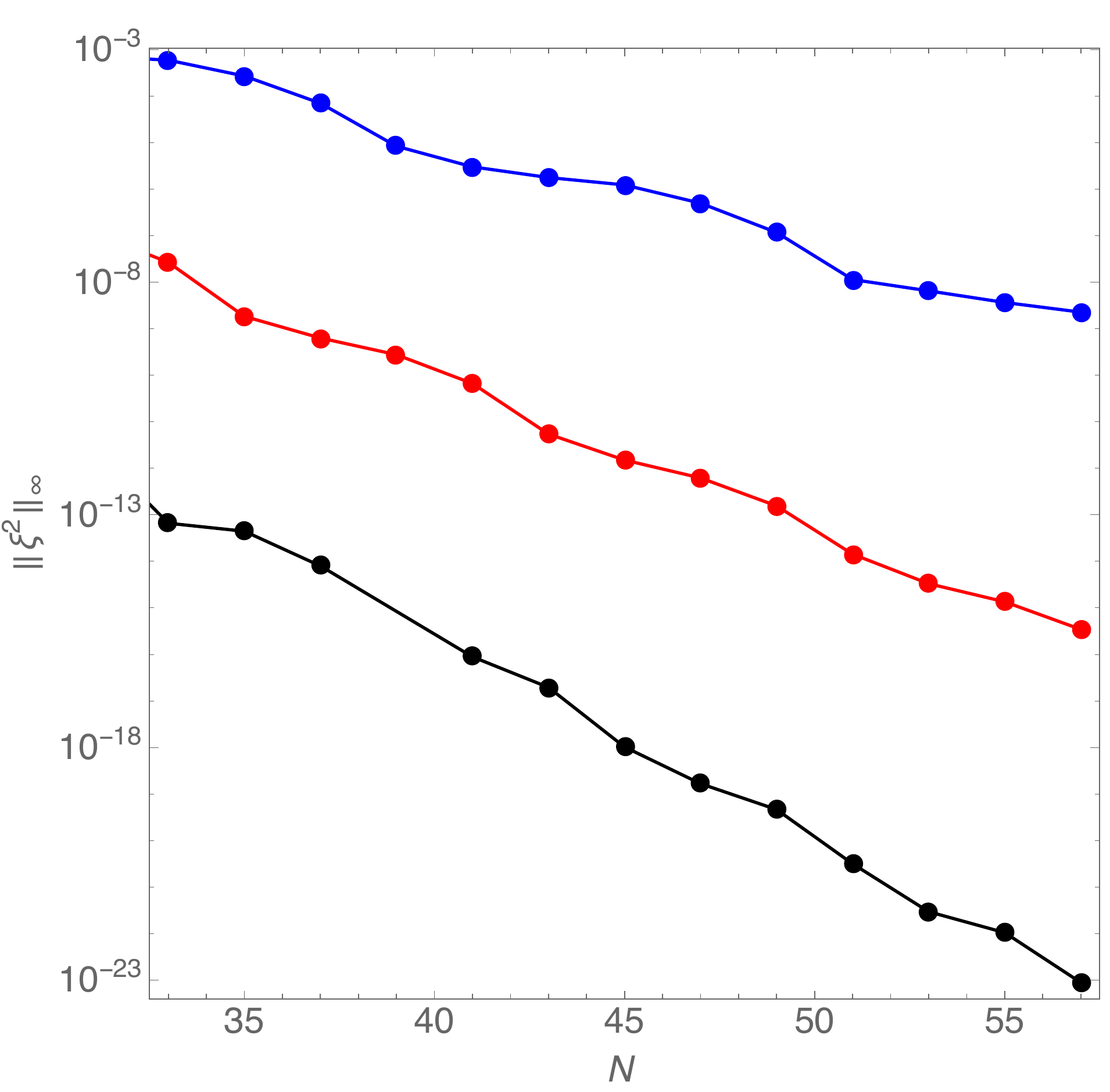}
\caption{Left panel: Convergence test for the lumpy AdS$_7$ black holes. This is a log plot of the error \eqref{RRR:conv} as a function of the number of grid points $N$. From bottom to top, these are for $\alpha = 2.570,\, 2.495,$ and $2.420$ (with $y_+=0.3$). These are the green squares in Fig.~\ref{Fig:RRRdSJ}.  Right panel: Convergence test of the infinite norm of the DeTurck vector $\left\| \xi^2\right\|_\infty$ as a function of the grid points $N$. The colour code is the same as the left panel.}\label{fig:converge}
\label{Fig:convtest}
\end{figure}

We also check the numerical convergence of our code. Since we are using pseudospectral collocation methods, we expect to find exponential convergence as the number of points $N$ is increased. We check this exponential convergence using the energy of the solution since this is the quantity that is  prone to higher numerical errors. This is because it requires taking a third-order derivative, see \eqref{RRR:lumpyThermo} (the angular momentum requires a first derivative and the entropy is read directly from the metric functions at the horizon). In the left panel of Fig.~\ref{Fig:convtest} we plot the relative error quantity
\begin{equation}\label{RRR:conv}
\chi_N \equiv {\biggl |}1-\frac{||E_N||_\infty}{||E_{N+2}||_\infty}{\biggr |}
\end{equation}
as a function of $N$ for three different lumpy black holes, namely for  black holes with $\alpha = 2.570,\, 2.495,$ and $2.420$ (and $y_+=0.3$; these are the solutions with a green square borderline in Fig.~\ref{Fig:RRRdSJ}). Note that this is a log plot so the a straight line indicates that the code indeed converges exponentially. 

Another useful quantity to test convergence is the norm of the DeTurck vector $\xi$. For the boundary value problems considered here, the solutions found are necessarily solutions of the vacuum Einstein equations in the gauge $\xi^\mu=0$ (ie there are no Ricci solitons), so the norm of the DeTurck vector is a measure of how well the gauge condition is satisfied.  On the right panel of Fig.~\ref{fig:converge}, we take the same three lumpy solutions of the left panel and we show the log plot of the largest value of the norm of the DeTurck vector  $\xi^2$ as a function of the grid points $N$.  We again see exponential convergence. In the worst cases the DeTurck norm vanishes with an error smaller than $10^{-8}$.

\appendix
\section{Collocation Methods}\label{appendix:collocation}

In this review, we have described the numerical methods up until a linear equation is obtained (either through linear perturbation theory or Newton-Raphson).  We have not given more details there because at this point in the computation, there is a vast number of well-documented numerical methods available, and the numerical practitioner is encouraged to use their own method of choice.  But for the reader who is not well-versed in these methods, we give here a particular class of methods that can be used. The methods we describe here are collocation methods, see \emph{e.g.} \cite{CanutoBook,Trefethen,Boyd,Grandclement:2007sb}. 

\subsection{Differentiation Matrices}
To illustrate collocation methods, let us begin with a single coordinate $x$. The general strategy is to place a number of points on the domain, $x_i$ called collocation points.  Given the value of a function at these points $f_i\equiv f(x_i)$, its derivatives at these points $f^{(n)}_i\equiv f^{(n)}(x_i)$ are approximated by differentiating an interpolation function (usually based on polynomial or trigonometric interpolation).  This operation reduces the derivative to a matrix operation $f^{(n)}_i\approx D^{(n)}_{ij}f_j$, where $D^{(n)}$ is the $n$-th order differentiation matrix.  This allows us to discretise linear differential equations, and reduce them to standard linear algebra problems.  The various different collocation methods differ only in their choice of collocation points $x_i$, and the type of functions to use for interpolation. 

The simplest collocation method is the finite difference family.  The collocation points are evenly spaced $x_0,\ldots, x_N$, with $x_{i+1}-x_i=\delta x$.  Finite differences uses polynomial interpolation.  

Let us consider second-order central differencing.  At each point $x_i$, a second-order polynomial is constructed from the locations $x_{i-1}$, $x_i$, $x_{i+1}$, and the function values $f_{i-1}$, $f_i$, $f_{i+1}$. This polynomial is then differentiated (up to twice) at $x_i$ to approximate $f^{(1)}$ and $f^{(2)}$.  The result of this is
\be\label{centraldifferencing}
f^{(1)}_i=\frac{f_{i+1}-f_{i-1}}{2\delta x}+O(\delta x^2),\qquad f^{(2)}_i=\frac{f_{i+1}-2f_i+f_{i-1}}{\delta x^2}+O(\delta x^2)\;.
\ee
By construction, third and higher order derivatives always vanish, and so cannot be computed with this method.    If the domain is periodic, one can identify $f_0=f_N$ and $f_{1}=f_{N+1}$.  Otherwise, central differences are not well-defined on the edges.  Then at $x_0$, we can interpolate with the points $x_0$, $x_1$, and $x_2$, and at $x_N$ we can use the points $x_N$, $x_{N-1}$, and $x_{N-2}$.  These are called (second-order) forward differencing and backward differencing, respectively.  The result of forward and backward difference at the point $x_i$ is given by
\be
f^{(1)}_i=\pm\frac{-f_{i\pm2}+4f_{i\pm1}-3f_i}{2\delta x}+O(\delta x^2),\qquad f^{(2)}_i=\frac{f_{i\pm2}-2f_{i\pm1}+f_i}{\delta x^2}+O(\delta x^2)\;.
\ee

From these formulas, we can construct a differentiation matrix such as
\be
{D^{(1)}}=
\frac{1}{2\delta x}\begin{bmatrix}
  -3 & 4 & -1  \\
   -1 & 0 & 1  \\
       & -1 & 0 & 1  \\
       &      &    &    &\ddots\\
       &      &    &    &           &-1&0&1 \\
        &      &    &    &           &   &-1&0&1 \\
         &      &    &    &          &   & 1&-4&3 
 \end{bmatrix}\;,
\ee
which is the first differentiation matrix for second-order finite differences.  

Higher-order differences can be derived by using higher order polynomials, which uses more neighbouring collocation points to construct the interpolant.  It is typical to use even-order differencing so that interpolants can be centred on collocation points.  Again, some form of forward and backward difference must be used near the edges. The fourth-order central differencing is given by
\begin{equation}
f_i^{(1)} = \frac{1}{12\,\delta x}(f_{i-2}-8 f_{i-1}+8 f_{i+1}-f_{i+2})\,.
\end{equation}
There are equivalent forward and backward differencing formulae as well.

For functions that are sufficiently differentiable, finite difference methods converge in a power-law fashion, with the power being equal to the order of the finite differencing.  Higher order methods converge more quickly, but yield differentiation matrices that are less sparse.  

Let us now describe pseudospectral collocation.  The strategy is to use all available points to form an interpolation function. This increasing interpolation order with grid size yields (if done properly) exponential convergence, but dense matrices.

For periodic domains, trigonometric interpolation functions are used on an evenly-spaced grid.   For simplicity, we will take the domain to lie in $(0,2\pi]$.  The matrices for other domains can be obtained by a scaling.  We will use an equidistant grid with $x_1=\delta x,\ldots, x_N=2\pi$.  By convention, we take $N$ to be even.  There are equivalent formulae for odd $N$, but in practice a single grid point makes little difference. Fitting a minimum-order trigonometric interpolant with the entire grid yields a differentiation matrix that is a Toeplitz matrix where the last column given by
\be
D^{(1)}_{iN}=
    \left\{
     \begin{array}{lr}
       0 & :i=0 \mod N\\
       \half(-1)^i\cot(i\delta x/2) & :i\neq 0 \mod N
     \end{array}
   \right.
   .
\ee
The entire matrix can be presented as
\be
{D^{(1)}}=
\begin{bmatrix}
  	          0                 &             &              &           &               &-\half\cot(\delta x/2)\\
   -\half\cot(\delta x/2) &\ddots &            & \ddots &           & \half\cot(2\delta x/2)  \\
    \half\cot(2\delta x/2)&            & \ddots&            &               &-\half\cot(3\delta x/2)\\
   -\half\cot(3\delta x/2)&             &            &\ddots&           &  \vdots \\
      \vdots                       &            & \ddots&           &\ddots& \half\cot(\delta x/2)\\
      \half\cot(\delta x/2) &              &            &              &          &  0
 \end{bmatrix}\;.
\ee
The second derivative matrix is again a Toeplitz matrix with last column defined by 
\be
D^{(2)}_{iN}=
    \left\{
     \begin{array}{lr}
       \frac{\pi^2}{3\delta x}-\frac{1}{6} & :i=0 \mod N\\
      -\frac{(-1)^i}{2\sin^2(i\delta x/2)} & :i\neq 0 \mod N
     \end{array}
   \right.\;.
\ee

Let us now consider non periodic domains $x\in[x_-,x_+]$.  The strategy of taking high-order polynomial interpolates will not work on equidistant grids because of the Runge phenomenon.  Even when interpolating a smooth, non-oscillatory function, large oscillations in the interpolant can appear near the edges of the interval and grow with increasing interpolation order.  This will spoil the accuracy to this method.  The solution is to use unevenly spaced grids that cluster near the edges.  There are several such grids available, but the most commonly used grid for pseudospectral collocation uses the Chebyshev-Gauss-Lobatto (often shorted to Chebyshev or CGL) collocation points:
\be\label{eq:cheb}
x_j=\frac{x_++x_-}{2}+\frac{x_+-x_-}{2}\cos\left(\frac{j\pi}{N}\right),\qquad j=0,1,\ldots, N\;.
\ee
Note that by convention, this gives $x_0=x_+$ and $x_N=x_-$, so grid points are ordered in reverse fashion.  A pseudospectral method on these points has been proven to achieve exponential convergence.  Now deriving differentiation matrices from polynomial interpolants on the whole grid gives us
\begin{align}\label{eq:chebD}
D^{(1)}_{jj}&=\sum_{k\neq j}\frac{1}{x_j-x_k}\nonumber\\
D^{(1)}_{ij}&=\frac{a_i}{a_j(x_i-x_j)}\;,\qquad (i\neq j)
\end{align}
where 
\begin{equation}
a_j=\prod_{k\neq j}(x_j-x_k)\;.
\end{equation}
Higher derivatives can be also be derived from interpolants, but in practice taking powers of first-derivative matrices $D^{(n)}=(D^{(1)})^n$ is equally good.

In this review, the pseudospectral collocation method on a Chebyshev grid was our method of choice since its  rapid exponential convergence allows accurate results with minimal computational resources.  A limitation of pseudospectral methods is that the functions need to be smooth in order to achieve this exponential convergence.  It also yields dense matrices which are more difficult to solve in a linear system.  Moreover, the method is global in that the entire grid is used to compute derivatives rather than neighbouring points.  This makes it more difficult to isolate troublesome areas if something is not working.  

If we are solving a (say, two-dimensional) PDE instead of an ODE, then we have a vector $f_I$ derived from $f_{ij}$ instead of $f_i$, where $ij$ label collocation points on a direct product grid.  We must also derive new partial derivative operators $D_{IJ}f_I$.  These operators can be obtained just as before, but now using higher dimensional interpolates.  Alternatively, we could make this task easier by a choice of the map between $I$ and $ij$ indices.   Suppose $i\in\{1,\ldots,n_x\}$ and $j\in\{1,\ldots,n_y\}$, then one possibility is to map between the indices $i,j$ and $I$ using
\be
I(i,j)=(i-1)n_y+j,\qquad (i(I),j(I))=\left(\left\lceil\frac{I}{n_y}\right\rceil,[(I-1)\,\mrm{mod}\, n_y]+1\right)\;.
\ee
This is also known as co-lexicographic ordering.  For example, 
\be
\begin{pmatrix}
	f_{11}	&f_{12}	&f_{13}\\
	f_{21}	&f_{22}	&f_{23}\\
 \end{pmatrix}\longleftrightarrow
 \begin{pmatrix}
	f_{11}\\
	f_{12}\\
	f_{13}\\
	f_{21}\\
	f_{22}\\
	f_{23}\\
 \end{pmatrix}
 \;.
\ee

Now that we have $f_{ij}$ written as a vector $f_I$, we can find the corresponding differentiation matrix.  To do this, we use a Kronecker product.  Let $D_x$ and $D_y$ be the differentiation matrices on the grids $x_i$ and $y_i$, respectively, and let $I_k$ be the $k\times k$ identity matrix.  Then the full derivatives $D_X$ and $D_Y$ acting on $f_I$ are given by
\be
D_X=I_{n_y}\otimes D_{x},\qquad D_Y=D_y\otimes I_{n_x}\;.
\ee
Let us demonstrate this with an example.  Let's take a product of Chebyshev grids with $N=1$ in $x$ and $N=2$ in $y$, both in the domain from $[-1,1]$.  The differentiation matrices are
\be
D_x=\begin{bmatrix}
	\frac{1}{2}&-\frac{1}{2} \\
	\frac{1}{2}&-\frac{1}{2} \\
	 \end{bmatrix},\qquad
D_y=\begin{bmatrix}
	\frac{3}{2}&-2&\frac{1}{2} \\
	\frac{1}{2}&0&-\frac{1}{2} \\
	-\frac{1}{2}&2&-\frac{3}{2}
 \end{bmatrix}\;.
\ee
Then the differentiation matrices for the product grid are
\begin{align}
D_X&=I_{n_y}\otimes D_{x}=
\left[\begin{array}{ccc|ccc}
\frac{1}{2}&&&-\frac{1}{2}&\\
&\frac{1}{2}&&&-\frac{1}{2}&\\
&&\frac{1}{2} &&&-\frac{1}{2}\\\hline
\frac{1}{2}&&&-\frac{1}{2}&\\
&\frac{1}{2}&&&-\frac{1}{2}&\\
&&\frac{1}{2} &&&-\frac{1}{2}
\end{array}\right]\\
D_Y&=D_y\otimes I_{n_x}=
\left[\begin{array}{ccc|ccc}
\frac{3}{2}&-2&\frac{1}{2} &&&\\
\frac{1}{2}&0&-\frac{1}{2} &&&\\
-\frac{1}{2}&2&-\frac{3}{2} &&&\\\hline
&&&\frac{3}{2}&-2&\frac{1}{2} \\
&&&\frac{1}{2}&0&-\frac{1}{2} \\
&&&-\frac{1}{2}&2&-\frac{3}{2} \\
\end{array}\right]\;.
\end{align}
Higher derivatives can be obtained as products of $D_X$ and $D_Y$, or by taking Kronecker products of higher derivative matrices.  The process for higher dimensional derivatives is similar.  

\subsection{Discretisation of Linear Equations}

Now that we have a means of computing derivatives, we can now attempt to solve linear PDEs.  For now, let us consider an ODE for one function.  Consider a second-order linear differential equation:
\be\label{linearform}
\Big[\delta E_2[x,f]\partial^2+\delta E_1[x,f]\partial+\delta E_0[x,f]\Big]\delta f=-E[x,f]\;,
\ee
where the $\delta E_i[x,f]$ and $E$ are arbitrary scalar functions of the coordinate $x$ and the numerically known function $f$.  $\delta E_i[x,f]$ and $E$ may also involve derivatives of $f$. This is the form for linear equation that one obtains from Newton-Raphson.   For eigenvalue problems, once a (polynomial) eigenvalue problem is reduced to a linear pencil, we have a problem of the form $(A+\lambda B)\delta f=0$, where $A$ and $B$ takes the form of the left hand side operator above, perhaps without the extra dependence on the function $f$. 

Our task is to discretise the differential operator and convert it to a matrix, and convert $\delta f$ and $E$ to vectors.  Let $x_i$ be the collocation points, and $f_i$ be the given function $f$ on these points.  The derivatives of $f$ can then be computed by multiplying by a differentiation matrix $f^{(n)}_i=D^{(n)}_{ij}f_j$.  This allows us to evaluate the $\delta E$'s and $E$.  Now, replace the $\delta E$ by a diagonal matrix by evaluation at the collocation points, and replace $\partial^2$ and $\partial$ with differentiation matrices. Also replace $\delta f$ with an unknown vector $\delta f_j$, and $E$ with a known vector $E_i$ by evaluation.  Then \eqref{linearform} becomes 
\be\label{matrixequation}
\delta E_{ij}\delta f_j=-E_i\;,
\ee
which is a linear system that can be solved for $\delta f_j$.  

This equation defines a linear system.  If we were using periodic boundary conditions, the linear system should be solvable.  For a non-periodic domain, the operator might not be invertible, so no solution would be found.  The reason is because we have not yet implemented boundary conditions.  To do so, we simply repeat the same process for the linear boundary conditions $\delta B_\pm[x_{\pm},f]\delta f=-B_{\pm}$, where $\delta B$ is some differential operator.  We will end up with the equivalent of \eqref{matrixequation}
\be\label{matrixequationbc}
\delta B_{\pm,j}\delta f_j=-B_\pm\;.
\ee
Note that we have lost the index $i$ because the boundary equations are only applied to a point in the domain.  Now we replace the top and bottom rows of \eqref{matrixequation} by \eqref{matrixequationbc}.  The linear system we must solve is then
\be
M_{ij}\delta f_j=-V_i
\ee
where
\begin{align}
M_{ij}&=\left\{
     \begin{array}{lr}
       \left[\frac{\delta B_\pm}{\delta f}\right]_{ij} & : x_i=x_\pm\\
       \left[\frac{\delta E}{\delta f}\right]_{\pm, j} & : x_i\neq x_\pm
     \end{array}
   \right.\nonumber\\
V_i&=\left\{
     \begin{array}{lr}
       B_\pm & : x_i= x_\pm\\
      E_i & : x_i\neq x_\pm
     \end{array}
   \right.
\end{align}

Standard algorithms for this include gaussian elimination, LU decomposition, and iterative methods like GMRES.  Standard packages for this include \emph{Mathematica}'s `LinearSolve', and LAPACK (or UMFPACK for sparse matrices).  

For a linear pencil, this process will yield a matrix eigenvalue problem of the form
\be
(A_{ij}+\lambda B_{ij})\delta f_j=0\;,
\ee
which is a \emph{generalised eigenvalue problem}.  We again, may have to replace the top and bottom rows of this equation with the equivalent boundary conditions.  There are many standard algorithms (such as QZ factorisation) and packages (\emph{Mathematica}'s `Eigensystem', or LAPACK) for solving this system.  

Let us mention two generalisations to the process outlined here.  If we have a system of $k$ coupled differential equations with several functions, we can replace the vectors $f_i$ and $\delta f_i$ with one that is $k$ times as long that includes all of the functions by joining them one after another.  The size of the operators increase to accommodate this.  The rest of the procedure follows in a straightforward manner.  

In higher dimensions (cohomogeneity), everything follows as before, but implementing the boundary conditions is more difficult.  Differentiation matrices are replaced by their multidimensional counterparts. We also can no longer replace just the first and last rows in the matrix, every row that corresponds to a boundary point.  For a two dimensional PDE, boundary conditions are defined on an edge of the domain rather than at a single point, so many more rows would have to be replaced in \eqref{matrixequation}.  Furthermore, there are corners which belong to two boundaries.  There, we have a choice of boundary conditions.  One can just impose the simplest one or some linear combination of both.  Keeping track of boundary points makes implementation more difficult.  It may help to have a routine that converts between $I$ indices to $ij$ indices. 

\subsection{Integration}

Let us demonstrate how to take an integral using collocation methods.  Consider the integral
\be
I(x)=\int_{x_{-}}^xf(y)\dd y\;,
\ee
Differentiating this, we find
\be
I'(x)=f(x)\;,\qquad I(x_-)=0\;,
\ee
which is a first-order ODE with a boundary condition. Discretising, we get
\be
D_{ij}I_j=f_i\;.
\ee 
We then replace one of the rows of the above equation to impose $I(x_-)=0$, and solve the linear system for $I_j$.  From this, we can read off the integral. Let us point out that there are alternate ways to approximate integrals.  One method is to use Gaussian quadrature. 

\subsection{Example Boundary Value Problem}

Let us now solve an example boundary value problem.  Consider the nonlinear ODE and boundary conditions
\be\label{eq:exampleode}
q''(x)-e^{\frac{q(x)}{2}}=0,\qquad q(-1)=0, \quad \text{and}\quad q'(1)-e^{q(1)}+1=0.
\ee
This equation, for any boundary conditions, admits a general solution
\begin{equation}\label{eq:exact}
q(x) = 4 \log\left\{\frac{2 A }{\cos\left[A(x+B)\right]}\right\},
\end{equation}
where both $A$ and $B$ are integration constants. For our particular boundary conditions, there are exactly two real solutions, corresponding to the doublets $\{A,B\}\approx\{0.483794,0.472301\}$ and $\{A,B\}\approx\{0.110278, -11.2273\}$, which can be determined numerically by a simple root-finding algorithm. Our goal is to solve the above ODE from scratch and show agreement with this semi-analytic result.

For this, we solve use Newton-Raphson which we explained in \ref{subsec:newtonraphson}.  Following the procedure, we linearise the equations \eqref{eq:exampleode} (\emph{i.e.} set $q\rightarrow q+\epsilon\delta q$, differentiate with respect to $\epsilon$, then take $\epsilon\to0$) to get the Newton-Raphson equations
\begin{align}\label{eq:linexampleode}
\left[\partial^2-\frac{1}{2}e^{q(x)/2}\right]\delta q(x)&=-[q''(x)-e^{\frac{q(x)}{2}}]\nonumber\\
\delta q(-1)&=-q(-1)\nonumber\\
\left[\partial-e^{q(x)}\right]\delta q(x)\bigg |_{x=1}&=-[q'(1)-e^{q(1)}+1]\;.
\end{align}

Our next task is to discretise the above equations.  Let us use Chebyshev collocation points $x_i$ (computed from \eqref{eq:cheb}), which have $x_0=1$, and $x_N=-1$.  These points yield the differentiation matrices $D^{(1)}_{ij}$ from \eqref{eq:chebD}, and $D^{(2)}=[D^{(1)}]^2$ which have rank $N+1$. Given the function $q$, define $q_i\equiv q(x_i)$.  Its derivatives can be obtained through $q'_i=D^{(1)}_{ij}q_j$, and $q''_i=D^{(2)}_{ij}q_j$. Then the discretised version of the above equations \eqref{eq:linexampleode} is
\begin{subequations}\label{eq:linexampleodediscrete}
\begin{align}
\left[D^{(2)}-\mathrm{Diag}\left(\frac{1}{2}e^{q_k/2}\right)\right]_{ij}\delta q_j&=-[q''_i-e^{\frac{q_i}{2}}]\label{eq:linexampleodediscretea}\\
\delta q_N&=-q_N\label{eq:linexampleodediscreteb}\\
\left[D^{(1)}-\mathrm{Diag}\left(e^{q_k}\right)\right]_{0j}\delta q_j&=-[q'_0-e^{q_0}+1]\;,\label{eq:linexampleodediscretec}
\end{align}
\end{subequations}
where $\mathrm{Diag}(f_k)$ is a diagonal matrix with diagonal entries given by $f_k$.  Then we construct the linear system $M_{ij}\delta q_j=V_j$, where the $0$th row is given by \eqref{eq:linexampleodediscretec}, the $N$th row is given by \eqref{eq:linexampleodediscreteb}, and the remaining rows are given by \eqref{eq:linexampleodediscretea}. This linear system can be solved for $\delta q_i$ using LU decomposition (which is implemented in \emph{Mathematica}'s `LinearSolve' and LAPACK).  

The numerical algorithm then proceeds as follows.  Given a seed $q$, evaluate $q_i$, $q'_i$, and $q''_i$.  Construct the linear system $M_{ij}\delta q_j=V_j$ as we have just described, and solve it to obtain $\delta q_i$.  Then update $q_i\rightarrow q_i+\delta q_i$. Repeat this processes as necessary until convergence or failure.  Our criteria for convergence is $\mathrm{Max}[\delta q_i]<10^{-10}$.

\begin{figure}[h]
\centering
\subfigure[Seed $q=0$.]{
\includegraphics[width = 5.5 cm]{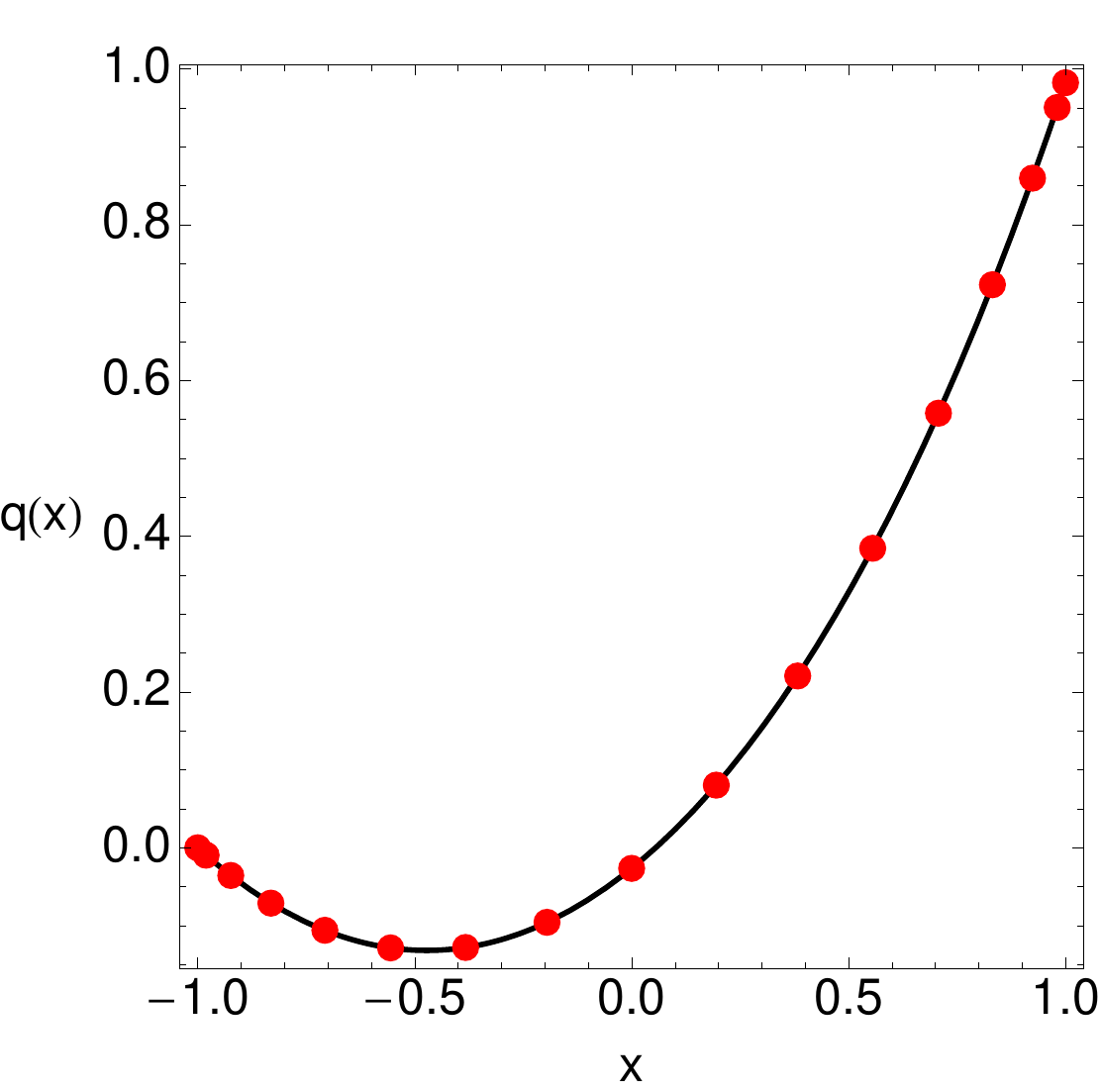}
\label{figs:example1a}
}
\subfigure[Seed $q=-1$.]{
\includegraphics[width = 5.5 cm]{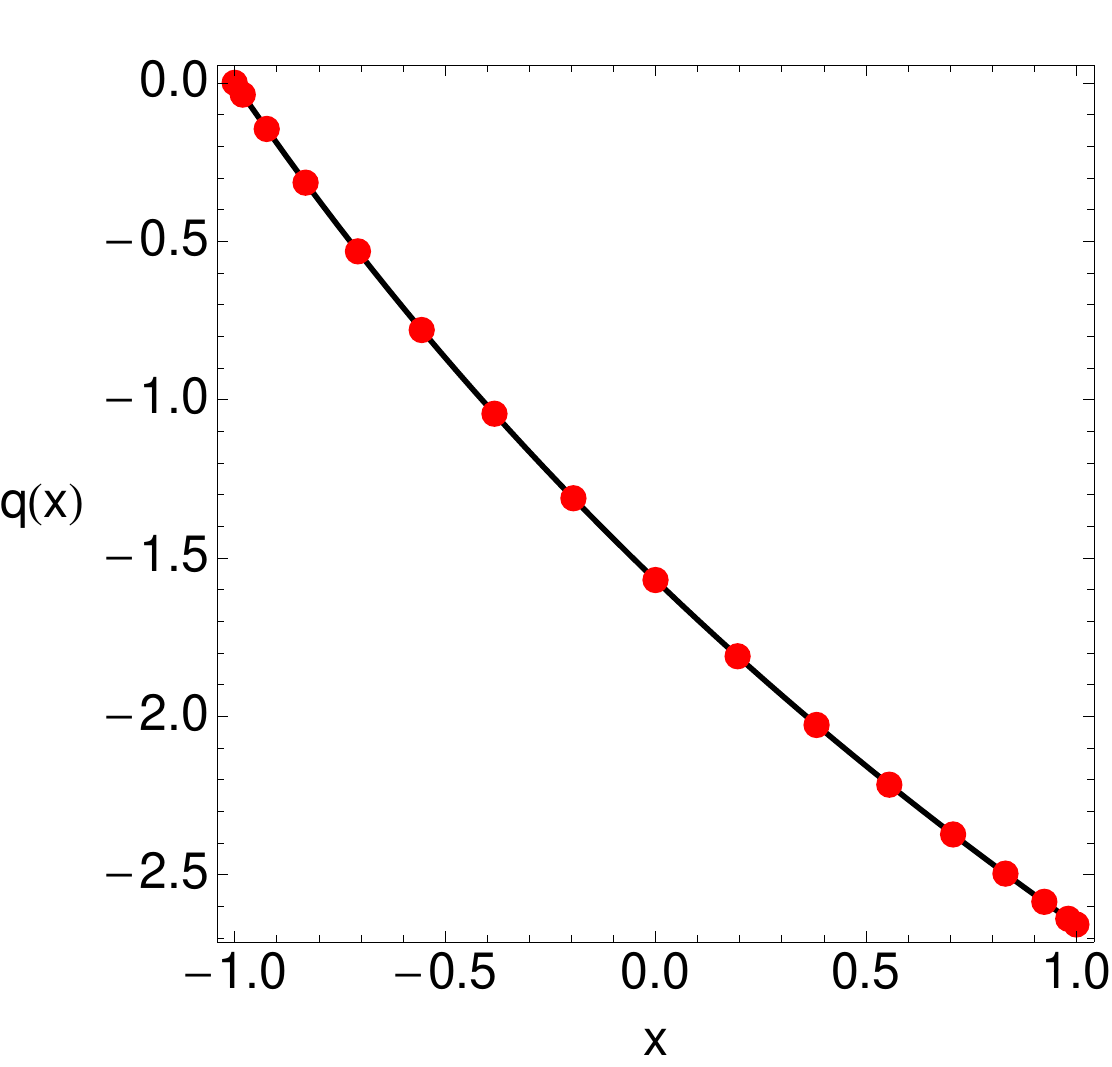}
\label{figs:example1b}
}
\caption{\label{figs:example1tot}Graphical representation of both semi-analytic and numerical solutions of (\ref{eq:exampleode}). The red dots represent the numerical result and the solid black line the semi-analytic result.}
\end{figure}

We can obtain both solutions by a suitable choice of starting seed.  The result of using the seed $q=0$ is shown in Fig.~\ref{figs:example1a}, together with the semi-analytical curve for $N=16$. For the seed $q=-1$, we obtain the other solution shown in Fig.~\ref{figs:example1b}.  There is good agreement between the numerical and semi-analytical results.

\section{Computing Conserved Charges and Thermodynamic Potentials}\label{appendix:computecharges}

The output of numerical codes are typically the metric and matter fields that describe a gravitational solution.
From this output, we can compute gauge invariant quantities and, in particular, the asymptotic conserved charges (energy $E$, angular momentum $J$, electric charge $Q$) and other thermodynamic quantities (entropy $S$, temperature $T_H$ and free energies). In this appendix we briefly review how these can be computed.  

For asymptotically flat spacetimes, the conserved asymptotic quantities are computed using the familiar Arnowitt-Deser-Misner (ADM) formulation \cite{Arnowitt:1961zz,Arnowitt:1962hi,Myers:1986un}. This computation is by now a standard one and is described also in detail, for example, in Section 3.1 of \cite{Emparan:2008eg}. Alternatively, we can use the Komar integrals  \eqref{KomarCharges} to compute the conserved charges \cite{Komar:1959}.

Asymptotically flat black holes further obey the Smarr relation \cite{Gibbons:2004ai}, 
\begin{equation}\label{smarr}
\frac{d-3}{d-2}M=T_{H}S_H+\Omega_H J\,,
\end{equation}
and the first law of thermodynamics \cite{Bardeen:1973gs}, 
\begin{equation}\label{1stlaw}
\dd M=T_H\dd S_H+\Omega_H\dd J.
\end{equation}
These relations are supplemented with an electromagnetic contribution proportional to $\Phi_H Q$ and $\Phi_H \dd Q$, respectively,  if a Maxwell potential is present.
These thermodynamic laws (and the vanishing of the DeTurck vector) are good monitors of numerical accuracy. 

Typically, the theory at hand will have different phases. It is then relevant to ask which of the phases is the dominant one. The answer depends on the thermodynamical ensemble. In the microcanonical ensemble, $E$ and $J$ are fixed, and solutions with higher $S$ dominate. In the canonical ensemble, we keep $\{T_H,J,Q\}$ fixed, and the lowest Helmoltz free energy ${\cal F}=E-T_H S$ dominates. Finally, in the grand canonical ensemble  $\{T_H,\Omega_H,\Phi_H\}$ are kept fixed and the lowest Gibbs free energy ${\cal G}=E-T_H S-\Omega_H J -\Phi_H Q$ dominates.

In AdS, the first law of thermodynamics still holds but the Smarr law requires more care. Computing conserved charges is now less straightforward, but there are still well-defined formalisms. One of them is known as `{\it holographic renormalization}' \cite{deHaro:2000xn,Henningson:1998gx}. Recall that  asymptotically AdS$_{d}$  gravitational fields must fall off according to a specific power law as detailed in \cite{Henneaux:1985tv,Ashtekar:1984zz}. Refs. \cite{deHaro:2000xn,Henningson:1998gx} have revisited these boundary conditions using Fefferman-Graham coordinates \cite{FG:1985,Graham:1999jg,Anderson:2004yi}. Taking the asymptotic boundary to be at the radial position $z=0$, the Fefferman-Graham expansion of the metric away from the boundary requires that $g_{zz}=L^2/z^2$ and $g_{zb}=0$ at each order, where $L$ is the AdS radius and $x^b$ are the coordinates on the boundary $z=0$. 

For an even dimension $d$, this reads\footnote{For odd $d$, the asymptotic expansion \eqref{introFG} contains a logarithmic term $z^{d-1} \log z^2 \tilde{g}_{ab}^{(d-1)}$ and the holographic stress tensor has an extra contribution proportional to the conformal anomalies of the boundary CFT \cite{deHaro:2000xn,Henningson:1998gx}. We omit this case in this general discussion but we discuss the details of the $d=7$ case in the application example of section \ref{Sec:RRR}.} 
\begin{eqnarray}\label{introFG}
&&\dd s^2 = \frac{L^2}{z^2}\left[\dd z^2+g_{ab}(z,x) \dd x^{a}\dd x^{b}\right]\,, \\
&& g_{ab}{\bigl |}_{z\to 0}=g_{ab}^{(0)}(x)+\cdots+z^{d-1} g_{ab}^{(d-1)}(x)+\cdots 
\qquad \hbox{with} \quad \langle T_{ab}(x)\rangle\equiv \frac{d-1}{16\pi G_N}\,g_{ab}^{(d-1)}(x),\nonumber
\end{eqnarray}
where  $G_N$ is Newton's constant, and $g^{(0)}(x)$, $g^{(d-1)}(x)$ are the two free coefficients of the expansion.  In the absence of matter, the first dots include only even powers of $z$ (smaller than $d-1$) and depend only on $g^{(0)}$, while the second dots depend on the two independent terms $g^{(0)}, g^{(d-1)} $. $g^{(0)}$ and $g^{(d-1)}$ are therefore the undetermined coefficients of the series expansion off the boundary.  $g^{(0)}$ represents the boundary metric.  Note that the Fefferman-Graham expansion \eqref{introFG} is not unique unless a conformal frame for $g^{(0)}$ is chosen. $g^{(d-1)}$ is related to the stress tensor $\langle T_{ab}(x)\rangle$.  Typical boundary conditions in AdS fix the (conformal class) of metrics $g^{(0)}$, and then reads off the holographic stress tensor $\langle T_{ab}(x)\rangle$, but other boundary conditions are possible.\footnote{Other boundary conditions that might be called asymptotically globally AdS (and that promote the boundary graviton to a dynamical field) were proposed in \cite{Andrade:2011dg}. However, they turn out to lead to ghosts (modes with negative kinetic energy) and thus make the energy unbounded below \cite{Compere:2008us}.} Alternatively, we also can compute the holographic stress tensor in terms of the intrinsic geometry and extrinsic curvature following the formalism of  \cite{Balasubramanian:1999re}. 

Pulling back the holographic stress tensor, $\langle T_{ij} \rangle$ to a $(d-1)$-dimensional spatial hypersurface and contracting it with the Killing vector that generates time (rotational) translations we can compute the energy and angular momentum of the solutions (see \eqref{RRR:lumpyThermo0}) and thus the thermodynamics of the system which obey the first law \eqref{1stlaw}  \cite{Caldarelli:1999xj,Gibbons:2004ai,Papadimitriou:2005ii}. 

Alternatively, we can use the Astekhar-Das formalism \cite{Ashtekar:1999jx} to compute the conserved quantities in AdS (e.g., see a detailed description of an application in \cite{Dias:2011at}). 

So far we have been assuming that our solution is asymptotically AdS$_d$. However, in the context of holography, we often have dual theories that are formulated on a gravitational background that asymptotes to a direct product spacetime ${\cal M}_p \times X^q$ where ${\cal M}_p$ asymptotes to AdS$_p$ and $X^q$ is a compact manifold. For example, in the original AdS/CFT correspondence ${\cal M}_p=$AdS$_5$ and $X^q=S^5$  \cite{Maldacena:1997re,Gubser:1998bc,Witten:1998qj}. Upon Kaluza-Klein dimensional reduction, the dual QFT is formulated on the holographic boundary of AdS$_p$ and the information about $X$ becomes encoded on some Kaluza-Klein gravitons and scalar fields that are generated by the dimensional reduction and live on the boundary of the asymptotically AdS$_p$ space. A gauge invariant formalism developed in \cite{Skenderis:2006uy} (following influential work of \cite{Kim:1985ez,Gunaydin:1984fk,Lee:1998bxa,Lee:1999pj,Arutyunov:1999en}, see also applications in \cite{Skenderis:2006di,Skenderis:2007yb,Dias:2015pda})  known as `{\it Kaluza-Klein holography}'  allows one to find the dimensionally reduced gauge invariant fields, and use these  fields to read the holographic stress tensor and the expectation values of the Kaluza-Klein gravitons and scalar fields using the standard  holographic renormalisation  procedure of  \cite{deHaro:2000xn,Henningson:1998gx} or  \cite{Balasubramanian:1999re}.

\bibliography{refs}{}

\providecommand{\href}[2]{#2}\begingroup\raggedright\begin{thebibliography}{100}

\bibitem{Sperhake:2014wpa}
U.~Sperhake, {\it {The numerical relativity breakthrough for binary black
  holes}},  {\em Class. Quant. Grav.} {\bf 32} (2015), no.~12 124011,
  [\href{http://xxx.lanl.gov/abs/1411.3997}{{\tt arXiv:1411.3997}}].

\bibitem{Winicour:2005ge}
J.~Winicour, {\it {Characteristic evolution and matching}},  {\em Living Rev.
  Rel.} {\bf 15} (2012) 2, [\href{http://xxx.lanl.gov/abs/gr-qc/0102085}{{\tt
  gr-qc/0102085}}]. [Living Rev. Rel.15,2(2012)].

\bibitem{Chesler:2013lia}
P.~M. Chesler and L.~G. Yaffe, {\it {Numerical solution of gravitational
  dynamics in asymptotically anti-de Sitter spacetimes}},  {\em JHEP} {\bf 07}
  (2014) 086, [\href{http://xxx.lanl.gov/abs/1309.1439}{{\tt
  arXiv:1309.1439}}].

\bibitem{Cook:2000vr}
G.~B. Cook, {\it {Initial data for numerical relativity}},  {\em Living Rev.
  Rel.} {\bf 3} (2000) 5, [\href{http://xxx.lanl.gov/abs/gr-qc/0007085}{{\tt
  gr-qc/0007085}}].

\bibitem{Baumgarte:2002jm}
T.~W. Baumgarte and S.~L. Shapiro, {\it {Numerical relativity and compact
  binaries}},  {\em Phys. Rept.} {\bf 376} (2003) 41--131,
  [\href{http://xxx.lanl.gov/abs/gr-qc/0211028}{{\tt gr-qc/0211028}}].

\bibitem{Gourgoulhon:2007ue}
E.~Gourgoulhon, {\it {3+1 formalism and bases of numerical relativity}},
  \href{http://xxx.lanl.gov/abs/gr-qc/0703035}{{\tt gr-qc/0703035}}.

\bibitem{AlcubierreBook}
M.~Alcubierre, {\em Introduction to 3+1 Numerical Relativity}.
\newblock Oxford University Press, Oxford, 2008.

\bibitem{BaumgarteShapiroBook}
T.~W. Baumgarte and S.~L. Shapiro, {\em Numerical Relativity}.
\newblock Cambridge University Press, 2010.

\bibitem{Centrella:2010mx}
J.~Centrella, J.~G. Baker, B.~J. Kelly, and J.~R. van Meter, {\it {Black-hole
  binaries, gravitational waves, and numerical relativity}},  {\em Rev. Mod.
  Phys.} {\bf 82} (2010) 3069, [\href{http://xxx.lanl.gov/abs/1010.5260}{{\tt
  arXiv:1010.5260}}].

\bibitem{Heller:2012je}
M.~P. Heller, R.~A. Janik, and P.~Witaszczyk, {\it {A numerical relativity
  approach to the initial value problem in asymptotically Anti-de Sitter
  spacetime for plasma thermalization - an ADM formulation}},  {\em Phys. Rev.}
  {\bf D85} (2012) 126002, [\href{http://xxx.lanl.gov/abs/1203.0755}{{\tt
  arXiv:1203.0755}}].

\bibitem{Pretorius:2004jg}
F.~Pretorius, {\it {Numerical relativity using a generalized harmonic
  decomposition}},  {\em Class. Quant. Grav.} {\bf 22} (2005) 425--452,
  [\href{http://xxx.lanl.gov/abs/gr-qc/0407110}{{\tt gr-qc/0407110}}].

\bibitem{Gundlach:2005eh}
C.~Gundlach, J.~M. Martin-Garcia, G.~Calabrese, and I.~Hinder, {\it {Constraint
  damping in the Z4 formulation and harmonic gauge}},  {\em Class. Quant.
  Grav.} {\bf 22} (2005) 3767--3774,
  [\href{http://xxx.lanl.gov/abs/gr-qc/0504114}{{\tt gr-qc/0504114}}].

\bibitem{Lindblom:2005qh}
L.~Lindblom, M.~A. Scheel, L.~E. Kidder, R.~Owen, and O.~Rinne, {\it {A New
  generalized harmonic evolution system}},  {\em Class. Quant. Grav.} {\bf 23}
  (2006) S447--S462, [\href{http://xxx.lanl.gov/abs/gr-qc/0512093}{{\tt
  gr-qc/0512093}}].

\bibitem{Palenzuela:2006wp}
C.~Palenzuela, I.~Olabarrieta, L.~Lehner, and S.~L. Liebling, {\it {Head-on
  collisions of boson stars}},  {\em Phys. Rev.} {\bf D75} (2007) 064005,
  [\href{http://xxx.lanl.gov/abs/gr-qc/0612067}{{\tt gr-qc/0612067}}].

\bibitem{Bantilan:2012vu}
H.~Bantilan, F.~Pretorius, and S.~S. Gubser, {\it {Simulation of Asymptotically
  AdS5 Spacetimes with a Generalized Harmonic Evolution Scheme}},  {\em Phys.
  Rev.} {\bf D85} (2012) 084038, [\href{http://xxx.lanl.gov/abs/1201.2132}{{\tt
  arXiv:1201.2132}}].

\bibitem{Hilditch:2015aba}
D.~Hilditch, A.~Weyhausen, and B.~Bruegmann, {\it {A Pseudospectral Method for
  Gravitational Wave Collapse}},  \href{http://xxx.lanl.gov/abs/1504.0473}{{\tt
  arXiv:1504.0473}}.

\bibitem{HorowitzBook2012}
{G.T. Horowitz, et. al}, {\em Black Holes in Higher Dimensions}.
\newblock Cambridge University Press, 2012.

\bibitem{Maldacena:1997re}
J.~M. Maldacena, {\it {The Large N limit of superconformal field theories and
  supergravity}},  {\em Int. J. Theor. Phys.} {\bf 38} (1999) 1113--1133,
  [\href{http://xxx.lanl.gov/abs/hep-th/9711200}{{\tt hep-th/9711200}}]. [Adv.
  Theor. Math. Phys.2,231(1998)].

\bibitem{Gubser:1998bc}
S.~S. Gubser, I.~R. Klebanov, and A.~M. Polyakov, {\it {Gauge theory
  correlators from noncritical string theory}},  {\em Phys. Lett.} {\bf B428}
  (1998) 105--114, [\href{http://xxx.lanl.gov/abs/hep-th/9802109}{{\tt
  hep-th/9802109}}].

\bibitem{Witten:1998qj}
E.~Witten, {\it {Anti-de Sitter space and holography}},  {\em Adv. Theor. Math.
  Phys.} {\bf 2} (1998) 253--291,
  [\href{http://xxx.lanl.gov/abs/hep-th/9802150}{{\tt hep-th/9802150}}].

\bibitem{Aharony:1999ti}
O.~Aharony, S.~S. Gubser, J.~M. Maldacena, H.~Ooguri, and Y.~Oz, {\it {Large N
  field theories, string theory and gravity}},  {\em Phys. Rept.} {\bf 323}
  (2000) 183--386, [\href{http://xxx.lanl.gov/abs/hep-th/9905111}{{\tt
  hep-th/9905111}}].

\bibitem{Maldacena:2011ut}
J.~Maldacena, {\it {The Gauge/gravity duality}},
  \href{http://xxx.lanl.gov/abs/1106.6073}{{\tt arXiv:1106.6073}}.

\bibitem{Hubeny:2014bla}
V.~E. Hubeny, {\it {The AdS/CFT Correspondence}},  {\em Class. Quant. Grav.}
  {\bf 32} (2015), no.~12 124010,
  [\href{http://xxx.lanl.gov/abs/1501.0000}{{\tt arXiv:1501.0000}}].

\bibitem{CasalderreySolana:2011us}
J.~Casalderrey-Solana, H.~Liu, D.~Mateos, K.~Rajagopal, and U.~A. Wiedemann,
  {\it {Gauge/String Duality, Hot QCD and Heavy Ion Collisions}},
  \href{http://xxx.lanl.gov/abs/1101.0618}{{\tt arXiv:1101.0618}}.

\bibitem{Erdmengerbook}
J.~Erdmenger and M.~Ammon, {\em Gauge/Gravity Duality - Foundations and
  Applications}.
\newblock Cambridge University Press, 2015.

\bibitem{Horowitz:2010gk}
G.~T. Horowitz, {\it {Introduction to Holographic Superconductors}},  {\em
  Lect. Notes Phys.} {\bf 828} (2011) 313--347,
  [\href{http://xxx.lanl.gov/abs/1002.1722}{{\tt arXiv:1002.1722}}].

\bibitem{Hubeny:2010ry}
V.~E. Hubeny and M.~Rangamani, {\it {A Holographic view on physics out of
  equilibrium}},  {\em Adv. High Energy Phys.} {\bf 2010} (2010) 297916,
  [\href{http://xxx.lanl.gov/abs/1006.3675}{{\tt arXiv:1006.3675}}].

\bibitem{Hartnoll:2011fn}
S.~A. Hartnoll, {\it {Horizons, holography and condensed matter}},
  \href{http://xxx.lanl.gov/abs/1106.4324}{{\tt arXiv:1106.4324}}.

\bibitem{Marolf:2013ioa}
D.~Marolf, M.~Rangamani, and T.~Wiseman, {\it {Holographic thermal field theory
  on curved spacetimes}},  {\em Class. Quant. Grav.} {\bf 31} (2014) 063001,
  [\href{http://xxx.lanl.gov/abs/1312.0612}{{\tt arXiv:1312.0612}}].

\bibitem{Kramerbook}
H.~Stephani, D.~Kramer, M.~MacCallum, C.~Hoenselaers, and E.~Herlt, {\em Exact
  solutions of Einstein's field equations}.
\newblock Cambridge University Press, 2003.

\bibitem{Starobinsky:1973scalar}
A.~A. Starobinsky and S.~M. Churilov, {\it {Amplification of waves during
  reflection from a rotating black hole}},  {\em Sov. Phys. JETP} {\bf 37}
  (1973) 28.

\bibitem{Starobinsky:1973}
A.~A. Starobinsky and S.~M. Churilov, {\it {Amplification of electromagnetic
  and gravitational waves scattered by a rotating black hole}},  {\em Sov.
  Phys. JETP} {\bf 38} (1973) 1.

\bibitem{Unruh:1976fm}
W.~G. Unruh, {\it {Absorption Cross-Section of Small Black Holes}},  {\em Phys.
  Rev.} {\bf D14} (1976) 3251--3259.

\bibitem{Detweiler:1980uk}
S.~L. Detweiler, {\it {Klein-Gordon equation and rotating black holes}},  {\em
  Phys. Rev.} {\bf D22} (1980) 2323--2326.

\bibitem{Maldacena:1997ih}
J.~M. Maldacena and A.~Strominger, {\it {Universal low-energy dynamics for
  rotating black holes}},  {\em Phys. Rev.} {\bf D56} (1997) 4975--4983,
  [\href{http://xxx.lanl.gov/abs/hep-th/9702015}{{\tt hep-th/9702015}}].

\bibitem{Dias:2007nj}
O.~J.~C. Dias, R.~Emparan, and A.~Maccarrone, {\it {Microscopic theory of black
  hole superradiance}},  {\em Phys. Rev.} {\bf D77} (2008) 064018,
  [\href{http://xxx.lanl.gov/abs/0712.0791}{{\tt arXiv:0712.0791}}].

\bibitem{DEathbook}
P.~D. D'Eath, {\em Black Holes: Gravitational Interactions}.
\newblock Series Oxford Mathematical Monographs, Oxford University Press, 1996.

\bibitem{Basu:2010uz}
P.~Basu, J.~Bhattacharya, S.~Bhattacharyya, R.~Loganayagam, S.~Minwalla, and
  V.~Umesh, {\it {Small Hairy Black Holes in Global AdS Spacetime}},  {\em
  JHEP} {\bf 10} (2010) 045, [\href{http://xxx.lanl.gov/abs/1003.3232}{{\tt
  arXiv:1003.3232}}].

\bibitem{Bhattacharyya:2010yg}
S.~Bhattacharyya, S.~Minwalla, and K.~Papadodimas, {\it {Small Hairy Black
  Holes in $AdS_5 x S^5$}},  {\em JHEP} {\bf 11} (2011) 035,
  [\href{http://xxx.lanl.gov/abs/1005.1287}{{\tt arXiv:1005.1287}}].

\bibitem{Dias:2011tj}
O.~J.~C. Dias, P.~Figueras, S.~Minwalla, P.~Mitra, R.~Monteiro, and J.~E.
  Santos, {\it {Hairy black holes and solitons in global $AdS_5$}},  {\em JHEP}
  {\bf 08} (2012) 117, [\href{http://xxx.lanl.gov/abs/1112.4447}{{\tt
  arXiv:1112.4447}}].

\bibitem{Dias:2011at}
O.~J.~C. Dias, G.~T. Horowitz, and J.~E. Santos, {\it {Black holes with only
  one Killing field}},  {\em JHEP} {\bf 07} (2011) 115,
  [\href{http://xxx.lanl.gov/abs/1105.4167}{{\tt arXiv:1105.4167}}].

\bibitem{Stotyn:2011ns}
S.~Stotyn, M.~Park, P.~McGrath, and R.~B. Mann, {\it {Black Holes and Boson
  Stars with One Killing Field in Arbitrary Odd Dimensions}},  {\em Phys. Rev.}
  {\bf D85} (2012) 044036, [\href{http://xxx.lanl.gov/abs/1110.2223}{{\tt
  arXiv:1110.2223}}].

\bibitem{Donos:2012ra}
A.~Donos and S.~A. Hartnoll, {\it {Universal linear in temperature resistivity
  from black hole superradiance}},  {\em Phys. Rev.} {\bf D86} (2012) 124046,
  [\href{http://xxx.lanl.gov/abs/1208.4102}{{\tt arXiv:1208.4102}}].

\bibitem{Cardoso:2013pza}
V.~Cardoso, O.~J.~C. Dias, G.~S. Hartnett, L.~Lehner, and J.~E. Santos, {\it
  {Holographic thermalization, quasinormal modes and superradiance in
  Kerr-AdS}},  {\em JHEP} {\bf 04} (2014) 183,
  [\href{http://xxx.lanl.gov/abs/1312.5323}{{\tt arXiv:1312.5323}}].

\bibitem{Iizuka:2015vsa}
N.~Iizuka, A.~Ishibashi, and K.~Maeda, {\it {A rotating hairy AdS$_{3}$ black
  hole with the metric having only one Killing vector field}},  {\em JHEP} {\bf
  08} (2015) 112, [\href{http://xxx.lanl.gov/abs/1505.0039}{{\tt
  arXiv:1505.0039}}].

\bibitem{Bhattacharyya:2008jc}
S.~Bhattacharyya, V.~E. Hubeny, S.~Minwalla, and M.~Rangamani, {\it {Nonlinear
  Fluid Dynamics from Gravity}},  {\em JHEP} {\bf 02} (2008) 045,
  [\href{http://xxx.lanl.gov/abs/0712.2456}{{\tt arXiv:0712.2456}}].

\bibitem{Bhattacharyya:2008xc}
S.~Bhattacharyya, V.~E. Hubeny, R.~Loganayagam, G.~Mandal, S.~Minwalla,
  T.~Morita, M.~Rangamani, and H.~S. Reall, {\it {Local Fluid Dynamical Entropy
  from Gravity}},  {\em JHEP} {\bf 06} (2008) 055,
  [\href{http://xxx.lanl.gov/abs/0803.2526}{{\tt arXiv:0803.2526}}].

\bibitem{Hubeny:2011hd}
V.~E. Hubeny, S.~Minwalla, and M.~Rangamani, {\it {The fluid/gravity
  correspondence}},  {\em in Black Holes in Higher Dimensions, CUP 2012, Ed.
  Gary T. Horowitz} (2012) [\href{http://xxx.lanl.gov/abs/1107.5780}{{\tt
  arXiv:1107.5780}}].

\bibitem{Caldarelli:2008pz}
M.~M. Caldarelli, R.~Emparan, and M.~J. Rodriguez, {\it {Black Rings in
  (Anti)-deSitter space}},  {\em JHEP} {\bf 11} (2008) 011,
  [\href{http://xxx.lanl.gov/abs/0806.1954}{{\tt arXiv:0806.1954}}].

\bibitem{Emparan:2009cs}
R.~Emparan, T.~Harmark, V.~Niarchos, and N.~A. Obers, {\it {World-Volume
  Effective Theory for Higher-Dimensional Black Holes}},  {\em Phys. Rev.
  Lett.} {\bf 102} (2009) 191301,
  [\href{http://xxx.lanl.gov/abs/0902.0427}{{\tt arXiv:0902.0427}}].

\bibitem{Emparan:2007wm}
R.~Emparan, T.~Harmark, V.~Niarchos, N.~A. Obers, and M.~J. Rodriguez, {\it
  {The Phase Structure of Higher-Dimensional Black Rings and Black Holes}},
  {\em JHEP} {\bf 10} (2007) 110,
  [\href{http://xxx.lanl.gov/abs/0708.2181}{{\tt arXiv:0708.2181}}].

\bibitem{Emparan:2009at}
R.~Emparan, T.~Harmark, V.~Niarchos, and N.~A. Obers, {\it {Essentials of
  Blackfold Dynamics}},  \href{http://xxx.lanl.gov/abs/0910.1601}{{\tt
  arXiv:0910.1601}}.

\bibitem{Caldarelli:2010xz}
M.~M. Caldarelli, R.~Emparan, and B.~Van~Pol, {\it {Higher-dimensional Rotating
  Charged Black Holes}},  {\em JHEP} {\bf 04} (2011) 013,
  [\href{http://xxx.lanl.gov/abs/1012.4517}{{\tt arXiv:1012.4517}}].

\bibitem{Armas:2010hz}
J.~Armas and N.~A. Obers, {\it {Blackfolds in (Anti)-de Sitter Backgrounds}},
  {\em Phys. Rev.} {\bf D83} (2011) 084039,
  [\href{http://xxx.lanl.gov/abs/1012.5081}{{\tt arXiv:1012.5081}}].

\bibitem{Armas:2011uf}
J.~Armas, J.~Camps, T.~Harmark, and N.~A. Obers, {\it {The Young Modulus of
  Black Strings and the Fine Structure of Blackfolds}},  {\em JHEP} {\bf 02}
  (2012) 110, [\href{http://xxx.lanl.gov/abs/1110.4835}{{\tt
  arXiv:1110.4835}}].

\bibitem{Camps:2012hw}
J.~Camps and R.~Emparan, {\it {Derivation of the blackfold effective theory}},
  {\em JHEP} {\bf 03} (2012) 038,
  [\href{http://xxx.lanl.gov/abs/1201.3506}{{\tt arXiv:1201.3506}}]. [Erratum:
  JHEP06,155(2012)].

\bibitem{Armas:2012bk}
J.~Armas, T.~Harmark, N.~A. Obers, M.~Orselli, and A.~V. Pedersen, {\it
  {Thermal Giant Gravitons}},  {\em JHEP} {\bf 11} (2012) 123,
  [\href{http://xxx.lanl.gov/abs/1207.2789}{{\tt arXiv:1207.2789}}].

\bibitem{Armas:2013hsa}
J.~Armas, {\it {How Fluids Bend: the Elastic Expansion for Higher-Dimensional
  Black Holes}},  {\em JHEP} {\bf 09} (2013) 073,
  [\href{http://xxx.lanl.gov/abs/1304.7773}{{\tt arXiv:1304.7773}}].

\bibitem{Armas:2014bia}
J.~Armas and T.~Harmark, {\it {Black Holes and Biophysical (Mem)-branes}},
  {\em Phys. Rev.} {\bf D90} (2014), no.~12 124022,
  [\href{http://xxx.lanl.gov/abs/1402.6330}{{\tt arXiv:1402.6330}}].

\bibitem{Emparan:2011br}
R.~Emparan, {\it {Blackfolds}},  {\em in Black Holes in Higher Dimensions, CUP
  2012, Ed. Gary T. Horowitz} (2012)
  [\href{http://xxx.lanl.gov/abs/1106.2021}{{\tt arXiv:1106.2021}}].

\bibitem{Emparan:2013moa}
R.~Emparan, R.~Suzuki, and K.~Tanabe, {\it {The large D limit of General
  Relativity}},  {\em JHEP} {\bf 06} (2013) 009,
  [\href{http://xxx.lanl.gov/abs/1302.6382}{{\tt arXiv:1302.6382}}].

\bibitem{Emparan:2013xia}
R.~Emparan, D.~Grumiller, and K.~Tanabe, {\it {Large-D gravity and low-D
  strings}},  {\em Phys. Rev. Lett.} {\bf 110} (2013), no.~25 251102,
  [\href{http://xxx.lanl.gov/abs/1303.1995}{{\tt arXiv:1303.1995}}].

\bibitem{Emparan:2013oza}
R.~Emparan and K.~Tanabe, {\it {Holographic superconductivity in the large D
  expansion}},  {\em JHEP} {\bf 01} (2014) 145,
  [\href{http://xxx.lanl.gov/abs/1312.1108}{{\tt arXiv:1312.1108}}].

\bibitem{Emparan:2014cia}
R.~Emparan and K.~Tanabe, {\it {Universal quasinormal modes of large D black
  holes}},  {\em Phys. Rev.} {\bf D89} (2014), no.~6 064028,
  [\href{http://xxx.lanl.gov/abs/1401.1957}{{\tt arXiv:1401.1957}}].

\bibitem{Emparan:2014jca}
R.~Emparan, R.~Suzuki, and K.~Tanabe, {\it {Instability of rotating black
  holes: large D analysis}},  {\em JHEP} {\bf 06} (2014) 106,
  [\href{http://xxx.lanl.gov/abs/1402.6215}{{\tt arXiv:1402.6215}}].

\bibitem{Emparan:2014aba}
R.~Emparan, R.~Suzuki, and K.~Tanabe, {\it {Decoupling and non-decoupling
  dynamics of large D black holes}},  {\em JHEP} {\bf 07} (2014) 113,
  [\href{http://xxx.lanl.gov/abs/1406.1258}{{\tt arXiv:1406.1258}}].

\bibitem{Emparan:2015hwa}
R.~Emparan, T.~Shiromizu, R.~Suzuki, K.~Tanabe, and T.~Tanaka, {\it {Effective
  theory of Black Holes in the 1/D expansion}},  {\em JHEP} {\bf 06} (2015)
  159, [\href{http://xxx.lanl.gov/abs/1504.0648}{{\tt arXiv:1504.0648}}].

\bibitem{Emparan:2015rva}
R.~Emparan, R.~Suzuki, and K.~Tanabe, {\it {Quasinormal modes of (Anti-)de
  Sitter black holes in the 1/D expansion}},  {\em JHEP} {\bf 04} (2015) 085,
  [\href{http://xxx.lanl.gov/abs/1502.0282}{{\tt arXiv:1502.0282}}].

\bibitem{Bhattacharyya:2015dva}
S.~Bhattacharyya, A.~De, S.~Minwalla, R.~Mohan, and A.~Saha, {\it {A membrane
  paradigm at large D}},  \href{http://xxx.lanl.gov/abs/1504.0661}{{\tt
  arXiv:1504.0661}}.

\bibitem{Suzuki:2015iha}
R.~Suzuki and K.~Tanabe, {\it {Stationary black holes: Large $D$ analysis}},
  \href{http://xxx.lanl.gov/abs/1505.0128}{{\tt arXiv:1505.0128}}.

\bibitem{Tanabe:2015hda}
K.~Tanabe, {\it {Black rings at large D}},
  \href{http://xxx.lanl.gov/abs/1510.0220}{{\tt arXiv:1510.0220}}.

\bibitem{Okawa:2013afa}
H.~Okawa, {\it {Initial Conditions for Numerical Relativity -- Introduction to
  numerical methods for solving elliptic PDEs}},  {\em Int. J. Mod. Phys.} {\bf
  A28} (2013) 1340016, [\href{http://xxx.lanl.gov/abs/1308.3502}{{\tt
  arXiv:1308.3502}}].

\bibitem{Okawa:2014nda}
H.~Okawa, H.~Witek, and V.~Cardoso, {\it {Black holes and fundamental fields in
  Numerical Relativity: initial data construction and evolution of bound
  states}},  {\em Phys. Rev.} {\bf D89} (2014), no.~10 104032,
  [\href{http://xxx.lanl.gov/abs/1401.1548}{{\tt arXiv:1401.1548}}].

\bibitem{Headrick:2009pv}
M.~Headrick, S.~Kitchen, and T.~Wiseman, {\it {A New approach to static
  numerical relativity, and its application to Kaluza-Klein black holes}},
  {\em Class. Quant. Grav.} {\bf 27} (2010) 035002,
  [\href{http://xxx.lanl.gov/abs/0905.1822}{{\tt arXiv:0905.1822}}].

\bibitem{DeTurck1983}
D.~M. DeTurck, {\it {Deforming metrics in the direction of their Ricci
  tensors}},  {\em J. Differential Geom.} {\bf 18} (1983) 157.

\bibitem{DeTurck2003}
D.~M. DeTurck, {\it {Deforming metrics in the direction of their Ricci tensors
  (Improved version)}},  {\em Series in Geometry and Topology,} {\bf 37} (2003)
  163.

\bibitem{Hamilton1982}
R.~S. Hamilton, {\it {Three-manifolds with positive Ricci curvature}},  {\em J.
  Differential Geom.} {\bf 17} (1982) 255.

\bibitem{Thurston1982}
W.~P. Thurston, {\it {Three-Dimensional Manifolds, Kleinian Groups and
  Hyperbolic Geometry.}},  {\em Bull. Amer. Math. Soc.} {\bf 6} (1982) 357.

\bibitem{Perelman2002}
G.~Perelman, {\it {The entropy formula for the Ricci flow and its geometric
  applications}},  \href{http://xxx.lanl.gov/abs/math/0211159}{{\tt
  math/0211159}}.

\bibitem{Perelman2003}
G.~Perelman, {\it {Ricci flow with surgery on three-manifolds}},
  \href{http://xxx.lanl.gov/abs/math/0303109}{{\tt math/0303109}}.

\bibitem{Morgan2005}
J.~W. Morgan, {\it {Recent progress on the Poincar\'e conjecture and the
  classification of 3-manifolds}},  {\em Bull. Amer. Math. Soc. (N.S.)} {\bf
  42} (2005) 57.

\bibitem{Topping2006}
P.~M. Topping, {\it {Lectures on the Ricci flow}},  {\em L.M.S. Lecture note
  series 325 C.U.P. (2006)} (2006)
  [\href{http://xxx.lanl.gov/abs/http://www.maths.warwick.ac.uk/~topping/RFnotes.html}{{\tt
  http://www.maths.warwick.ac.uk/~topping/RFnotes.html}}].

\bibitem{ChowKnopfBook}
B.~Chow and D.~Knopf, {\em {The Ricci flow: an introduction}}, vol.~110.
\newblock Mathematical Surveys and Monographs, American Mathematical Society,
  Providence, RI, 2004, 2004.

\bibitem{Wiseman:2011by}
T.~Wiseman, {\em {Numerical construction of static and stationary black
  holes}}.
\newblock in Black Holes in Higher Dimensions, CUP 2012, Ed. Gary T. Horowitz,
  2011.

\bibitem{Wiseman:2002zc}
T.~Wiseman, {\it {Static axisymmetric vacuum solutions and nonuniform black
  strings}},  {\em Class. Quant. Grav.} {\bf 20} (2003) 1137--1176,
  [\href{http://xxx.lanl.gov/abs/hep-th/0209051}{{\tt hep-th/0209051}}].

\bibitem{Kol:2003ja}
B.~Kol and T.~Wiseman, {\it {Evidence that highly nonuniform black strings have
  a conical waist}},  {\em Class. Quant. Grav.} {\bf 20} (2003) 3493--3504,
  [\href{http://xxx.lanl.gov/abs/hep-th/0304070}{{\tt hep-th/0304070}}].

\bibitem{Kudoh:2004hs}
H.~Kudoh and T.~Wiseman, {\it {Connecting black holes and black strings}},
  {\em Phys. Rev. Lett.} {\bf 94} (2005) 161102,
  [\href{http://xxx.lanl.gov/abs/hep-th/0409111}{{\tt hep-th/0409111}}].

\bibitem{Kudoh:2003ki}
H.~Kudoh and T.~Wiseman, {\it {Properties of Kaluza-Klein black holes}},  {\em
  Prog. Theor. Phys.} {\bf 111} (2004) 475--507,
  [\href{http://xxx.lanl.gov/abs/hep-th/0310104}{{\tt hep-th/0310104}}].

\bibitem{Aharony:2005bm}
O.~Aharony, S.~Minwalla, and T.~Wiseman, {\it {Plasma-balls in large N gauge
  theories and localized black holes}},  {\em Class. Quant. Grav.} {\bf 23}
  (2006) 2171--2210, [\href{http://xxx.lanl.gov/abs/hep-th/0507219}{{\tt
  hep-th/0507219}}].

\bibitem{Kleihaus:1997mn}
B.~Kleihaus and J.~Kunz, {\it {Static axially symmetric Einstein Yang-Mills
  dilaton solutions: 1. Regular solutions}},  {\em Phys. Rev.} {\bf D57} (1998)
  834--856, [\href{http://xxx.lanl.gov/abs/gr-qc/9707045}{{\tt
  gr-qc/9707045}}].

\bibitem{Kleihaus:1997ws}
B.~Kleihaus and J.~Kunz, {\it {Static axially symmetric Einstein Yang-Mills
  dilaton solutions. 2. Black hole solutions}},  {\em Phys. Rev.} {\bf D57}
  (1998) 6138--6157, [\href{http://xxx.lanl.gov/abs/gr-qc/9712086}{{\tt
  gr-qc/9712086}}].

\bibitem{Kleihaus:2000kg}
B.~Kleihaus and J.~Kunz, {\it {Rotating hairy black holes}},  {\em Phys. Rev.
  Lett.} {\bf 86} (2001) 3704--3707,
  [\href{http://xxx.lanl.gov/abs/gr-qc/0012081}{{\tt gr-qc/0012081}}].

\bibitem{Hartmann:2001ic}
B.~Hartmann, B.~Kleihaus, and J.~Kunz, {\it {Axially symmetric monopoles and
  black holes in Einstein-Yang-Mills-Higgs theory}},  {\em Phys. Rev.} {\bf
  D65} (2002) 024027, [\href{http://xxx.lanl.gov/abs/hep-th/0108129}{{\tt
  hep-th/0108129}}].

\bibitem{Kleihaus:2002ee}
B.~Kleihaus, J.~Kunz, and F.~Navarro-Lerida, {\it {Rotating Einstein-Yang-Mills
  black holes}},  {\em Phys. Rev.} {\bf D66} (2002) 104001,
  [\href{http://xxx.lanl.gov/abs/gr-qc/0207042}{{\tt gr-qc/0207042}}].

\bibitem{Kleihaus:2005me}
B.~Kleihaus, J.~Kunz, and M.~List, {\it {Rotating boson stars and Q-balls}},
  {\em Phys. Rev.} {\bf D72} (2005) 064002,
  [\href{http://xxx.lanl.gov/abs/gr-qc/0505143}{{\tt gr-qc/0505143}}].

\bibitem{Kleihaus:2006ee}
B.~Kleihaus, J.~Kunz, and E.~Radu, {\it {New nonuniform black string
  solutions}},  {\em JHEP} {\bf 06} (2006) 016,
  [\href{http://xxx.lanl.gov/abs/hep-th/0603119}{{\tt hep-th/0603119}}].

\bibitem{Kalisch:2015via}
M.~Kalisch and M.~Ansorg, {\it {Highly Deformed Non-uniform Black Strings in
  Six Dimensions}},  in {\em {14th Marcel Grossmann Meeting on Recent
  Developments in Theoretical and Experimental General Relativity,
  Astrophysics, and Relativistic Field Theories (MG14) Rome, Italy, July 12-18,
  2015}}, 2015.
\newblock \href{http://xxx.lanl.gov/abs/1509.0308}{{\tt arXiv:1509.0308}}.

\bibitem{Press:1992zz}
W.~H. Press, S.~A. Teukolsky, W.~T. Vetterling, and B.~P. Flannery, {\it
  {Numerical Recipes in FORTRAN: The Art of Scientific Computing}}, .

\bibitem{Grandclement:2007sb}
P.~Grandclement and J.~Novak, {\it {Spectral methods for numerical
  relativity}},  {\em Living Rev. Rel.} {\bf 12} (2009) 1,
  [\href{http://xxx.lanl.gov/abs/0706.2286}{{\tt arXiv:0706.2286}}].

\bibitem{HagemanYoungBook}
L.~A. Hageman and D.~M. Young, {\em Applied Iterative Methods}.
\newblock Dover Publications, 2012.

\bibitem{VargaBook}
R.~S. Varga, {\em Matrix Iterative Analysis}.
\newblock Springer Series in Computational Mathematics, 2009.

\bibitem{CanutoBook}
C.~{Canuto}, M.~Y. {Hussaini}, A.~{Quarteroni}, and T.~A. {Zang}, {\em
  {Spectral Methods}}.
\newblock Springer-Verlag, 2006.

\bibitem{Trefethen}
L.~N. Trefethen, {\em Spectral Methods in MATLAB}.
\newblock SIAM, Philadelphia, 2000.

\bibitem{Boyd}
J.~P. Boyd, {\em Chebyshev and Fourier Spectral Methods}.
\newblock Dover Books on Mathematics, 2001.

\bibitem{Harmark:2007md}
T.~Harmark, V.~Niarchos, and N.~A. Obers, {\it {Instabilities of black strings
  and branes}},  {\em Class. Quant. Grav.} {\bf 24} (2007) R1--R90,
  [\href{http://xxx.lanl.gov/abs/hep-th/0701022}{{\tt hep-th/0701022}}].

\bibitem{Horowitz:2011cq}
G.~T. Horowitz and T.~Wiseman, {\em {General black holes in Kaluza-Klein
  theory}}.
\newblock in Black Holes in Higher Dimensions, CUP 2012, Ed. Gary T. Horowitz,
  2011.

\bibitem{Cardoso:2006ks}
V.~Cardoso and O.~J.~C. Dias, {\it {Rayleigh-Plateau and Gregory-Laflamme
  instabilities of black strings}},  {\em Phys. Rev. Lett.} {\bf 96} (2006)
  181601, [\href{http://xxx.lanl.gov/abs/hep-th/0602017}{{\tt
  hep-th/0602017}}].

\bibitem{Caldarelli:2008mv}
M.~M. Caldarelli, O.~J.~C. Dias, R.~Emparan, and D.~Klemm, {\it {Black Holes as
  Lumps of Fluid}},  {\em JHEP} {\bf 04} (2009) 024,
  [\href{http://xxx.lanl.gov/abs/0811.2381}{{\tt arXiv:0811.2381}}].

\bibitem{Caldarelli:2008ze}
M.~M. Caldarelli, O.~J.~C. Dias, and D.~Klemm, {\it {Dyonic AdS black holes
  from magnetohydrodynamics}},  {\em JHEP} {\bf 03} (2009) 025,
  [\href{http://xxx.lanl.gov/abs/0812.0801}{{\tt arXiv:0812.0801}}].

\bibitem{Camps:2010br}
J.~Camps, R.~Emparan, and N.~Haddad, {\it {Black Brane Viscosity and the
  Gregory-Laflamme Instability}},  {\em JHEP} {\bf 05} (2010) 042,
  [\href{http://xxx.lanl.gov/abs/1003.3636}{{\tt arXiv:1003.3636}}].

\bibitem{Caldarelli:2012hy}
M.~M. Caldarelli, J.~Camps, B.~Gout\'eraux, and K.~Skenderis, {\it
  {AdS/Ricci-flat correspondence and the Gregory-Laflamme instability}},  {\em
  Phys. Rev.} {\bf D87} (2013), no.~6 061502,
  [\href{http://xxx.lanl.gov/abs/1211.2815}{{\tt arXiv:1211.2815}}].

\bibitem{Caldarelli:2013aaa}
M.~M. Caldarelli, J.~Camps, B.~Gout\'eraux, and K.~Skenderis, {\it
  {AdS/Ricci-flat correspondence}},  {\em JHEP} {\bf 04} (2014) 071,
  [\href{http://xxx.lanl.gov/abs/1312.7874}{{\tt arXiv:1312.7874}}].

\bibitem{Gross:1982cv}
D.~J. Gross, M.~J. Perry, and L.~G. Yaffe, {\it {Instability of Flat Space at
  Finite Temperature}},  {\em Phys. Rev.} {\bf D25} (1982) 330--355.

\bibitem{Horowitz:1991cd}
G.~T. Horowitz and A.~Strominger, {\it {Black strings and P-branes}},  {\em
  Nucl. Phys.} {\bf B360} (1991) 197--209.

\bibitem{Gregory:1993vy}
R.~Gregory and R.~Laflamme, {\it {Black strings and p-branes are unstable}},
  {\em Phys. Rev. Lett.} {\bf 70} (1993) 2837--2840,
  [\href{http://xxx.lanl.gov/abs/hep-th/9301052}{{\tt hep-th/9301052}}].

\bibitem{Gregory:1994bj}
R.~Gregory and R.~Laflamme, {\it {The Instability of charged black strings and
  p-branes}},  {\em Nucl. Phys.} {\bf B428} (1994) 399--434,
  [\href{http://xxx.lanl.gov/abs/hep-th/9404071}{{\tt hep-th/9404071}}].

\bibitem{Horowitz:2001cz}
G.~T. Horowitz and K.~Maeda, {\it {Fate of the black string instability}},
  {\em Phys. Rev. Lett.} {\bf 87} (2001) 131301,
  [\href{http://xxx.lanl.gov/abs/hep-th/0105111}{{\tt hep-th/0105111}}].

\bibitem{Gubser:2001ac}
S.~S. Gubser, {\it {On non-uniform black branes}},  {\em Class. Quant. Grav.}
  {\bf 19} (2002) 4825--4844,
  [\href{http://xxx.lanl.gov/abs/hep-th/0110193}{{\tt hep-th/0110193}}].

\bibitem{Reall:2001ag}
H.~S. Reall, {\it {Classical and thermodynamic stability of black branes}},
  {\em Phys. Rev.} {\bf D64} (2001) 044005,
  [\href{http://xxx.lanl.gov/abs/hep-th/0104071}{{\tt hep-th/0104071}}].

\bibitem{Kol:2002xz}
B.~Kol, {\it {Topology change in general relativity, and the black hole black
  string transition}},  {\em JHEP} {\bf 10} (2005) 049,
  [\href{http://xxx.lanl.gov/abs/hep-th/0206220}{{\tt hep-th/0206220}}].

\bibitem{Harmark:2002tr}
T.~Harmark and N.~A. Obers, {\it {Black holes on cylinders}},  {\em JHEP} {\bf
  05} (2002) 032, [\href{http://xxx.lanl.gov/abs/hep-th/0204047}{{\tt
  hep-th/0204047}}].

\bibitem{Harmark:2003yz}
T.~Harmark, {\it {Small black holes on cylinders}},  {\em Phys. Rev.} {\bf D69}
  (2004) 104015, [\href{http://xxx.lanl.gov/abs/hep-th/0310259}{{\tt
  hep-th/0310259}}].

\bibitem{Gorbonos:2004uc}
D.~Gorbonos and B.~Kol, {\it {A Dialogue of multipoles: Matched asymptotic
  expansion for caged black holes}},  {\em JHEP} {\bf 06} (2004) 053,
  [\href{http://xxx.lanl.gov/abs/hep-th/0406002}{{\tt hep-th/0406002}}].

\bibitem{Sorkin:2004qq}
E.~Sorkin, {\it {A Critical dimension in the black string phase transition}},
  {\em Phys. Rev. Lett.} {\bf 93} (2004) 031601,
  [\href{http://xxx.lanl.gov/abs/hep-th/0402216}{{\tt hep-th/0402216}}].

\bibitem{Asnin:2006ip}
V.~Asnin, B.~Kol, and M.~Smolkin, {\it {Analytic evidence for continuous self
  similarity of the critical merger solution}},  {\em Class. Quant. Grav.} {\bf
  23} (2006) 6805--6827, [\href{http://xxx.lanl.gov/abs/hep-th/0607129}{{\tt
  hep-th/0607129}}].

\bibitem{Dias:2007hg}
O.~J.~C. Dias, T.~Harmark, R.~C. Myers, and N.~A. Obers, {\it {Multi-black hole
  configurations on the cylinder}},  {\em Phys. Rev.} {\bf D76} (2007) 104025,
  [\href{http://xxx.lanl.gov/abs/0706.3645}{{\tt arXiv:0706.3645}}].

\bibitem{Monteiro:2009tc}
R.~Monteiro, M.~J. Perry, and J.~E. Santos, {\it {Thermodynamic instability of
  rotating black holes}},  {\em Phys. Rev.} {\bf D80} (2009) 024041,
  [\href{http://xxx.lanl.gov/abs/0903.3256}{{\tt arXiv:0903.3256}}].

\bibitem{Dias:2009iu}
O.~J.~C. Dias, P.~Figueras, R.~Monteiro, J.~E. Santos, and R.~Emparan, {\it
  {Instability and new phases of higher-dimensional rotating black holes}},
  {\em Phys. Rev.} {\bf D80} (2009) 111701,
  [\href{http://xxx.lanl.gov/abs/0907.2248}{{\tt arXiv:0907.2248}}].

\bibitem{Dias:2010maa}
O.~J.~C. Dias, P.~Figueras, R.~Monteiro, and J.~E. Santos, {\it {Ultraspinning
  instability of rotating black holes}},  {\em Phys. Rev.} {\bf D82} (2010)
  104025, [\href{http://xxx.lanl.gov/abs/1006.1904}{{\tt arXiv:1006.1904}}].

\bibitem{Dias:2010eu}
O.~J.~C. Dias, P.~Figueras, R.~Monteiro, H.~S. Reall, and J.~E. Santos, {\it
  {An instability of higher-dimensional rotating black holes}},  {\em JHEP}
  {\bf 05} (2010) 076, [\href{http://xxx.lanl.gov/abs/1001.4527}{{\tt
  arXiv:1001.4527}}].

\bibitem{Dias:2011jg}
O.~J.~C. Dias, R.~Monteiro, and J.~E. Santos, {\it {Ultraspinning instability:
  the missing link}},  {\em JHEP} {\bf 08} (2011) 139,
  [\href{http://xxx.lanl.gov/abs/1106.4554}{{\tt arXiv:1106.4554}}].

\bibitem{Lehner:2010pn}
L.~Lehner and F.~Pretorius, {\it {Black Strings, Low Viscosity Fluids, and
  Violation of Cosmic Censorship}},  {\em Phys. Rev. Lett.} {\bf 105} (2010)
  101102, [\href{http://xxx.lanl.gov/abs/1006.5960}{{\tt arXiv:1006.5960}}].

\bibitem{Lehner:2011wc}
L.~Lehner and F.~Pretorius, {\it {Final State of Gregory-Laflamme
  Instability}},  \href{http://xxx.lanl.gov/abs/1106.5184}{{\tt
  arXiv:1106.5184}}.

\bibitem{Figueras:2012xj}
P.~Figueras, K.~Murata, and H.~S. Reall, {\it {Stable non-uniform black strings
  below the critical dimension}},  {\em JHEP} {\bf 11} (2012) 071,
  [\href{http://xxx.lanl.gov/abs/1209.1981}{{\tt arXiv:1209.1981}}].

\bibitem{Prestidge:1999uq}
T.~Prestidge, {\it {Dynamic and thermodynamic stability and negative modes in
  Schwarzschild-anti-de Sitter}},  {\em Phys. Rev.} {\bf D61} (2000) 084002,
  [\href{http://xxx.lanl.gov/abs/hep-th/9907163}{{\tt hep-th/9907163}}].

\bibitem{Dias:2010gk}
O.~J.~C. Dias, P.~Figueras, R.~Monteiro, and J.~E. Santos, {\it {Ultraspinning
  instability of anti-de Sitter black holes}},  {\em JHEP} {\bf 12} (2010) 067,
  [\href{http://xxx.lanl.gov/abs/1011.0996}{{\tt arXiv:1011.0996}}].

\bibitem{Sorkin:2006wp}
E.~Sorkin, {\it {Nonuniform black strings in various dimensions}},  {\em Phys.
  Rev.} {\bf D74} (2006) 104027,
  [\href{http://xxx.lanl.gov/abs/gr-qc/0608115}{{\tt gr-qc/0608115}}].

\bibitem{Penrose:1969pc}
R.~Penrose, {\it {Gravitational collapse: The role of general relativity}},
  {\em Riv. Nuovo Cim.} {\bf 1} (1969) 252--276. [Gen. Rel.
  Grav.34,1141(2002)].

\bibitem{HawkingEllisBook}
S.~Hawking and G.~Ellis, {\em The Large Scale Structure of Space-Time}.
\newblock Cambridge University Press, Cambridge, 1973.

\bibitem{waldbook}
R.~M. Wald, {\em General Relativity}.
\newblock University Of Chicago Press, 1984.

\bibitem{Christodoulou:1999}
D.~Christodoulou, {\it {On the global initial value problem and the issue of
  singularities}},  {\em Class. Quantum Gravity} {\bf 97} (1999) A23.

\bibitem{Senovilla:2014gza}
J.~M.~M. Senovilla and D.~Garfinkle, {\it {The 1965 Penrose singularity
  theorem}},  {\em Class. Quant. Grav.} {\bf 32} (2015), no.~12 124008,
  [\href{http://xxx.lanl.gov/abs/1410.5226}{{\tt arXiv:1410.5226}}].

\bibitem{Emparan:2015gva}
R.~Emparan, R.~Suzuki, and K.~Tanabe, {\it {Evolution and End Point of the
  Black String Instability: Large D Solution}},  {\em Phys. Rev. Lett.} {\bf
  115} (2015), no.~9 091102, [\href{http://xxx.lanl.gov/abs/1506.0677}{{\tt
  arXiv:1506.0677}}].

\bibitem{Heusler:1996}
M.~Heusler, {\em {Black hole uniqueness theorems}}.
\newblock Cambridge University Press, 1996.

\bibitem{Chrusciel:2012jk}
P.~T. Chrusciel, J.~L. Costa, and M.~Heusler, {\it {Stationary Black Holes:
  Uniqueness and Beyond}},  {\em Living Rev. Rel.} {\bf 15} (2012) 7,
  [\href{http://xxx.lanl.gov/abs/1205.6112}{{\tt arXiv:1205.6112}}].

\bibitem{Robinson:2004zz}
D.~Robinson, {\it {Four decades of black holes uniqueness theorems}},  {\em in
  {``The Kerr Spacetime: Rotating Black Holes in General Relativity."}}
  (Editors: D. L. Wiltshire, M. Visser, S. M. Scott (Cambridge University
  Press, 2009).).

\bibitem{Hollands:2012xy}
S.~Hollands and A.~Ishibashi, {\it {Black hole uniqueness theorems in higher
  dimensional spacetimes}},  {\em Class. Quant. Grav.} {\bf 29} (2012) 163001,
  [\href{http://xxx.lanl.gov/abs/1206.1164}{{\tt arXiv:1206.1164}}].

\bibitem{Emparan:2006mm}
R.~Emparan and H.~S. Reall, {\it {Black Rings}},  {\em Class. Quant. Grav.}
  {\bf 23} (2006) R169, [\href{http://xxx.lanl.gov/abs/hep-th/0608012}{{\tt
  hep-th/0608012}}].

\bibitem{Emparan:2008eg}
R.~Emparan and H.~S. Reall, {\it {Black Holes in Higher Dimensions}},  {\em
  Living Rev. Rel.} {\bf 11} (2008) 6,
  [\href{http://xxx.lanl.gov/abs/0801.3471}{{\tt arXiv:0801.3471}}].

\bibitem{Ionescu:2015dna}
A.~Ionescu and S.~Klainerman, {\it {Rigidity Results in General Relativity: a
  Review}},  \href{http://xxx.lanl.gov/abs/1501.0158}{{\tt arXiv:1501.0158}}.

\bibitem{Hawking:1971vc}
S.~W. Hawking, {\it {Black holes in general relativity}},  {\em Commun. Math.
  Phys.} {\bf 25} (1972) 152--166.

\bibitem{HawkingEllis:1973}
S.~Hawking and G.~Ellis, {\em {The large scale Structure of space-time}}.
\newblock Cambridge University Press, 1973, 1973.

\bibitem{SudarskyW:1992}
D.~Sudarsky and R.~M. Wald, {\it {Extrema of mass, stationarity, and staticity,
  and solutions to the Einstein Yang-Mills equations}},  {\em Phys. Rev. D}
  {\bf 46} (1992) 1453.

\bibitem{Chrusciel:1993cv}
P.~T. Chrusciel and R.~M. Wald, {\it {Maximal hypersurfaces in asymptotically
  stationary space-times}},  {\em Commun. Math. Phys.} {\bf 163} (1994)
  561--604, [\href{http://xxx.lanl.gov/abs/gr-qc/9304009}{{\tt
  gr-qc/9304009}}].

\bibitem{Friedrich:1998wq}
H.~Friedrich, I.~Racz, and R.~M. Wald, {\it {On the rigidity theorem for
  space-times with a stationary event horizon or a compact Cauchy horizon}},
  {\em Commun. Math. Phys.} {\bf 204} (1999) 691--707,
  [\href{http://xxx.lanl.gov/abs/gr-qc/9811021}{{\tt gr-qc/9811021}}].

\bibitem{Hollands:2008wn}
S.~Hollands and A.~Ishibashi, {\it {On the `Stationary Implies Axisymmetric'
  Theorem for Extremal Black Holes in Higher Dimensions}},  {\em Commun. Math.
  Phys.} {\bf 291} (2009) 403--441,
  [\href{http://xxx.lanl.gov/abs/0809.2659}{{\tt arXiv:0809.2659}}].

\bibitem{Chrusciel:2008js}
P.~T. Chrusciel and J.~Lopes~Costa, {\it {On uniqueness of stationary vacuum
  black holes}},  {\em Asterisque} {\bf 321} (2008) 195--265,
  [\href{http://xxx.lanl.gov/abs/0806.0016}{{\tt arXiv:0806.0016}}].

\bibitem{Israel:1967wq}
W.~Israel, {\it {Event horizons in static vacuum space-times}},  {\em Phys.
  Rev.} {\bf 164} (1967) 1776--1779.

\bibitem{Israel:1967za}
W.~Israel, {\it {Event horizons in static electrovac space-times}},  {\em
  Commun. Math. Phys.} {\bf 8} (1968) 245--260.

\bibitem{Carter:1971zc}
B.~Carter, {\it {Axisymmetric Black Hole Has Only Two Degrees of Freedom}},
  {\em Phys. Rev. Lett.} {\bf 26} (1971) 331--333.

\bibitem{Wald:1971iw}
R.~M. Wald, {\it {Final states of gravitational collapse}},  {\em Phys. Rev.
  Lett.} {\bf 26} (1971) 1653--1655.

\bibitem{Carter:1973}
B.~Carter, {\em {Black Hole Equilibrium States: II General Theory of Stationary
  Black Hole States}}.
\newblock Black Holes (1972 Les Houches Summer School), Ed. B. \& C. DeWitt,
  1973, 1972.

\bibitem{Carter:1986}
B.~Carter, {\em {Mathematical Foundations of the Theory of Relativistic Stellar
  and Black Hole Configurations}}.
\newblock Gravitation in Astrophysics: Carg\`ese 1986, Ed. Carter, B. and
  Hartle, J.B (Plenum Press, New York, 1987), 1987.

\bibitem{Robinson:1975bv}
D.~C. Robinson, {\it {Uniqueness of the Kerr black hole}},  {\em Phys. Rev.
  Lett.} {\bf 34} (1975) 905--906.

\bibitem{Mazur:1982}
P.~O. Mazur, {\it {Proof of uniqueness of the Kerr-Newman black hole
  solution}},  {\em J. Phys.} {\bf A15} (1982) 3173--3180.

\bibitem{Bunting:1983}
G.~L. Bunting, {\it {Proof of the uniqueness conjecture for black holes}},
  {\em PhD Thesis, Univ. of New England, Armidale, N.S.W.} (1983).

\bibitem{Amsel:2009et}
A.~J. Amsel, G.~T. Horowitz, D.~Marolf, and M.~M. Roberts, {\it {Uniqueness of
  Extremal Kerr and Kerr-Newman Black Holes}},  {\em Phys. Rev.} {\bf D81}
  (2010) 024033, [\href{http://xxx.lanl.gov/abs/0906.2367}{{\tt
  arXiv:0906.2367}}].

\bibitem{Figueras:2009ci}
P.~Figueras and J.~Lucietti, {\it {On the uniqueness of extremal vacuum black
  holes}},  {\em Class. Quant. Grav.} {\bf 27} (2010) 095001,
  [\href{http://xxx.lanl.gov/abs/0906.5565}{{\tt arXiv:0906.5565}}].

\bibitem{Newman:1965my}
E.~T. Newman, R.~Couch, K.~Chinnapared, A.~Exton, A.~Prakash, et~al., {\it
  {Metric of a Rotating, Charged Mass}},  {\em J.Math.Phys.} {\bf 6} (1965)
  918--919.

\bibitem{Adamo:2014baa}
T.~Adamo and E.~Newman, {\it {The Kerr-Newman metric: A Review}},
  \href{http://xxx.lanl.gov/abs/1410.6626}{{\tt arXiv:1410.6626}}.

\bibitem{Kerr:1963ud}
R.~P. Kerr, {\it {Gravitational field of a spinning mass as an example of
  algebraically special metrics}},  {\em Phys. Rev. Lett.} {\bf 11} (1963)
  237--238.

\bibitem{Kerr:2007dk}
R.~P. Kerr, {\it {Discovering the Kerr and Kerr-Schild metrics}},  in {\em
  {Kerr Fest: Black Holes in Astrophysics, General Relativity and Quantum
  Gravity Christchurch, New Zealand, August 26-28, 2004}}, 2007.
\newblock \href{http://xxx.lanl.gov/abs/0706.1109}{{\tt arXiv:0706.1109}}.

\bibitem{Teukolsky:2014vca}
S.~A. Teukolsky, {\it {The Kerr Metric}},  {\em Class. Quant. Grav.} {\bf 32}
  (2015), no.~12 124006, [\href{http://xxx.lanl.gov/abs/1410.2130}{{\tt
  arXiv:1410.2130}}].

\bibitem{Reissner:1916}
B.~Reissner, {\it {\"Uber die Eigengravitation des elektrischen Feldes nach der
  Einsteinschen Theorie}},  {\em Annalen Phys.} {\bf 106} (1916) 355.

\bibitem{Nordstrom:1918}
{Nordstr\"om, G.}, {\it {On the Energy of the Gravitational Field in Einstein's
  Theory}},  {\em Verhandl. Koninkl. Ned. Akad. Wetenschap., Afdel. Natuurk.,
  Amsterdam} {\bf 26} (1918) 1201.

\bibitem{Schwarzschild:1916}
K.~Schwarzschild, {\it {On the gravitational field of a mass point according to
  Einstein's theory}},  {\em Sitzungsber.Preuss.Akad.Wiss.Berlin (Math.Phys.)
  1916 (1916) 189} {\bf 1916} (1916) 189,
  [\href{http://xxx.lanl.gov/abs/physics/9905030}{{\tt physics/9905030}}].

\bibitem{Majumdar:1947eu}
S.~D. Majumdar, {\it {A class of exact solutions of Einstein's field
  equations}},  {\em Phys. Rev.} {\bf 72} (1947) 390--398.

\bibitem{Papaetrou:1947ib}
A.~Papapetrou, {\it {A Static solution of the equations of the gravitational
  field for an arbitrary charge distribution}},  {\em Proc. Roy. Irish
  Acad.(Sect. A)} {\bf A51} (1947) 191--204.

\bibitem{Chrusciel:1994qa}
P.~T. Chrusciel and N.~S. Nadirashvili, {\it {All electrovacuum
  Majumdar-Papapetrou space-times with nonsingular black holes}},  {\em Class.
  Quant. Grav.} {\bf 12} (1995) L17--L23,
  [\href{http://xxx.lanl.gov/abs/gr-qc/9412044}{{\tt gr-qc/9412044}}].

\bibitem{Heusler:1996ex}
M.~Heusler, {\it {On the uniqueness of the Papapetrou-Majumdar metric}},  {\em
  Class. Quant. Grav.} {\bf 14} (1997) L129--L134,
  [\href{http://xxx.lanl.gov/abs/gr-qc/9607001}{{\tt gr-qc/9607001}}].

\bibitem{WheelerRuffiniNohair:1971}
R.~Ruffini and J.~A. Wheeler, {\it {Introducing the black hole}},  {\em Physics
  Today} {\bf 24} (1971) 30.

\bibitem{Bekenstein:PhysTod1980}
J.~D. {Bekenstein}, {\it {Black-hole thermodynamics}},  {\em Physics Today}
  {\bf 33} (1980) 24--31.

\bibitem{Bekenstein:1996pn}
J.~D. Bekenstein, {\it {Black hole hair: 25 - years after}},  in {\em {Physics.
  Proceedings, 2nd International A.D. Sakharov Conference, Moscow, Russia, May
  20-24, 1996}}, 1996.
\newblock \href{http://xxx.lanl.gov/abs/gr-qc/9605059}{{\tt gr-qc/9605059}}.

\bibitem{Wald:1999vt}
R.~M. Wald, {\it {The thermodynamics of black holes}},  {\em Living Rev. Rel.}
  {\bf 4} (2001) 6, [\href{http://xxx.lanl.gov/abs/gr-qc/9912119}{{\tt
  gr-qc/9912119}}].

\bibitem{Reall:2002bh}
H.~S. Reall, {\it {Higher dimensional black holes and supersymmetry}},  {\em
  Phys. Rev.} {\bf D68} (2003) 024024,
  [\href{http://xxx.lanl.gov/abs/hep-th/0211290}{{\tt hep-th/0211290}}].
  [Erratum: Phys. Rev.D70,089902(2004)].

\bibitem{Chase:1970}
J.~E. Chase, {\it {Event horizons in static scalar-vacuum space-times}},  {\em
  Commun. Math. Phys.} {\bf 19} (1970) 276.

\bibitem{Penney:1968zz}
R.~Penney, {\it {Axially Symmetric Zero-Mass Meson Solutions of Einstein
  Equations}},  {\em Phys. Rev.} {\bf 174} (1968) 1578--1579.

\bibitem{Bekenstein:1972ny}
J.~D. Bekenstein, {\it {Transcendence of the law of baryon-number conservation
  in black hole physics}},  {\em Phys. Rev. Lett.} {\bf 28} (1972) 452--455.

\bibitem{Bekenstein:1971hc}
J.~D. Bekenstein, {\it {Nonexistence of baryon number for static black holes}},
   {\em Phys. Rev.} {\bf D5} (1972) 1239--1246.

\bibitem{Bekenstein:1972ky}
J.~D. Bekenstein, {\it {Nonexistence of baryon number for black holes. ii}},
  {\em Phys. Rev.} {\bf D5} (1972) 2403--2412.

\bibitem{Teitelboim:1972qx}
C.~Teitelboim, {\it {Nonmeasurability of the quantum numbers of a black hole}},
   {\em Phys. Rev.} {\bf D5} (1972) 2941--2954.

\bibitem{Hartle1972}
J.~B. Hartle, {\em Magic Without Magic}.
\newblock edited by J. Klauder, Freeman, San Francisco, 1972.

\bibitem{Heusler:1992ss}
M.~Heusler, {\it {A No hair theorem for selfgravitating nonlinear sigma
  models}},  {\em J. Math. Phys.} {\bf 33} (1992) 3497--3502.

\bibitem{Bekenstein:1995un}
J.~D. Bekenstein, {\it {Novel 'no scalar hair' theorem for black holes}},  {\em
  Phys. Rev.} {\bf D51} (1995) 6608--6611.

\bibitem{Sudarsky:1995zg}
D.~Sudarsky, {\it {A Simple proof of a no hair theorem in Einstein Higgs
  theory,}},  {\em Class. Quant. Grav.} {\bf 12} (1995) 579--584.

\bibitem{Hertog:2006rr}
T.~Hertog, {\it {Towards a Novel no-hair Theorem for Black Holes}},  {\em Phys.
  Rev.} {\bf D74} (2006) 084008,
  [\href{http://xxx.lanl.gov/abs/gr-qc/0608075}{{\tt gr-qc/0608075}}].

\bibitem{Bizon:1994dh}
P.~Bizon, {\it {Gravitating solitons and hairy black holes}},  {\em Acta Phys.
  Polon.} {\bf B25} (1994) 877--898,
  [\href{http://xxx.lanl.gov/abs/gr-qc/9402016}{{\tt gr-qc/9402016}}].

\bibitem{Volkov:1998cc}
M.~S. Volkov and D.~V. Gal'tsov, {\it {Gravitating nonAbelian solitons and
  black holes with Yang-Mills fields}},  {\em Phys. Rept.} {\bf 319} (1999)
  1--83, [\href{http://xxx.lanl.gov/abs/hep-th/9810070}{{\tt hep-th/9810070}}].

\bibitem{Ashtekar:2000nx}
A.~Ashtekar, A.~Corichi, and D.~Sudarsky, {\it {Hairy black holes, horizon mass
  and solitons}},  {\em Class. Quant. Grav.} {\bf 18} (2001) 919--940,
  [\href{http://xxx.lanl.gov/abs/gr-qc/0011081}{{\tt gr-qc/0011081}}].

\bibitem{Gibbons:1982ih}
G.~W. Gibbons, {\it {Antigravitating Black Hole Solitons with Scalar Hair in
  N=4 Supergravity}},  {\em Nucl. Phys.} {\bf B207} (1982) 337--349.

\bibitem{Gibbons:1987ps}
G.~W. Gibbons and K.~Maeda, {\it {Black Holes and Membranes in Higher
  Dimensional Theories with Dilaton Fields}},  {\em Nucl. Phys.} {\bf B298}
  (1988) 741.

\bibitem{Garfinkle:1990qj}
D.~Garfinkle, G.~T. Horowitz, and A.~Strominger, {\it {Charged black holes in
  string theory}},  {\em Phys. Rev.} {\bf D43} (1991) 3140. [Erratum: Phys.
  Rev.D45,3888(1992)].

\bibitem{Lee:1991qs}
K.-M. Lee, V.~P. Nair, and E.~J. Weinberg, {\it {A Classical instability of
  Reissner-Nordstrom solutions and the fate of magnetically charged black
  holes}},  {\em Phys. Rev. Lett.} {\bf 68} (1992) 1100--1103,
  [\href{http://xxx.lanl.gov/abs/hep-th/9111045}{{\tt hep-th/9111045}}].

\bibitem{Achucarro:1995nu}
A.~Achucarro, R.~Gregory, and K.~Kuijken, {\it {Abelian Higgs hair for black
  holes}},  {\em Phys. Rev.} {\bf D52} (1995) 5729--5742,
  [\href{http://xxx.lanl.gov/abs/gr-qc/9505039}{{\tt gr-qc/9505039}}].

\bibitem{Volkov:1989fi}
M.~S. Volkov and D.~V. Galtsov, {\it {NonAbelian Einstein Yang-Mills black
  holes}},  {\em JETP Lett.} {\bf 50} (1989) 346--350. [Pisma Zh. Eksp. Teor.
  Fiz.50,312(1989)].

\bibitem{Volkov:1990sva}
M.~S. Volkov and D.~V. Galtsov, {\it {Black holes in Einstein Yang-Mills
  theory. (In Russian)}},  {\em Sov. J. Nucl. Phys.} {\bf 51} (1990) 747--753.
  [Yad. Fiz.51,1171(1990)].

\bibitem{Bizon:1990sr}
P.~Bizon, {\it {Colored black holes}},  {\em Phys. Rev. Lett.} {\bf 64} (1990)
  2844--2847.

\bibitem{Kuenzle:1990is}
H.~P. Kuenzle and A.~K.~M. Masood-ul Alam, {\it {Spherically symmetric static
  SU(2) Einstein Yang-Mills fields}},  {\em J. Math. Phys.} {\bf 31} (1990)
  928--935.

\bibitem{Breitenlohner:1991aa}
P.~Breitenlohner, P.~Forgacs, and D.~Maison, {\it {Gravitating monopole
  solutions}},  {\em Nucl. Phys.} {\bf B383} (1992) 357--376.

\bibitem{Lavrelashvili:1992ia}
G.~V. Lavrelashvili and D.~Maison, {\it {Regular and black hole solutions of
  Einstein Yang-Mills Dilaton theory}},  {\em Nucl. Phys.} {\bf B410} (1993)
  407--422.

\bibitem{Greene:1992fw}
B.~R. Greene, S.~D. Mathur, and C.~M. O'Neill, {\it {Eluding the no hair
  conjecture: Black holes in spontaneously broken gauge theories}},  {\em Phys.
  Rev.} {\bf D47} (1993) 2242--2259,
  [\href{http://xxx.lanl.gov/abs/hep-th/9211007}{{\tt hep-th/9211007}}].

\bibitem{Bizon:1992gb}
P.~Bizon and T.~Chmaj, {\it {Gravitating skyrmions}},  {\em Phys. Lett.} {\bf
  B297} (1992) 55--62.

\bibitem{Droz:1991cx}
S.~Droz, M.~Heusler, and N.~Straumann, {\it {New black hole solutions with
  hair}},  {\em Phys. Lett.} {\bf B268} (1991) 371--376.

\bibitem{Winstanley:2008ac}
E.~Winstanley, {\it {Classical Yang-Mills black hole hair in anti-de Sitter
  space}},  {\em Lect. Notes Phys.} {\bf 769} (2009) 49--87,
  [\href{http://xxx.lanl.gov/abs/0801.0527}{{\tt arXiv:0801.0527}}].

\bibitem{Torii:1998ir}
T.~Torii, K.~Maeda, and M.~Narita, {\it {No scalar hair conjecture in
  asymptotic de Sitter space-time}},  {\em Phys. Rev.} {\bf D59} (1999) 064027,
  [\href{http://xxx.lanl.gov/abs/gr-qc/9809036}{{\tt gr-qc/9809036}}].

\bibitem{Torii:2001pg}
T.~Torii, K.~Maeda, and M.~Narita, {\it {Scalar hair on the black hole in
  asymptotically anti-de Sitter space-time}},  {\em Phys. Rev.} {\bf D64}
  (2001) 044007.

\bibitem{Zloshchastiev:2004ny}
K.~G. Zloshchastiev, {\it {On co-existence of black holes and scalar field}},
  {\em Phys. Rev. Lett.} {\bf 94} (2005) 121101,
  [\href{http://xxx.lanl.gov/abs/hep-th/0408163}{{\tt hep-th/0408163}}].

\bibitem{Martinez:2004nb}
C.~Martinez, R.~Troncoso, and J.~Zanelli, {\it {Exact black hole solution with
  a minimally coupled scalar field}},  {\em Phys. Rev.} {\bf D70} (2004)
  084035, [\href{http://xxx.lanl.gov/abs/hep-th/0406111}{{\tt
  hep-th/0406111}}].

\bibitem{Martinez:2006an}
C.~Martinez and R.~Troncoso, {\it {Electrically charged black hole with scalar
  hair}},  {\em Phys. Rev.} {\bf D74} (2006) 064007,
  [\href{http://xxx.lanl.gov/abs/hep-th/0606130}{{\tt hep-th/0606130}}].

\bibitem{Gubser:2008px}
S.~S. Gubser, {\it {Breaking an Abelian gauge symmetry near a black hole
  horizon}},  {\em Phys.Rev.} {\bf D78} (2008) 065034,
  [\href{http://xxx.lanl.gov/abs/0801.2977}{{\tt arXiv:0801.2977}}].

\bibitem{Hartnoll:2008vx}
S.~A. Hartnoll, C.~P. Herzog, and G.~T. Horowitz, {\it {Building a Holographic
  Superconductor}},  {\em Phys.Rev.Lett.} {\bf 101} (2008) 031601,
  [\href{http://xxx.lanl.gov/abs/0803.3295}{{\tt arXiv:0803.3295}}].

\bibitem{Hartnoll:2008kx}
S.~A. Hartnoll, C.~P. Herzog, and G.~T. Horowitz, {\it {Holographic
  Superconductors}},  {\em JHEP} {\bf 0812} (2008) 015,
  [\href{http://xxx.lanl.gov/abs/0810.1563}{{\tt arXiv:0810.1563}}].

\bibitem{Faulkner:2010gj}
T.~Faulkner, G.~T. Horowitz, and M.~M. Roberts, {\it {Holographic quantum
  criticality from multi-trace deformations}},  {\em JHEP} {\bf 1104} (2011)
  051, [\href{http://xxx.lanl.gov/abs/1008.1581}{{\tt arXiv:1008.1581}}].

\bibitem{Dias:2010ma}
O.~J.~C. Dias, R.~Monteiro, H.~S. Reall, and J.~E. Santos, {\it {A Scalar field
  condensation instability of rotating anti-de Sitter black holes}},  {\em
  JHEP} {\bf 11} (2010) 036, [\href{http://xxx.lanl.gov/abs/1007.3745}{{\tt
  arXiv:1007.3745}}].

\bibitem{Herdeiro:2014goa}
C.~A.~R. Herdeiro and E.~Radu, {\it {Kerr black holes with scalar hair}},  {\em
  Phys. Rev. Lett.} {\bf 112} (2014) 221101,
  [\href{http://xxx.lanl.gov/abs/1403.2757}{{\tt arXiv:1403.2757}}].

\bibitem{Dias:2015rxy}
O.~J.~C. Dias, J.~E. Santos, and B.~Way, {\it {Black holes with a single
  Killing vector field: black resonators}},
  \href{http://xxx.lanl.gov/abs/1505.0479}{{\tt arXiv:1505.0479}}.

\bibitem{Breitenlohner:1982jf}
P.~Breitenlohner and D.~Z. Freedman, {\it {Stability in Gauged Extended
  Supergravity}},  {\em Annals Phys.} {\bf 144} (1982) 249.

\bibitem{Hartnoll:2009sz}
S.~A. Hartnoll, {\it {Lectures on holographic methods for condensed matter
  physics}},  {\em Class. Quant. Grav.} {\bf 26} (2009) 224002,
  [\href{http://xxx.lanl.gov/abs/0903.3246}{{\tt arXiv:0903.3246}}].

\bibitem{Herzog:2009xv}
C.~P. Herzog, {\it {Lectures on Holographic Superfluidity and
  Superconductivity}},  {\em J. Phys.} {\bf A42} (2009) 343001,
  [\href{http://xxx.lanl.gov/abs/0904.1975}{{\tt arXiv:0904.1975}}].

\bibitem{2010uqpt.book..701H}
S.~Hartnoll, {\em {Quantum Critical Dynamics from Black Holes}}.
\newblock CRC Press, Edited by Lincoln Carr, pp.~701-723, 2010.

\bibitem{McGreevy:2009xe}
J.~McGreevy, {\it {Holographic duality with a view toward many-body physics}},
  {\em Adv. High Energy Phys.} {\bf 2010} (2010) 723105,
  [\href{http://xxx.lanl.gov/abs/0909.0518}{{\tt arXiv:0909.0518}}].

\bibitem{DonosGauntlett:CQGreview2015}
A.~Donos and J.~P. Gauntlett, {\it {Holographic lattices, (Topical Review)}},
  {\em Class. Quantum Grav., to appear} (2015).

\bibitem{Horowitz:2011dz}
G.~T. Horowitz, J.~E. Santos, and B.~Way, {\it {A Holographic Josephson
  Junction}},  {\em Phys. Rev. Lett.} {\bf 106} (2011) 221601,
  [\href{http://xxx.lanl.gov/abs/1101.3326}{{\tt arXiv:1101.3326}}].

\bibitem{Horowitz:2012ky}
G.~T. Horowitz, J.~E. Santos, and D.~Tong, {\it {Optical Conductivity with
  Holographic Lattices}},  {\em JHEP} {\bf 07} (2012) 168,
  [\href{http://xxx.lanl.gov/abs/1204.0519}{{\tt arXiv:1204.0519}}].

\bibitem{GarciaGarcia:2012zd}
A.~M. Garcia-Garcia, J.~E. Santos, and B.~Way, {\it {Holographic Description of
  Finite Size Effects in Strongly Coupled Superconductors}},  {\em Phys. Rev.}
  {\bf B86} (2012) 064526, [\href{http://xxx.lanl.gov/abs/1204.4189}{{\tt
  arXiv:1204.4189}}].

\bibitem{Horowitz:2012gs}
G.~T. Horowitz, J.~E. Santos, and D.~Tong, {\it {Further Evidence for
  Lattice-Induced Scaling}},  {\em JHEP} {\bf 11} (2012) 102,
  [\href{http://xxx.lanl.gov/abs/1209.1098}{{\tt arXiv:1209.1098}}].

\bibitem{Donos:2012yu}
A.~Donos, J.~P. Gauntlett, J.~Sonner, and B.~Withers, {\it {Competing orders in
  M-theory: superfluids, stripes and metamagnetism}},  {\em JHEP} {\bf 03}
  (2013) 108, [\href{http://xxx.lanl.gov/abs/1212.0871}{{\tt
  arXiv:1212.0871}}].

\bibitem{Horowitz:2013jaa}
G.~T. Horowitz and J.~E. Santos, {\it {General Relativity and the Cuprates}},
  {\em JHEP} {\bf 06} (2013) 087,
  [\href{http://xxx.lanl.gov/abs/1302.6586}{{\tt arXiv:1302.6586}}].

\bibitem{Donos:2013wia}
A.~Donos, {\it {Striped phases from holography}},  {\em JHEP} {\bf 05} (2013)
  059, [\href{http://xxx.lanl.gov/abs/1303.7211}{{\tt arXiv:1303.7211}}].

\bibitem{Withers:2013loa}
B.~Withers, {\it {Black branes dual to striped phases}},  {\em Class. Quant.
  Grav.} {\bf 30} (2013) 155025, [\href{http://xxx.lanl.gov/abs/1304.0129}{{\tt
  arXiv:1304.0129}}].

\bibitem{Withers:2013kva}
B.~Withers, {\it {The moduli space of striped black branes}},
  \href{http://xxx.lanl.gov/abs/1304.2011}{{\tt arXiv:1304.2011}}.

\bibitem{Ling:2013aya}
Y.~Ling, C.~Niu, J.-P. Wu, Z.-Y. Xian, and H.-b. Zhang, {\it {Holographic
  Fermionic Liquid with Lattices}},  {\em JHEP} {\bf 07} (2013) 045,
  [\href{http://xxx.lanl.gov/abs/1304.2128}{{\tt arXiv:1304.2128}}].

\bibitem{Chesler:2013qla}
P.~Chesler, A.~Lucas, and S.~Sachdev, {\it {Conformal field theories in a
  periodic potential: results from holography and field theory}},  {\em Phys.
  Rev.} {\bf D89} (2014), no.~2 026005,
  [\href{http://xxx.lanl.gov/abs/1308.0329}{{\tt arXiv:1308.0329}}].

\bibitem{Ling:2013nxa}
Y.~Ling, C.~Niu, J.-P. Wu, and Z.-Y. Xian, {\it {Holographic Lattice in
  Einstein-Maxwell-Dilaton Gravity}},  {\em JHEP} {\bf 11} (2013) 006,
  [\href{http://xxx.lanl.gov/abs/1309.4580}{{\tt arXiv:1309.4580}}].

\bibitem{Horowitz:2013mia}
G.~T. Horowitz, N.~Iqbal, and J.~E. Santos, {\it {Simple holographic model of
  nonlinear conductivity}},  {\em Phys. Rev.} {\bf D88} (2013), no.~12 126002,
  [\href{http://xxx.lanl.gov/abs/1309.5088}{{\tt arXiv:1309.5088}}].

\bibitem{Dias:2013bwa}
O.~J.~C. Dias, G.~T. Horowitz, N.~Iqbal, and J.~E. Santos, {\it {Vortices in
  holographic superfluids and superconductors as conformal defects}},  {\em
  JHEP} {\bf 04} (2014) 096, [\href{http://xxx.lanl.gov/abs/1311.3673}{{\tt
  arXiv:1311.3673}}].

\bibitem{Hartnoll:2014cua}
S.~A. Hartnoll and J.~E. Santos, {\it {Disordered horizons: Holography of
  randomly disordered fixed points}},  {\em Phys. Rev. Lett.} {\bf 112} (2014)
  231601, [\href{http://xxx.lanl.gov/abs/1402.0872}{{\tt arXiv:1402.0872}}].

\bibitem{Hartnoll:2014gaa}
S.~A. Hartnoll and J.~E. Santos, {\it {Cold planar horizons are floppy}},  {\em
  Phys. Rev.} {\bf D89} (2014), no.~12 126002,
  [\href{http://xxx.lanl.gov/abs/1403.4612}{{\tt arXiv:1403.4612}}].

\bibitem{Ling:2014saa}
Y.~Ling, C.~Niu, J.~Wu, Z.~Xian, and H.-b. Zhang, {\it {Metal-insulator
  Transition by Holographic Charge Density Waves}},  {\em Phys. Rev. Lett.}
  {\bf 113} (2014) 091602, [\href{http://xxx.lanl.gov/abs/1404.0777}{{\tt
  arXiv:1404.0777}}].

\bibitem{Mefford:2014gia}
E.~Mefford and G.~T. Horowitz, {\it {Simple holographic insulator}},  {\em
  Phys. Rev.} {\bf D90} (2014), no.~8 084042,
  [\href{http://xxx.lanl.gov/abs/1406.4188}{{\tt arXiv:1406.4188}}].

\bibitem{Withers:2014sja}
B.~Withers, {\it {Holographic Checkerboards}},  {\em JHEP} {\bf 09} (2014) 102,
  [\href{http://xxx.lanl.gov/abs/1407.1085}{{\tt arXiv:1407.1085}}].

\bibitem{Donos:2014yya}
A.~Donos and J.~P. Gauntlett, {\it {The thermoelectric properties of
  inhomogeneous holographic lattices}},  {\em JHEP} {\bf 01} (2015) 035,
  [\href{http://xxx.lanl.gov/abs/1409.6875}{{\tt arXiv:1409.6875}}].

\bibitem{Hartnoll:2015faa}
S.~A. Hartnoll, D.~M. Ramirez, and J.~E. Santos, {\it {Emergent scale
  invariance of disordered horizons}},  {\em JHEP} {\bf 09} (2015) 160,
  [\href{http://xxx.lanl.gov/abs/1504.0332}{{\tt arXiv:1504.0332}}].

\bibitem{Rangamani:2015hka}
M.~Rangamani, M.~Rozali, and D.~Smyth, {\it {Spatial Modulation and
  Conductivities in Effective Holographic Theories}},  {\em JHEP} {\bf 07}
  (2015) 024, [\href{http://xxx.lanl.gov/abs/1505.0517}{{\tt
  arXiv:1505.0517}}].

\bibitem{Langley:2015exa}
B.~W. Langley, G.~Vanacore, and P.~W. Phillips, {\it {Absence of Power-Law
  Mid-Infrared Conductivity in Gravitational Crystals}},
  \href{http://xxx.lanl.gov/abs/1506.0676}{{\tt arXiv:1506.0676}}.

\bibitem{Hartnoll:2015rza}
S.~A. Hartnoll, D.~M. Ramirez, and J.~E. Santos, {\it {Thermal conductivity at
  a disordered quantum critical point}},
  \href{http://xxx.lanl.gov/abs/1508.0443}{{\tt arXiv:1508.0443}}.

\bibitem{Figueras:2014lka}
P.~Figueras and S.~Tunyasuvunakool, {\it {Localized Plasma Balls}},  {\em JHEP}
  {\bf 06} (2014) 025, [\href{http://xxx.lanl.gov/abs/1404.0018}{{\tt
  arXiv:1404.0018}}].

\bibitem{Witten:1998zw}
E.~Witten, {\it {Anti-de Sitter space, thermal phase transition, and
  confinement in gauge theories}},  {\em Adv. Theor. Math. Phys.} {\bf 2}
  (1998) 505--532, [\href{http://xxx.lanl.gov/abs/hep-th/9803131}{{\tt
  hep-th/9803131}}].

\bibitem{Lahiri:2007ae}
S.~Lahiri and S.~Minwalla, {\it {Plasmarings as dual black rings}},  {\em JHEP}
  {\bf 05} (2008) 001, [\href{http://xxx.lanl.gov/abs/0705.3404}{{\tt
  arXiv:0705.3404}}].

\bibitem{Costa:2014wya}
M.~S. Costa, L.~Greenspan, J.~Penedones, and J.~Santos, {\it {Thermodynamics of
  the BMN matrix model at strong coupling}},  {\em JHEP} {\bf 03} (2015) 069,
  [\href{http://xxx.lanl.gov/abs/1411.5541}{{\tt arXiv:1411.5541}}].

\bibitem{Horowitz:2014gva}
G.~T. Horowitz, N.~Iqbal, J.~E. Santos, and B.~Way, {\it {Hovering Black Holes
  from Charged Defects}},  {\em Class. Quant. Grav.} {\bf 32} (2015) 105001,
  [\href{http://xxx.lanl.gov/abs/1412.1830}{{\tt arXiv:1412.1830}}].

\bibitem{Janik:2015oja}
R.~A. Janik, J.~Jankowski, and P.~Witkowski, {\it {Conformal defects in
  supergravity - backreacted Dirac delta sources}},  {\em JHEP} {\bf 07} (2015)
  050, [\href{http://xxx.lanl.gov/abs/1503.0845}{{\tt arXiv:1503.0845}}].

\bibitem{Hickling:2015ooa}
A.~Hickling, {\it {Bulk Duals for Generic Static, Scale-Invariant Holographic
  CFT States}},  {\em Class. Quant. Grav.} {\bf 32} (2015), no.~17 175011,
  [\href{http://xxx.lanl.gov/abs/1504.0372}{{\tt arXiv:1504.0372}}].

\bibitem{Kichakova:2014fta}
O.~Kichakova, J.~Kunz, E.~Radu, and Y.~Shnir, {\it {Non-Abelian fields in
  AdS$_4$ spacetime: Axially symmetric, composite configurations}},  {\em Phys.
  Rev.} {\bf D90} (2014), no.~12 124012,
  [\href{http://xxx.lanl.gov/abs/1409.1894}{{\tt arXiv:1409.1894}}].

\bibitem{Yoshida:1997nd}
S.~Yoshida and Y.~Eriguchi, {\it {New static axisymmetric and nonvacuum
  solutions in general relativity: Equilibrium solutions of boson stars}},
  {\em Phys. Rev.} {\bf D55} (1997) 1994--2001.

\bibitem{Yoshida:1997qf}
S.~Yoshida and Y.~Eriguchi, {\it {Rotating boson stars in general relativity}},
   {\em Phys. Rev.} {\bf D56} (1997) 762--771.

\bibitem{Lee:1991ax}
T.~D. Lee and Y.~Pang, {\it {Nontopological solitons}},  {\em Phys. Rept.} {\bf
  221} (1992) 251--350.

\bibitem{Schunck:2003kk}
F.~E. Schunck and E.~W. Mielke, {\it {General relativistic boson stars}},  {\em
  Class. Quant. Grav.} {\bf 20} (2003) R301--R356,
  [\href{http://xxx.lanl.gov/abs/0801.0307}{{\tt arXiv:0801.0307}}].

\bibitem{Liebling:2012fv}
S.~L. Liebling and C.~Palenzuela, {\it {Dynamical Boson Stars}},  {\em Living
  Rev. Rel.} {\bf 15} (2012) 6, [\href{http://xxx.lanl.gov/abs/1202.5809}{{\tt
  arXiv:1202.5809}}].

\bibitem{Friedberg:1986tp}
R.~Friedberg, T.~D. Lee, and Y.~Pang, {\it {Mini-soliton Stars}},  {\em Phys.
  Rev.} {\bf D35} (1987) 3640.

\bibitem{Seidel:1991zh}
E.~Seidel and W.~M. Suen, {\it {Oscillating soliton stars}},  {\em Phys. Rev.
  Lett.} {\bf 66} (1991) 1659--1662.

\bibitem{Maliborski:2013jca}
M.~Maliborski and A.~Rostworowski, {\it {Time-Periodic Solutions in an Einstein
  AdS-Massless-Scalar-Field System}},  {\em Phys. Rev. Lett.} {\bf 111} (2013)
  051102, [\href{http://xxx.lanl.gov/abs/1303.3186}{{\tt arXiv:1303.3186}}].

\bibitem{Kaup:1968zz}
D.~J. Kaup, {\it {Klein-Gordon Geon}},  {\em Phys. Rev.} {\bf 172} (1968)
  1331--1342.

\bibitem{Ruffini:1969qy}
R.~Ruffini and S.~Bonazzola, {\it {Systems of selfgravitating particles in
  general relativity and the concept of an equation of state}},  {\em Phys.
  Rev.} {\bf 187} (1969) 1767--1783.

\bibitem{Astefanesei:2003qy}
D.~Astefanesei and E.~Radu, {\it {Boson stars with negative cosmological
  constant}},  {\em Nucl. Phys.} {\bf B665} (2003) 594--622,
  [\href{http://xxx.lanl.gov/abs/gr-qc/0309131}{{\tt gr-qc/0309131}}].

\bibitem{Horowitz:2010jq}
G.~T. Horowitz and B.~Way, {\it {Complete Phase Diagrams for a Holographic
  Superconductor/Insulator System}},  {\em JHEP} {\bf 11} (2010) 011,
  [\href{http://xxx.lanl.gov/abs/1007.3714}{{\tt arXiv:1007.3714}}].

\bibitem{Buchel:2013uba}
A.~Buchel, S.~L. Liebling, and L.~Lehner, {\it {Boson stars in AdS spacetime}},
   {\em Phys. Rev.} {\bf D87} (2013), no.~12 123006,
  [\href{http://xxx.lanl.gov/abs/1304.4166}{{\tt arXiv:1304.4166}}].

\bibitem{Jetzer:1989av}
P.~Jetzer and J.~J. van~der Bij, {\it {Charged boson stars}},  {\em Phys.
  Lett.} {\bf B227} (1989) 341.

\bibitem{Gentle:2011kv}
S.~A. Gentle, M.~Rangamani, and B.~Withers, {\it {A Soliton Menagerie in AdS}},
   {\em JHEP} {\bf 05} (2012) 106,
  [\href{http://xxx.lanl.gov/abs/1112.3979}{{\tt arXiv:1112.3979}}].

\bibitem{Schunck:1996book}
F.~E. Schunck and E.~W. Mielke, {\em in Relativity and Scientific Computing,
  edited by F. W. Hehl, R. A. Puntigam, and H. Ruder, pp. 138-151}.
\newblock Springer, Berlin, 1996.

\bibitem{Schunck:1996he}
F.~E. Schunck and E.~W. Mielke, {\it {Rotating boson star as an effective mass
  torus in general relativity}},  {\em Phys. Lett.} {\bf A249} (1998) 389--394.

\bibitem{Astefanesei:2003rw}
D.~Astefanesei and E.~Radu, {\it {Rotating boson stars in (2+1)-dimensions}},
  {\em Phys. Lett.} {\bf B587} (2004) 7--15,
  [\href{http://xxx.lanl.gov/abs/gr-qc/0310135}{{\tt gr-qc/0310135}}].

\bibitem{Hartmann:2010pm}
B.~Hartmann, B.~Kleihaus, J.~Kunz, and M.~List, {\it {Rotating Boson Stars in 5
  Dimensions}},  {\em Phys. Rev.} {\bf D82} (2010) 084022,
  [\href{http://xxx.lanl.gov/abs/1008.3137}{{\tt arXiv:1008.3137}}].

\bibitem{Bartnik:1988am}
R.~Bartnik and J.~Mckinnon, {\it {Particle - Like Solutions of the Einstein
  Yang-Mills Equations}},  {\em Phys. Rev. Lett.} {\bf 61} (1988) 141--144.

\bibitem{Lichnerowicz:1955}
A.~Lichnerowicz, {\it {Th\'eories Relativistes de la Gravitation et de
  l\'Electromagn\'etisme}},  {\em Masson, Paris,} (1955).

\bibitem{Coleman:1975}
S.~Coleman, {\em in New Phenomena in Subnuclear Physics, ed. by A. Zichichi}.
\newblock Plenum, New York, 1975.

\bibitem{Deser:1976wq}
S.~Deser, {\it {Absence of Static Solutions in Source-Free Yang-Mills Theory}},
   {\em Phys. Lett.} {\bf B64} (1976) 463.

\bibitem{Lee:1991vy}
K.-M. Lee, V.~P. Nair, and E.~J. Weinberg, {\it {Black holes in magnetic
  monopoles}},  {\em Phys. Rev.} {\bf D45} (1992) 2751--2761,
  [\href{http://xxx.lanl.gov/abs/hep-th/9112008}{{\tt hep-th/9112008}}].

\bibitem{Ortiz:1991eu}
M.~E. Ortiz, {\it {Curved space magnetic monopoles}},  {\em Phys. Rev.} {\bf
  D45} (1992) 2586--2589.

\bibitem{Aichelburg:1992st}
P.~C. Aichelburg and P.~Bizon, {\it {Magnetically charged black holes and their
  stability}},  {\em Phys. Rev.} {\bf D48} (1993) 607--615,
  [\href{http://xxx.lanl.gov/abs/gr-qc/9212009}{{\tt gr-qc/9212009}}].

\bibitem{Kastor:1992qy}
D.~Kastor and J.~H. Traschen, {\it {Horizons inside classical lumps}},  {\em
  Phys. Rev.} {\bf D46} (1992) 5399--5403,
  [\href{http://xxx.lanl.gov/abs/hep-th/9207070}{{\tt hep-th/9207070}}].

\bibitem{Heusler:1992av}
M.~Heusler, S.~Droz, and N.~Straumann, {\it {Linear stability of Einstein
  Skyrme black holes}},  {\em Phys. Lett.} {\bf B285} (1992) 21--26.

\bibitem{Pena:1997cy}
I.~Pena and D.~Sudarsky, {\it {Do collapsed boson stars result in new types of
  black holes?}},  {\em Class. Quant. Grav.} {\bf 14} (1997) 3131--3134.

\bibitem{Brito:2015pxa}
R.~Brito, V.~Cardoso, C.~A.~R. Herdeiro, and E.~Radu, {\it {Proca Stars:
  gravitating Bose-Einstein condensates of massive spin 1 particles}},
  \href{http://xxx.lanl.gov/abs/1508.0539}{{\tt arXiv:1508.0539}}.

\bibitem{Wheeler:1955zz}
J.~A. Wheeler, {\it {Geons}},  {\em Phys. Rev.} {\bf 97} (1955) 511--536.

\bibitem{Brill:1957fx}
D.~R. Brill and J.~A. Wheeler, {\it {Interaction of neutrinos and gravitational
  fields}},  {\em Rev. Mod. Phys.} {\bf 29} (1957) 465--479.

\bibitem{Ernst:1957zza}
F.~J. Ernst, {\it {Variational Calculations in Geon Theory}},  {\em Phys. Rev.}
  {\bf 105} (1957) 1662--1664.

\bibitem{Ernst:1957zz}
F.~J. Ernst, {\it {Linear and Toroidal Geons}},  {\em Phys. Rev.} {\bf 105}
  (1957) 1665--1670.

\bibitem{Misner:1957mt}
C.~W. Misner and J.~A. Wheeler, {\it {Classical physics as geometry:
  Gravitation, electromagnetism, unquantized charge, and mass as properties of
  curved empty space}},  {\em Annals Phys.} {\bf 2} (1957) 525--603.

\bibitem{Melvin:1963qx}
M.~A. Melvin, {\it {Pure magnetic and electric geons}},  {\em Phys. Lett.} {\bf
  8} (1964) 65--70.

\bibitem{Brill:1964zz}
D.~R. Brill and J.~B. Hartle, {\it {Method of the Self-Consistent Field in
  General Relativity and its Application to the Gravitational Geon}},  {\em
  Phys. Rev.} {\bf 135} (1964) B271--B278.

\bibitem{Melvin:1965zza}
M.~A. Melvin, {\it {Dynamics of Cylindrical Electromagnetic Universes}},  {\em
  Phys. Rev.} {\bf 139} (1965) B225--B243.

\bibitem{Dias:2011ss}
O.~J.~C. Dias, G.~T. Horowitz, and J.~E. Santos, {\it {Gravitational Turbulent
  Instability of Anti-de Sitter Space}},  {\em Class. Quant. Grav.} {\bf 29}
  (2012) 194002, [\href{http://xxx.lanl.gov/abs/1109.1825}{{\tt
  arXiv:1109.1825}}].

\bibitem{Horowitz:2014hja}
G.~T. Horowitz and J.~E. Santos, {\it {Geons and the Instability of Anti-de
  Sitter Spacetime}},  \href{http://xxx.lanl.gov/abs/1408.5906}{{\tt
  arXiv:1408.5906}}.

\bibitem{Kunduri:2006qa}
H.~K. Kunduri, J.~Lucietti, and H.~S. Reall, {\it {Gravitational perturbations
  of higher dimensional rotating black holes: Tensor perturbations}},  {\em
  Phys. Rev.} {\bf D74} (2006) 084021,
  [\href{http://xxx.lanl.gov/abs/hep-th/0606076}{{\tt hep-th/0606076}}].

\bibitem{Ridgway:1995ke}
S.~A. Ridgway and E.~J. Weinberg, {\it {Static black hole solutions without
  rotational symmetry}},  {\em Phys. Rev.} {\bf D52} (1995) 3440--3456,
  [\href{http://xxx.lanl.gov/abs/gr-qc/9503035}{{\tt gr-qc/9503035}}].

\bibitem{Ridgway:1995ac}
S.~A. Ridgway and E.~J. Weinberg, {\it {Are all static black hole solutions
  spherically symmetric?}},  {\em Gen. Rel. Grav.} {\bf 27} (1995) 1017--1021,
  [\href{http://xxx.lanl.gov/abs/gr-qc/9504003}{{\tt gr-qc/9504003}}].

\bibitem{Lemos:1994xp}
J.~P.~S. Lemos, {\it {Cylindrical black hole in general relativity}},  {\em
  Phys. Lett.} {\bf B353} (1995) 46--51,
  [\href{http://xxx.lanl.gov/abs/gr-qc/9404041}{{\tt gr-qc/9404041}}].

\bibitem{Mann:1996gj}
R.~B. Mann, {\it {Pair production of topological anti-de Sitter black holes}},
  {\em Class. Quant. Grav.} {\bf 14} (1997) L109--L114,
  [\href{http://xxx.lanl.gov/abs/gr-qc/9607071}{{\tt gr-qc/9607071}}].

\bibitem{Cai:1996eg}
R.-G. Cai and Y.-Z. Zhang, {\it {Black plane solutions in four-dimensional
  space-times}},  {\em Phys. Rev.} {\bf D54} (1996) 4891--4898,
  [\href{http://xxx.lanl.gov/abs/gr-qc/9609065}{{\tt gr-qc/9609065}}].

\bibitem{Vanzo:1997gw}
L.~Vanzo, {\it {Black holes with unusual topology}},  {\em Phys. Rev.} {\bf
  D56} (1997) 6475--6483, [\href{http://xxx.lanl.gov/abs/gr-qc/9705004}{{\tt
  gr-qc/9705004}}].

\bibitem{Mann:1997iz}
R.~B. Mann, {\it {Topological black holes: Outside looking in}},
  \href{http://xxx.lanl.gov/abs/gr-qc/9709039}{{\tt gr-qc/9709039}}. [Annals
  Israel Phys. Soc.13,311(1997)].

\bibitem{Birmingham:1998nr}
D.~Birmingham, {\it {Topological black holes in Anti-de Sitter space}},  {\em
  Class. Quant. Grav.} {\bf 16} (1999) 1197--1205,
  [\href{http://xxx.lanl.gov/abs/hep-th/9808032}{{\tt hep-th/9808032}}].

\bibitem{Hubeny:2009ru}
V.~E. Hubeny, D.~Marolf, and M.~Rangamani, {\it {Hawking radiation in large N
  strongly-coupled field theories}},  {\em Class. Quant. Grav.} {\bf 27} (2010)
  095015, [\href{http://xxx.lanl.gov/abs/0908.2270}{{\tt arXiv:0908.2270}}].

\bibitem{Fitzpatrick:2006cd}
A.~L. Fitzpatrick, L.~Randall, and T.~Wiseman, {\it {On the existence and
  dynamics of braneworld black holes}},  {\em JHEP} {\bf 11} (2006) 033,
  [\href{http://xxx.lanl.gov/abs/hep-th/0608208}{{\tt hep-th/0608208}}].

\bibitem{Hubeny:2009kz}
V.~E. Hubeny, D.~Marolf, and M.~Rangamani, {\it {Black funnels and droplets
  from the AdS C-metrics}},  {\em Class. Quant. Grav.} {\bf 27} (2010) 025001,
  [\href{http://xxx.lanl.gov/abs/0909.0005}{{\tt arXiv:0909.0005}}].

\bibitem{Hubeny:2009rc}
V.~E. Hubeny, D.~Marolf, and M.~Rangamani, {\it {Hawking radiation from AdS
  black holes}},  {\em Class. Quant. Grav.} {\bf 27} (2010) 095018,
  [\href{http://xxx.lanl.gov/abs/0911.4144}{{\tt arXiv:0911.4144}}].

\bibitem{Caldarelli:2011wa}
M.~M. Caldarelli, O.~J.~C. Dias, R.~Monteiro, and J.~E. Santos, {\it {Black
  funnels and droplets in thermal equilibrium}},  {\em JHEP} {\bf 05} (2011)
  116, [\href{http://xxx.lanl.gov/abs/1102.4337}{{\tt arXiv:1102.4337}}].

\bibitem{Haehl:2012tw}
F.~M. Haehl, {\it {The Schwarzschild-Black String AdS Soliton: Instability and
  Holographic Heat Transport}},  {\em Class. Quant. Grav.} {\bf 30} (2013)
  055002, [\href{http://xxx.lanl.gov/abs/1210.5763}{{\tt arXiv:1210.5763}}].

\bibitem{Haddad:2013tha}
N.~Haddad, {\it {Hawking Radiation from Small Black Holes at Strong Coupling
  and Large N}},  {\em Class. Quant. Grav.} {\bf 30} (2013) 195002,
  [\href{http://xxx.lanl.gov/abs/1306.0086}{{\tt arXiv:1306.0086}}].

\bibitem{Emparan:2013fha}
R.~Emparan and M.~Martinez, {\it {Black String Flow}},  {\em JHEP} {\bf 09}
  (2013) 068, [\href{http://xxx.lanl.gov/abs/1307.2276}{{\tt
  arXiv:1307.2276}}].

\bibitem{Figueras:2011va}
P.~Figueras, J.~Lucietti, and T.~Wiseman, {\it {Ricci solitons, Ricci flow, and
  strongly coupled CFT in the Schwarzschild Unruh or Boulware vacua}},  {\em
  Class. Quant. Grav.} {\bf 28} (2011) 215018,
  [\href{http://xxx.lanl.gov/abs/1104.4489}{{\tt arXiv:1104.4489}}].

\bibitem{Fischetti:2013hja}
S.~Fischetti and J.~E. Santos, {\it {Rotating Black Droplet}},  {\em JHEP} {\bf
  07} (2013) 156, [\href{http://xxx.lanl.gov/abs/1304.1156}{{\tt
  arXiv:1304.1156}}].

\bibitem{Figueras:2013jja}
P.~Figueras and S.~Tunyasuvunakool, {\it {CFTs in rotating black hole
  backgrounds}},  {\em Class. Quant. Grav.} {\bf 30} (2013) 125015,
  [\href{http://xxx.lanl.gov/abs/1304.1162}{{\tt arXiv:1304.1162}}].

\bibitem{Santos:2012he}
J.~E. Santos and B.~Way, {\it {Black Funnels}},  {\em JHEP} {\bf 12} (2012)
  060, [\href{http://xxx.lanl.gov/abs/1208.6291}{{\tt arXiv:1208.6291}}].

\bibitem{Santos:2014yja}
J.~E. Santos and B.~Way, {\it {Black Droplets}},  {\em JHEP} {\bf 08} (2014)
  072, [\href{http://xxx.lanl.gov/abs/1405.2078}{{\tt arXiv:1405.2078}}].

\bibitem{Emparan:1999wa}
R.~Emparan, G.~T. Horowitz, and R.~C. Myers, {\it {Exact description of black
  holes on branes}},  {\em JHEP} {\bf 01} (2000) 007,
  [\href{http://xxx.lanl.gov/abs/hep-th/9911043}{{\tt hep-th/9911043}}].

\bibitem{Figueras:2011gd}
P.~Figueras and T.~Wiseman, {\it {Gravity and large black holes in
  Randall-Sundrum II braneworlds}},  {\em Phys. Rev. Lett.} {\bf 107} (2011)
  081101, [\href{http://xxx.lanl.gov/abs/1105.2558}{{\tt arXiv:1105.2558}}].

\bibitem{Abdolrahimi:2012qi}
S.~Abdolrahimi, C.~Cattoen, D.~N. Page, and S.~Yaghoobpour-Tari, {\it {Large
  Randall-Sundrum II Black Holes}},  {\em Phys. Lett.} {\bf B720} (2013)
  405--409, [\href{http://xxx.lanl.gov/abs/1206.0708}{{\tt arXiv:1206.0708}}].

\bibitem{Abdolrahimi:2012pb}
S.~Abdolrahimi, C.~Cattoen, D.~N. Page, and S.~Yaghoobpour-Tari, {\it {Spectral
  methods in general relativity and large Randall-Sundrum II black holes}},
  {\em JCAP} {\bf 1306} (2013) 039,
  [\href{http://xxx.lanl.gov/abs/1212.5623}{{\tt arXiv:1212.5623}}].

\bibitem{Carter:1968ks}
B.~Carter, {\it {Hamilton-Jacobi and Schrodinger separable solutions of
  Einstein's equations}},  {\em Commun. Math. Phys.} {\bf 10} (1968) 280.

\bibitem{Myers:1986un}
R.~C. Myers and M.~J. Perry, {\it {Black Holes in Higher Dimensional
  Space-Times}},  {\em Annals Phys.} {\bf 172} (1986) 304.

\bibitem{Myers:2011yc}
R.~C. Myers, {\em {Myers-Perry black holes}}.
\newblock in Black Holes in Higher Dimensions, CUP 2012, Ed. Gary T. Horowitz,
  2011.

\bibitem{Hawking:1998kw}
S.~W. Hawking, C.~J. Hunter, and M.~Taylor, {\it {Rotation and the AdS / CFT
  correspondence}},  {\em Phys. Rev.} {\bf D59} (1999) 064005,
  [\href{http://xxx.lanl.gov/abs/hep-th/9811056}{{\tt hep-th/9811056}}].

\bibitem{Gibbons:2004uw}
G.~W. Gibbons, H.~Lu, D.~N. Page, and C.~N. Pope, {\it {The General Kerr-de
  Sitter metrics in all dimensions}},  {\em J. Geom. Phys.} {\bf 53} (2005)
  49--73, [\href{http://xxx.lanl.gov/abs/hep-th/0404008}{{\tt
  hep-th/0404008}}].

\bibitem{Emparan:2001wn}
R.~Emparan and H.~S. Reall, {\it {A Rotating black ring solution in
  five-dimensions}},  {\em Phys. Rev. Lett.} {\bf 88} (2002) 101101,
  [\href{http://xxx.lanl.gov/abs/hep-th/0110260}{{\tt hep-th/0110260}}].

\bibitem{Pomeransky:2006bd}
A.~A. Pomeransky and R.~A. Sen'kov, {\it {Black ring with two angular
  momenta}},  \href{http://xxx.lanl.gov/abs/hep-th/0612005}{{\tt
  hep-th/0612005}}.

\bibitem{Elvang:2007rd}
H.~Elvang and P.~Figueras, {\it {Black Saturn}},  {\em JHEP} {\bf 05} (2007)
  050, [\href{http://xxx.lanl.gov/abs/hep-th/0701035}{{\tt hep-th/0701035}}].

\bibitem{Iguchi:2007is}
H.~Iguchi and T.~Mishima, {\it {Black di-ring and infinite nonuniqueness}},
  {\em Phys. Rev.} {\bf D75} (2007) 064018,
  [\href{http://xxx.lanl.gov/abs/hep-th/0701043}{{\tt hep-th/0701043}}].
  [Erratum: Phys. Rev.D78,069903(2008)].

\bibitem{Izumi:2007qx}
K.~Izumi, {\it {Orthogonal black di-ring solution}},  {\em Prog. Theor. Phys.}
  {\bf 119} (2008) 757--774, [\href{http://xxx.lanl.gov/abs/0712.0902}{{\tt
  arXiv:0712.0902}}].

\bibitem{Galloway:2005mf}
G.~J. Galloway and R.~Schoen, {\it {A Generalization of Hawking's black hole
  topology theorem to higher dimensions}},  {\em Commun. Math. Phys.} {\bf 266}
  (2006) 571--576, [\href{http://xxx.lanl.gov/abs/gr-qc/0509107}{{\tt
  gr-qc/0509107}}].

\bibitem{Kunduri:2014kja}
H.~K. Kunduri and J.~Lucietti, {\it {Supersymmetric Black Holes with Lens-Space
  Topology}},  {\em Phys. Rev. Lett.} {\bf 113} (2014), no.~21 211101,
  [\href{http://xxx.lanl.gov/abs/1408.6083}{{\tt arXiv:1408.6083}}].

\bibitem{Kleihaus:2012xh}
B.~Kleihaus, J.~Kunz, and E.~Radu, {\it {Black rings in six dimensions}},  {\em
  Phys. Lett.} {\bf B718} (2013) 1073--1077,
  [\href{http://xxx.lanl.gov/abs/1205.5437}{{\tt arXiv:1205.5437}}].

\bibitem{Dias:2014cia}
O.~J.~C. Dias, J.~E. Santos, and B.~Way, {\it {Rings, Ripples, and Rotation:
  Connecting Black Holes to Black Rings}},  {\em JHEP} {\bf 07} (2014) 045,
  [\href{http://xxx.lanl.gov/abs/1402.6345}{{\tt arXiv:1402.6345}}].

\bibitem{Figueras:2014dta}
P.~Figueras and S.~Tunyasuvunakool, {\it {Black rings in global anti-de Sitter
  space}},  {\em JHEP} {\bf 03} (2015) 149,
  [\href{http://xxx.lanl.gov/abs/1412.5680}{{\tt arXiv:1412.5680}}].

\bibitem{Emparan:2009vd}
R.~Emparan, T.~Harmark, V.~Niarchos, and N.~A. Obers, {\it {New Horizons for
  Black Holes and Branes}},  {\em JHEP} {\bf 04} (2010) 046,
  [\href{http://xxx.lanl.gov/abs/0912.2352}{{\tt arXiv:0912.2352}}].

\bibitem{Armas:2015kra}
J.~Armas and M.~Blau, {\it {Blackfolds, Plane Waves and Minimal Surfaces}},
  {\em JHEP} {\bf 07} (2015) 156,
  [\href{http://xxx.lanl.gov/abs/1503.0883}{{\tt arXiv:1503.0883}}].

\bibitem{Armas:2015nea}
J.~Armas and M.~Blau, {\it {New Geometries for Black Hole Horizons}},  {\em
  JHEP} {\bf 07} (2015) 048, [\href{http://xxx.lanl.gov/abs/1504.0139}{{\tt
  arXiv:1504.0139}}].

\bibitem{Emparan:2014pra}
R.~Emparan, P.~Figueras, and M.~Martinez, {\it {Bumpy black holes}},  {\em
  JHEP} {\bf 12} (2014) 072, [\href{http://xxx.lanl.gov/abs/1410.4764}{{\tt
  arXiv:1410.4764}}].

\bibitem{Hollands:2006rj}
S.~Hollands, A.~Ishibashi, and R.~M. Wald, {\it {A Higher Dimensional
  Stationary Rotating Black Hole Must be Axisymmetric}},  {\em Commun. Math.
  Phys.} {\bf 271} (2007) 699--722,
  [\href{http://xxx.lanl.gov/abs/gr-qc/0605106}{{\tt gr-qc/0605106}}].

\bibitem{Moncrief:2008mr}
V.~Moncrief and J.~Isenberg, {\it {Symmetries of Higher Dimensional Black
  Holes}},  {\em Class.Quant.Grav.} {\bf 25} (2008) 195015,
  [\href{http://xxx.lanl.gov/abs/0805.1451}{{\tt arXiv:0805.1451}}].

\bibitem{Morisawa:2004tc}
Y.~Morisawa and D.~Ida, {\it {A Boundary value problem for the five-dimensional
  stationary rotating black holes}},  {\em Phys. Rev.} {\bf D69} (2004) 124005,
  [\href{http://xxx.lanl.gov/abs/gr-qc/0401100}{{\tt gr-qc/0401100}}].

\bibitem{Hollands:2007aj}
S.~Hollands and S.~Yazadjiev, {\it {Uniqueness theorem for 5-dimensional black
  holes with two axial Killing fields}},  {\em Commun. Math. Phys.} {\bf 283}
  (2008) 749--768, [\href{http://xxx.lanl.gov/abs/0707.2775}{{\tt
  arXiv:0707.2775}}].

\bibitem{Harmark:2009dh}
T.~Harmark, {\it {Domain Structure of Black Hole Space-Times}},  {\em Phys.
  Rev.} {\bf D80} (2009) 024019, [\href{http://xxx.lanl.gov/abs/0904.4246}{{\tt
  arXiv:0904.4246}}].

\bibitem{Durkee:2010ea}
M.~Durkee and H.~S. Reall, {\it {Perturbations of near-horizon geometries and
  instabilities of Myers-Perry black holes}},  {\em Phys. Rev.} {\bf D83}
  (2011) 104044, [\href{http://xxx.lanl.gov/abs/1012.4805}{{\tt
  arXiv:1012.4805}}].

\bibitem{Khlebnikov:2010yt}
S.~Khlebnikov, M.~Kruczenski, and G.~Michalogiorgakis, {\it {Shock waves in
  strongly coupled plasmas}},  {\em Phys. Rev.} {\bf D82} (2010) 125003,
  [\href{http://xxx.lanl.gov/abs/1004.3803}{{\tt arXiv:1004.3803}}].

\bibitem{Khlebnikov:2011ka}
S.~Khlebnikov, M.~Kruczenski, and G.~Michalogiorgakis, {\it {Shock waves in
  strongly coupled plasmas II}},  {\em JHEP} {\bf 07} (2011) 097,
  [\href{http://xxx.lanl.gov/abs/1105.1355}{{\tt arXiv:1105.1355}}].

\bibitem{Hubeny:2011yk}
V.~E. Hubeny, {\it {Holographic insights and puzzles}},  {\em Fortsch. Phys.}
  {\bf 59} (2011) 586--601, [\href{http://xxx.lanl.gov/abs/1103.1999}{{\tt
  arXiv:1103.1999}}].

\bibitem{Fischetti:2012ps}
S.~Fischetti and D.~Marolf, {\it {Flowing Funnels: Heat sources for field
  theories and the AdS$_3$ dual of CFT$_2$ Hawking radiation}},  {\em Class.
  Quant. Grav.} {\bf 29} (2012) 105004,
  [\href{http://xxx.lanl.gov/abs/1202.5069}{{\tt arXiv:1202.5069}}].

\bibitem{Figueras:2012rb}
P.~Figueras and T.~Wiseman, {\it {Stationary holographic plasma quenches and
  numerical methods for non-Killing horizons}},  {\em Phys. Rev. Lett.} {\bf
  110} (2013) 171602, [\href{http://xxx.lanl.gov/abs/1212.4498}{{\tt
  arXiv:1212.4498}}].

\bibitem{Fischetti:2012vt}
S.~Fischetti, D.~Marolf, and J.~E. Santos, {\it {AdS flowing black funnels:
  Stationary AdS black holes with non-Killing horizons and heat transport in
  the dual CFT}},  {\em Class. Quant. Grav.} {\bf 30} (2013) 075001,
  [\href{http://xxx.lanl.gov/abs/1212.4820}{{\tt arXiv:1212.4820}}].

\bibitem{Aretakis:2011ha}
S.~Aretakis, {\it {Stability and Instability of Extreme Reissner-Nordstr\'om
  Black Hole Spacetimes for Linear Scalar Perturbations I}},  {\em Commun.
  Math. Phys.} {\bf 307} (2011) 17--63,
  [\href{http://xxx.lanl.gov/abs/1110.2007}{{\tt arXiv:1110.2007}}].

\bibitem{Aretakis:2011hc}
S.~Aretakis, {\it {Stability and Instability of Extreme Reissner-Nordstrom
  Black Hole Spacetimes for Linear Scalar Perturbations II}},  {\em Annales
  Henri Poincare} {\bf 12} (2011) 1491--1538,
  [\href{http://xxx.lanl.gov/abs/1110.2009}{{\tt arXiv:1110.2009}}].

\bibitem{Aretakis:2011gz}
S.~Aretakis, {\it {Decay of Axisymmetric Solutions of the Wave Equation on
  Extreme Kerr Backgrounds}},  {\em J. Funct. Anal.} {\bf 263} (2012)
  2770--2831, [\href{http://xxx.lanl.gov/abs/1110.2006}{{\tt
  arXiv:1110.2006}}].

\bibitem{Aretakis:2012ei}
S.~Aretakis, {\it {Horizon Instability of Extremal Black Holes}},
  \href{http://xxx.lanl.gov/abs/1206.6598}{{\tt arXiv:1206.6598}}.

\bibitem{Aretakis:2012bm}
S.~Aretakis, {\it {A note on instabilities of extremal black holes under scalar
  perturbations from afar}},  {\em Class. Quant. Grav.} {\bf 30} (2013) 095010,
  [\href{http://xxx.lanl.gov/abs/1212.1103}{{\tt arXiv:1212.1103}}].

\bibitem{Aretakis:2013dpa}
S.~Aretakis, {\it {Nonlinear instability of scalar fields on extremal black
  holes}},  {\em Phys. Rev.} {\bf D87} (2013) 084052,
  [\href{http://xxx.lanl.gov/abs/1304.4616}{{\tt arXiv:1304.4616}}].

\bibitem{Lucietti:2012sf}
J.~Lucietti and H.~S. Reall, {\it {Gravitational instability of an extreme Kerr
  black hole}},  {\em Phys. Rev.} {\bf D86} (2012) 104030,
  [\href{http://xxx.lanl.gov/abs/1208.1437}{{\tt arXiv:1208.1437}}].

\bibitem{Murata:2013daa}
K.~Murata, H.~S. Reall, and N.~Tanahashi, {\it {What happens at the horizon(s)
  of an extreme black hole?}},  {\em Class. Quant. Grav.} {\bf 30} (2013)
  235007, [\href{http://xxx.lanl.gov/abs/1307.6800}{{\tt arXiv:1307.6800}}].

\bibitem{Regge:1957td}
T.~Regge and J.~A. Wheeler, {\it {Stability of a Schwarzschild singularity}},
  {\em Phys. Rev.} {\bf 108} (1957) 1063--1069.

\bibitem{Zerilli:1970se}
F.~J. Zerilli, {\it {Effective potential for even parity Regge-Wheeler
  gravitational perturbation equations}},  {\em Phys. Rev. Lett.} {\bf 24}
  (1970) 737--738.

\bibitem{Newman:1961qr}
E.~Newman and R.~Penrose, {\it {An Approach to gravitational radiation by a
  method of spin coefficients}},  {\em J.Math.Phys.} {\bf 3} (1962) 566--578.

\bibitem{Teukolsky:1972my}
S.~A. Teukolsky, {\it {Rotating black holes - separable wave equations for
  gravitational and electromagnetic perturbations}},  {\em Phys. Rev. Lett.}
  {\bf 29} (1972) 1114--1118.

\bibitem{Geroch:1973am}
R.~P. Geroch, A.~Held, and R.~Penrose, {\it {A space-time calculus based on
  pairs of null directions}},  {\em J. Math. Phys.} {\bf 14} (1973) 874--881.

\bibitem{Teukolsky:1973ha}
S.~A. Teukolsky, {\it {Perturbations of a rotating black hole. I. Fundamental
  equations for gravitational electromagnetic and neutrino field
  perturbations}},  {\em Astrophys. J.} {\bf 185} (1973) 635--647.

\bibitem{Chandrasekhar:1978a}
S.~Chandrasekhar, {\it {The Gravitational Perturbations of the Kerr Black Hole.
  I. The Perturbations in the Quantities which Vanish in the Stationary
  State}},  {\em Proc. R. Soc. Lond. A} {\bf 358} (1978) 421.

\bibitem{Chandrasekhar:1978b}
S.~Chandrasekhar, {\it {The Gravitational Perturbations of the Kerr Black Hole.
  II. The Perturbations in the Quantities which are Finite in the Stationary
  State}},  {\em Proc. R. Soc. Lond. A} {\bf 358} (1978) 441.

\bibitem{Chandrasekhar:1985kt}
S.~Chandrasekhar, {\em {The mathematical theory of black holes}}.
\newblock Oxford, UK: Clarendon (1992) 646 p., 1985.

\bibitem{Leaver:1985ax}
E.~Leaver, {\it {An Analytic representation for the quasi normal modes of Kerr
  black holes}},  {\em Proc.Roy.Soc.Lond.} {\bf A402} (1985) 285--298.

\bibitem{Whiting:1988vc}
B.~F. Whiting, {\it {Mode Stability of the Kerr Black Hole}},  {\em J. Math.
  Phys.} {\bf 30} (1989) 1301.

\bibitem{Berti:2009kk}
E.~Berti, V.~Cardoso, and A.~O. Starinets, {\it {Quasinormal modes of black
  holes and black branes}},  {\em Class. Quant. Grav.} {\bf 26} (2009) 163001,
  [\href{http://xxx.lanl.gov/abs/0905.2975}{{\tt arXiv:0905.2975}}].

\bibitem{Pani:2013ija}
P.~Pani, E.~Berti, and L.~Gualtieri, {\it {Gravitoelectromagnetic Perturbations
  of Kerr-Newman Black Holes: Stability and Isospectrality in the Slow-Rotation
  Limit}},  {\em Phys.Rev.Lett.} {\bf 110} (2013), no.~24 241103,
  [\href{http://xxx.lanl.gov/abs/1304.1160}{{\tt arXiv:1304.1160}}].

\bibitem{Pani:2013wsa}
P.~Pani, E.~Berti, and L.~Gualtieri, {\it {Scalar, Electromagnetic and
  Gravitational Perturbations of Kerr-Newman Black Holes in the Slow-Rotation
  Limit}},  {\em Phys.Rev.} {\bf D88} (2013) 064048,
  [\href{http://xxx.lanl.gov/abs/1307.7315}{{\tt arXiv:1307.7315}}].

\bibitem{Mark:2014aja}
Z.~Mark, H.~Yang, A.~Zimmerman, and Y.~Chen, {\it {The Quasinormal Modes of
  Weakly Charged Kerr-Newman Spacetimes}},
  \href{http://xxx.lanl.gov/abs/1409.5800}{{\tt arXiv:1409.5800}}.

\bibitem{Zilhao:2014wqa}
M.~Zilhão, V.~Cardoso, C.~Herdeiro, L.~Lehner, and U.~Sperhake, {\it {Testing
  the nonlinear stability of Kerr-Newman black holes}},  {\em Phys. Rev.} {\bf
  D90} (2014), no.~12 124088, [\href{http://xxx.lanl.gov/abs/1410.0694}{{\tt
  arXiv:1410.0694}}].

\bibitem{Dias:2015wqa}
O.~J.~C. Dias, M.~Godazgar, and J.~E. Santos, {\it {Linear Mode Stability of
  the Kerr-Newman Black Hole and Its Quasinormal Modes}},  {\em Phys. Rev.
  Lett.} {\bf 114} (2015), no.~15 151101,
  [\href{http://xxx.lanl.gov/abs/1501.0462}{{\tt arXiv:1501.0462}}].

\bibitem{Tangherlini:1963bw}
F.~R. Tangherlini, {\it {Schwarzschild field in n dimensions and the
  dimensionality of space problem}},  {\em Nuovo Cim.} {\bf 27} (1963)
  636--651.

\bibitem{Kodama:2003jz}
H.~Kodama and A.~Ishibashi, {\it {A Master equation for gravitational
  perturbations of maximally symmetric black holes in higher dimensions}},
  {\em Prog. Theor. Phys.} {\bf 110} (2003) 701--722,
  [\href{http://xxx.lanl.gov/abs/hep-th/0305147}{{\tt hep-th/0305147}}].

\bibitem{Ishibashi:2003ap}
A.~Ishibashi and H.~Kodama, {\it {Stability of higher dimensional Schwarzschild
  black holes}},  {\em Prog. Theor. Phys.} {\bf 110} (2003) 901--919,
  [\href{http://xxx.lanl.gov/abs/hep-th/0305185}{{\tt hep-th/0305185}}].

\bibitem{Kodama:2003kk}
H.~Kodama and A.~Ishibashi, {\it {Master equations for perturbations of
  generalized static black holes with charge in higher dimensions}},  {\em
  Prog. Theor. Phys.} {\bf 111} (2004) 29--73,
  [\href{http://xxx.lanl.gov/abs/hep-th/0308128}{{\tt hep-th/0308128}}].

\bibitem{Murata:2008yx}
K.~Murata and J.~Soda, {\it {Stability of Five-dimensional Myers-Perry Black
  Holes with Equal Angular Momenta}},  {\em Prog. Theor. Phys.} {\bf 120}
  (2008) 561--579, [\href{http://xxx.lanl.gov/abs/0803.1371}{{\tt
  arXiv:0803.1371}}].

\bibitem{Kodama:2009bf}
H.~Kodama, R.~A. Konoplya, and A.~Zhidenko, {\it {Gravitational stability of
  simply rotating Myers-Perry black holes: Tensorial perturbations}},  {\em
  Phys. Rev.} {\bf D81} (2010) 044007,
  [\href{http://xxx.lanl.gov/abs/0904.2154}{{\tt arXiv:0904.2154}}].

\bibitem{Kovtun:2005ev}
P.~K. Kovtun and A.~O. Starinets, {\it {Quasinormal modes and holography}},
  {\em Phys. Rev.} {\bf D72} (2005) 086009,
  [\href{http://xxx.lanl.gov/abs/hep-th/0506184}{{\tt hep-th/0506184}}].

\bibitem{Friess:2006kw}
J.~J. Friess, S.~S. Gubser, G.~Michalogiorgakis, and S.~S. Pufu, {\it
  {Expanding plasmas and quasinormal modes of anti-de Sitter black holes}},
  {\em JHEP} {\bf 04} (2007) 080,
  [\href{http://xxx.lanl.gov/abs/hep-th/0611005}{{\tt hep-th/0611005}}].

\bibitem{Michalogiorgakis:2006jc}
G.~Michalogiorgakis and S.~S. Pufu, {\it {Low-lying gravitational modes in the
  scalar sector of the global AdS(4) black hole}},  {\em JHEP} {\bf 02} (2007)
  023, [\href{http://xxx.lanl.gov/abs/hep-th/0612065}{{\tt hep-th/0612065}}].

\bibitem{Dias:2013sdc}
O.~J.~C. Dias and J.~E. Santos, {\it {Boundary Conditions for Kerr-AdS
  Perturbations}},  {\em JHEP} {\bf 10} (2013) 156,
  [\href{http://xxx.lanl.gov/abs/1302.1580}{{\tt arXiv:1302.1580}}].

\bibitem{Emparan:2003sy}
R.~Emparan and R.~C. Myers, {\it {Instability of ultra-spinning black holes}},
  {\em JHEP} {\bf 09} (2003) 025,
  [\href{http://xxx.lanl.gov/abs/hep-th/0308056}{{\tt hep-th/0308056}}].

\bibitem{Shibata:2009ad}
M.~Shibata and H.~Yoshino, {\it {Nonaxisymmetric instability of rapidly
  rotating black hole in five dimensions}},  {\em Phys. Rev.} {\bf D81} (2010)
  021501, [\href{http://xxx.lanl.gov/abs/0912.3606}{{\tt arXiv:0912.3606}}].

\bibitem{Shibata:2010wz}
M.~Shibata and H.~Yoshino, {\it {Bar-mode instability of rapidly spinning black
  hole in higher dimensions: Numerical simulation in general relativity}},
  {\em Phys. Rev.} {\bf D81} (2010) 104035,
  [\href{http://xxx.lanl.gov/abs/1004.4970}{{\tt arXiv:1004.4970}}].

\bibitem{Dias:2014eua}
O.~J.~C. Dias, G.~S. Hartnett, and J.~E. Santos, {\it {Quasinormal modes of
  asymptotically flat rotating black holes}},  {\em Class. Quant. Grav.} {\bf
  31} (2014), no.~24 245011, [\href{http://xxx.lanl.gov/abs/1402.7047}{{\tt
  arXiv:1402.7047}}].

\bibitem{Arcioni:2004ww}
G.~Arcioni and E.~Lozano-Tellechea, {\it {Stability and critical phenomena of
  black holes and black rings}},  {\em Phys. Rev.} {\bf D72} (2005) 104021,
  [\href{http://xxx.lanl.gov/abs/hep-th/0412118}{{\tt hep-th/0412118}}].

\bibitem{Elvang:2006dd}
H.~Elvang, R.~Emparan, and A.~Virmani, {\it {Dynamics and stability of black
  rings}},  {\em JHEP} {\bf 12} (2006) 074,
  [\href{http://xxx.lanl.gov/abs/hep-th/0608076}{{\tt hep-th/0608076}}].

\bibitem{Figueras:2011he}
P.~Figueras, K.~Murata, and H.~S. Reall, {\it {Black hole instabilities and
  local Penrose inequalities}},  {\em Class. Quant. Grav.} {\bf 28} (2011)
  225030, [\href{http://xxx.lanl.gov/abs/1107.5785}{{\tt arXiv:1107.5785}}].

\bibitem{Santos:2015iua}
J.~E. Santos and B.~Way, {\it {Neutral Black Rings in Five Dimensions are
  Unstable}},  {\em Phys. Rev. Lett.} {\bf 114} (2015), no.~22 221101,
  [\href{http://xxx.lanl.gov/abs/1503.0072}{{\tt arXiv:1503.0072}}].

\bibitem{Banks:1998dd}
T.~Banks, M.~R. Douglas, G.~T. Horowitz, and E.~J. Martinec, {\it {AdS dynamics
  from conformal field theory}},
  \href{http://xxx.lanl.gov/abs/hep-th/9808016}{{\tt hep-th/9808016}}.

\bibitem{Peet:1998cr}
A.~W. Peet and S.~F. Ross, {\it {Microcanonical phases of string theory on
  AdS(m)$\times S^n$}},  {\em JHEP} {\bf 9812} (1998) 020,
  [\href{http://xxx.lanl.gov/abs/hep-th/9810200}{{\tt hep-th/9810200}}].

\bibitem{Hubeny:2002xn}
V.~E. Hubeny and M.~Rangamani, {\it {Unstable horizons}},  {\em JHEP} {\bf
  0205} (2002) 027, [\href{http://xxx.lanl.gov/abs/hep-th/0202189}{{\tt
  hep-th/0202189}}].

\bibitem{Dias:2015pda}
O.~J.~C. Dias, J.~E. Santos, and B.~Way, {\it {Lumpy AdS$_{5}\times$S$^{5}$
  black holes and black belts}},  {\em JHEP} {\bf 04} (2015) 060,
  [\href{http://xxx.lanl.gov/abs/1501.0657}{{\tt arXiv:1501.0657}}].

\bibitem{Buchel:2015gxa}
A.~Buchel and L.~Lehner, {\it {Small black holes in $AdS_5\times S^5$}},  {\em
  Class. Quant. Grav.} {\bf 32} (2015), no.~14 145003,
  [\href{http://xxx.lanl.gov/abs/1502.0157}{{\tt arXiv:1502.0157}}].

\bibitem{Emparan:2011ve}
R.~Emparan and N.~Haddad, {\it {Self-similar critical geometries at horizon
  intersections and mergers}},  {\em JHEP} {\bf 10} (2011) 064,
  [\href{http://xxx.lanl.gov/abs/1109.1983}{{\tt arXiv:1109.1983}}].

\bibitem{PauBarM}
P.~Figueras, M.~Kunesch, and S.~Tunyasuvunakool, {\it {(to appear)}}, .

\bibitem{Hartnett:2012np}
G.~S. Hartnett and G.~T. Horowitz, {\it {Geons and Spin-2 Condensates in the
  AdS Soliton}},  {\em JHEP} {\bf 01} (2013) 010,
  [\href{http://xxx.lanl.gov/abs/1210.1606}{{\tt arXiv:1210.1606}}].

\bibitem{Murata:2010dx}
K.~Murata, S.~Kinoshita, and N.~Tanahashi, {\it {Non-equilibrium Condensation
  Process in a Holographic Superconductor}},  {\em JHEP} {\bf 07} (2010) 050,
  [\href{http://xxx.lanl.gov/abs/1005.0633}{{\tt arXiv:1005.0633}}].

\bibitem{Penrose:1971uk}
R.~Penrose and R.~Floyd, {\it {Extraction of rotational energy from a black
  hole}},  {\em Nature} {\bf 229} (1971) 177--179.

\bibitem{Zeldovich:1971}
Y.~B. Zeldovich, {\it {Generation of Waves by a Rotating Body}},  {\em JETP
  Lett.} {\bf 14} (1971) 180.

\bibitem{Zeldovich:1972}
Y.~B. Zeldovich, {\it {Amplification of cylindrical electromagnetic waves
  reflected from a rotating body}},  {\em Sov. Phys. JETP} {\bf 35} (1972)
  1085.

\bibitem{Teukolsky:1974yv}
S.~A. Teukolsky and W.~H. Press, {\it {Perturbations of a rotating black hole.
  III - Interaction of the hole with gravitational and electromagnetic
  radiation}},  {\em Astrophys. J.} {\bf 193} (1974) 443--461.

\bibitem{Brito:2015oca}
R.~Brito, V.~Cardoso, and P.~Pani, {\it {Superradiance}},  {\em Lect. Notes
  Phys.} {\bf 906} (2015) [\href{http://xxx.lanl.gov/abs/1501.0657}{{\tt
  arXiv:1501.0657}}].

\bibitem{Furuhashi:2004jk}
H.~Furuhashi and Y.~Nambu, {\it {Instability of massive scalar fields in
  Kerr-Newman space-time}},  {\em Prog. Theor. Phys.} {\bf 112} (2004)
  983--995, [\href{http://xxx.lanl.gov/abs/gr-qc/0402037}{{\tt
  gr-qc/0402037}}].

\bibitem{PhysRevD.73.084013}
J.~L. Hovdebo and R.~C. Myers, {\it Black rings, boosted strings, and
  gregory-laflamme instability},  {\em Phys. Rev. D} {\bf 73} (Apr, 2006)
  084013.

\bibitem{Press:1972zz}
W.~H. Press and S.~A. Teukolsky, {\it {Floating Orbits, Superradiant Scattering
  and the Black-hole Bomb}},  {\em Nature} {\bf 238} (1972) 211--212.

\bibitem{King:1977}
A.~R. King, {\it Black-hole magnetostatics},  {\em Mathematical Proceedings of
  the Cambridge Philosophical Society} {\bf 81} (1, 1977) 149--156.

\bibitem{Cardoso:2004nk}
V.~Cardoso, O.~J.~C. Dias, J.~P.~S. Lemos, and S.~Yoshida, {\it {The Black hole
  bomb and superradiant instabilities}},  {\em Phys. Rev.} {\bf D70} (2004)
  044039, [\href{http://xxx.lanl.gov/abs/hep-th/0404096}{{\tt
  hep-th/0404096}}]. [Erratum: Phys. Rev.D70,049903(2004)].

\bibitem{Hod:2009cp}
S.~Hod and O.~Hod, {\it {Analytic treatment of the black-hole bomb}},  {\em
  Phys. Rev.} {\bf D81} (2010) 061502,
  [\href{http://xxx.lanl.gov/abs/0910.0734}{{\tt arXiv:0910.0734}}].

\bibitem{Rosa:2009ei}
J.~G. Rosa, {\it {The Extremal black hole bomb}},  {\em JHEP} {\bf 06} (2010)
  015, [\href{http://xxx.lanl.gov/abs/0912.1780}{{\tt arXiv:0912.1780}}].

\bibitem{Hod:2009cw}
S.~Hod and O.~Hod, {\it {Comment on `The Extremal black hole bomb'}},
  \href{http://xxx.lanl.gov/abs/0912.2761}{{\tt arXiv:0912.2761}}.

\bibitem{Witek:2010qc}
H.~Witek, V.~Cardoso, C.~Herdeiro, A.~Nerozzi, U.~Sperhake, and M.~Zilhao, {\it
  {Black holes in a box: towards the numerical evolution of black holes in
  AdS}},  {\em Phys. Rev.} {\bf D82} (2010) 104037,
  [\href{http://xxx.lanl.gov/abs/1004.4633}{{\tt arXiv:1004.4633}}].

\bibitem{Lee:2011ez}
J.-P. Lee, {\it {Superradiance by mini black holes with mirror}},  {\em JHEP}
  {\bf 01} (2012) 091, [\href{http://xxx.lanl.gov/abs/1107.5641}{{\tt
  arXiv:1107.5641}}].

\bibitem{Dolan:2012yt}
S.~R. Dolan, {\it {Superradiant instabilities of rotating black holes in the
  time domain}},  {\em Phys. Rev.} {\bf D87} (2013), no.~12 124026,
  [\href{http://xxx.lanl.gov/abs/1212.1477}{{\tt arXiv:1212.1477}}].

\bibitem{Herdeiro:2013pia}
C.~A.~R. Herdeiro, J.~C. Degollado, and H.~F. Rúnarsson, {\it {Rapid growth of
  superradiant instabilities for charged black holes in a cavity}},  {\em Phys.
  Rev.} {\bf D88} (2013) 063003, [\href{http://xxx.lanl.gov/abs/1305.5513}{{\tt
  arXiv:1305.5513}}].

\bibitem{Degollado:2013bha}
J.~C. Degollado and C.~A.~R. Herdeiro, {\it {Time evolution of superradiant
  instabilities for charged black holes in a cavity}},  {\em Phys. Rev.} {\bf
  D89} (2014), no.~6 063005, [\href{http://xxx.lanl.gov/abs/1312.4579}{{\tt
  arXiv:1312.4579}}].

\bibitem{Hod:2013fvl}
S.~Hod, {\it {Analytic treatment of the charged black-hole-mirror bomb in the
  highly explosive regime}},  {\em Phys. Rev.} {\bf D88} (2013), no.~6 064055,
  [\href{http://xxx.lanl.gov/abs/1310.6101}{{\tt arXiv:1310.6101}}].

\bibitem{Hod:2014pza}
S.~Hod, {\it {Onset of superradiant instabilities in the composed
  Kerr-black-hole-mirror bomb}},  {\em Phys. Lett.} {\bf B736} (2014) 398--402,
  [\href{http://xxx.lanl.gov/abs/1412.6108}{{\tt arXiv:1412.6108}}].

\bibitem{Li:2014gfg}
R.~Li, J.-K. Zhao, and Y.-M. Zhang, {\it {Superradiant Instability of
  D-Dimensional Reissner-Nordstr\"om Black Hole Mirror System}},  {\em Commun.
  Theor. Phys.} {\bf 63} (2015), no.~5 569--574,
  [\href{http://xxx.lanl.gov/abs/1404.6309}{{\tt arXiv:1404.6309}}].

\bibitem{Aliev:2014aba}
A.~N. Aliev, {\it {Superradiance and black hole bomb in five-dimensional
  minimal ungauged supergravity}},  {\em JCAP} {\bf 1411} (2014), no.~11 029,
  [\href{http://xxx.lanl.gov/abs/1408.4269}{{\tt arXiv:1408.4269}}].

\bibitem{Li:2014fna}
R.~Li and J.~Zhao, {\it {Numerical study of superradiant instability for
  charged stringy black hole mirror system}},  {\em Phys. Lett.} {\bf B740}
  (2015) 317--321, [\href{http://xxx.lanl.gov/abs/1412.1527}{{\tt
  arXiv:1412.1527}}].

\bibitem{DiMenza2015}
L.~{Di Menza} and J.-P. {Nicolas}, {\it {Superradiance on the
  Reissner-Nordstrom metric}},  {\em Classical and Quantum Gravity} {\bf 32}
  (July, 2015) 145013, [\href{http://xxx.lanl.gov/abs/1411.3988}{{\tt
  arXiv:1411.3988}}].

\bibitem{Dolan:2015dha}
S.~R. Dolan, S.~Ponglertsakul, and E.~Winstanley, {\it {Stability of black
  holes in Einstein-charged scalar field theory in a cavity}},
  \href{http://xxx.lanl.gov/abs/1507.0215}{{\tt arXiv:1507.0215}}.

\bibitem{Aliev:2015wla}
A.~N. Aliev, {\it {Superradiance and instability of small rotating charged AdS
  black holes in all dimensions}},
  \href{http://xxx.lanl.gov/abs/1503.0860}{{\tt arXiv:1503.0860}}.

\bibitem{Delice:2015zga}
{Delice, \"Ozg\"ur and Durut, T\"urk\"uler}, {\it {Superradiance Instability of
  Small Rotating AdS Black Holes in Arbitrary Dimensions}},  {\em Phys. Rev.}
  {\bf D92} (2015), no.~2 024053,
  [\href{http://xxx.lanl.gov/abs/1503.0581}{{\tt arXiv:1503.0581}}].

\bibitem{Damour:1976kh}
T.~Damour, N.~Deruelle, and R.~Ruffini, {\it {On Quantum Resonances in
  Stationary Geometries}},  {\em Lett. Nuovo Cim.} {\bf 15} (1976) 257--262.

\bibitem{Zouros:1979iw}
T.~J.~M. Zouros and D.~M. Eardley, {\it {Instabilities of massive scalar
  perturbations of a rotating black hole}},  {\em Annals Phys.} {\bf 118}
  (1979) 139--155.

\bibitem{Dolan:2007mj}
S.~R. Dolan, {\it {Instability of the massive Klein-Gordon field on the Kerr
  spacetime}},  {\em Phys. Rev.} {\bf D76} (2007) 084001,
  [\href{http://xxx.lanl.gov/abs/0705.2880}{{\tt arXiv:0705.2880}}].

\bibitem{Strafuss:2004qc}
M.~J. Strafuss and G.~Khanna, {\it {Massive scalar field instability in Kerr
  spacetime}},  {\em Phys. Rev.} {\bf D71} (2005) 024034,
  [\href{http://xxx.lanl.gov/abs/gr-qc/0412023}{{\tt gr-qc/0412023}}].

\bibitem{Cardoso:2011xi}
V.~Cardoso, S.~Chakrabarti, P.~Pani, E.~Berti, and L.~Gualtieri, {\it {Floating
  and sinking: The Imprint of massive scalars around rotating black holes}},
  {\em Phys. Rev. Lett.} {\bf 107} (2011) 241101,
  [\href{http://xxx.lanl.gov/abs/1109.6021}{{\tt arXiv:1109.6021}}].

\bibitem{Yoshino:2012kn}
H.~Yoshino and H.~Kodama, {\it {Bosenova collapse of axion cloud around a
  rotating black hole}},  {\em Prog. Theor. Phys.} {\bf 128} (2012) 153--190,
  [\href{http://xxx.lanl.gov/abs/1203.5070}{{\tt arXiv:1203.5070}}].

\bibitem{Pani:2012bp}
P.~Pani, V.~Cardoso, L.~Gualtieri, E.~Berti, and A.~Ishibashi, {\it
  {Perturbations of slowly rotating black holes: massive vector fields in the
  Kerr metric}},  {\em Phys. Rev.} {\bf D86} (2012) 104017,
  [\href{http://xxx.lanl.gov/abs/1209.0773}{{\tt arXiv:1209.0773}}].

\bibitem{Witek:2012tr}
H.~Witek, V.~Cardoso, A.~Ishibashi, and U.~Sperhake, {\it {Superradiant
  instabilities in astrophysical systems}},  {\em Phys. Rev.} {\bf D87} (2013),
  no.~4 043513, [\href{http://xxx.lanl.gov/abs/1212.0551}{{\tt
  arXiv:1212.0551}}].

\bibitem{Hawking:1999dp}
S.~W. Hawking and H.~S. Reall, {\it {Charged and rotating AdS black holes and
  their CFT duals}},  {\em Phys. Rev.} {\bf D61} (2000) 024014,
  [\href{http://xxx.lanl.gov/abs/hep-th/9908109}{{\tt hep-th/9908109}}].

\bibitem{Cardoso:2004hs}
V.~Cardoso and O.~J.~C. Dias, {\it {Small Kerr-anti-de Sitter black holes are
  unstable}},  {\em Phys. Rev.} {\bf D70} (2004) 084011,
  [\href{http://xxx.lanl.gov/abs/hep-th/0405006}{{\tt hep-th/0405006}}].

\bibitem{Dold:2015cqa}
D.~Dold, {\it {Unstable mode solutions to the Klein-Gordon equation in
  Kerr-anti-de Sitter spacetimes}},
  \href{http://xxx.lanl.gov/abs/1509.0497}{{\tt arXiv:1509.0497}}.

\bibitem{Dafermos:2008en}
M.~Dafermos and I.~Rodnianski, {\it {Lectures on black holes and linear
  waves}},  {\em Clay Math. Proc.} {\bf 17} (2013) 97--205,
  [\href{http://xxx.lanl.gov/abs/0811.0354}{{\tt arXiv:0811.0354}}].

\bibitem{Dafermos:2010hb}
M.~Dafermos and I.~Rodnianski, {\it {Decay for solutions of the wave equation
  on Kerr exterior spacetimes I-II: The cases $|a| \ll M$ or axisymmetry}},
  \href{http://xxx.lanl.gov/abs/1010.5132}{{\tt arXiv:1010.5132}}.

\bibitem{Shlapentokh-Rothman:2013ysa}
Y.~Shlapentokh-Rothman, {\it {Exponentially growing finite energy solutions for
  the Klein-Gordon equation on sub-extremal Kerr spacetimes}},  {\em Commun.
  Math. Phys.} {\bf 329} (2014) 859--891,
  [\href{http://xxx.lanl.gov/abs/1302.3448}{{\tt arXiv:1302.3448}}].

\bibitem{Dafermos:2014cua}
M.~Dafermos, I.~Rodnianski, and Y.~Shlapentokh-Rothman, {\it {Decay for
  solutions of the wave equation on Kerr exterior spacetimes III: The full
  subextremal case $|a| < M$}},  \href{http://xxx.lanl.gov/abs/1402.7034}{{\tt
  arXiv:1402.7034}}.

\bibitem{Dafermos:2014jwa}
M.~Dafermos, I.~Rodnianski, and Y.~Shlapentokh-Rothman, {\it {A scattering
  theory for the wave equation on Kerr black hole exteriors}},
  \href{http://xxx.lanl.gov/abs/1412.8379}{{\tt arXiv:1412.8379}}.

\bibitem{Ishibashi:2004wx}
A.~Ishibashi and R.~M. Wald, {\it {Dynamics in nonglobally hyperbolic static
  space-times. 3. Anti-de Sitter space-time}},  {\em Class. Quant. Grav.} {\bf
  21} (2004) 2981--3014, [\href{http://xxx.lanl.gov/abs/hep-th/0402184}{{\tt
  hep-th/0402184}}].

\bibitem{Vasy:2009}
A.~{Vasy}, {\it {The wave equation on asymptotically Anti-de Sitter spaces}},
  {\em ArXiv e-prints} (Nov., 2009)
  [\href{http://xxx.lanl.gov/abs/0911.5440}{{\tt arXiv:0911.5440}}].

\bibitem{Holzegel:2009ye}
G.~Holzegel, {\it {On the massive wave equation on slowly rotating Kerr-AdS
  spacetimes}},  {\em Commun. Math. Phys.} {\bf 294} (2010) 169--197,
  [\href{http://xxx.lanl.gov/abs/0902.0973}{{\tt arXiv:0902.0973}}].

\bibitem{Holzegel:2011uu}
G.~Holzegel and J.~Smulevici, {\it {Decay properties of Klein-Gordon fields on
  Kerr-AdS spacetimes}},  {\em Commun. Pure Appl. Math.} {\bf 66} (2013)
  1751--1802, [\href{http://xxx.lanl.gov/abs/1110.6794}{{\tt
  arXiv:1110.6794}}].

\bibitem{Holzegel:2011qj}
G.~Holzegel, {\it {Well-posedness for the massive wave equation on
  asymptotically anti-de Sitter spacetimes}},
  \href{http://xxx.lanl.gov/abs/1103.0710}{{\tt arXiv:1103.0710}}.

\bibitem{Holzegel:2012wt}
G.~H. Holzegel and C.~M. Warnick, {\it {Boundedness and growth for the massive
  wave equation on asymptotically anti-de Sitter black holes}},  {\em J. Funct.
  Anal.} {\bf 266} (2014), no.~4 2436--2485,
  [\href{http://xxx.lanl.gov/abs/1209.3308}{{\tt arXiv:1209.3308}}].

\bibitem{Warnick:2012fi}
C.~M. Warnick, {\it {The Massive wave equation in asymptotically AdS
  spacetimes}},  {\em Commun. Math. Phys.} {\bf 321} (2013) 85--111,
  [\href{http://xxx.lanl.gov/abs/1202.3445}{{\tt arXiv:1202.3445}}].

\bibitem{2012arXiv1212.1907G}
O.~{Gannot}, {\it {Quasinormal modes for Schwarzschild-AdS black holes:
  exponential convergence to the real axis}},  {\em ArXiv e-prints} (Dec.,
  2012) [\href{http://xxx.lanl.gov/abs/1212.1907}{{\tt arXiv:1212.1907}}].

\bibitem{Holzegel:2013kna}
G.~Holzegel and J.~Smulevici, {\it {Quasimodes and a Lower Bound on the Uniform
  Energy Decay Rate for Kerr-AdS Spacetimes}},
  \href{http://xxx.lanl.gov/abs/1303.5944}{{\tt arXiv:1303.5944}}.

\bibitem{Gannot:2014boa}
O.~Gannot, {\it {A global definition of quasinormal modes for Kerr-AdS Black
  Holes}},  \href{http://xxx.lanl.gov/abs/1407.6686}{{\tt arXiv:1407.6686}}.

\bibitem{Holzegel:2015swa}
G.~Holzegel, J.~Luk, J.~Smulevici, and C.~Warnick, {\it {Asymptotic properties
  of linear field equations in anti-de Sitter space}},
  \href{http://xxx.lanl.gov/abs/1502.0496}{{\tt arXiv:1502.0496}}.

\bibitem{friedman1978}
J.~L. Friedman, {\it Ergosphere instability},  {\em Comm. Math. Phys.} {\bf 63}
  (1978), no.~3 243--255.

\bibitem{cominsschutz1978}
N.~Comins and B.~Schutz, {\it {On the ergoregion instability}},  {\em Proc. R.
  Soc. Lond. A} {\bf 364} (1978) 211.

\bibitem{1996MNRAS.282..580Y}
S.~Yoshida and Y.~Eriguchi, {\it {Ergoregion instability revisited - a new and
  general method for numerical analysis of stability}},  {\em Mon. Not. Roy.
  Astron. Soc.} {\bf 282} (1996) 580.

\bibitem{Cardoso:2005gj}
V.~Cardoso, O.~J.~C. Dias, J.~L. Hovdebo, and R.~C. Myers, {\it {Instability of
  non-supersymmetric smooth geometries}},  {\em Phys. Rev.} {\bf D73} (2006)
  064031, [\href{http://xxx.lanl.gov/abs/hep-th/0512277}{{\tt
  hep-th/0512277}}].

\bibitem{Chowdhury:2007jx}
B.~D. Chowdhury and S.~D. Mathur, {\it {Radiation from the non-extremal
  fuzzball}},  {\em Class. Quant. Grav.} {\bf 25} (2008) 135005,
  [\href{http://xxx.lanl.gov/abs/0711.4817}{{\tt arXiv:0711.4817}}].

\bibitem{Cardoso:2007az}
V.~Cardoso, P.~Pani, M.~Cadoni, and M.~Cavaglia, {\it {Ergoregion instability
  of ultracompact astrophysical objects}},  {\em Phys. Rev.} {\bf D77} (2008)
  124044, [\href{http://xxx.lanl.gov/abs/0709.0532}{{\tt arXiv:0709.0532}}].

\bibitem{waldResonator}
S.~Green, S.~Hollands, A.~Ishibashi, and R.~M. Wald, {\it {(to appear)}}, .

\bibitem{East:2013mfa}
W.~E. East, F.~M. Ramazanolu, and F.~Pretorius, {\it {Black Hole Superradiance
  in Dynamical Spacetime}},  {\em Phys. Rev.} {\bf D89} (2014), no.~6 061503,
  [\href{http://xxx.lanl.gov/abs/1312.4529}{{\tt arXiv:1312.4529}}].

\bibitem{Niehoff:2015oga}
B.~E. Niehoff, J.~E. Santos, and B.~Way, {\it {Towards a violation of cosmic
  censorship}},  \href{http://xxx.lanl.gov/abs/1510.0070}{{\tt
  arXiv:1510.0070}}.

\bibitem{Friedrich:1986}
H.~Friedrich, {\it {On the existence of n-geodesically complete or future
  complete solutions of Einsteins field equations with smooth asymptotic
  structure}},  {\em Commun. Math. Phys.} {\bf 107} (1986) 587.

\bibitem{Christodoulou:1993uv}
D.~Christodoulou and S.~Klainerman, {\em {The Global nonlinear stability of the
  Minkowski space}}.
\newblock {Princeton University Press}, 1993.

\bibitem{Lindblad:2004ue}
H.~Lindblad and I.~Rodnianski, {\it {The Global stability of the Minkowski
  space-time in harmonic gauge}},
  \href{http://xxx.lanl.gov/abs/math/0411109}{{\tt math/0411109}}.

\bibitem{ChoquetBruhat:2006jc}
Y.~Choquet-Bruhat, P.~T. Chrusciel, and J.~Loizelet, {\it {Global solutions of
  the Einstein-Maxwell equations in higher dimensions}},  {\em Class. Quant.
  Grav.} {\bf 23} (2006) 7383--7394,
  [\href{http://xxx.lanl.gov/abs/gr-qc/0608108}{{\tt gr-qc/0608108}}].

\bibitem{Ringstrom:2015jza}
H.~Ringström, {\it {Origins and development of the Cauchy problem in general
  relativity}},  {\em Class. Quant. Grav.} {\bf 32} (2015), no.~12 124003.

\bibitem{DafermosHolzegel2006}
M.~Dafermos and G.~Holzegel, {\it Dynamic instability of solitons in 4+1
  dimensional gravity with negative cosmological constant},  in {\em Seminar at
  DAMTP}, University of Cambridge, 2006.
\newblock \href{http://xxx.lanl.gov/abs/Available at:
  {https://www.dpmms.cam.ac.uk/$\sim$md384/ADSinstability.pdf}}{{\tt Available
  at: {https://www.dpmms.cam.ac.uk/$\sim$md384/ADSinstability.pdf}}}.

\bibitem{Bizon:2011gg}
P.~Bizon and A.~Rostworowski, {\it {On weakly turbulent instability of anti-de
  Sitter space}},  {\em Phys. Rev. Lett.} {\bf 107} (2011) 031102,
  [\href{http://xxx.lanl.gov/abs/1104.3702}{{\tt arXiv:1104.3702}}].

\bibitem{Choptuik:1992jv}
M.~W. Choptuik, {\it {Universality and scaling in gravitational collapse of a
  massless scalar field}},  {\em Phys. Rev. Lett.} {\bf 70} (1993) 9--12.

\bibitem{Carrasco:2012nf}
F.~Carrasco, L.~Lehner, R.~C. Myers, O.~Reula, and A.~Singh, {\it {Turbulent
  flows for relativistic conformal fluids in 2+1 dimensions}},  {\em Phys.
  Rev.} {\bf D86} (2012) 126006, [\href{http://xxx.lanl.gov/abs/1210.6702}{{\tt
  arXiv:1210.6702}}].

\bibitem{Adams:2012pj}
A.~Adams, P.~M. Chesler, and H.~Liu, {\it {Holographic Vortex Liquids and
  Superfluid Turbulence}},  {\em Science} {\bf 341} (2013) 368--372,
  [\href{http://xxx.lanl.gov/abs/1212.0281}{{\tt arXiv:1212.0281}}].

\bibitem{Adams:2013vsa}
A.~Adams, P.~M. Chesler, and H.~Liu, {\it {Holographic turbulence}},  {\em
  Phys. Rev. Lett.} {\bf 112} (2014), no.~15 151602,
  [\href{http://xxx.lanl.gov/abs/1307.7267}{{\tt arXiv:1307.7267}}].

\bibitem{Green:2013zba}
S.~R. Green, F.~Carrasco, and L.~Lehner, {\it {Holographic Path to the
  Turbulent Side of Gravity}},  {\em Phys. Rev.} {\bf X4} (2014), no.~1 011001,
  [\href{http://xxx.lanl.gov/abs/1309.7940}{{\tt arXiv:1309.7940}}].

\bibitem{Chesler:2014pka}
P.~M. Chesler and A.~Lucas, {\it {Vortex annihilation and inverse cascades in
  two dimensional superfluid turbulence}},
  \href{http://xxx.lanl.gov/abs/1411.2610}{{\tt arXiv:1411.2610}}.

\bibitem{Dias:2012tq}
O.~J.~C. Dias, G.~T. Horowitz, D.~Marolf, and J.~E. Santos, {\it {On the
  Nonlinear Stability of Asymptotically Anti-de Sitter Solutions}},  {\em
  Class. Quant. Grav.} {\bf 29} (2012) 235019,
  [\href{http://xxx.lanl.gov/abs/1208.5772}{{\tt arXiv:1208.5772}}].

\bibitem{Buchel:2012uh}
A.~Buchel, L.~Lehner, and S.~L. Liebling, {\it {Scalar Collapse in AdS}},  {\em
  Phys. Rev.} {\bf D86} (2012) 123011,
  [\href{http://xxx.lanl.gov/abs/1210.0890}{{\tt arXiv:1210.0890}}].

\bibitem{Bizon:2013xha}
P.~Bizon and J.~Jamuna, {\it {Globally regular instability of $AdS_3$}},  {\em
  Phys. Rev. Lett.} {\bf 111} (2013), no.~4 041102,
  [\href{http://xxx.lanl.gov/abs/1306.0317}{{\tt arXiv:1306.0317}}].

\bibitem{Maliborski:2012gx}
M.~Maliborski, {\it {Instability of Flat Space Enclosed in a Cavity}},  {\em
  Phys. Rev. Lett.} {\bf 109} (2012) 221101,
  [\href{http://xxx.lanl.gov/abs/1208.2934}{{\tt arXiv:1208.2934}}].

\bibitem{Maliborski:2013ula}
M.~Maliborski and A.~Rostworowski, {\it {A comment on "Boson stars in AdS"}},
  \href{http://xxx.lanl.gov/abs/1307.2875}{{\tt arXiv:1307.2875}}.

\bibitem{Baier:2013gsa}
R.~Baier, S.~A. Stricker, and O.~Taanila, {\it {Critical scalar field collapse
  in AdS$_{3}$: an analytical approach}},  {\em Class. Quant. Grav.} {\bf 31}
  (2014) 025007, [\href{http://xxx.lanl.gov/abs/1309.1629}{{\tt
  arXiv:1309.1629}}].

\bibitem{Jalmuzna:2013rwa}
J.~Jamuna, {\it {Three-dimensional Gravity and Instability of AdS$_{3}$}},
  {\em Acta Phys. Polon.} {\bf B44} (2013), no.~12 2603--2620,
  [\href{http://xxx.lanl.gov/abs/1311.7409}{{\tt arXiv:1311.7409}}].

\bibitem{Basu:2012gg}
P.~Basu, D.~Das, S.~R. Das, and T.~Nishioka, {\it {Quantum Quench Across a Zero
  Temperature Holographic Superfluid Transition}},  {\em JHEP} {\bf 03} (2013)
  146, [\href{http://xxx.lanl.gov/abs/1211.7076}{{\tt arXiv:1211.7076}}].

\bibitem{Friedrich:2014raa}
H.~Friedrich, {\it {On the AdS stability problem}},  {\em Class. Quant. Grav.}
  {\bf 31} (2014) 105001, [\href{http://xxx.lanl.gov/abs/1401.7172}{{\tt
  arXiv:1401.7172}}].

\bibitem{Maliborski:2014rma}
M.~Maliborski and A.~Rostworowski, {\it {What drives AdS spacetime unstable?}},
   {\em Phys. Rev.} {\bf D89} (2014), no.~12 124006,
  [\href{http://xxx.lanl.gov/abs/1403.5434}{{\tt arXiv:1403.5434}}].

\bibitem{Abajo-Arrastia:2014fma}
J.~Abajo-Arrastia, E.~da~Silva, E.~Lopez, J.~Mas, and A.~Serantes, {\it
  {Holographic Relaxation of Finite Size Isolated Quantum Systems}},  {\em
  JHEP} {\bf 05} (2014) 126, [\href{http://xxx.lanl.gov/abs/1403.2632}{{\tt
  arXiv:1403.2632}}].

\bibitem{Balasubramanian:2014cja}
V.~Balasubramanian, A.~Buchel, S.~R. Green, L.~Lehner, and S.~L. Liebling, {\it
  {Holographic Thermalization, Stability of Anti-de Sitter Space, and the
  Fermi-Pasta-Ulam Paradox}},  {\em Phys. Rev. Lett.} {\bf 113} (2014), no.~7
  071601, [\href{http://xxx.lanl.gov/abs/1403.6471}{{\tt arXiv:1403.6471}}].

\bibitem{Bizon:2014bya}
P.~Bizon and A.~Rostworowski, {\it {Comment on Holographic Thermalization,
  Stability of Anti-de Sitter Space, and the Fermi-Pasta-Ulam Paradox?}},  {\em
  Phys. Rev. Lett.} {\bf 115} (2015), no.~4 049101,
  [\href{http://xxx.lanl.gov/abs/1410.2631}{{\tt arXiv:1410.2631}}].

\bibitem{Balasubramanian:2015uua}
V.~Balasubramanian, A.~Buchel, S.~R. Green, L.~Lehner, and S.~L. Liebling, {\it
  {Reply to Comment on Holographic Thermalization, Stability of Anti-de Sitter
  Space, and the Fermi-Pasta-Ulam Paradox?}},  {\em Phys. Rev. Lett.} {\bf 115}
  (2015), no.~4 049102, [\href{http://xxx.lanl.gov/abs/1506.0790}{{\tt
  arXiv:1506.0790}}].

\bibitem{daSilva:2014zva}
E.~da~Silva, E.~Lopez, J.~Mas, and A.~Serantes, {\it {Collapse and Revival in
  Holographic Quenches}},  {\em JHEP} {\bf 04} (2015) 038,
  [\href{http://xxx.lanl.gov/abs/1412.6002}{{\tt arXiv:1412.6002}}].

\bibitem{Craps:2014vaa}
B.~Craps, O.~Evnin, and J.~Vanhoof, {\it {Renormalization group, secular term
  resummation and AdS (in)stability}},  {\em JHEP} {\bf 10} (2014) 48,
  [\href{http://xxx.lanl.gov/abs/1407.6273}{{\tt arXiv:1407.6273}}].

\bibitem{Okawa:2014nea}
H.~Okawa, V.~Cardoso, and P.~Pani, {\it {Study of the nonlinear instability of
  confined geometries}},  {\em Phys. Rev.} {\bf D90} (2014), no.~10 104032,
  [\href{http://xxx.lanl.gov/abs/1409.0533}{{\tt arXiv:1409.0533}}].

\bibitem{Deppe:2014oua}
N.~Deppe, A.~Kolly, A.~Frey, and G.~Kunstatter, {\it {Stability of AdS in
  Einstein Gauss Bonnet Gravity}},  {\em Phys. Rev. Lett.} {\bf 114} (2015)
  071102, [\href{http://xxx.lanl.gov/abs/1410.1869}{{\tt arXiv:1410.1869}}].

\bibitem{Dimitrakopoulos:2014ada}
F.~V. Dimitrakopoulos, B.~Freivogel, M.~Lippert, and I.-S. Yang, {\it {Position
  space analysis of the AdS (in)stability problem}},  {\em JHEP} {\bf 08}
  (2015) 077, [\href{http://xxx.lanl.gov/abs/1410.1880}{{\tt
  arXiv:1410.1880}}].

\bibitem{Buchel:2014xwa}
A.~Buchel, S.~R. Green, L.~Lehner, and S.~L. Liebling, {\it {Conserved
  quantities and dual turbulent cascades in anti-de Sitter spacetime}},  {\em
  Phys. Rev.} {\bf D91} (2015), no.~6 064026,
  [\href{http://xxx.lanl.gov/abs/1412.4761}{{\tt arXiv:1412.4761}}].

\bibitem{Craps:2014jwa}
B.~Craps, O.~Evnin, and J.~Vanhoof, {\it {Renormalization, averaging,
  conservation laws and AdS (in)stability}},  {\em JHEP} {\bf 01} (2015) 108,
  [\href{http://xxx.lanl.gov/abs/1412.3249}{{\tt arXiv:1412.3249}}].

\bibitem{Yang:2015jha}
I.-S. Yang, {\it {Missing top of the AdS resonance structure}},  {\em Phys.
  Rev.} {\bf D91} (2015), no.~6 065011,
  [\href{http://xxx.lanl.gov/abs/1501.0099}{{\tt arXiv:1501.0099}}].

\bibitem{Okawa:2015xma}
H.~Okawa, J.~C. Lopes, and V.~Cardoso, {\it {Collapse of massive fields in
  anti-de Sitter spacetime}},  \href{http://xxx.lanl.gov/abs/1504.0520}{{\tt
  arXiv:1504.0520}}.

\bibitem{Bizon:2015pfa}
P.~Bizon, M.~Maliborski, and A.~Rostworowski, {\it {Resonant Dynamics and the
  Instability of Anti-de Sitter Spacetime}},  {\em Phys. Rev. Lett.} {\bf 115}
  (2015), no.~8 081103, [\href{http://xxx.lanl.gov/abs/1506.0351}{{\tt
  arXiv:1506.0351}}].

\bibitem{Green:2015dsa}
S.~R. Green, A.~Maillard, L.~Lehner, and S.~L. Liebling, {\it {Islands of
  stability and recurrence times in AdS}},
  \href{http://xxx.lanl.gov/abs/1507.0826}{{\tt arXiv:1507.0826}}.

\bibitem{Deppe:2015qsa}
N.~Deppe and A.~R. Frey, {\it {Classes of Stable Initial Data for Massless and
  Massive Scalars in Anti-de Sitter Spacetime}},
  \href{http://xxx.lanl.gov/abs/1508.0270}{{\tt arXiv:1508.0270}}.

\bibitem{Craps:2015iia}
B.~Craps, O.~Evnin, and J.~Vanhoof, {\it {Ultraviolet asymptotics and singular
  dynamics of AdS perturbations}},
  \href{http://xxx.lanl.gov/abs/1508.0494}{{\tt arXiv:1508.0494}}.

\bibitem{Craps:2015xya}
B.~Craps, O.~Evnin, P.~Jai-akson, and J.~Vanhoof, {\it {Ultraviolet asymptotics
  for quasiperiodic AdS4 perturbations}},
  \href{http://xxx.lanl.gov/abs/1508.0547}{{\tt arXiv:1508.0547}}.

\bibitem{Evnin:2015gma}
O.~Evnin and C.~Krishnan, {\it {A Hidden Symmetry of AdS Resonances}},  {\em
  Phys. Rev.} {\bf D91} (2015), no.~12 126010,
  [\href{http://xxx.lanl.gov/abs/1502.0374}{{\tt arXiv:1502.0374}}].

\bibitem{Menon:2015oda}
D.~S. Menon and V.~Suneeta, {\it {Necessary conditions for an AdS-type
  instability}},  \href{http://xxx.lanl.gov/abs/1509.0023}{{\tt
  arXiv:1509.0023}}.

\bibitem{Bizon:CQG2015}
P.~Bizon and A.~Rostworowski, {\it {Stability of AdS, (Topical Review)}},  {\em
  Class. Quantum. Grav., to appear} (2015).

\bibitem{deDonder:1921}
T.~de~Donder, {\it {La gravifique Einsteinienne}},  {\em Annales de
  l'Observatoire Royal de Belgique} (1921).

\bibitem{Gander1989815}
W.~Gander, G.~H. Golub, and U.~von Matt, {\it A constrained eigenvalue
  problem},  {\em Linear Algebra and its Applications} {\bf 114–115} (1989)
  815 -- 839. Special Issue Dedicated to Alan J. Hoffman.

\bibitem{Godazgar:2012zq}
M.~Godazgar and H.~S. Reall, {\it {Peeling of the Weyl tensor and gravitational
  radiation in higher dimensions}},  {\em Phys. Rev.} {\bf D85} (2012) 084021,
  [\href{http://xxx.lanl.gov/abs/1201.4373}{{\tt arXiv:1201.4373}}].

\bibitem{Monteiro:2009ke}
R.~Monteiro, M.~J. Perry, and J.~E. Santos, {\it {Semiclassical instabilities
  of Kerr-AdS black holes}},  {\em Phys. Rev.} {\bf D81} (2010) 024001,
  [\href{http://xxx.lanl.gov/abs/0905.2334}{{\tt arXiv:0905.2334}}].

\bibitem{FouresBruhat:1952zz}
Y.~Foures-Bruhat, {\it {Theoreme d'existence pour certains systemes derivees
  partielles non lineaires}},  {\em Acta Mat.} {\bf 88} (1952) 141--225.

\bibitem{Bruhat:1967}
Y.~Bruhat, {\it {Cauchy problem}},  {\em An Introduction to Current Research.
  Ed. L Witten (New York: Wiley)} (1967) 130.

\bibitem{Fischer:1972}
A.~E. Fischer and J.~E. Marsden, {\it {The Einstein evolution equations as a
  first-order quasi-linear symmetric hyperbolic system}},  {\em Commun. Math.
  Phys.} {\bf 28} (1972) 1.

\bibitem{ChoquetBruhat:1969cb}
Y.~Choquet-Bruhat and R.~P. Geroch, {\it {Global aspects of the Cauchy problem
  in general relativity}},  {\em Commun. Math. Phys.} {\bf 14} (1969) 329--335.

\bibitem{LerayBook}
J.~Leray, {\em {Hyperbolic differential equations}}.
\newblock The Institute for Advanced Study, Princeton, N. J., 1953., 1953.

\bibitem{Adam:2011dn}
A.~Adam, S.~Kitchen, and T.~Wiseman, {\it {A numerical approach to finding
  general stationary vacuum black holes}},  {\em Class.Quant.Grav.} {\bf 29}
  (2012) 165002, [\href{http://xxx.lanl.gov/abs/1105.6347}{{\tt
  arXiv:1105.6347}}].

\bibitem{Friedrich}
H.~Friedrich, {\it {On the hyperbolicity of Einsteins and other gauge field
  equations}},  {\em Commun. Math. Phys.} {\bf 100} (1985) 525.

\bibitem{Garfinkle:2001ni}
D.~Garfinkle, {\it {Harmonic coordinate method for simulating generic
  singularities}},  {\em Phys. Rev.} {\bf D65} (2002) 044029,
  [\href{http://xxx.lanl.gov/abs/gr-qc/0110013}{{\tt gr-qc/0110013}}].

\bibitem{Szilagyi:2001fy}
B.~Szilagyi, B.~G. Schmidt, and J.~Winicour, {\it {Boundary conditions in
  linearized harmonic gravity}},  {\em Phys. Rev.} {\bf D65} (2002) 064015,
  [\href{http://xxx.lanl.gov/abs/gr-qc/0106026}{{\tt gr-qc/0106026}}].

\bibitem{Szilagyi:2002kv}
B.~Szilagyi and J.~Winicour, {\it {Well posed initial boundary evolution in
  general relativity}},  {\em Phys. Rev.} {\bf D68} (2003) 041501,
  [\href{http://xxx.lanl.gov/abs/gr-qc/0205044}{{\tt gr-qc/0205044}}].

\bibitem{Pretorius:2005gq}
F.~Pretorius, {\it {Evolution of binary black hole spacetimes}},  {\em Phys.
  Rev. Lett.} {\bf 95} (2005) 121101,
  [\href{http://xxx.lanl.gov/abs/gr-qc/0507014}{{\tt gr-qc/0507014}}].

\bibitem{Szilagyi:2006qy}
B.~Szilagyi, D.~Pollney, L.~Rezzolla, J.~Thornburg, and J.~Winicour, {\it {An
  Explicit harmonic code for black-hole evolution using excision}},  {\em
  Class. Quant. Grav.} {\bf 24} (2007) S275--S293,
  [\href{http://xxx.lanl.gov/abs/gr-qc/0612150}{{\tt gr-qc/0612150}}].

\bibitem{Pretorius:2007nq}
F.~Pretorius, {\it {Binary Black Hole Coalescence}},
  \href{http://xxx.lanl.gov/abs/0710.1338}{{\tt arXiv:0710.1338}}.

\bibitem{Szilagyi:2009qz}
B.~Szilagyi, L.~Lindblom, and M.~A. Scheel, {\it {Simulations of Binary Black
  Hole Mergers Using Spectral Methods}},  {\em Phys. Rev.} {\bf D80} (2009)
  124010, [\href{http://xxx.lanl.gov/abs/0909.3557}{{\tt arXiv:0909.3557}}].

\bibitem{Friedan1985}
D.~H. Friedan, {\it {Nonlinear models in two + epsilon dimensions}},  {\em Ann.
  Phys.} {\bf 163} (1985) 318.

\bibitem{Garfinkle:2003an}
D.~Garfinkle and J.~Isenberg, {\it {Critical behavior in Ricci flow}},
  \href{http://xxx.lanl.gov/abs/math/0306129}{{\tt math/0306129}}.

\bibitem{Headrick:2006ti}
M.~Headrick and T.~Wiseman, {\it {Ricci flow and black holes}},  {\em Class.
  Quant. Grav.} {\bf 23} (2006) 6683--6708,
  [\href{http://xxx.lanl.gov/abs/hep-th/0606086}{{\tt hep-th/0606086}}].

\bibitem{Holzegel:2007zz}
G.~Holzegel, C.~Warnick, and T.~Schmelzer, {\it {Ricci flows connecting
  Taub-Bolt and Taub-NUT metrics}},  {\em Class. Quant. Grav.} {\bf 24} (2007)
  6201--6217.

\bibitem{Holzegel:2007ud}
G.~Holzegel, T.~Schmelzer, and C.~Warnick, {\it {Ricci Flow of Biaxial Bianchi
  IX Metrics}},  \href{http://xxx.lanl.gov/abs/0706.1694}{{\tt
  arXiv:0706.1694}}.

\bibitem{Headrick:2007fk}
M.~Headrick and T.~Wiseman, {\it {Numerical K\"ahler-Ricci soliton on the
  second del Pezzo}},  \href{http://xxx.lanl.gov/abs/0706.2329}{{\tt
  arXiv:0706.2329}}.

\bibitem{Rozali:2013fna}
M.~Rozali, J.~B. Stang, and M.~van Raamsdonk, {\it {Holographic Baryons from
  Oblate Instantons}},  {\em JHEP} {\bf 02} (2014) 044,
  [\href{http://xxx.lanl.gov/abs/1309.7037}{{\tt arXiv:1309.7037}}].

\bibitem{Press:1973zz}
W.~H. Press and S.~A. Teukolsky, {\it {Perturbations of a Rotating Black Hole.
  II. Dynamical Stability of the Kerr Metric}},  {\em Astrophys. J.} {\bf 185}
  (1973) 649--674.

\bibitem{Dias:2009ex}
O.~J.~C. Dias, H.~S. Reall, and J.~E. Santos, {\it {Kerr-CFT and gravitational
  perturbations}},  {\em JHEP} {\bf 08} (2009) 101,
  [\href{http://xxx.lanl.gov/abs/0906.2380}{{\tt arXiv:0906.2380}}].

\bibitem{Kunduri:2007vf}
H.~K. Kunduri, J.~Lucietti, and H.~S. Reall, {\it {Near-horizon symmetries of
  extremal black holes}},  {\em Class. Quant. Grav.} {\bf 24} (2007)
  4169--4190, [\href{http://xxx.lanl.gov/abs/0705.4214}{{\tt
  arXiv:0705.4214}}].

\bibitem{Figueras:2008qh}
P.~Figueras, H.~K. Kunduri, J.~Lucietti, and M.~Rangamani, {\it {Extremal
  vacuum black holes in higher dimensions}},  {\em Phys. Rev.} {\bf D78} (2008)
  044042, [\href{http://xxx.lanl.gov/abs/0803.2998}{{\tt arXiv:0803.2998}}].

\bibitem{Kunduri:2008rs}
H.~K. Kunduri and J.~Lucietti, {\it {A Classification of near-horizon
  geometries of extremal vacuum black holes}},  {\em J. Math. Phys.} {\bf 50}
  (2009) 082502, [\href{http://xxx.lanl.gov/abs/0806.2051}{{\tt
  arXiv:0806.2051}}].

\bibitem{Kunduri:2008tk}
H.~K. Kunduri and J.~Lucietti, {\it {Uniqueness of near-horizon geometries of
  rotating extremal AdS(4) black holes}},  {\em Class. Quant. Grav.} {\bf 26}
  (2009) 055019, [\href{http://xxx.lanl.gov/abs/0812.1576}{{\tt
  arXiv:0812.1576}}].

\bibitem{Kunduri:2009ud}
H.~K. Kunduri and J.~Lucietti, {\it {Static near-horizon geometries in five
  dimensions}},  {\em Class. Quant. Grav.} {\bf 26} (2009) 245010,
  [\href{http://xxx.lanl.gov/abs/0907.0410}{{\tt arXiv:0907.0410}}].

\bibitem{Kunduri:2013ana}
H.~K. Kunduri and J.~Lucietti, {\it {Classification of near-horizon geometries
  of extremal black holes}},  {\em Living Rev. Rel.} {\bf 16} (2013) 8,
  [\href{http://xxx.lanl.gov/abs/1306.2517}{{\tt arXiv:1306.2517}}].

\bibitem{Horowitz:2009ij}
G.~T. Horowitz and M.~M. Roberts, {\it {Zero Temperature Limit of Holographic
  Superconductors}},  {\em JHEP} {\bf 0911} (2009) 015,
  [\href{http://xxx.lanl.gov/abs/0908.3677}{{\tt arXiv:0908.3677}}].

\bibitem{Horowitz:2002ym}
G.~T. Horowitz and K.~Maeda, {\it {Inhomogeneous near extremal black branes}},
  {\em Phys. Rev.} {\bf D65} (2002) 104028,
  [\href{http://xxx.lanl.gov/abs/hep-th/0201241}{{\tt hep-th/0201241}}].

\bibitem{Mezincescu:1984ev}
L.~Mezincescu and P.~K. Townsend, {\it {Stability at a Local Maximum in Higher
  Dimensional Anti-de Sitter Space and Applications to Supergravity}},  {\em
  Annals Phys.} {\bf 160} (1985) 406.

\bibitem{Klebanov:1999tb}
I.~R. Klebanov and E.~Witten, {\it {AdS / CFT correspondence and symmetry
  breaking}},  {\em Nucl.Phys.} {\bf B556} (1999) 89--114,
  [\href{http://xxx.lanl.gov/abs/hep-th/9905104}{{\tt hep-th/9905104}}].

\bibitem{Hertog:2004rz}
T.~Hertog and G.~T. Horowitz, {\it {Towards a big crunch dual}},  {\em JHEP}
  {\bf 07} (2004) 073, [\href{http://xxx.lanl.gov/abs/hep-th/0406134}{{\tt
  hep-th/0406134}}].

\bibitem{Witten:2001ua}
E.~Witten, {\it {Multitrace operators, boundary conditions, and AdS / CFT
  correspondence}},  \href{http://xxx.lanl.gov/abs/hep-th/0112258}{{\tt
  hep-th/0112258}}.

\bibitem{Sever:2002fk}
A.~Sever and A.~Shomer, {\it {A Note on multitrace deformations and AdS/CFT}},
  {\em JHEP} {\bf 0207} (2002) 027,
  [\href{http://xxx.lanl.gov/abs/hep-th/0203168}{{\tt hep-th/0203168}}].

\bibitem{Faulkner:2010fh}
T.~Faulkner, G.~T. Horowitz, and M.~M. Roberts, {\it {New stability results for
  Einstein scalar gravity}},  {\em Class. Quant. Grav.} {\bf 27} (2010) 205007,
  [\href{http://xxx.lanl.gov/abs/1006.2387}{{\tt arXiv:1006.2387}}].

\bibitem{Hubeny:2004cn}
V.~E. Hubeny, X.~Liu, M.~Rangamani, and S.~Shenker, {\it {Comments on cosmic
  censorship in AdS / CFT}},  {\em JHEP} {\bf 12} (2004) 067,
  [\href{http://xxx.lanl.gov/abs/hep-th/0403198}{{\tt hep-th/0403198}}].

\bibitem{Monteiro:2008wr}
R.~Monteiro and J.~E. Santos, {\it {Negative modes and the thermodynamics of
  Reissner- Nordstr\'om black holes}},  {\em Phys. Rev.} {\bf D79} (2009)
  064006, [\href{http://xxx.lanl.gov/abs/0812.1767}{{\tt arXiv:0812.1767}}].

\bibitem{Gubser:2000ec}
S.~S. Gubser and I.~Mitra, {\it {Instability of charged black holes in anti-de
  Sitter space}},  \href{http://xxx.lanl.gov/abs/hep-th/0009126}{{\tt
  hep-th/0009126}}.

\bibitem{Gubser:2000mm}
S.~S. Gubser and I.~Mitra, {\it {The evolution of unstable black holes in
  anti-de Sitter space}},  {\em JHEP} {\bf 08} (2001) 018,
  [\href{http://xxx.lanl.gov/abs/hep-th/0011127}{{\tt hep-th/0011127}}].

\bibitem{Caldarelli:1999xj}
M.~M. Caldarelli, G.~Cognola, and D.~Klemm, {\it {Thermodynamics of
  Kerr-Newman-AdS black holes and conformal field theories}},  {\em Class.
  Quant. Grav.} {\bf 17} (2000) 399--420,
  [\href{http://xxx.lanl.gov/abs/hep-th/9908022}{{\tt hep-th/9908022}}].

\bibitem{Gibbons:2004ai}
G.~W. Gibbons, M.~J. Perry, and C.~N. Pope, {\it {The First law of
  thermodynamics for Kerr-anti-de Sitter black holes}},  {\em Class. Quant.
  Grav.} {\bf 22} (2005) 1503--1526,
  [\href{http://xxx.lanl.gov/abs/hep-th/0408217}{{\tt hep-th/0408217}}].

\bibitem{Papadimitriou:2005ii}
I.~Papadimitriou and K.~Skenderis, {\it {Thermodynamics of asymptotically
  locally AdS spacetimes}},  {\em JHEP} {\bf 08} (2005) 004,
  [\href{http://xxx.lanl.gov/abs/hep-th/0505190}{{\tt hep-th/0505190}}].

\bibitem{deHaro:2000xn}
S.~de~Haro, S.~N. Solodukhin, and K.~Skenderis, {\it {Holographic
  reconstruction of space-time and renormalization in the AdS / CFT
  correspondence}},  {\em Commun. Math. Phys.} {\bf 217} (2001) 595--622,
  [\href{http://xxx.lanl.gov/abs/hep-th/0002230}{{\tt hep-th/0002230}}].

\bibitem{Henningson:1998gx}
M.~Henningson and K.~Skenderis, {\it {The Holographic Weyl anomaly}},  {\em
  JHEP} {\bf 07} (1998) 023,
  [\href{http://xxx.lanl.gov/abs/hep-th/9806087}{{\tt hep-th/9806087}}].

\bibitem{Arnowitt:1961zz}
R.~L. Arnowitt, S.~Deser, and C.~W. Misner, {\it {Coordinate invariance and
  energy expressions in general relativity}},  {\em Phys. Rev.} {\bf 122}
  (1961) 997.

\bibitem{Arnowitt:1962hi}
R.~L. Arnowitt, S.~Deser, and C.~W. Misner, {\it {The Dynamics of general
  relativity}},  {\em Gen. Rel. Grav.} {\bf 40} (2008) 1997--2027,
  [\href{http://xxx.lanl.gov/abs/gr-qc/0405109}{{\tt gr-qc/0405109}}].

\bibitem{Komar:1959}
A.~Komar, {\it {Covariant conservation laws in general relativity}},  {\em
  Phys. Rev.} {\bf 113} (1959) 934.

\bibitem{Bardeen:1973gs}
J.~M. Bardeen, B.~Carter, and S.~W. Hawking, {\it {The Four laws of black hole
  mechanics}},  {\em Commun. Math. Phys.} {\bf 31} (1973) 161--170.

\bibitem{Henneaux:1985tv}
M.~Henneaux and C.~Teitelboim, {\it {Asymptotically anti-De Sitter Spaces}},
  {\em Commun. Math. Phys.} {\bf 98} (1985) 391--424.

\bibitem{Ashtekar:1984zz}
A.~Ashtekar and A.~Magnon, {\it {Asymptotically anti-de Sitter space-times}},
  {\em Class. Quant. Grav.} {\bf 1} (1984) L39--L44.

\bibitem{FG:1985}
C.~Fefferman and C.~R. Graham, {\it {Conformal invariants, in The Mathematical
  Heritage of \'Elie Cartan (Lyon, 1984)}},  {\em Ast\'erisque} {\bf 98} (1985)
  95--116.

\bibitem{Graham:1999jg}
C.~R. Graham, {\it {Volume and area renormalizations for conformally compact
  Einstein metrics}},  in {\em {Proceedings, 19th Winter School on Geometry and
  Physics}}, 1999.
\newblock \href{http://xxx.lanl.gov/abs/math/9909042}{{\tt math/9909042}}.

\bibitem{Anderson:2004yi}
M.~T. Anderson, {\it {Geometric aspects of the AdS / CFT correspondence}},  in
  {\em {AdS/CFT correspondence: Einstein metrics and their conformal
  boundaries. Proceedings, 73rd Meeting of Theoretical Physicists and
  Mathematicians, Strasbourg, France, September 11-13, 2003}}, pp.~1--31, 2004.
\newblock \href{http://xxx.lanl.gov/abs/hep-th/0403087}{{\tt hep-th/0403087}}.

\bibitem{Andrade:2011dg}
T.~Andrade and D.~Marolf, {\it {AdS/CFT beyond the unitarity bound}},  {\em
  JHEP} {\bf 01} (2012) 049, [\href{http://xxx.lanl.gov/abs/1105.6337}{{\tt
  arXiv:1105.6337}}].

\bibitem{Compere:2008us}
G.~Compere and D.~Marolf, {\it {Setting the boundary free in AdS/CFT}},  {\em
  Class. Quant. Grav.} {\bf 25} (2008) 195014,
  [\href{http://xxx.lanl.gov/abs/0805.1902}{{\tt arXiv:0805.1902}}].

\bibitem{Balasubramanian:1999re}
V.~Balasubramanian and P.~Kraus, {\it {A Stress tensor for Anti-de Sitter
  gravity}},  {\em Commun. Math. Phys.} {\bf 208} (1999) 413--428,
  [\href{http://xxx.lanl.gov/abs/hep-th/9902121}{{\tt hep-th/9902121}}].

\bibitem{Ashtekar:1999jx}
A.~Ashtekar and S.~Das, {\it {Asymptotically Anti-de Sitter space-times:
  Conserved quantities}},  {\em Class. Quant. Grav.} {\bf 17} (2000) L17--L30,
  [\href{http://xxx.lanl.gov/abs/hep-th/9911230}{{\tt hep-th/9911230}}].

\bibitem{Skenderis:2006uy}
K.~Skenderis and M.~Taylor, {\it {Kaluza-Klein holography}},  {\em JHEP} {\bf
  0605} (2006) 057, [\href{http://xxx.lanl.gov/abs/hep-th/0603016}{{\tt
  hep-th/0603016}}].

\bibitem{Kim:1985ez}
H.~Kim, L.~Romans, and P.~van Nieuwenhuizen, {\it {The Mass Spectrum of Chiral
  N=2 D=10 Supergravity on $S^5$}},  {\em Phys.Rev.} {\bf D32} (1985) 389.

\bibitem{Gunaydin:1984fk}
M.~Gunaydin and N.~Marcus, {\it {The Spectrum of the $S^5$ Compactification of
  the Chiral N=2, D=10 Supergravity and the Unitary Supermultiplets of U(2,
  2/4)}},  {\em Class.Quant.Grav.} {\bf 2} (1985) L11.

\bibitem{Lee:1998bxa}
S.~Lee, S.~Minwalla, M.~Rangamani, and N.~Seiberg, {\it {Three point functions
  of chiral operators in D = 4, N=4 SYM at large N}},  {\em
  Adv.Theor.Math.Phys.} {\bf 2} (1998) 697--718,
  [\href{http://xxx.lanl.gov/abs/hep-th/9806074}{{\tt hep-th/9806074}}].

\bibitem{Lee:1999pj}
S.~Lee, {\it {AdS(5) / CFT(4) four point functions of chiral primary operators:
  Cubic vertices}},  {\em Nucl.Phys.} {\bf B563} (1999) 349--360,
  [\href{http://xxx.lanl.gov/abs/hep-th/9907108}{{\tt hep-th/9907108}}].

\bibitem{Arutyunov:1999en}
G.~Arutyunov and S.~Frolov, {\it {Some cubic couplings in type IIB supergravity
  on AdS(5) $\times S^5$ and three point functions in SYM(4) at large N}},
  {\em Phys.Rev.} {\bf D61} (2000) 064009,
  [\href{http://xxx.lanl.gov/abs/hep-th/9907085}{{\tt hep-th/9907085}}].

\bibitem{Skenderis:2006di}
K.~Skenderis and M.~Taylor, {\it {Holographic Coulomb branch vevs}},  {\em
  JHEP} {\bf 0608} (2006) 001,
  [\href{http://xxx.lanl.gov/abs/hep-th/0604169}{{\tt hep-th/0604169}}].

\bibitem{Skenderis:2007yb}
K.~Skenderis and M.~Taylor, {\it {Anatomy of bubbling solutions}},  {\em JHEP}
  {\bf 0709} (2007) 019, [\href{http://xxx.lanl.gov/abs/0706.0216}{{\tt
  arXiv:0706.0216}}].

\end{thebibliography}\endgroup
\bibliographystyle{JHEP}

\end{document}